# A Compact Model of Silicon-Based Nanowire Field Effect Transistor for Circuit Simulation and Design

Thesis submitted in partial fulfillment of the
Requirements for the degree of

**Master of Science in Technology**
in
**Microelectronics**
By

**Mayank Chakraverty**
(122510024)

Under the guidance of

Dr. Vaibhav Meshram
Professor and Head, Department of ECE
Jain University
Bangalore

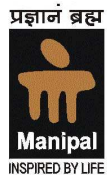

**MANIPAL UNIVERSITY, MANIPAL**

# A Compact Model of Silicon-Based Nanowire Field Effect Transistor for Circuit Simulation and Design

Thesis submitted in partial fulfillment of the
Requirements for the degree of

**Master of Science in Technology**

in

**Microelectronics**

By

**Mayank Chakraverty**

(122510024)

**Examiner 1**

Signature:

Name:

**Examiner 2**

Signature:

Name:

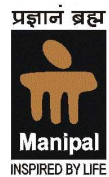

**MANIPAL UNIVERSITY, MANIPAL**

## CERTIFICATE

This is to certify that this thesis work titled

## A Compact Model of Silicon-Based Nanowire Field Effect Transistor for Circuit Simulation and Design

Is a bonafide record of the work done by

**Mayank Chakraverty**

Reg. No. 122510024

In partial fulfillment of the requirements for the award of the degree of **Master of Science in Technology in Microelectronics** under Manipal University, Manipal and the same has not been submitted elsewhere for the award for any other degree.

Dr. Vaibhav Meshram
Professor and Head,
Department of ECE,
Jain University
Bangalore

# Acknowledgment

At the very outset, I take the opportunity to thank Almighty for showering his choicest blessings on a little mortal like me. But for his I would not have overcome the several odds that came in my way.

I thank the Security, Governance, Risk & Compliance (SGRC) practice under GBS Business Unit and Semiconductor Research & Development Center under STG Business Unit, IBM, Bangalore, for giving me an opportunity to carry out my studies and successfully complete this project work along with my deliverables and assignments at office.

I take the opportunity to thank Manipal University, Manipal, India for providing their approval to go ahead with this project.

With deep sense of gratitude I thank all the faculty members of Centre for Nanotechnology Research and VLSI Division, VIT University, Vellore, India, for motivating me to take up this project and complete it in the field of Nanoelectronic devices within the specified deadline.

Words fail when it comes to thank my project guide Prof. Dr. Vaibhav Meshram, Professor, Department of ECE, Jain University, Bangalore, India. He not only instigated me into the project but also has rendered valuable guidance every time I sought his advice. His spontaneity in offering suggestions and timely guidance took us through my difficult times and I am eternally grateful for all that he has done for me. The innumerable suggestions he offered, the plentiful tips he gave me, the valuable knowledge, the sharing sessions I had with him will no doubt go a long way in shaping my career.

i

The project would not have seen the light of the day but for the unfailing support and inspiration given to us by my parents, the teaching and non teaching staffs and friends. Special thanks to my friends Mr. Kinshuk Gupta, Mr. Hitesh Kataria, Mr. Anil T and Mr. Karthik R of Microelectronics and VLSI CAD batches, Manipal University, Manipal, India, with regards to discussions on devices physics, technicalities in circuit design and many other aspects of transistor design and fabrication. Finally, I place my humble accolades to all those who helped me in one way or the other in completing this project successfully.

Mayank Chakraverty

Reg. No. 122510024

# Abstract

Over the past three decades, by reducing the transistor lengths and the gate oxide thickness along with decreasing the supply voltage, there has been a steady improvement in transistor performance with a reduction in transistor size and a reduction in cost per function. The more an IC is down scaled, the higher becomes its packing density, the higher its circuit speed, and the lower its power dissipation. However, as CMOS dimensions start approaching the nanometer regime (<100 nm), we start seeing new effects in the device performance arising from new physical phenomenon. To maintain the rate of improvement in device performance with continued down scaling, modifications to device designs and fabrication are required. In particular, a collection of undesirable problems arise that are collectively called "short channel effects" which impede the further downscaling of transistors. One of the major challenges in transistor scaling are the "short channel effects" which become more visible with gate lengths less than 100 nm. The dominant short channel effects that obstruct the downscaling of FETs include a decrease in threshold voltage with a reduction in channel length, Drain Induced Barrier Lowering (DIBL), drain punchthrough, degradation of sub threshold slope resulting in increased off-state leakage current, gate oxide tunneling resulting in high gate current and so on. Scaling of MOSFET transistors has led to increased performance due to the gate length reduction. However, intrinsic semiconductor properties, like electron and hole mobilities, for the silicon lattice cannot be scaled. Since they are unaffected by scaling, beyond the 90 nm technology node new innovations in transport have been sought to increase the channel carrier mobilities and the MOSFET performance. One approach, known as strained silicon technology, tunes the strain using silicon grown on an underlying SiGe layer. This strain in the Si causes a reduction in the effective electron (hole) mass and increased electron (hole) mobility. This increased mobility improves the transistor's switching delay and leads to a higher transconductance and larger drain current.

Shrinking of MOSFETs beyond 50-nm-technology node requires additional innovations to deal with barriers imposed by fundamental physics. The classical approach used to scale the conventional MOSFET starts to fail at such a small scale and new issues emerge such as short channel effects which are important to overcome to continue the scaling trend. The issues most often cited are:

1) current tunneling through thin gate oxide;

2) quantum mechanical tunneling of carriers from source to drain;

3) threshold voltage increase due to quantum confinement; and

4) random dopant induced fluctuations.

As the gate length is reduced to around 10 nm level, gate control over the channel region decreases and there is increased source-drain tunneling of electrons. This leads to increased off current and degradation in the sub threshold slope. The source-drain tunneling significantly degrades the sub threshold slope S at gate lengths less than 10nm, and increases the off state current. MOSFET's with gate lengths approaching 10 nm need to have a thinner channel layer to ensure adequate device turnoff. With new device designs like ultra thin body (UTB) FET's where the MOSFET is fabricated on a very thin silicon layer on an oxide substrate (SOI), it is imperative to have a body thickness below 10 nm to maintain electrostatic integrity. Due to quantum confinement effects in UTB-FETs we start to see a threshold voltage increase with reducing channel width. As the MOSFET scaling process continues the International Technology Roadmap for Semiconductors (ITRS) anticipates that the semiconductor industry would require channel lengths in the range of 10 nm by 2015. Besides the introduction of new materials and improving bulk MOSFET performance, newer device concepts are likely to be required to continue scaling into the sub-10nm gate length regime. Advanced MOSFET structures like ultra thin-body (UTB) FET, Dual-gate FET, FinFET, TriGate FET and Gate All Around (GAA) FET offer the opportunity to continue scaling beyond the bulk because they provide reduced short channel effects, a sharper subthreshold slope, and better carrier transport as channel doping is reduced. In UTB single gate MOSFETs, though a superior performance has been reported in terms of short channel effects, thicker self aligned source and drain structures are required to minimize parasitic source/drain series resistance. A dual-gate device FET (DG FET) structure allows for more aggressive device scaling as short channel effects are further suppressed by doubling the effective gate control. There have been several variations proposed for DGFET structure, but most of them suffer from process complexities, among these, the FinFET has emerged as the most practical design. The FinFET is a double gate FET since the gate oxide is thin on the vertical sidewall but thick on the top. The fin width is an important parameter for the device as it determines the body thickness and short channel effects depend on it. For effective gate control it is required that the fin width be half the gate length or less. Because of the vertical nature of a FinFET channel, it has (110) oriented surfaces when fabricated on a standard (100) wafer. This crystal orientation leads to enhanced hole mobility but degraded electron mobility. The Tri-Gate and Omega-Gate FETs are multi-gate transistors having three sided gate structures. Omega-Gate FET has the gate extending into the substrate on the sides creating an effective fourth gate which provides better gate control than a Tri-Gate FET. Therefore, there is a need to circumvent the short channel effects dominantly encountered while downscaling transistors. Silicon nanowire transistor is a potential candidate among the other non classical device structures to completely circumvent short channel effects, thereby enabling the downscaling of devices to the next level as it is reported to obstruct the prevailing short channel effects in downscaled MOSFETs. Cylindrical gate-all-around (GAA) devices are considered as the most efficient structure, with better Short Channel Effects (SCE) immunity and enhanced mobility.

As the conventional silicon metal-oxide-semiconductor field-effect transistor (MOSFET) approaches its scaling limits; many novel device structures are being extensively explored. Among them, the silicon nanowire transistor (SNWT) has attracted broad attention from both the semiconductor industry and academia. To understand device physics in depth and to assess the performance limits of SNWTs, simulation is becoming increasingly important. The objectives of this report are: 1) to theoretically explore the essential physics of SNWTs (e.g., electrostatics, transport and band structure) by performing computer-based simulations, and 2) to assess the performance limits and scaling potentials of SNWTs and to address the SNWT design issues. The computer based simulations carried out are essentially based on DFT using NEGF formalism. A silicon nanowire has been modeled as PN diode (Zener Diode), PIN diode, PIP & NIN diode configurations by selectively doping the nanowire and simulated by biasing one end of the nanowire to ground and sweeping the other end of the nanowire from -1 V to 1 V to obtain the electrical characteristics of the respective diodes. In order to determine the effectiveness of the modeled diodes in silicon nanowire, the same diodes have been modeled using a germanium nanowire by selective doping and simulated in the same manner to obtain the electrical characteristics of the germanium nanowire based diodes which has been used as a reference to analyze the characteristics obtained using silicon nanowire. The modeled diodes are extremely small in dimension when compared to the conventional bulk silicon and germanium based diodes.

The final results are expected to show that SNWTs provide better scaling capability than planar MOSFETs. A microscopic, quantum treatment of surface roughness scattering (SRS) in SNWTs is also planned to be accomplished, and it is expected to show that SRS is less important in SNWTs with small diameters than in planar MOSFETs. Finally, band structure effects in SNWTs with small diameters have to be examined by using an empirical tight binding model, and a channel orientation optimization has to be done for both silicon nanowire field effect transistors.

# Index

| CHAPTER NO. | TITLE | PAGE NUMBERS |
|:---:|:---:|:---:|

# LIST OF TABLES

# LIST OF FIGURES

xvi

# 1. Introduction

The NW transistor is one candidate which has the potential to overcome the problems caused by short channel effects (SCEs) in SOI MOSFETs and has gained significant attention from both device and circuit developers. In addition to the effective suppression of SCEs due to the improved gate strength, the multi-gate Silicon nanowire (SiNW) FETs show excellent current drive and have the merit that they are compatible with conventional CMOS processes.

## 1.1 Overview of Nano Wires

Following the discovery of Carbon Nanotubes (CNTs) by Iijima [1], there has been great interest in the synthesis and characterization of other One Dimensional (1D) structures, which include nanowires (NWs), nanorods and nanobelts [2]. Inorganic NWs can act as active components in devices, as revealed by recent investigations. In the last 4–5 years, NWs of various inorganic materials have been synthesized and characterized [11]. A NW is an object with a 1D aspect in which the ratio of the length to the width is greater than 10 and the width does not exceed a few tens of nanometers [26].

Today, this definition has been extended to atomic and molecular wires, which have proved to exhibit very interesting physical properties without necessarily having the geometrical characteristics, defined earlier. NWs represent the smallest dimension for efficient transport of electrons and excitons. Thus, they will be used as interconnects and critical devices in nanoelectronics and nano-optoelectronics [3]. They are especially attractive for nano-science studies as well as for nanotechnology applications. NWs, compared to other low dimensional systems, have two quantum confined directions, while still leaving one unconfined direction for electrical conduction [4]. Today, different types of nanowires (NWs) are being investigated and are being fabricated.

## 1.1.1 Types of NWs

NWs can be prepared from metals, semiconductors, organic molecules, etc. and offer prospects in mechanical, electronic, optical, or medical applications. Therefore, depending on the materials they are made, NWs are classified as: metallic NWs, semiconductor NWs and molecular NWs [5]. They can also be classified as elemental, metal oxide, metal nitride, metal carbide and metal Chalcogenide. Thus, NWs of elements, oxides, nitrides, carbides and chalcogenides have been generated by employing several strategies [6]. Fig. 1.1 shows the images of different semiconductor NWs imaged using high-angle annular dark-field scanning transmission electron microscopy and Table 1.1 summarizes the different types of NWs together with their specific examples.

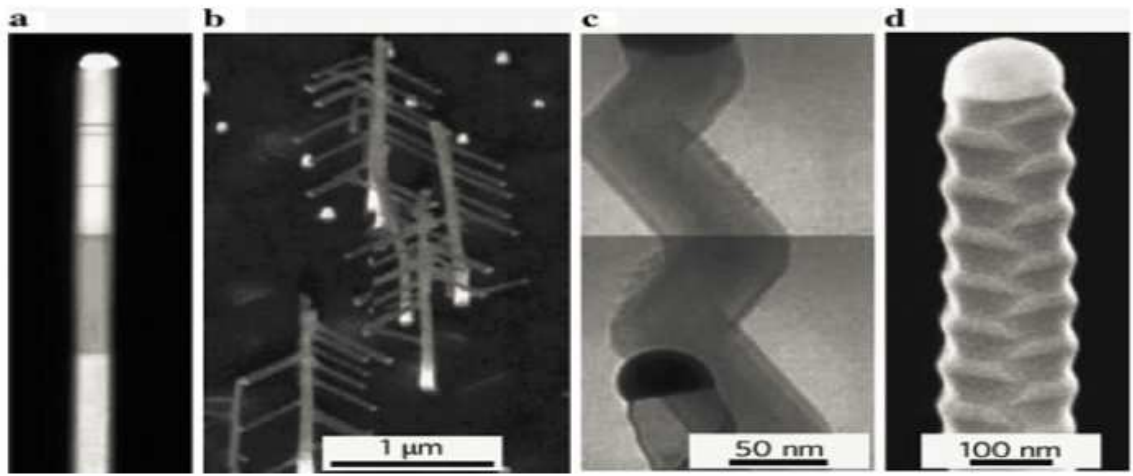

**Fig.1.1** Images of different semiconductor NWs. (a) Axially modulated InAs/In NW (30 nm indiameter), (b) GaP NWs, (c) SiNWs, (d) Periodically twinned InP NW [28].

Although many different types of semiconductor NWs have been investigated, SiNWs have become prototypical NWs. This is because they can be readily prepared, the Si/SiO$_2$ interface is chemically stable, and SiNWs are utilized in a number of device demonstrations that have well-known silicon technology- based counterparts [7].

**Table 1.1** The different types of NWs

| Types of NWs | Specific Examples |
|---|---|
| Elemental | Ge, B, In, Sn, Pb, Sb, Bi, Se, Te, Au, Ag, Fe, Co, Ni and Cu |
| Metal Oxide | MgO, Al$_2$O$_3$ ,Ga $_2$O  In$_2$O$_3$, SnO$_2$, SiO$_2$, GeO$_2$, TiO$_2$, MnO$_2$, Mn$_3$O$_4$, Cu$_x$O and ZnO |
| Metal Nitride | BN, AlN, GaN, InN, Si$_3$N$_4$ and Si$_2$N$_2$O |
| Metal Carbide | BC and SiC |
| Metal Chalcogenide | CdS, CdSe, CdTe, PbS, PbSe, Bismuth Chalcogenides, CuS, CuSe, ZnS and ZnSe, NbS$_2$ and NbSe$_2$ |

## 1.1.2 Electrical Properties of NWs

The effects of size on electrical conductivity of nanostructures and nanomaterials are complex, since they are based on distinct mechanisms. These mechanisms can be generally grouped into four categories: surface scattering including grain boundary scattering, quantized conduction including ballistic conduction, Coulomb charging, tunneling and widening of the band gap. In addition, increased perfection such as reduced impurity, structural defects and dislocations, would affect the electrical conductivity of nanostructures and nanomaterials [8].

When the size of a material is smaller than the de Broglie wavelength, electrons and holes are spatially confined and electric dipoles are formed. In addition, discrete electronic energy level would be formed in all materials. Similar to a particle in a box, the energy separation between adjacent levels increases with decreasing dimensions. Also, the electron Density of States (DOS) depends dramatically on the dimensionality of nanostructures [21]. Whereas for bulk systems a square-root dependence of energy prevails, a staircase behavior is a characteristic for Two Dimensional (2D) quantum well structures, spikes are found in 1D quantum wires (QWs), and discrete features appear in Zero Dimensional (0D) quantum dots [28]. The DOS for Three Dimensional (3D) (bulk), 2D (quantum well), 1D (QW) and 0D (quantum dots) is illustrated in Fig. 1.2.

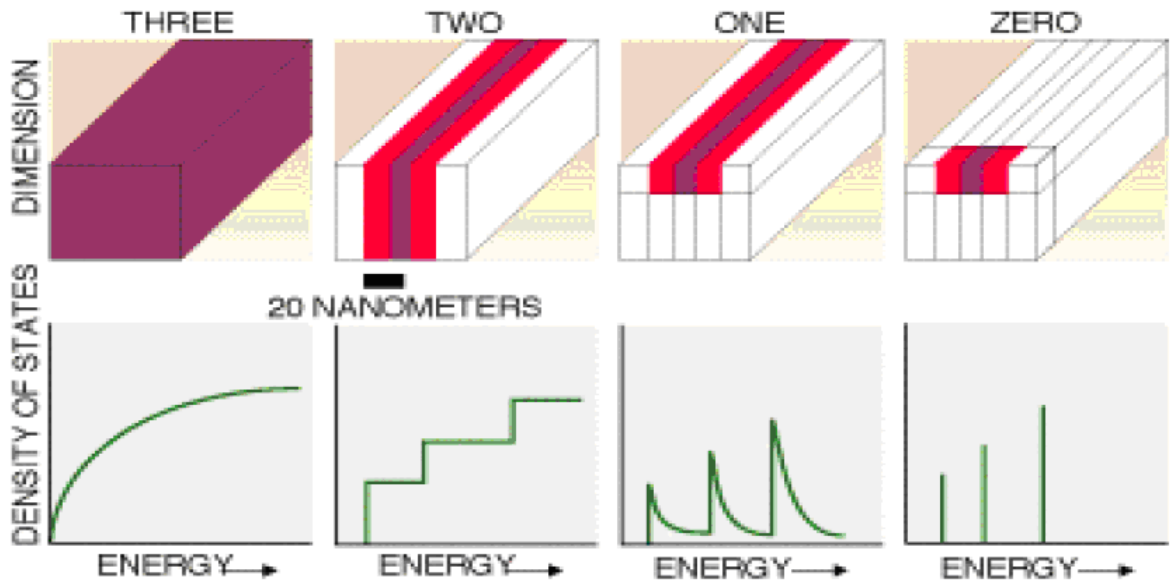

**Fig. 1.2** Electron density of states for 3D, 2D, 1D and 0D structures [20]

The other mechanism that occurs in quantum wires is ballistic conduction. It occurs when the length of device is smaller than the electron mean-free path. In this case, each transverse waveguide mode or conducting channel contributes $G_o$ = $2e^2/h$ = 12.9 kW to the total conductance. Another important aspect of ballistic transport is that no energy is dissipated in the conduction, and there exist no elastic scattering. The latter requires the absence of impurity and defects. When elastic scattering occurs, the transmission coefficients, and thus the electrical conductance will be reduced, which is then no longer precisely quantized [9]. Finally, tunneling conduction is a phenomenon that affects the electrical conductivity of NWs. Tunneling involves charge transport through an insulating medium separating two conductors that are extremely closely spaced. It occurs when the electron wave functions from two conductors overlap inside the insulating material, at extremely thin thickness of the insulator [10].

## 1.1.3 Applications of NWs

In the early 1980s it was theoretically predicted that QWs may have applications in high performance transport devices due to their saw tooth like DOS [11]. Semiconductor NWs are emerging as a powerful class of materials that, through controlled growth and organization, are opening up novel opportunities for nanoscale electronic and photonic devices. From these QWs, logic gates such as inverters or oscillators can be built. QW FETs have emerged as powerful sensors for label free detection of biological and chemical species. Crossed NWs can also be used to fabricate nanoscale p-n diodes, for example, for band-edge emission Light Emitting Diodes (LEDs) at the nanoscale cross-points [11].

SiNWs have been demonstrated as one of the promising building blocks for future nano-devices such as FETs, solar cells, sensors and lithium battery [12]. Generally, NWs have been pursued for their intrinsic ability to make smaller devices for several years. Vertically grown NWs have been shown in several materials, and hetero structures have been embedded within these wires to create quantum dots and resonant-tunneling diodes [13].

## 1.2 SiNW FETs

NW FETs have been proposed and now studied by many research groups around the world. This is because, they are promising candidate to sustain the relentless progress in scaling for CMOS devices [14]. Several key factors have contributed to the boom of NW research. First, semiconductor NWs can be prepared in high-yield with reproducible electronic properties as required for Large Scale integrated (LSI) systems. Second, compared with "top–down" nanofabricated device structures, "bottom–up" synthesized NW materials offer well controlled size; that is at or beyond the limits of lithography. In addition, the crystalline structure, smooth surfaces and the ability to produce radial and axial NW hetero structures can reduce scattering. These results in higher carrier mobility compared with nanofabricated samples with similar size. Finally, the body thickness (diameter) of NWs can be controlled down to well below 10 nm. Therefore, electrical integrity of NW-based electronics can be maintained even as the gate length is aggressively scaled. This is a feature that has become increasingly difficult to achieve in conventional MOSFETs [15]. GAA SiNW FETs have attracted significant interest because of their excellent electrostatic integrity even at the nanoscale. Various types of SiNW FETs are being explored as a promising candidate for future transistors replacing planar MOSFETs in logic and Dynamic Random Access Memory (DRAM) applications, and their fabrication is being studied either from top-down or bottom-up approaches [16]. The schematic of a GAA SiNW FET is shown in Fig. 1.3.

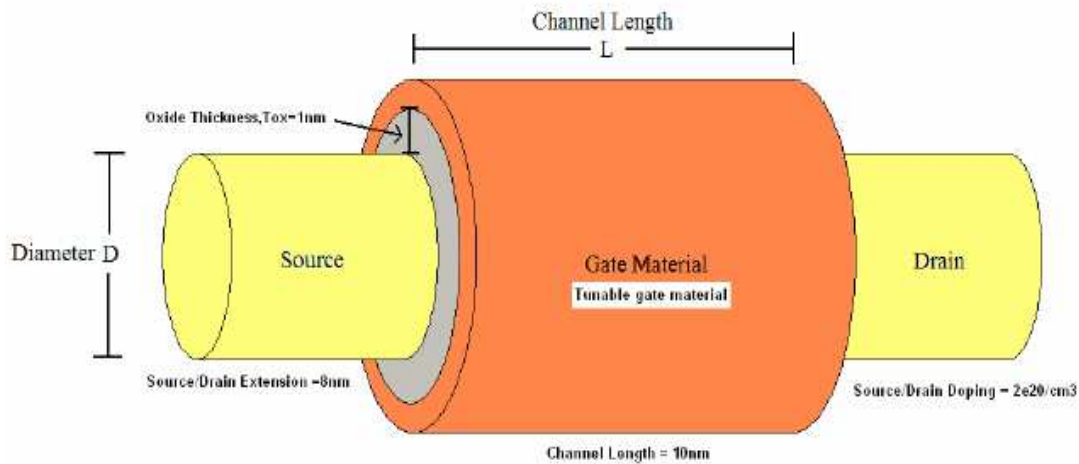

**Fig.1.3** Schematic of GAA SiNW FET [12]

SiNW FETs, unlike planar MOSFETs, have metal source and drain contacts. That is, the source and drain contacts are made from metals instead of degenerately doped semiconductors. Typically, positive Schottky barriers are observed at the metal/semiconductor interface due to the combined effect of metal work function and Fermi level pinning by surface states [17]. As a result, the device performance is to a large degree affected by contact properties. Because of this property, application of annealing can lead to the formation of essentially ohmic contacts and dramatically increase on-state current and the apparent field effect mobility. This is illustrated in Fig. 1.5. A SiNW FET with aluminum source/drain contact is shown in Fig. 1.4. Fig. 1.5(a) and (b) show the output characteristics ($I_{ds}$ - $V_{ds}$) graph and the transfer characteristics graph ($I_{ds}$ - $V_{gs}$) graph respectively. From these graphs we can see that there is a huge increase of On-state current after annealing.

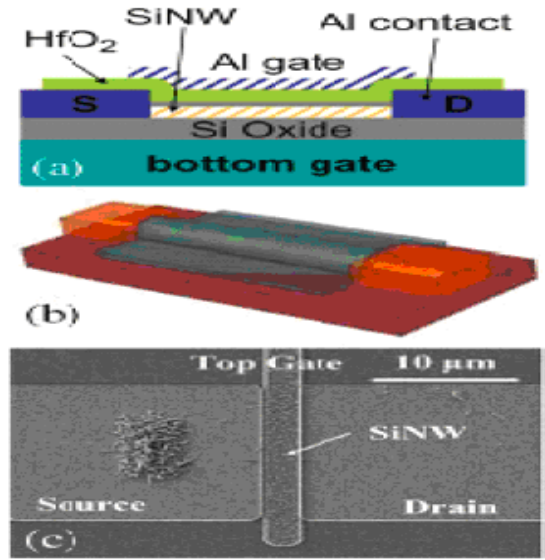

**Fig. 1.4** SiNW FET (a) Schematic drawing of device cross section along the length of the SiNW. (b) Three-dimensional schematic of top-gated SiNW FET. (c) Scanning electron beam micrograph of a typical top-gated SiNW FET with no gate-to-source/drain overlap [12].

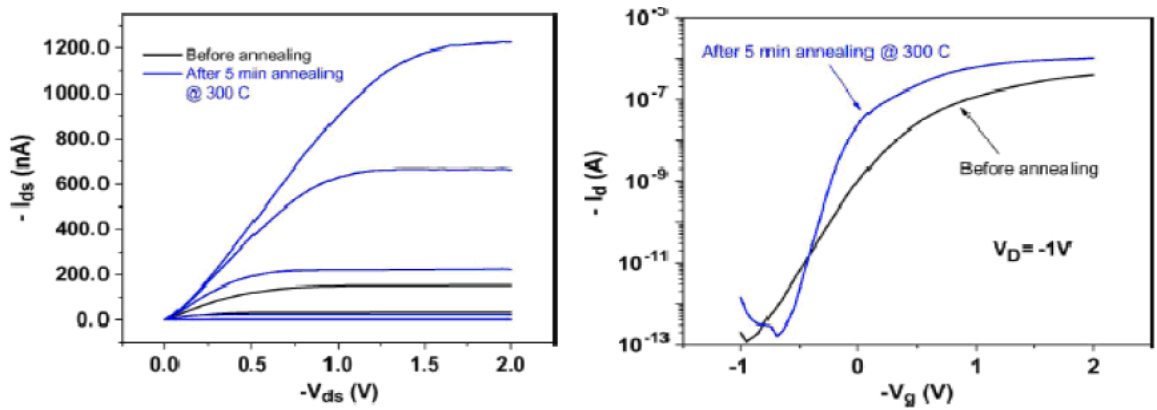

**Fig. 1.5** Electrical characterization of a typical top-gated SiNW FET before annealing and rapid thermal annealing for 300 sec at 300 $^0$C. (a) Ids – Vds curves at Vgs = +0.5 to –1.5 V in –0.5V steps. (b) Ids as a function of Vgs (before and after annealing) for Vds=-1V [12].

## 1.2.1 Transport Mechanisms in NW FETs

If the device length is smaller than the mean free path, it is very probable for carriers not to undergo any scattering event during their motion within the device and the transport is ballistic [15]. Since SiNW FETs have lengths in the order of nanometers their transport mechanism can be mainly described by ballistic transport. However, in general, electronic-carrier mobility in real systems can be affected by the scattering of carriers in a number of ways. These include, scattering by other carriers, by surfaces, by interfacial roughness, by acoustic phonons, optical phonons, impurities and by plasmons [18].

Significant theoretical and experimental work involving electronic transport in CNTs has helped in distinguishing ballistic and diffusive modes of transport. However, for free standing semiconductor NWs, experimental and theoretical consensus of carrier-scattering mechanisms that are most significant in single- and multi component coaxial semiconductor NWs is less clear. In some cases, carrier mobilities in SiNW FETs and gm values have been reported to exceed those associated with conventional Si planar technology devices [19]. Although many studies pointed out that transport in nanoscale transistors is ballistic, in practice there is some sort of scattering in these devices. In reference [14], scattering mechanisms affecting the performance of SiNW FETs are studied. The Non Equilibrium Green's Function (NEGF) formalism within parabolic Effective Mass Approximation (EMA) and the coupled mode space approach is applied in the analysis of this device. In addition, various scattering mechanisms are analyzed, namely, the Surface-Roughness (SR), the Remote-Coulomb Scattering (RCS), and phonon scatterings. SR and RCS are investigated by using a non-perturbative approach, considering devices with specific random realizations of rough $Si/SiO_2$ interfaces and fixed-charge center distributions at the high-κ/$SiO_2$ interface. A general analysis is thus statistically carried out on a set of device samples. The impact of the SR and RCS on the transport properties of SiNW FETs was analyzed with special attention devoted to the effective mobility. Effective mobility is found to be an important performance metric also in the quasi-ballistic regime, showing findings in accordance with the main semi-classical models. Finally, a global analysis of the interplay of the different scattering mechanisms has been performed. This shows interesting results on the mobility trend for devices scaled down to 10 nm channel length [20].

## 1.2.2 Operation Modes of SiNW FETs

The operation of NW FETs is similar to that of planar FETs and UTB FETs [21]. However due to its 2D confinement structure, transport characteristics in NW MOSFETs are different from those in planar and UTB MOSFETs. Also, NW FETs show some differences as compared to the conventional planar MOSFETs. Among these differences, the presence of ambipolar conduction in NW FETs is major one. This property is due to the fact that source and drain contacts are made from metals instead of degenerately doped semiconductors.

Ambipolar behavior in FETs is defined as both n- and p-type conduction in a single device at the appropriate gate bias conditions [29]. Ambipolar conduction in n-SiNW FETs is shown in Fig. 1.6. As we can see from this figure, there is both p-type and n-type conduction in this device, depending on the bias applied to the gate voltage. With the application of negative gate bias (for Vgs<0), there is strong p-type conduction in the device and with zero gate bias (Vgs=0V) application, there is a weak p-type conduction. Finally for a positive gate bias (Vgs>0), n-type conduction is seen in the device [22].

Generally, there are two types of operation modes for SiNW FETs. These are the Junction FET (JFET) like operation and the MOSFET like operation [23]. In JFET like operation of NW FETs, the NW region is treated as a bulk volume of charge whereas in MOSFET like operation, the NW channel is treated as a thin sheet of charge. Fig. 1.7a shows the JFET like operation of the SiNW FET. The MOSFET like operation is depicted in Fig. 1.7b.

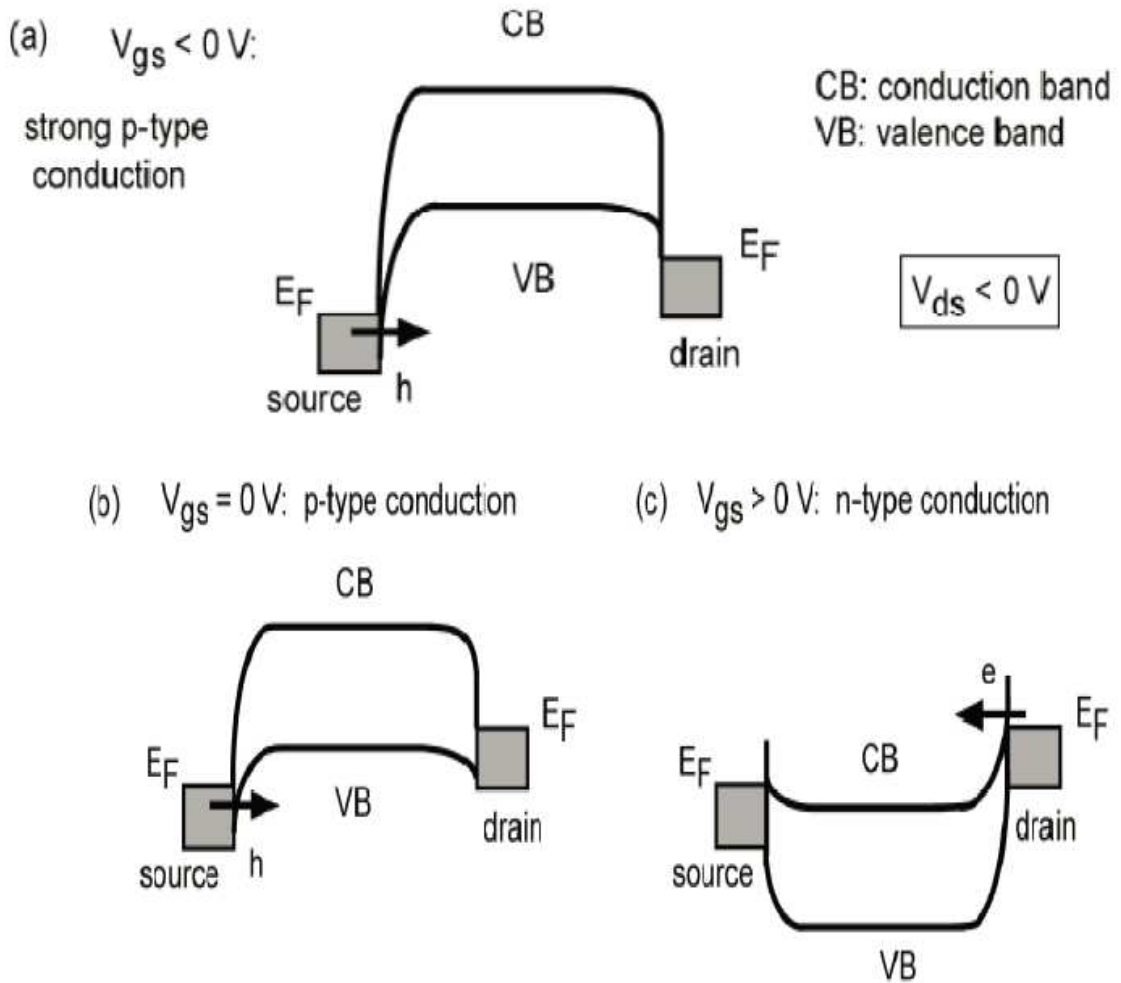

**Fig. 1.6** A schematic showing ambipolar conduction in n-SiNW FETs (a) strong p-type conduction for Vgs < 0 V. (b) p-type conduction at Vgs = 0 V. (c) n-type conduction for Vgs > 0

## 1.2.3 Review of Performance Evaluation of SiNW FETs

In order to assess the performance of SiNW FETs, we will compare the different parameters of this device with that of the planar MOSFETs. The simulation results of reference [24] are used for comparing the performance of SiNW FET with that of DG planar FET.

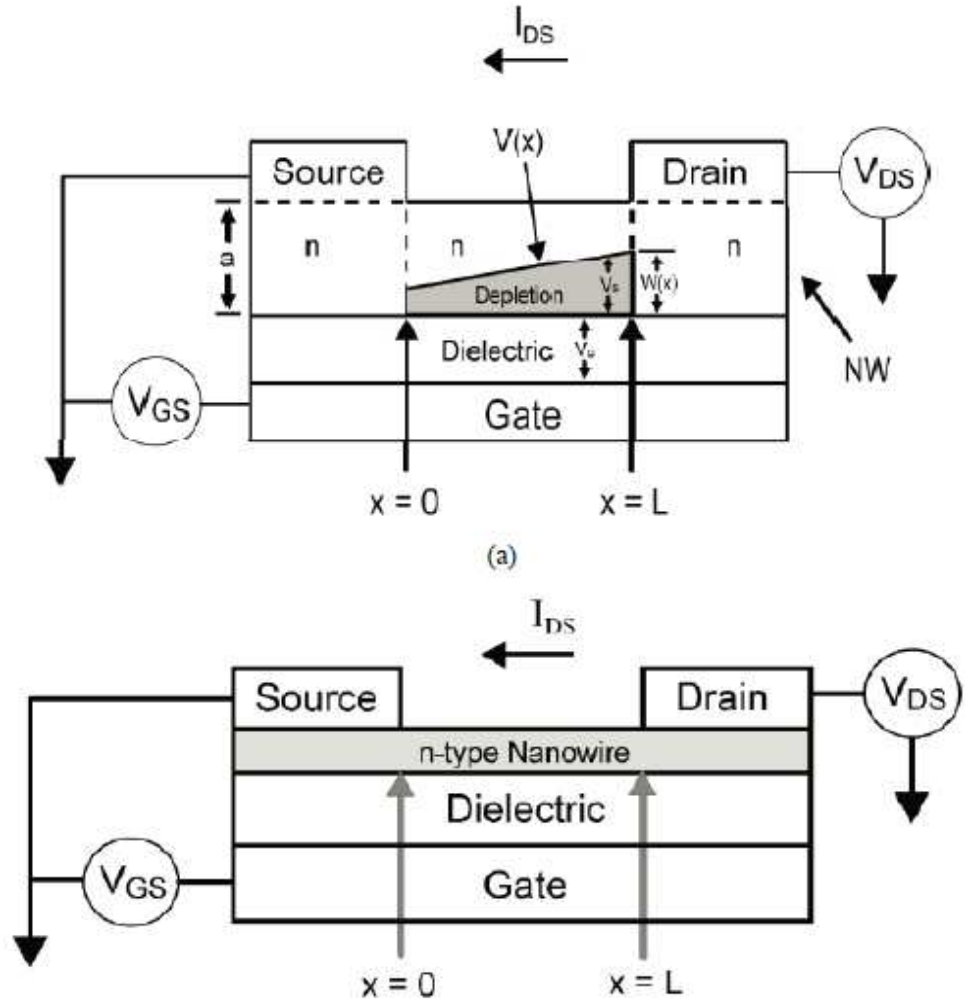

**Fig. 1.7** Diagram of a Bottom-Gated NW-based FET showing (a) JFET like operation of SiNW FETs (b) MOSFET like operation of SiNW FETs

In reference [24], Ids – Vgs graph of SiNW FET and planar DG FET is simulated. The simulation is done assuming ballistic quantum transport using two different models. These models are the EMA model and the Nonparabolicity model (NP). In addition, there are simplifications of these models included in the simulation namely, the MC which is a model for increasing the conduction mass alone and ALPHA model for increasing the NP mass alone. Neglecting both effects yields the EMA case while inclusion of both effects is still referred to as the NP case [40]. In this reference, simulation of the transfer characteristics is done for investigation of the impact of band structure effects on transfer characteristics of SiNW FET and planar FETs. However we use these simulation results for comparing the transfer characteristics and parameters of these two transistors.

The simulated NW FET has square GAA structure with a Si NW channel. The source and drain regions are both n-doped with a concentration of $N_D$ = 1020 $cm^{-3}$ and the lengths are $l_s = l_d$ = 10 nm. The gate contact surrounds the FET as shown in Fig. 1.8(a) (GAA). Like wise, the planar FET considered is a DG structure as shown in Fig. 1.8(b). Both transistors have a body thickness of $t_c$ and an oxide thickness $t_{ox}$. For the calculation of Threshold Voltage ($V_{th}$), $I_{on}$, and SS the values $I_{off}$ = 10–7 A, $\Delta V_{on}$ = 0.2 V, and $\Delta V_{SS}$ = 0.2 V, respectively, are employed.

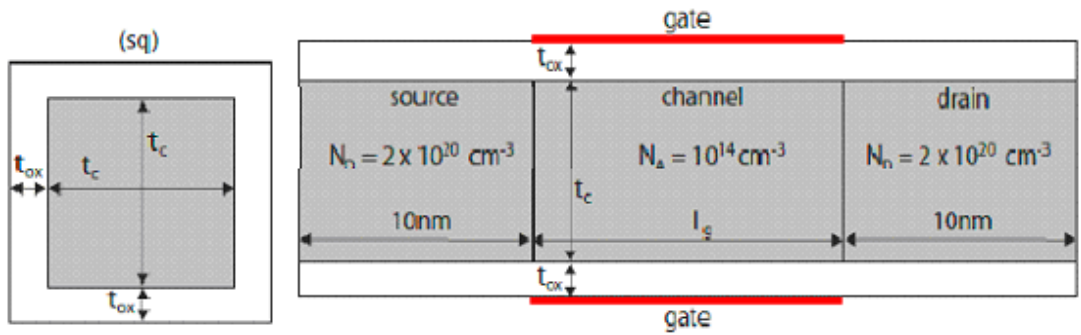

Fig. 1.8 (a) Profile specifications of the square (sq) NW and (b) DG FET with oxide thickness of $t_c$ = 0.6 nm

## 1.2.3.1 Comparison of the Off-state Current

The $I_{ds} - V_{gs}$ graphs of the SiNW and the planar transistor are given in Fig. 1.9 (a) and (b). The body thickness and NW diameter are taken to be equal to 2 nm and the gate length ($l_g$) is taken as 10 nm for both transistors. From the first graph (Fig. 1.9 (a)), Off-state current is found to be approximately 0.8525 pA/μm for the NW FET and from Fig. 1.9 (b) Ioff is found to be approximately 0.415 nA/μm for the planar FET. From these results we can conclude that the Off-state current of the planar FET is greater by a factor of 1000. This result implies that NW FETs have greater gate control than the planar FETs. This is because of the GAA structure of the NW FET compared to its planar counter part which has only a double gate.

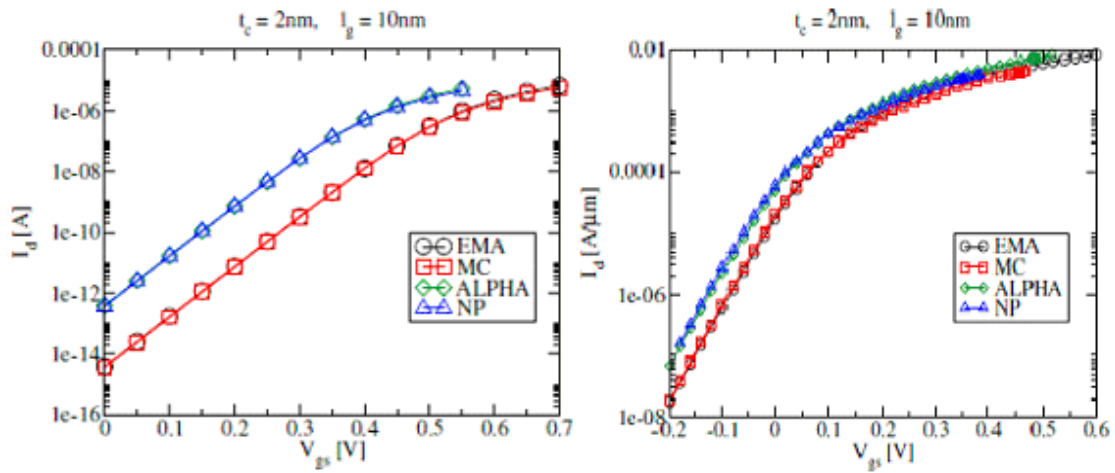

**Fig. 1.9** Transfer characteristics (sub threshold regime) of (a) sq NW transistor. (b) Planar FETs by various models

## 1.2.3.2 Comparison of Subthreshold Slope

Fig. 1.10 (a) and (b) show the graph of SS versus $t_C$ for a gate length of 10 nm for the NW and planar FETs respectively. From these graphs the SS is found as approximately 61.5 mV/dec for NW FET and 65 mV/dec for the planar FET at $t_C$ of 2 nm. Similarly SS at $t_C$ of 4 nm is found to be approximately 67.5 mV/dec and 77.5 mV/dec for the NW and the planar FETs respectively. From these values of SS we can see that the SS value of planar FETs is much higher than that of the NW FETs. In addition the SS value for the planar FET increases very rapidly as the channel thickness increases as compared to the NW FET.

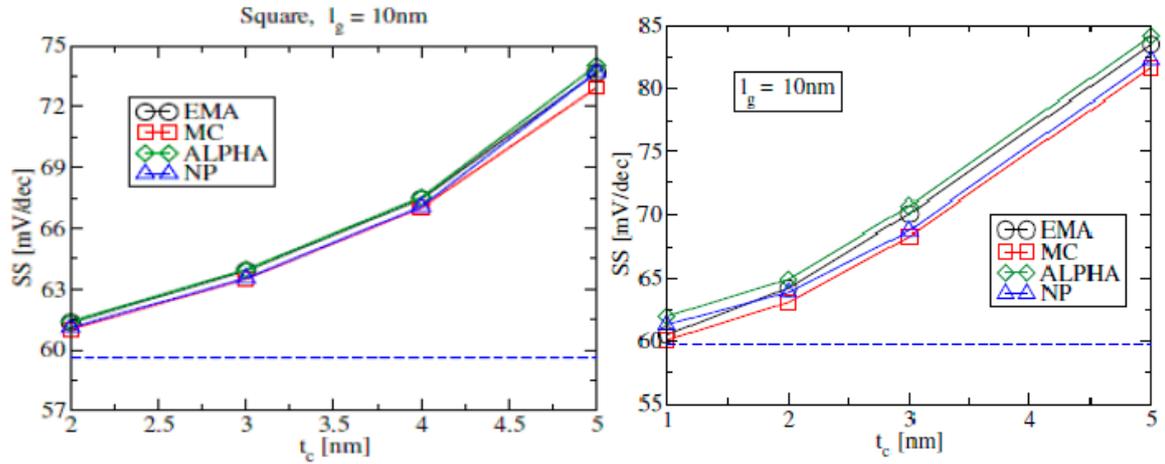

**Fig. 1.10** Subthreshold Slope versus body thickness of (a) NW FETs and (b) planar FETs

## 1.2.3.3 Comparison of Threshold Voltage

Fig. 1.11 (a) and (b) shows the graph of $V_{th}$ versus $t_C$ for the NW FET and for the planar FET at a gate length of 10 nm. The vth value for $t_C$ of 2 nm is approximately 0.348 V for the NW FET and - 0.058 V for the planar FET. Similarly, at $t_C$ of 3 nm, the value of $V_{th}$ is found to be approximately 0.097 V and -0.125 V for the NW and planar FETs respectively. Here again, we can deduce that the NW FET has very high gate control than that of the planar FET.

Therefore, we can conclude that SiNW FETs have higher performance when compared to conventional planar FETs. The advantages of SiNW FETs over planar FETs are summarized in the following subsection.

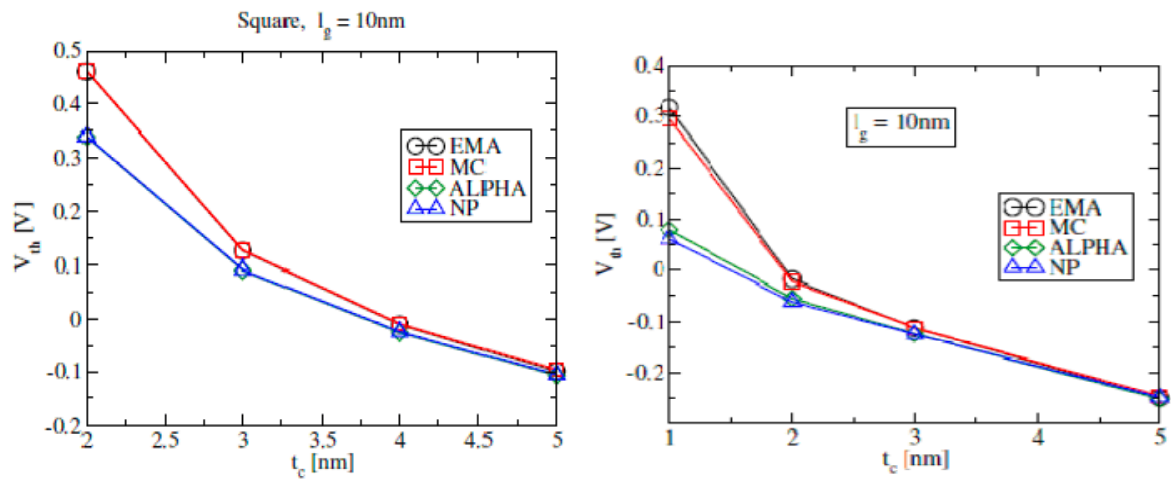

**Fig. 1.11** Threshold voltage versus body thickness of (a) NW FETs and (b) planar FETs

## 1.2.4 Advantages of SiNW FETs Over Planar FETs

SiNW FETs have several advantages as the candidate of main stream CMOS devices for 2020s. First of all, the ability of the suppression of the short channel effects and thus, the suppression of the off-leakage current of SiNW FETs are expected to be very good, because of the gate surrounding configuration. Secondly, Si NW FETs are expected to have high on current because of the following 3 reasons [25].

I. The nature of quasi-1D conduction of thin NW with small freedom of the carrier scattering angle [26]; because of the small freedom of the carrier scattering, its conduction will be high.

II. The use of multi-quantum channels for the conduction; the band structure of SiNWs is quite different from that of the bulk and many conduction sub-bands appear near the lowest sub-band [27]. Those sub-bands contribute to the conduction as the gate voltage increase.

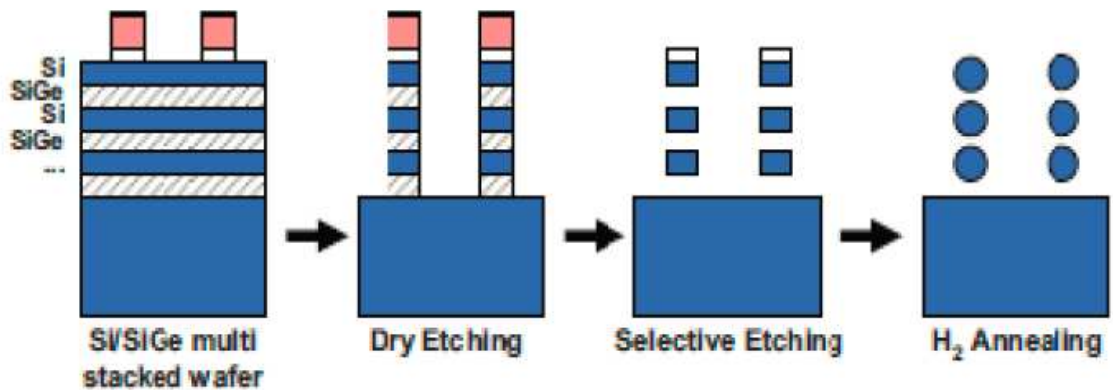

**Fig. 1.12** Fabrication of Multilayer NW

III. Multilayer NWs can be implemented easily by utilization of Si/Ge multi-layers as shown in Fig. 1.12 [28]. For the fabrication, basically, today's conventional Si CMOS IC production process can be used almost as it is to fabricate SiNW FET, although process tuning kind of developments are necessary. This is a very big advantage for the production to minimize the risk and cost of the new process technology development. Furthermore, the number of SiNW FET fabrication process will be smaller than that of today's planar CMOS. It is assumed that no channel implantation including that of halo is necessary because of good SCEs control of the NW structure, assuming that threshold voltage control can be done by the work function control of gate stack. In some future, metal or silicide source/drain is assumed to be introduced into ultra-short channel SiNW FETs because of the necessity of abrupt junction, resulting in the further elimination of source/drain doping [29].

The advantages of SiNW FETs over planar FETs are summarized in Fig. 1.13.

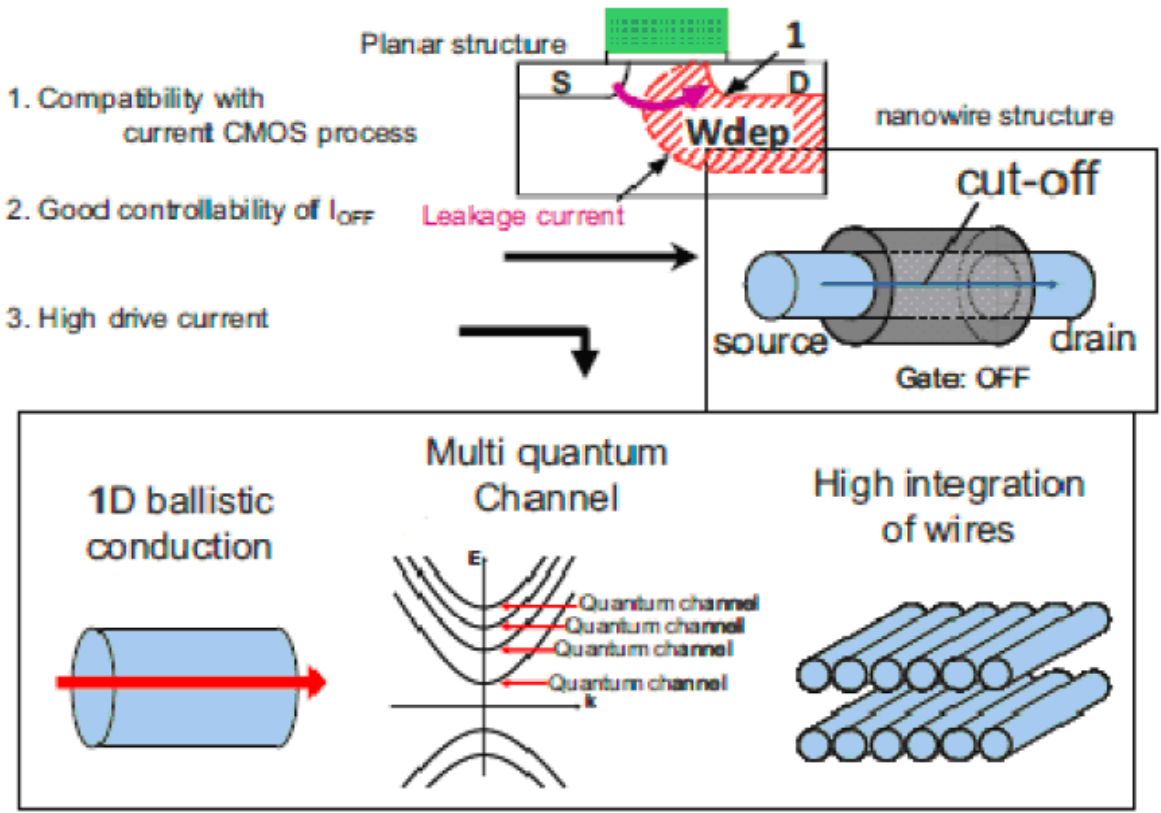

**Fig. 1.13** Advantages of SiNW FETs over Planar FETs

# 2. Literature Survey

The journey of the development of the metal-oxide-semiconductor field-effect transistor (MOSFET) started some seventy years ago. Fundamental concepts and design of a MOSFET were first described by Julius Edgar Lilienfeld of Brooklyn and Cedarhurst, NY in two of his first three patents [1] [2]. However it was not until 1960 that the first MOSFET was fabricated and demonstrated by Dawon Kahng and Martin Atalla at the Bell Labs [3]. Making use of a thermally grown silicon dioxide layer to passivate the surface, this device did not suffer from the high density of surface states (interface and oxide traps), which had affected the performance of the bipolar junction transistor (BJT) proposed in 1948 by Shockley and was considered one of the most important and significant technological advance for that period [4]. The Kahng and Atalla paper [3] also reported reduction and stabilization in the device's leakage current and that the transistor was easy to integrate with the planar fabrication process. This discovery led to MOS transistor integration and the development of the silicon MOSFET integrated circuit manufacturing industry. Later, in 1962 an important idea was conceived by Wanlass at Fairchild, which was the CMOS (Complementary MOS) concept of integrating p-channel and n-channel MOSFETs [5]. The development of a passivating oxide on the semiconductor surface was extremely important in reducing the power used by MOS transistors, which was one of the reasons MOS was lagging behind bipolar technology. By late 1964 the first commercial MOSFETs were announced by Fairchild and RCA.

The dramatic development of the MOSFET fabrication technology enabled the growth of the modern integrated circuit and computer industry [6]. MOS technology required fewer processing steps than bipolar technology, which translated into lower fabrication costs and higher yield. Also, while bipolar transistor technology could not be scaled down in size with harming transistor performance characteristics, MOSFETs could be scaled down without compromising on performance. This led to increased growth of MOSFET technology and by the end of 1970 it had taken over bipolar as the dominant choice for integrated circuits.

## 2.1 Moore's Law

Moore's Law is one of the ubiquitous concepts associated with dramatic rise of the semiconductor industry over last few decades. Gordon E Moore made an empirical observation in 1965 that '*The complexity for minimum component costs has increased at a rate of roughly a factor of two per year.....That means by 1975, the number of components per integrated circuit for minimum cost will be 65,000. I believe that such a large circuit can be built on a single wafer*' [7]. Though Moore's prediction made in 1965 envisioned only the next ten years, it has remained relevant over the last forty years.

Moore's Law is illustrated in Fig. 2.1 and shows the trend of increasing transistor count with each new microprocessor technology. The consequences of Moore's Law are ever increasing processing power while decreasing costs because of higher levels of transistor and circuit integration. The latest Intel Duo-Core processor now employs about one billion transistors. Through technology improvements and innovations like strained Si channels [8], high – k gate dielectric [9] [10], metal gates [10] and the development of novel device designs like ultra thin body (UTB) transistors [11], FinFETs and other Dual- Gate transistors [12], Tri-gate transistors [13], and Silicon nanowire gate all around (GAA) transistors [14], it is possible that Moore's Law will continue through next decade.

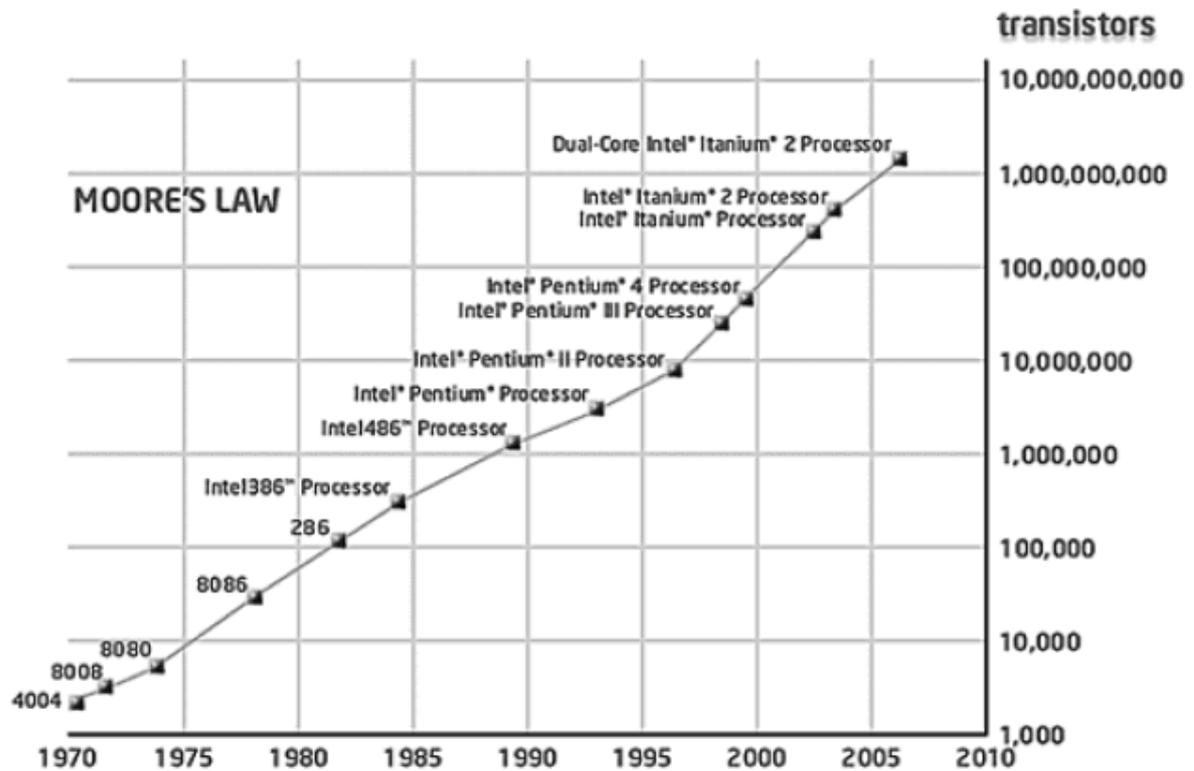

**Fig. 2.1** Increasing number transistors with each new microprocessor technology

## 2.2 MOS Transistor

The MOSFET is the basic building block for very large scale integration (VLSI) circuits and in microprocessors and dynamic memories. Since current in a MOSFET is primarily transported by carriers of one polarity (e.g., electrons in an n-channel device), the MOSFET is usually referred to as unipolar or majority carrier device.

The basic structure of a MOSFET is shown in Fig. 2.2. It is a four terminal device with the terminals designated as gate (G), source (S), drain (D), and substrate or body (B). An n-channel (p-channel) MOSFET consists of a p-type (n-type) substrate into which two n+ (p+) regions, the source and the drain are formed.

The gate electrode is usually made of heavily doped polysilicon and is separated from the semiconductor channel by a thin gate oxide layer. Application of a positive (negative) voltage at the gate terminal for n-channel (p-channel) MOSFET leads to accumulation of negative (positive) charge at the oxide-silicon interface, creating a channel for current conduction. The gate is capacitively coupled to the channel via the gate electric field and there is effectively no gate drain current. The transistor effect is achieved by modulating gate electric field using gate-source bias leading a MOS transistor to be popularly known as the *'field effect transistor'*.

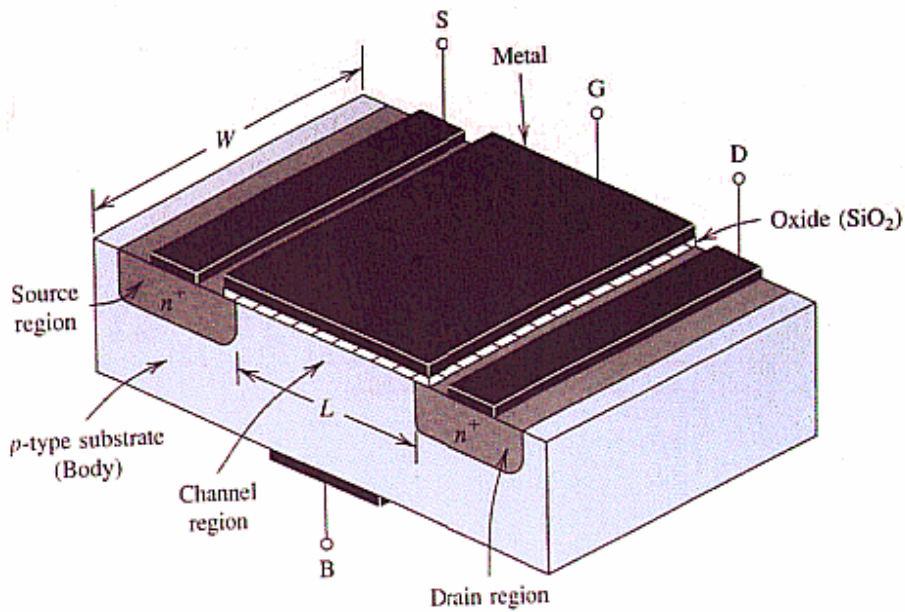

**Fig. 2.2** 3-D view of a basic n-channel MOSFET

## 2.3 Scaling of MOSFETS

It has been now forty five years since the invention of the MOSFET. During these years we have seen rapid and steady progress in the development of integrated-circuit (IC) technology. The main driving force behind this growth has been the down scaling of the MOSFET's dimensions, which was first proposed in 1974 by Dennard et al. Starting with gate lengths of 10μm in 1970, we have reached about 32nm in 2007 (currently known as 65nm technology node) corresponding to approximately a 15% reduction each year [16].

Two types of scaling mechanisms are commonly used, known as constant voltage scaling and constant electric field scaling as summarized in Fig. 2.3. Constant voltage scaling is purely a geometric approach where the power supply is kept constant while the transistor's dimensions are scaled down by a factor α. However, reducing the channel length and gate dielectric thickness increases electric field in the channel. This initially improves the mobility in the channel, but as the field starts to increase beyond 1MV/cm, it starts decreasing again due to saturation of the carrier velocity, which reduces the current gain. By contrast, the constant field scaling approach involves reducing the transistor's dimensions along with power voltage supply in order to maintain the electric field strength in the channel and ensure the same transistor physics and operation. As a result, the power per transistor decreases quadratically, as shown in Fig. 2.3, so the power density (P/L*W) remains constant instead of exploding as in constant voltage scaling approach. However the transistor's speed ($f_T$) only increases linearly. Current density also increases linearly ($I/W^2$) rather than quadratically. Because of its benefits, the computer industry essentially followed constant voltage scaling for the period from 1973 to 1993, best reducing the supply voltage only twice instead of continuously with each new generation. From 1993 to 2003, to the power supply voltage decreased with every new technology generation, although not as rapidly as the constant electric field scaling approach required [17]. Fig. 2.4 sums up the industry trends in device parameters with channel length scaling. It is apparent that the drain voltage (V) has not been

| Parameter | Symbol | Constant Field Scaling | Constant Voltage Scaling |
|---|---|---|---|
| Gate length | $L$ | $1/\alpha$ | $1/\alpha$ |
| Gate width | $W$ | $1/\alpha$ | $1/\alpha$ |
| Field | $\varepsilon$ | 1 | $\alpha$ |
| Oxide thickness | $t_{ox}$ | $1/\alpha$ | $1/\alpha$ |
| Substrate doping | $N_a$ | $\alpha^2$ | $\alpha^2$ |
| Gate capacitance | $C_G$ | $1/\alpha$ | $1/\alpha$ |
| Oxide capacitance | $C_{ox}$ | $\alpha$ | $\alpha$ |
| Transit time | $t_r$ | $1/\alpha^2$ | $1/\alpha^2$ |
| Transit frequency | $f_T$ | $\alpha$ | $\alpha^2$ |
| Voltage | $V$ | $1/\alpha$ | 1 |
| Current | $I$ | $1/\alpha$ | $\alpha$ |
| Power | $P$ | $1/\alpha^2$ | $\alpha$ |
| Power-delay | $P \vartriangle t$ | $1/\alpha^3$ | $1/\alpha$ |

**Fig. 2.3** Constant field scaling and constant voltage scaling parameters

Decreasing as fast as channel length (L), which means that the transistor's electric fields have been increasing as opposed to staying constant for a constant field scaling approach.

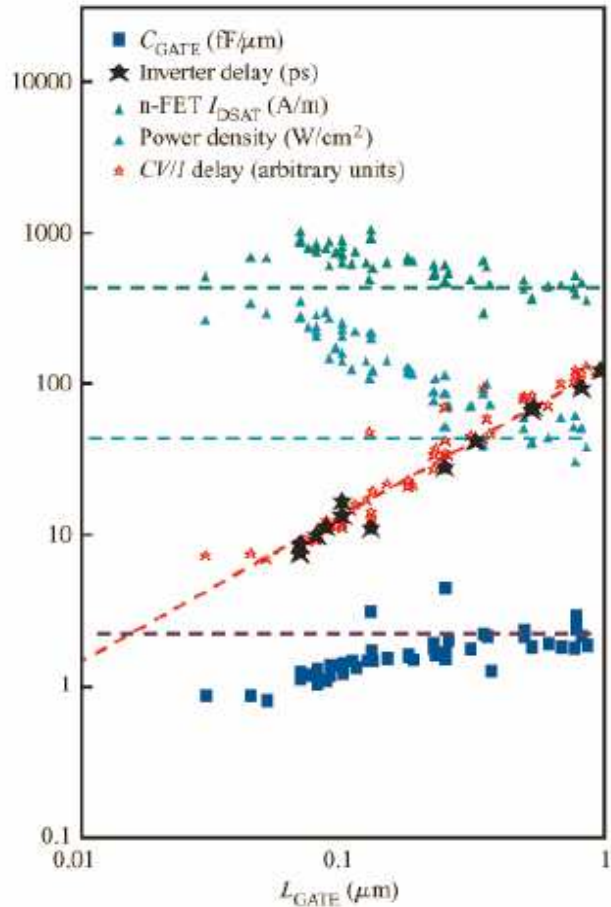

**Fig. 2.4** Industry-trend scaling (points) contrasted with classical scaling (dashed lines)

## 2.4 Scaling Issues and Approaches

Over the past three decades, by reducing the transistor lengths and the gate oxide thickness along with decreasing the supply voltage as seen in Fig. 2.5, there has been a steady improvement in transistor performance a reduction in transistor size and a reduction in cost per function. The more an IC is down scaled, the higher becomes its packing density, the higher its circuit speed, and the lower its power dissipation [15]. However, as CMOS dimensions start approaching the nanometer regime (<100 nm), we start seeing new effects in the device performance arising from new physical phenomenon. To maintain the rate of improvement in device performance with continued down scaling, modifications to device design and fabrication are required. In particular, a collection of undesirable phenomenon problems arises that are collectively called "short channel effects" which impedes further progress in transistor downscaling.

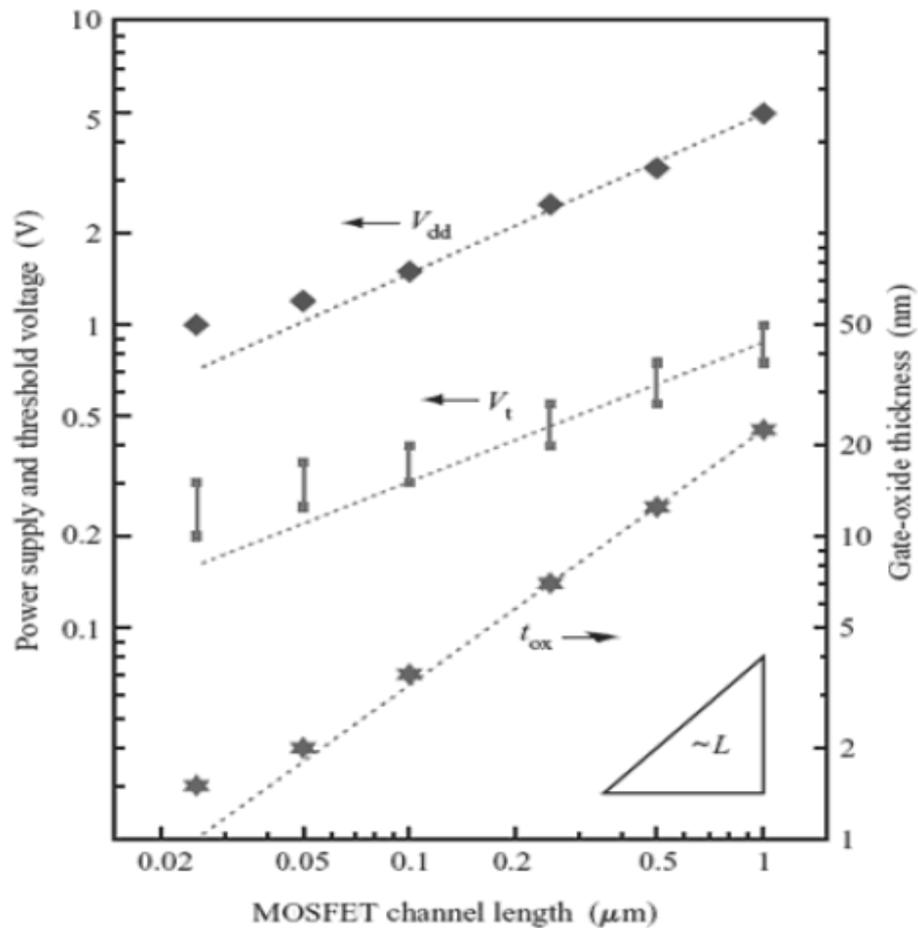

**Fig. 2.5** History and trends for supply voltage (Vdd), threshold voltage (V$_t$) and oxide thickness (tox) versus channel length for CMOS logic technologies

## 2.4.1 Short Channel Effects

One of the major challenges in transistor scaling is the "short channel effects" which become more visible with gate lengths less than 100 nm. One short channel effect is a decrease in threshold voltage as channel length is reduced. As MOSFETs shrink in dimensions, the source and drain regions move closer to each other. In a long channel device, the source and drain regions are far apart and sufficiently separated that their depletion regions have no effect on the potential in the central region under the gate as seen in Fig 2.6(a). In a short channel device, however, the source-drain distance is comparable to the MOS depletion width in the vertical direction, and the source-drain potential has a strong effect on the surface band bending under the gate [20] as seen in Fig 2.6(b). Since some of the charge is shared between the source and the drain in the channel, the net effective charge being controlled by gate is reduced, lowering the threshold voltage. Fig.2.6 (a) and (b) show simulation results of the potential profiles present in both the long channel and short channel devices, respectively.

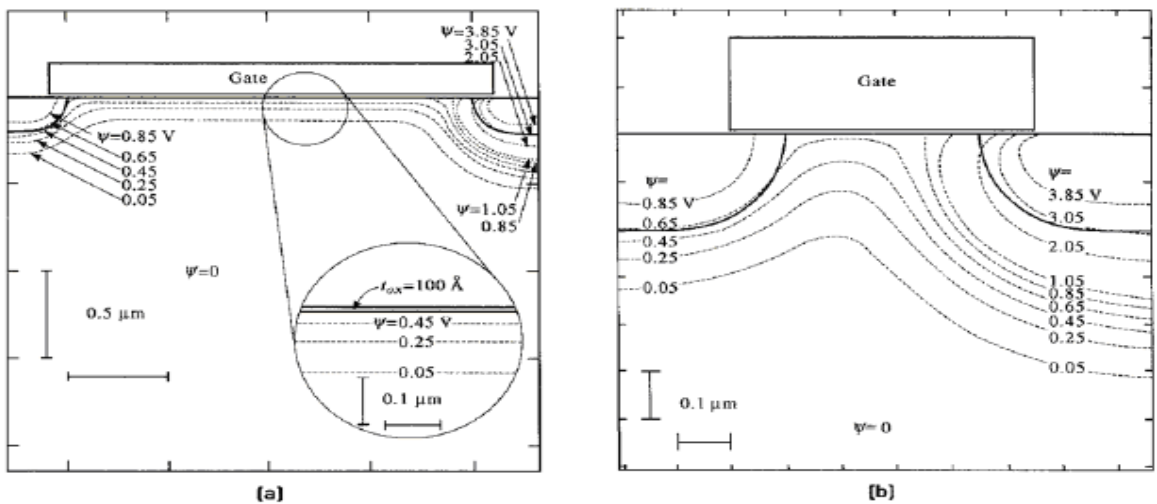

**Fig. 2.6** Potential contour for a (a) long channel device and a (b) short channel device

From the Fig.2.6, it can be seen that in short channel lengths the electric field under the gate has a two dimensional nature. Fig. 2.7 shows an example of the *threshold voltage roll-off* for n- and p-MOSFETs, which affects the circuit design as process variations in the gate length lead to variation in threshold voltage. Also, as we continue to down scale, new techniques are required to obtain adequate control and reproducibility in the threshold voltage for subsequent technology nodes.

A second short channel effect occurs when a high drain bias is applied to a short channel device; the barrier height at the source end of the channel is by the drain bias resulting which reduces the threshold voltage as seen in Fig. 2.8. This effect is known as *Drain Induced Barrier Lowering* (DIBL).

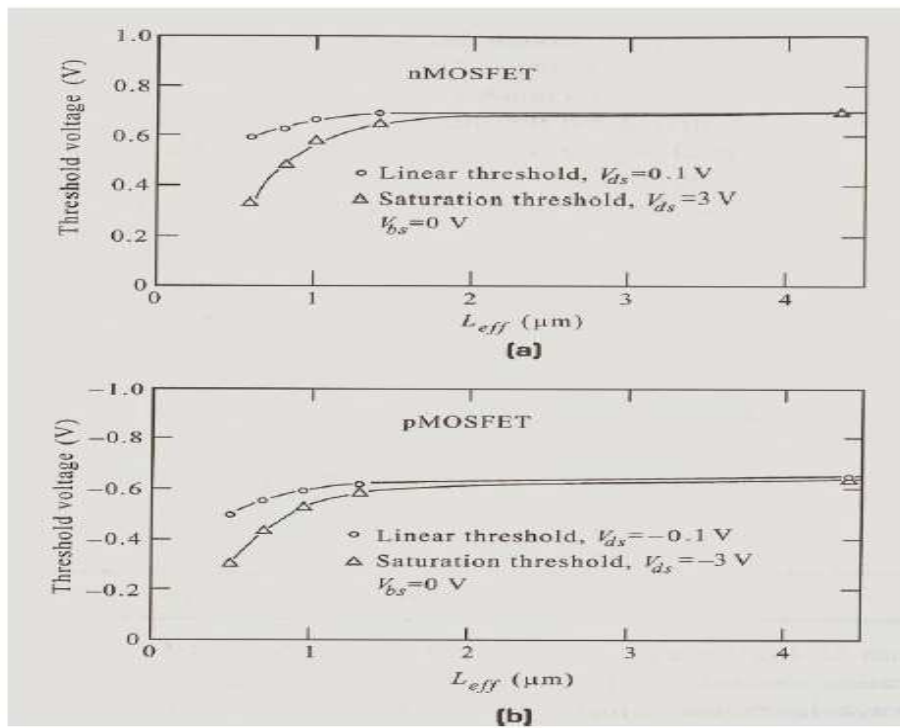

**Fig. 2.7** Short-channel threshold roll off: measured low- and high- drain threshold voltage of n- and p- MOSFET's versus channel

For the short channel device the drain voltage has an effect on the electron barrier height which appears as a variation in the threshold voltage with drain bias. This effect worsens with reduction in the channel length ultimately leading to the *punch through* condition when an excessive drain current flows and the gate loses all control.

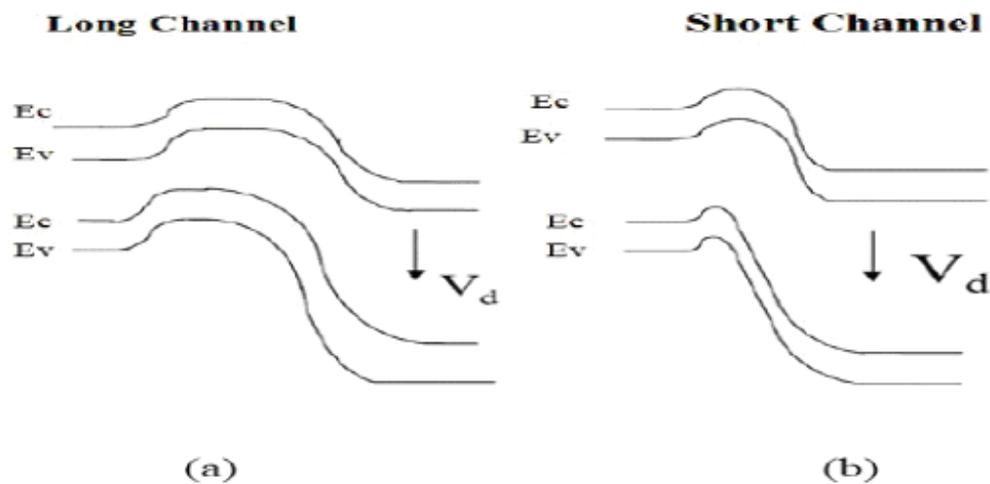

**Fig. 2.8** Band diagram for a (a) long channel and a (b) short channel device

Since the power supply voltage is also scaled down following the constant field scaling approach, there is a reduced gate voltage swing available so that the net effect of the DIBL phenomenon is to cause increased off-current leakage for the transistor.

A third short channel effect is associated with the transistor's turn-off characteristics. The subthreshold slope (SS) is defined as the change in gate voltage required produce a decade change in drain current (units of mV/dec). As we scale down the transistor with decreasing channel lengths, the subthreshold slope degrades (increases in magnitude from its theoretical minimum of 60mV/dec [20]). Ideally the sub threshold slope should be as small as possible to quickly turn on or off the transistor but the device's physics limits it to about 60 mV/dec. *Subthreshold slope degradation* causes increased off-state leakage current and induces another process variability in device and circuit fabrication.

## 2.4.2 Gate Oxide

Following the constant field scaling approach, the gate oxide ($SiO_2$) thickness has been scaled down in conjunction with gate length. This down scaling helps to keep short channel effects under control and to maintain electrostatic integrity as shown in Fig. 2.5 with the gate oxide thickness now of the order of 1 nm. However, an oxide thickness of 1.2 nm, which is used in the 90 nm logic technology comprises only a five atom thick oxide layer which means we are approaching a physical limit beyond which carrier tunneling current through the gate increases dramatically. As can be seen from Fig. 2.9, the *gate oxide tunneling current* increases exponentially as the gate oxide thickness decreases so that it approaches the drain on-current ($I_{on}$) at an oxide thickness of 1 nm. Since the gate oxide leakage current is an undesirable parasitic current, this is clearly an undesirable circumstance. Another issue associated with an excessively thin gate oxide is the loss of inversion charge, which leads to smaller gate capacitance and so smaller transconductance [19]. Quantum mechanics dictates that the peak of the inversion charge density lie at a small distance from the $Si$-$SiO_2$ surface. This decreases the depletion capacitance and effectively reduces the total gate capacitance. Similarly, a third effect, known as the polysilicon gate depletion effect, also occurs with a thinner gate oxide. A thin space charge forms in the heavily doped polysilicon gate near the gate oxide surface, which acts to reduce the overall gate capacitance [19]. For a polysilicon doping of 1020 /cm$^3$ and a 2-nm oxide, about 20% of the inversion charge is lost at 1.5 V gate voltage because of the combined effects of *polysilicon gate depletion* and *inversion layer quantization* [19].

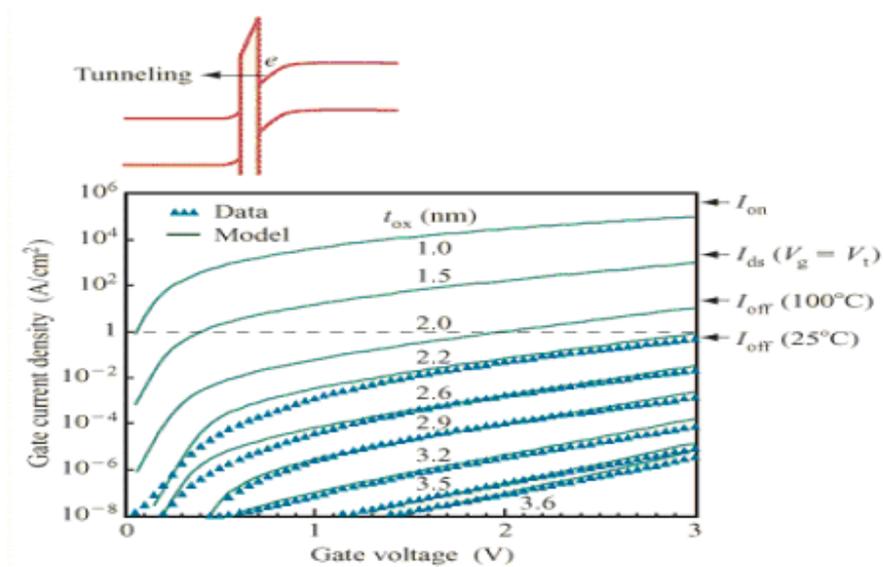

**Fig. 2.9** Calculated gate oxide tunneling current vs gate voltage for different oxide thickness

## 2.4.3 New Materials

Scaling of MOSFET transistors has led to increased performance due to the gate length reduction. However, intrinsic semiconductor properties, like electron and hole mobilities, for the silicon lattice cannot be scaled. Since they are unaffected by scaling, beyond the 90 nm technology node new innovations in transport have been sought to increase the channel carrier mobilities and the MOSFET performance. One approach has been to utilize strain using silicon grown on an underlying SiGe layer [21]. In strained silicon a layer of SixGe1-x is initially grown during epitaxial growth by adding a few Ge atoms near the wafer's crystalline surface.

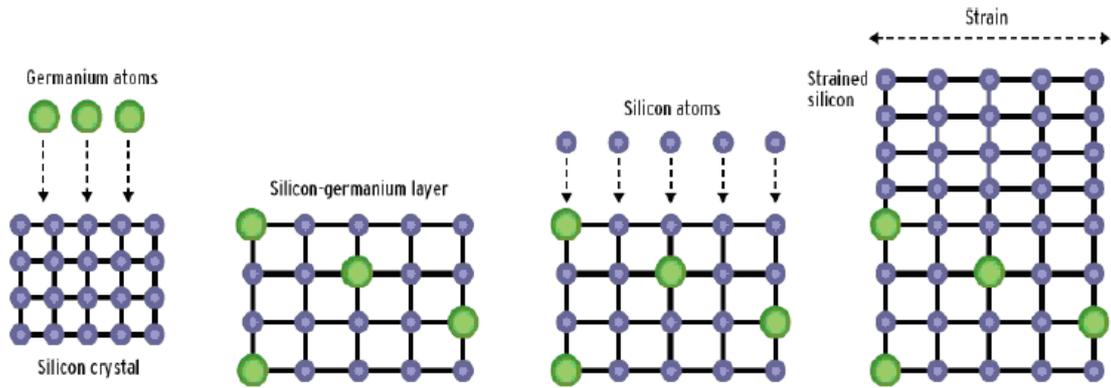

**Fig.2.10** Strained silicon grown over a silicon germanium (SiGe) layer [22]

Since Ge has a larger lattice constant (5.65 Å) than Si (5.4Å), the resulting crystal structure is larger so that a subsequent silicon layer grown on the SiGe is strained. The top Si layer grown over the $Si_xGe_{1-x}$ surface is in tensile strain as the Si atoms try to align according the to the expanded lattice as shown in

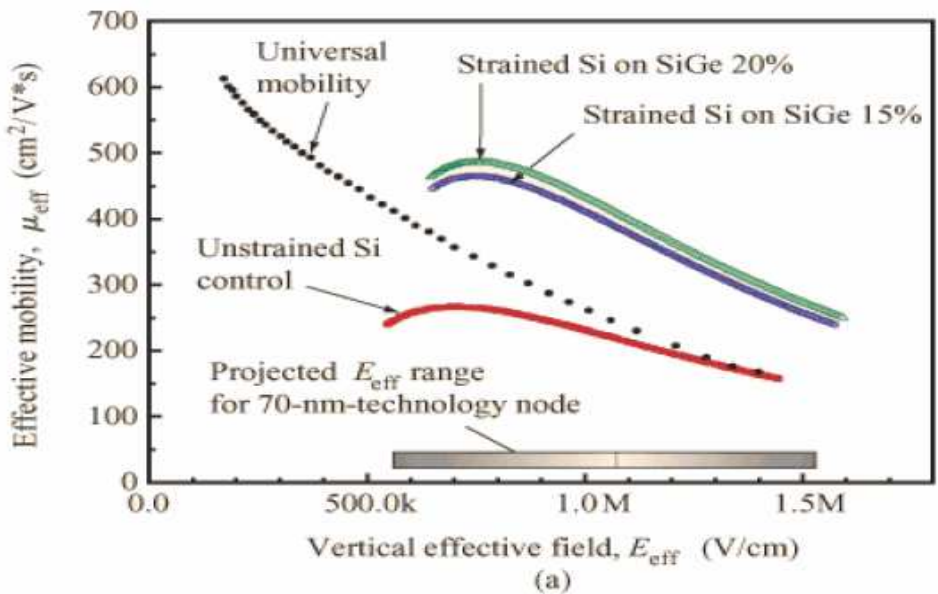

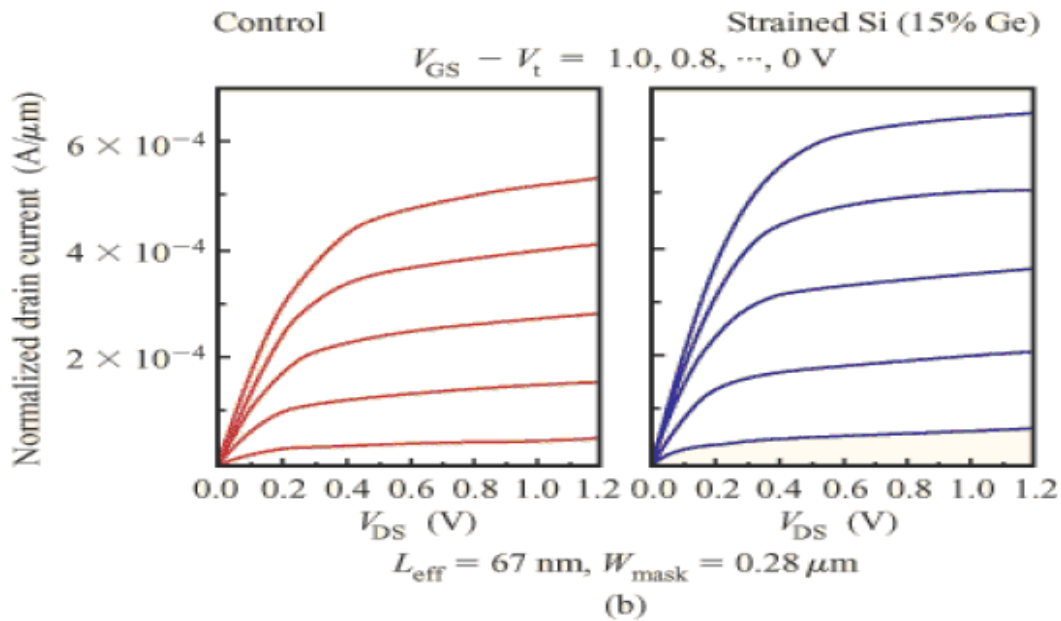

**Fig. 2.11** (a) Increased mobility and (b) drain current with strained Si technology [23]

Fig. 2.10. This strain in the Si causes a reduction in the effective electron (hole) mass and increased electron (hole) mobility as shown in Fig. 2.11(a). This increased mobility improves the transistor's switching delay and leads to a higher transconductance and larger drain current as can be seen from Fig. 2.11(b). Intel has reported an increase of 10- 20% in the MOSFET's drive current with *strained silicon technology* for a 50 nm long

channel [21].

Problems associated with thinning of the oxide layer presents one of the biggest challenges for continued MOSFET down scaling since the physical limits of oxide scaling are being approached with only few atomic layers for the current technology node. As a result, it is imperative to use an alternate gate dielectric to solve the increasing gate leakage current problem associated with thinning the gate oxide (SiO2). Using a *high-k dielectric* allows use of a thicker dielectric while providing same gate capacitance and so equivalent transistor performance. The gate capacitance for a parallel plate capacitor is $C = k \, \varepsilon_0 \, A/d$ where k is the native dielectric constant (=3.9 for $SiO_2$), $\varepsilon_0$ is permittivity of free space (=0.0885 pF/cm), A is the area of the capacitor and d is the dielectric thickness. To maintain the same gate capacitance, the same k/d ratio is needed. Thus, using a high-k dielectric allows a thicker dielectric given by $d_{high-k} = d_{ox} \times (K_{high-k})/3.9$. For a high-k dielectric with $k_{high-k}$=16 and $d_{ox}$=1nm, then the thickness for the high-k of $d_{high-k}$ ~4nm. After almost a decade of research, hafnium-oxide based materials such as $HfO_2$, $HfSi_xO_y$, $HfO_xN_y$, and $HfSi_xO_yNz$, ($k_{HfO2}$ ~ 25) have emerged as a leading candidate to replace $SiO_2$ gate dielectrics in advanced CMOS applications [24].

Downscaling the gate dielectric also requires replacing polysilicon as the gate electrode material. *Metal gate technology* involves no poly depletion effects and offers much better threshold voltage control [24]. Over last few years, research in metal gate technology has identified several promising candidates, such as W, Ti, Mo, Nb, Re, Ru and their binary or ternary derivatives such as WN, TiN, TaN, MoN, and TaSiN [25]. An alternative to metal gates is to fabricate fully silicided gates. It involves converting poly Si into silicides which are in direct contact with the gate dielectric after their fabrication, e.g. MoSi, WSi, TiSi, HfSi, PtSi, CoSi, and NiSi [24]. Recently, Intel announced plans to introduce *high-k dielectric* and *metal gates* for their 45nm technology node, which will be in production by the end of 2007 as shown in Fig. 2.12 [26].

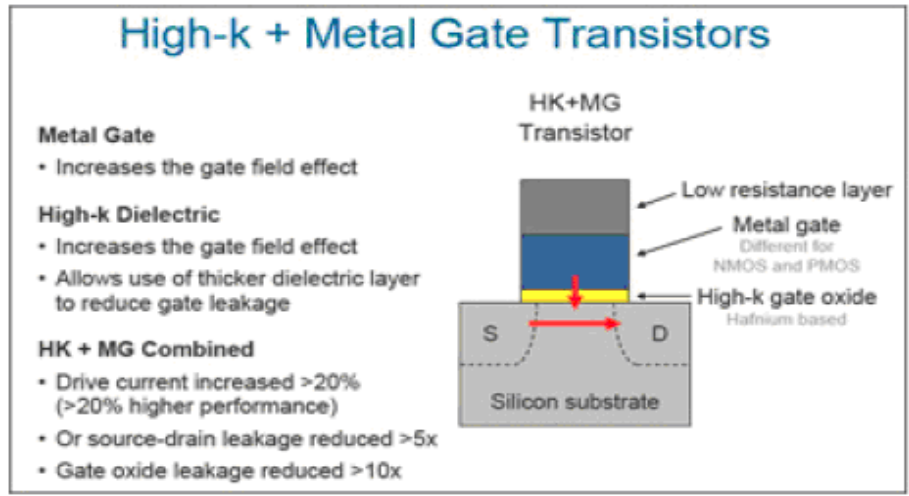

**Fig. 2.12** High-k and metal gate technology proposed by Intel (production year: 2007) [26]

## 2.5 Issues at the Nanoscale

Shrinking of MOSFETs beyond 50-nm-technology node requires additional innovations to deal with barriers imposed by fundamental physics. The classical approach used to scale the conventional MOSFET starts to fail at such a small scale and new issues emerge such as short channel effects which are important to overcome to continue the scaling trend. The issues most often cited are: 1) current tunneling through thin gate oxide (discussed above); 2) quantum mechanical tunneling of carriers from source to drain [27-28]; 3) threshold voltage increase due to quantum confinement [29]; and 4) random dopant induced fluctuations [26].

As the gate length is reduced to around 10 nm level, gate control over the channel region decreases and there is increased source-drain tunneling of electrons. This leads to increased off current and degradation in the sub threshold slope. It is still a debatable topic that whether source-drain tunneling or the device's electrostatics degradation will be the limiting factor for scaling. A simulation study by Lundstrom et al. revealed that source-drain tunneling might set the scaling limit well below 10 nm [28]. The source-drain tunneling significantly degrades the sub threshold slope S at gate lengths less than 10nm, and increases the off-state current.

MOSFETs with gate lengths approaching 10 nm need to have a thinner channel layer to ensure adequate device turnoff. With new device designs like ultra thin body (UTB) FETs where the MOSFET is fabricated on a very thin silicon layer on an oxide substrate (SOI), it is imperative to have a body thickness below 10 nm to maintain electrostatic integrity. Due to quantum confinement effects in UTB-FETs we start to see a threshold voltage increase with reducing channel width as shown in Fig. 2.13 [29]. The *quantum mechanical narrow channel effect* occurs because electrons in the inversion layer are not only located away from the surface but also occupy discrete energy levels in the channel. The lowest energy level is some finite energy higher above the bottom of the conduction band due to energy quantization due to lateral confinement. Hence, a larger surface potential is required to populate the inversion layer, which increases the threshold voltage [20].

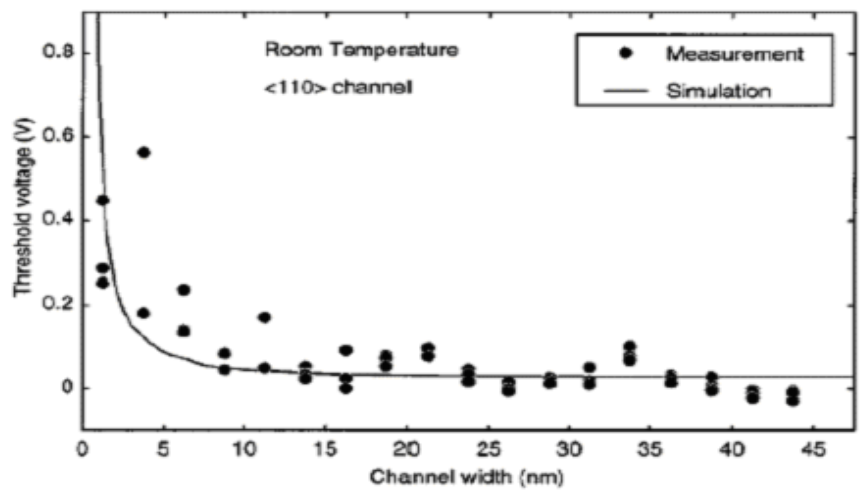

**Fig. 2.13** Threshold voltage increase with reducing SOI channel width

As can be seen in Fig. 2.13 for a thin SOI silicon film thickness, the quantum effect becomes important as the silicon thickness is reduced below ~5nm. Random fluctuation of the number of dopant atoms in the channel was predicted as a limiting factor in transistor scaling back in 1970's [21]. Recall that constant field scaling requires the substrate doping to rise at the rate of scaling factor α (see Fig. 2.3). However, in a deviation from constant field scaling, undoped channels have been adopted for FET's below 100 nm gate length due to statistical variation in the doping level called *random dopant induced fluctuation*. Fig. 2.14 shows variation in Id-Vg curves with random dopant fluctuation.

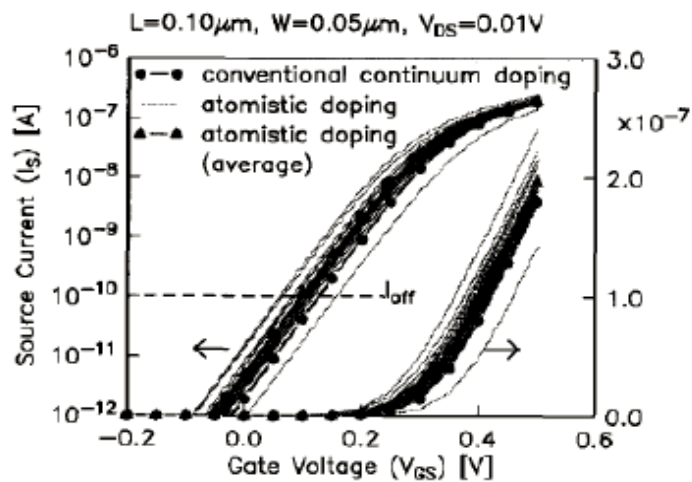

**Fig. 2.14** Variation in Id-Vg curves with different discrete dopant distribution for 24 devices. Solid dots indicate the conventionally doped device. It shows 20- 30mV variation along gate voltage, 30mV shift in sub threshold region and 15 mV shifts in linear region

compared with the conventionally doped device for 24 MOSFETs with different random "atom" distributions for W=50 nm, L=100 nm, $t_{ox}$=30 Å , and an average uniform substrate doping of $8.6 \times 10^{17}$ cm$^{-3}$ [25].   While channel doping has historically been used to adjust the MOSFET's threshold voltage metal gates with appropriate work-function are now used to adjust the threshold voltage

## 2.6 Non Classical Device Structures

As the MOSFET scaling process continues the International Technology Roadmap for Semiconductors (ITRS) anticipates that the semiconductor industry would require channel lengths in the range of 10 nm by 2015 [16]. Besides the introduction of new materials and improving bulk MOSFET performance, newer device concepts are likely to be required to continue scaling into the sub-10nm gate length region [25][22]. Advanced MOSFET structures like ultrathin-body (UTB) FET, Dual-gate FET, FinFET, TriGate FET and Gate All Around (GAA) FET offer the opportunity to continue scaling beyond the bulk because they provide reduced short channel effects, a sharper sub threshold slope, and better carrier transport as channel doping is reduced.

The basic concept of UTB MOSFET device is to have a thin silicon channel with an underlying, insulation oxide to eliminate leakage current paths through the substrate and to reduce parasitic capacitances to enhance the device's speed. Since most of the off – state current flows through the bottom of the body, it is desirable to replace the semiconductor substrate with an insulating dielectric. However thicker self aligned source and drain structures are required to minimize parasitic source/drain series resistance [23]. As an example, Fig. 2.15 (b) shows a TEM picture of a 40 nm gate length UTB n-MOSFET with 20 nm thick silicon body and 2.4 nm thick gate oxide [24]. The device shows superior short channel effects with a sub threshold slope of 87mV/dec.

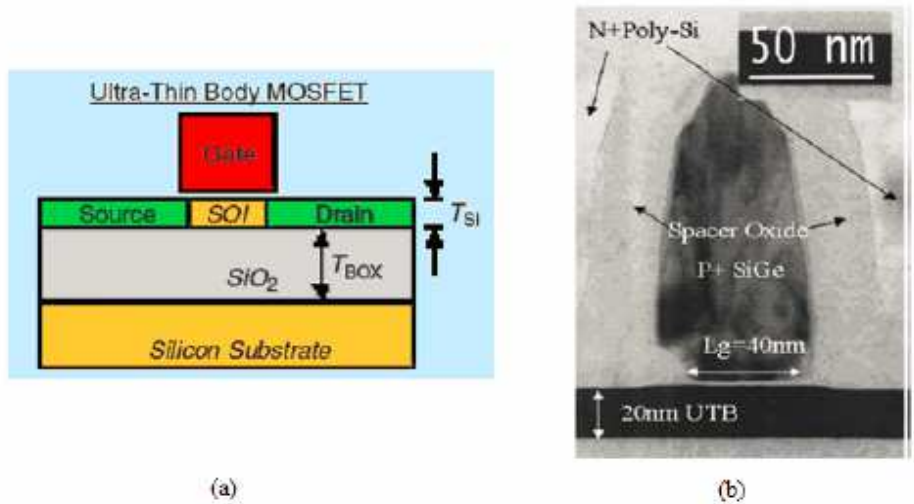

**Fig. 2.15** (a) Schematic diagram for UTB MOSFET (b) TEM image of a UTB Device

Raised poly-Si S/D contact regions (not shown) were employed to reduce the parasitic resistance of the device [24]. Drain currents of 400 µA/µm were achieved at a drain voltage of 1V and $V_g$-$V_t$=1.2 V.

A dual-gate device FET (DG FET) structure allows for more aggressive device scaling as short channel effects are further suppressed by doubling the effective gate control. There have been several variations proposed for DGFET structure, but most of them suffer from process complexities, among these, the FinFET has emerged as the most practical design as shown in the Fig. 2.16 [25]. The channel consists of a thin vertical fin around which the gate wraps on three sides. The FinFET is a double gate FET since the gate oxide is thin on the vertical sidewall but thick on the top. The fin width is an important parameter for the device as it determines the body thickness and short channel effects depend on it. For effective gate control it is required that the fin width be half the gate length or less [25]. Because of the vertical nature of a FinFET channel, it has (110) oriented surfaces when fabricated on a standard (100) wafer. This crystal orientation leads to enhanced hole mobility but degraded electron mobility [25]. The primary advantage of the FinFET over the planar MOSFET is that it offers reduced short channel effects such in Fig. 2.17. A DGFET offers more ideal sub threshold slope and better DIBL characteristics. As a result the channel length is scaled down, the silicon body thickness (fin width) must also be reduced, which poses a fabrication challenge.

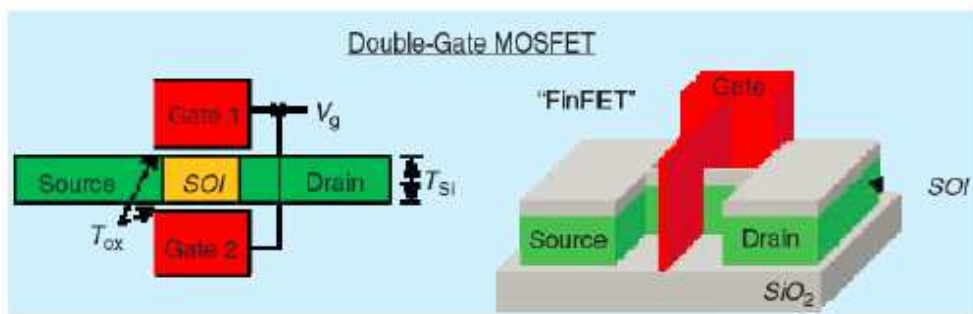

**Fig. 2.16** Dual-Gate FET structure (left) and a FinFET structure (right)

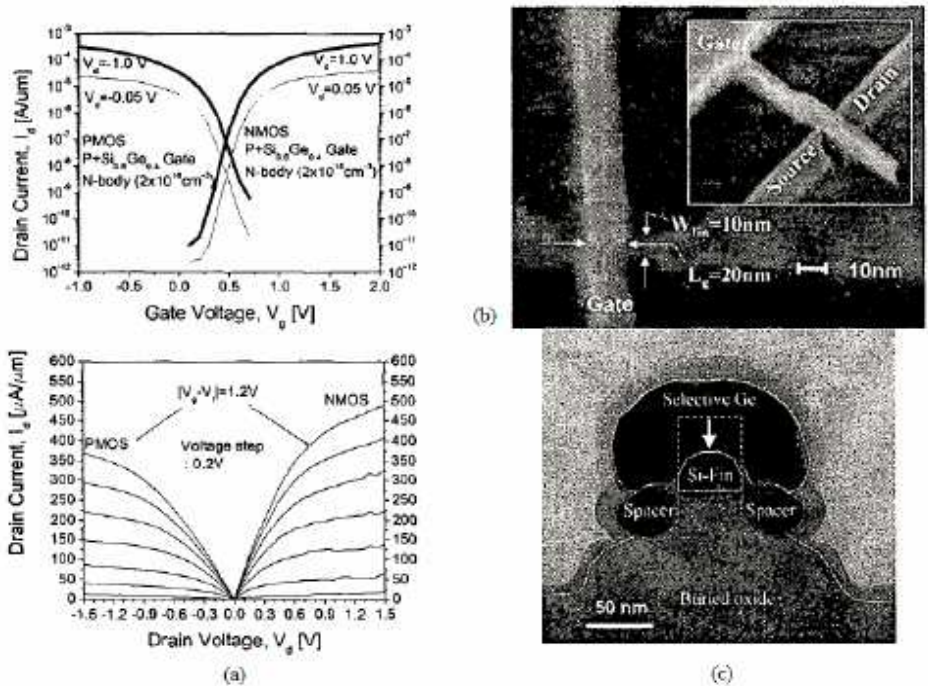

**Fig. 2.17** (a) $I_d$-$V_g$ (top) and $I_d$-$V_d$ curves (b) SEM image and gate profile (c) TEM image of a FinFET

The Tri-Gate and Omega-Gate FETs are multi-gate transistors having three sided gate structures as shown in Fig. 2.18. Omega-Gate FET has the gate extending into the substrate on the sides creating an effective fourth gate which provides better gate control than a Tri-Gate FET.

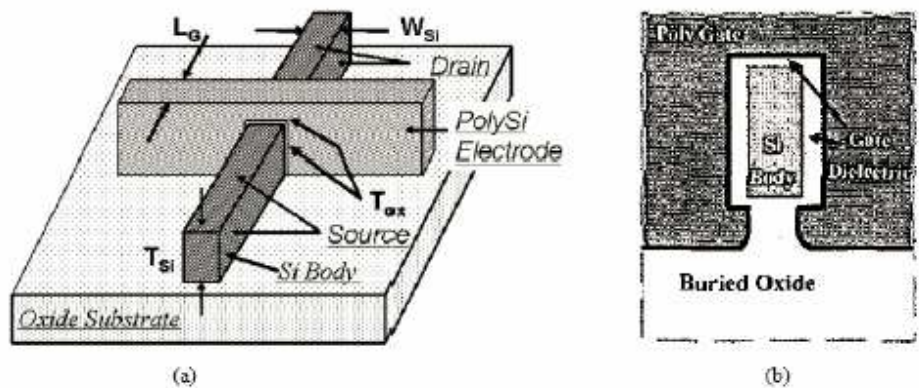

**Fig. 2.18** Schematic diagram for a (a) TriGate FET [26] (b) Omega-Gate FET

## 2.7 Silicon Nanowire Technology

Semiconductor nanowires are cylindrical single crystal structures with a diameter of a few nanometers that exhibit several interesting and novel properties because of their small one dimensions and confinement in two dimensions. The nanowire approach to nanoscale MOSFET fabrication offers the opportunity for ultimate scaling of the MOSFET using Gate All Around device structures (GAA-FET) as shown in Fig. 2.19. Since the addition of extra gates (two or more) improves MOSFET performance, including reducing short channel effects, the GAA FET is attractive for very short channel MOSFET.

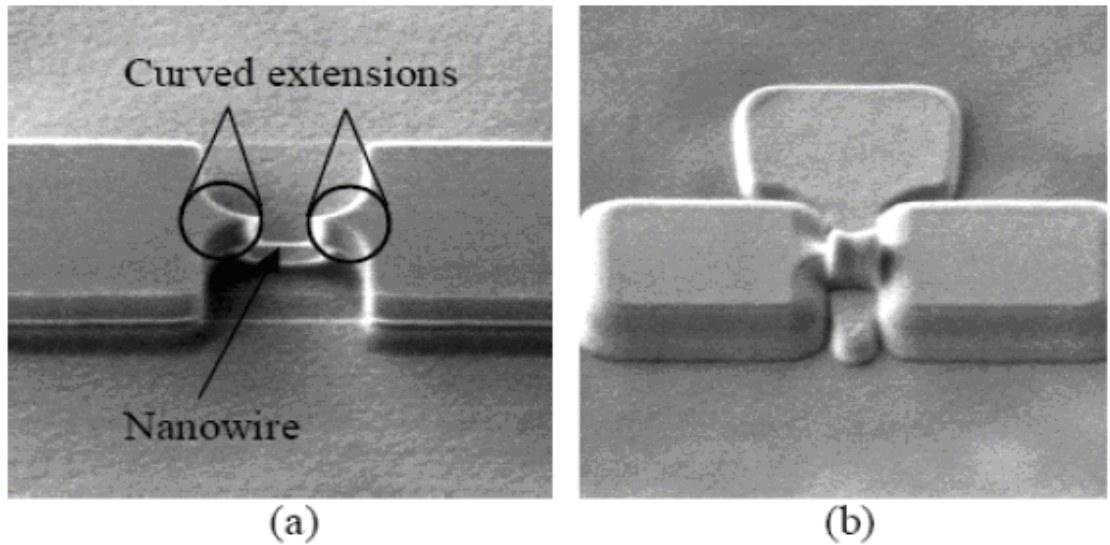

**Fig. 2.19** SEM image of (a) 200 nm long silicon nanowire and (b) after gate electrode definition ;
4 nm grown oxide followed by 130 nm amorphous silicon

A number of attempts have been made at fabricating discrete silicon nanowire transistors using both back gate [19] [20] and gate-all-around (GAA) device geometries. Shown in Fig. 2.20 is an example of a back gate nanowire MOSFET with a 5 nm silicon nanowire. Research groups have also successfully fabricated integrated circuits using nanowires as building blocks. Much of the recent (since 2000) research work on fabrication of nanowire devices has been carried out at Harvard University, USA. The research group led by Charles Lieber at Harvard has done some pioneering work in this area. They were the first to report controlled doping of silicon nanowire devices besides reports of diameter controlled synthesis of nanowires, multishell Si-Ge nanowire heterostructures [15] and also integrated circuits.

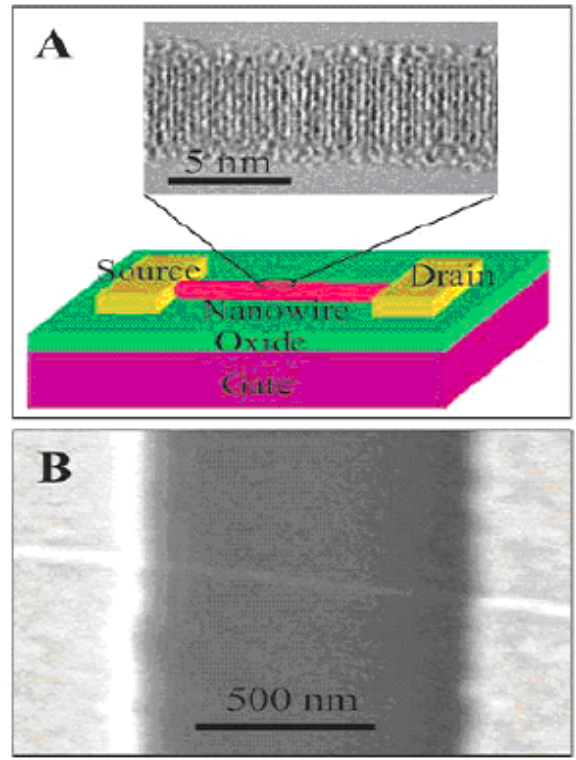

**Fig. 2.20** (a) Schematic of the back gate SiNW FET (b) TEM image of a 5 nm diameter SiNW

Fig. 2.20 shows one of the first silicon nanowire transistors, a back gated silicon nanowire transistor reported by Cui et al. The nanowire device was grown via vapor-liquid-solid (VLS) mechanism using gold nanoclusters as the catalysts as shown in Fig. 2.21. This fabrication method leads to high-quality single-crystal nanowire growth with well-controlled diameter by using well defined Au nanoclusters. Silane ($SiH_4$) was used as the vapor phase reactant as shown in the Fig. 2.21. The fabricated nanowires were deposited onto oxidized silicon substrates with electrodes separated by about by 800-2000 nm. Thermal annealing was done to improve the contact and passivate Si-SiO$_x$ traps as can be seen in the current measurements (Fig. 2.22).

An important observation from the results is that there is an improved mobility in nanowire devices. Hole mobility is estimated to be 1000 cm$^2$/V.s, which is considerably larger than bulk hole mobility (450 cm$^2$/V.s). Silicon nanowire devices also exhibit higher transconductance and more ideal sub threshold behavior. Table 2.1 shows the comparison between converted nanowire MOSFET data and a bulk silicon device. Note the improved on current, reduced off current, lower sub threshold slope and improved transconductance. Recently, there have been reports of successful fabrication of gate-all-around nanowire MOSFET devices in the literature [14]. The GAA structure is reported to lead to better gate control and better short channel performance.

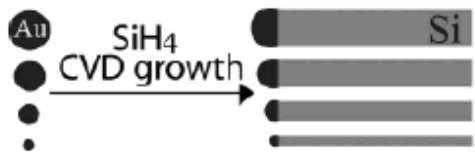

**Fig. 2.21** Size-controlled synthesis of SiNW from Au nanoclusters for diameter control

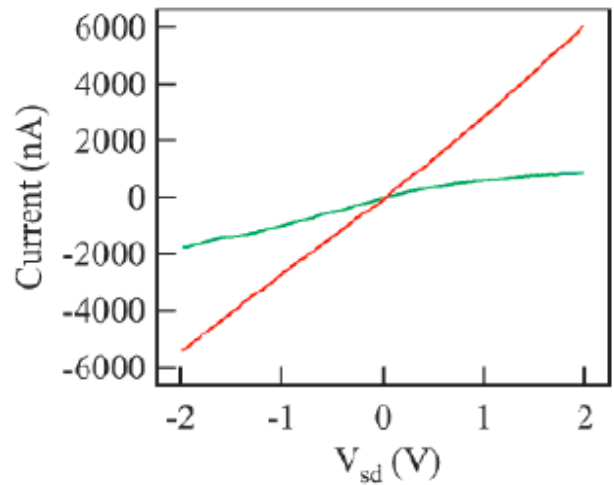

**Fig. 2.22** I$_d$-V$_d$ curve for the back gate silicon nanowire transistor. Red curve is after annealing

.

**Table 2.1** Nanowire Data Compared with Bulk Data for Silicon Nanowire MOSFETs [18]

| | nanowire raw data | nanowire coverted data | planar Si device |
|---|---|---|---|
| gate length (nm) | 800–2000 | 50 | 50 |
| gate oxide thickness (nm) | 600 | 1.5 | 1.5 |
| mobility (cm²/V s) | 230–1350 | 230–1350 | |
| $I_{on}$ ($\mu$A/$\mu$m) | 50–200 | 2000–5600 | 650 |
| $I_{off}$ (nA/$\mu$m) | 2–50 | 4–45 | 9 |
| subthreshold slope (mV/decade) | 174–609 | 60 | 70 |
| transconductance ($\mu$S/$\mu$m) | 17–100 | 2700–7500 | 650 |

Fig. 2.23 below shows a TEM image of a reported 200 nm long nanowire with 4 nm diameter and 9 nm thick oxide [14]. The fabrication process started with a p-type silicon-on-insulator (SOI) wafer. Active areas were etched out down to the buried oxide to form a silicon fin structure. The patterned silicon was then oxidized in dry $O_2$ which resulted in two nanowire cores, one at the bottom and another at the top of fin. The top nanowire was etched out and bottom one was released from the underlying oxide using a wet etch process. The release was followed by a 9 nm gate oxide and 130 nm α-Si deposition to form the gate dielectric and polysilicon gate electrode.

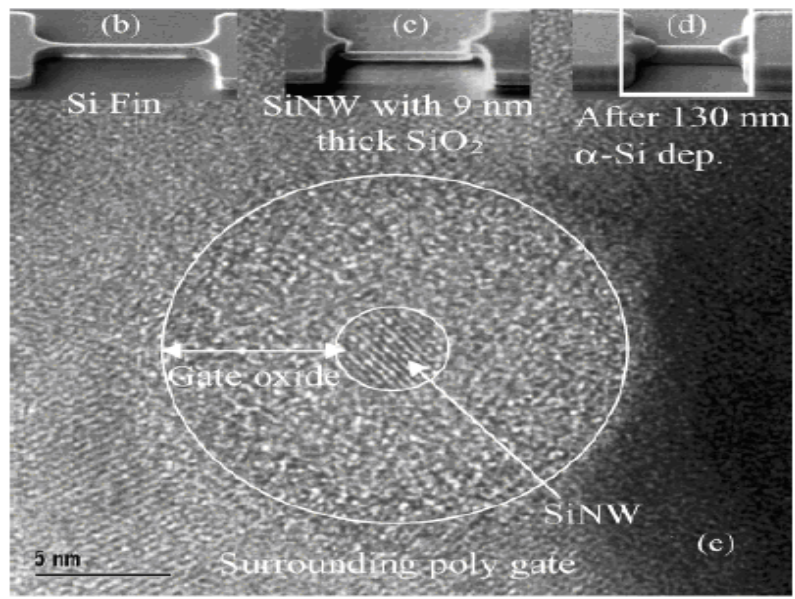

**Fig. 2.23** TEM image of fabricated GAA silicon nanowire transistor structure

The $I_d$-$V_d$ and $I_d$-$V_g$ curves for the device are shown in Fig. 2.24 for fabricated 5 nm diameter and 180 nm channel length SiNW FET shows ON-state currents of 1.5 and 1 mA/µm for n- and p-FETs, respectively, and OFF-state current < 1nA/µm at 1.2 V of operating voltage. The electron and hole mobilities were estimated to be ~ 750 cm$^2$/V.s and ~ 325 cm$^2$/V.s, respectively, for holes and electrons, which are lower than other reported nanowire results [29]. However, the subthreshold characteristics of n-FET were nearly ideal with SS ~ 63 mV/dec (66 mV/dec for p-FET) and the DIBL were also very good, ~ 10 mV/V (20mV/V for p-FET) even with a 9 nm thick gate oxide [14].

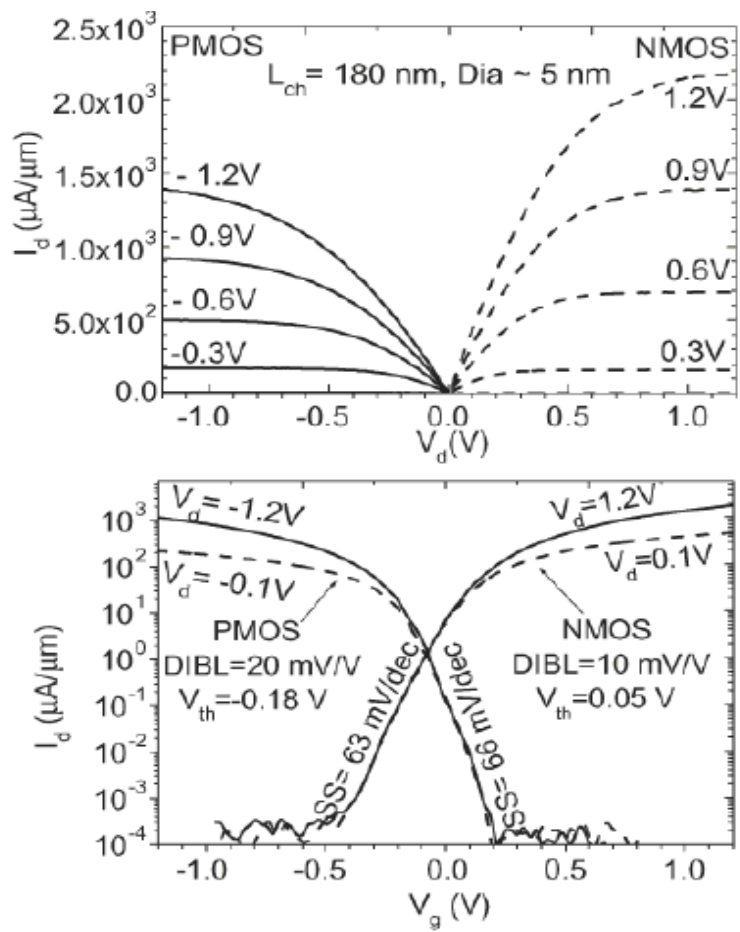

**Fig. 2.24** $I_d$-$V_d$ and $I_d$-$V_g$ curves for a 5 nm diameter GAA silicon nanowire transistor

# 3. Simulation of Silicon Nanowires

The work done till date has been presented in this section. An intrinsic silicon nanowire and an intrinsic germanium nanowire have been simulated using QuantumWise ATK version 12.8.2 by doping the nanowires in different ways to realize Zener diodes, PIN, PIP, NIN diodes respectively. The diameter of the nanowires simulated is taken as 3 nm.

## 3.1. Silicon Based Systems

In this section, all the simulations presented are based on silicon nanowire. Firstly, a silicon nanowire of the above mentioned diameter and geometry has been built using Virtual NanoLab tool present as part of QuantumWise ATK. The total number of atoms comprising the silicon nanowire is 56. The lattice constant of silicon is taken as 5.43 angstroms.

### 3.1.1 Simulation of Silicon Nanowire

The nanowire built in the previous section is simulated by applying a voltage sweeping from -1 V to 1 V at the right end of the nanowire while grounding the left end of the silicon nanowire to obtain the I-V curve, as shown in Fig. 3.1.

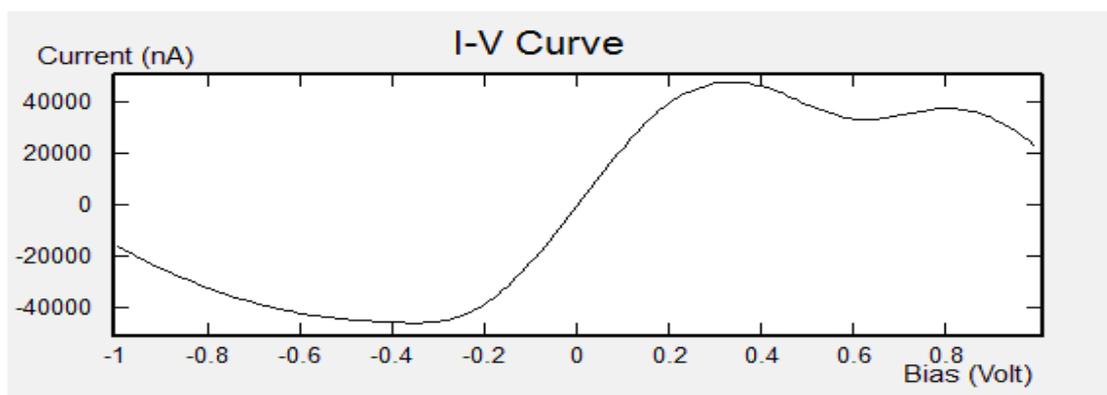

**Fig. 3.1** Electrical characteristics of Intrinsic Silicon Nanowire

The electrical characteristics shown in the figure above has been obtained for an undoped/intrinsic silicon nanowire, where the maximum value of current reported is of the order of 4000 nA. Negative resistance regions can be seen, thus enabling to use the device for microwave applications.

## 3.1.2 Uniform Doping of Silicon Nanowire to obtain PN Diode

The intrinsic silicon nanowire is doped atom by atom to investigate the effect of gradual doping on the electrical characteristics taking the characteristics of Fig. 3.1 as reference.    To ensure regularity in dopant distribution, the nanowire is doped in a way that after every two silicon atoms, there is a dopant atom sitting on the lattice. For N-doping, phosphorus is used as dopant while for P-doping, boron is used as the dopant.

To start with doped nanowire simulation, one atom of boron and one atom of phosphorus have been incorporated in the nanowire. Once doped, the nanowire is simulated to obtain the electrical characteristics. This is followed by the incorporation of two, three, four, five, six, seven, eight and nine dopant atoms, each of boron and phosphorus to obtain the electrical characteristics. The curves obtained as shown together in Fig. 3.2.  Thus, a P-N diode configuration has been obtained where both the P and N regions are heavily doped. This depicts the configuration of a Zener diode for which the electrical characteristics are shown in Fig. 3.2 (i). As shown in the figures below, there is a variation in the electrical characteristics with the incorporation of every additional dopant atom of boron and phosphorus. Negative resistance regions are evident in each of the nine plots. The Fig. 3.2 (j) shows a consolidated plot characterizing the Zener diode. It can be seen that the addition of dopants on either side of the nanowire progressively shifts the characteristics towards right, thereby making the curves flatter.

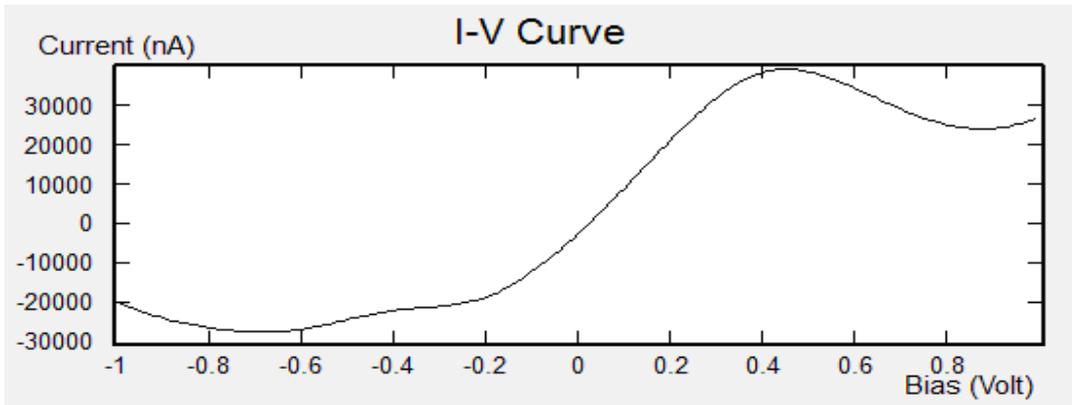

(a) Silicon nanowire with one boron and one phosphorus dopant atom

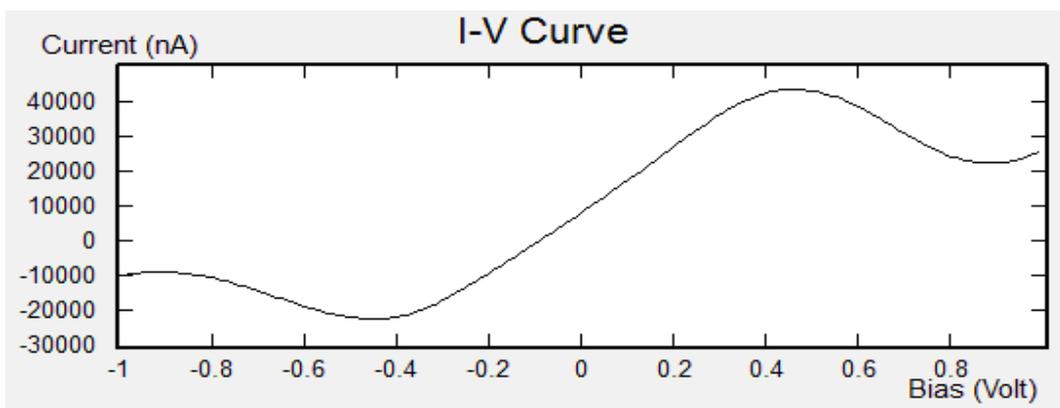

(b) Silicon nanowire with two boron and two phosphorus dopant atoms

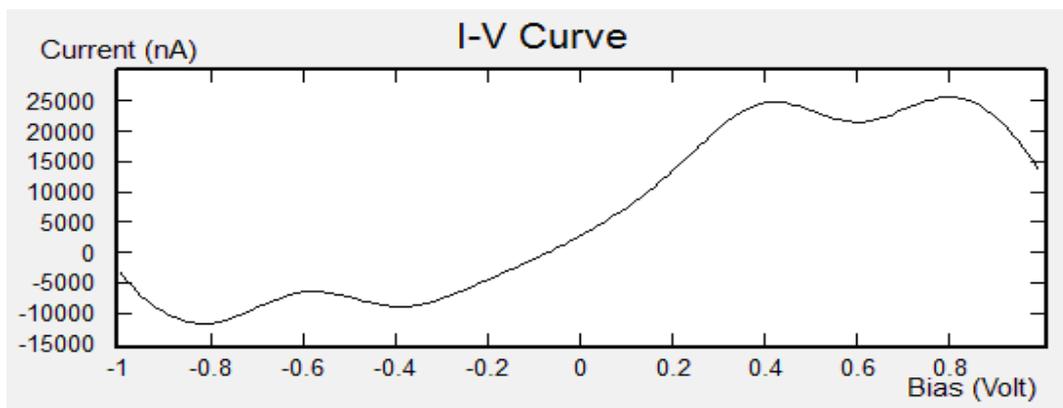

(c) Silicon nanowire with three boron and three phosphorus dopant atoms

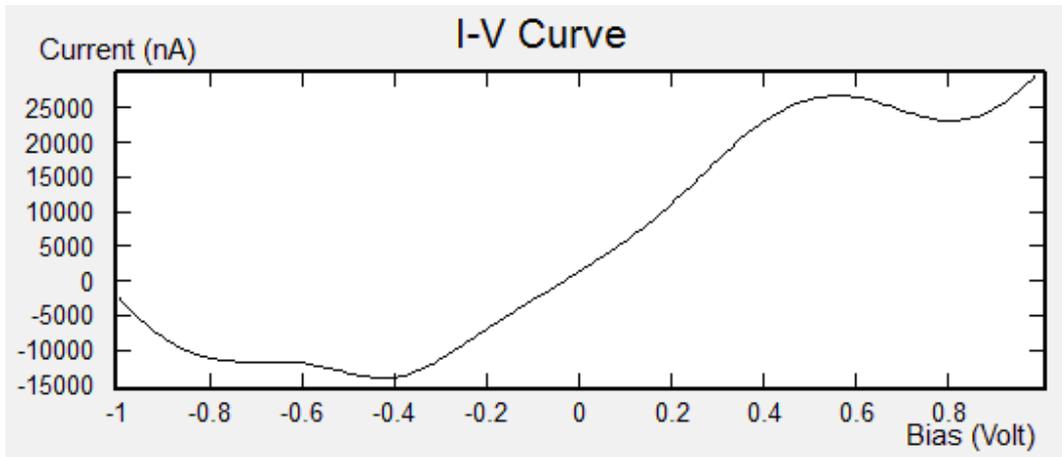

(d) Silicon nanowire with four boron and four phosphorus dopant atoms

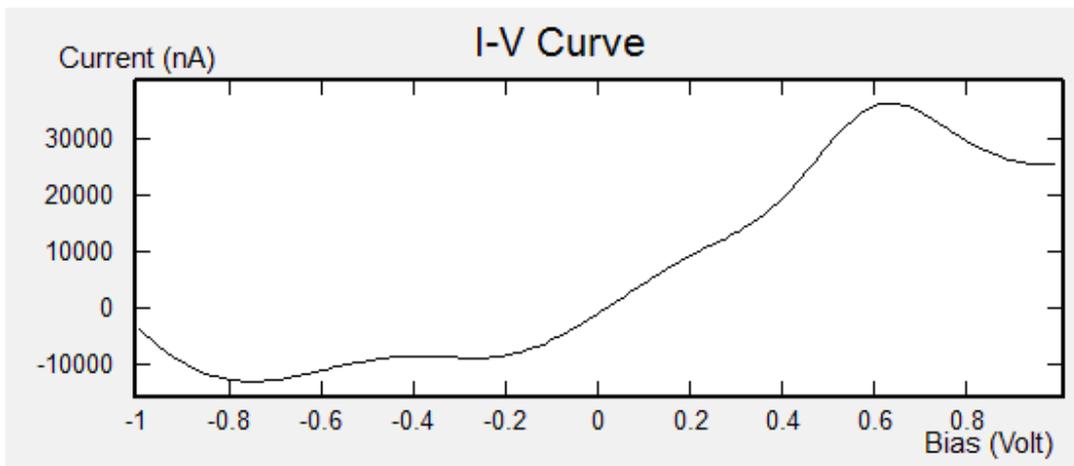

(e) Silicon nanowire with five boron and five phosphorus dopant atoms

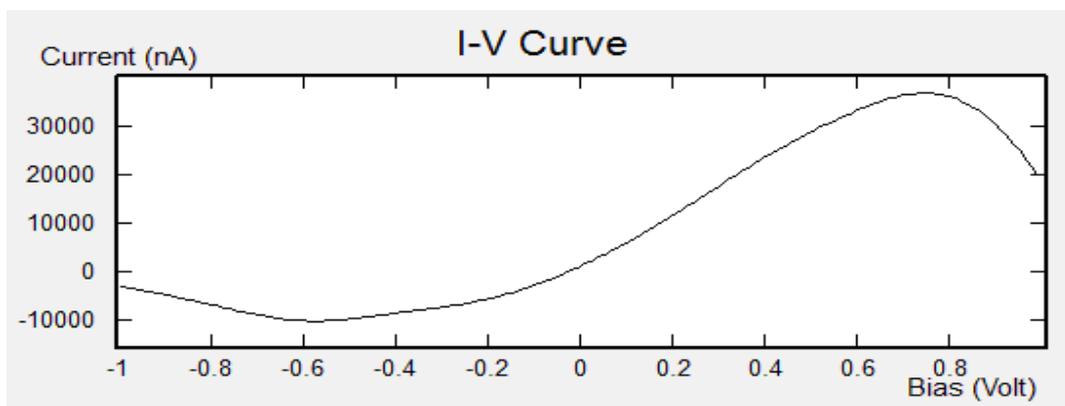

(f) Silicon nanowire with six boron and six phosphorus dopant atoms

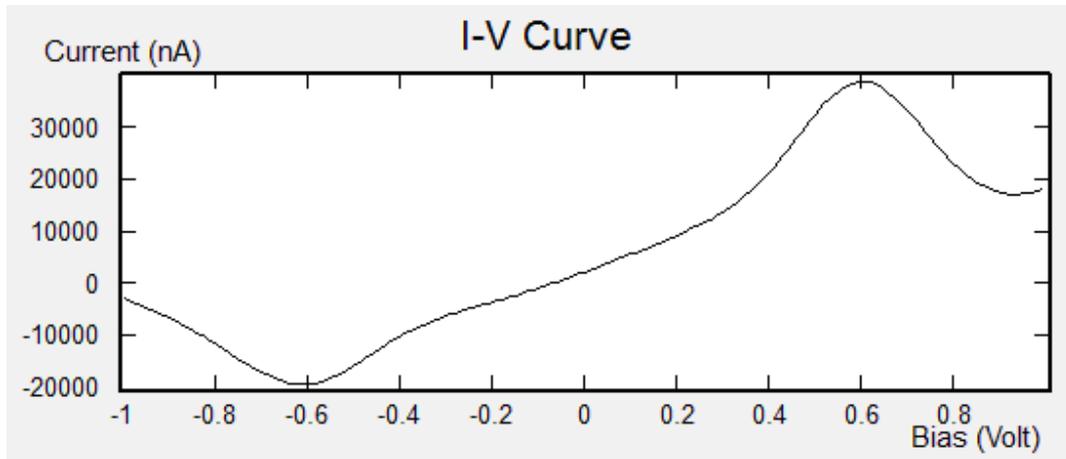

(g) Silicon nanowire with seven boron and seven phosphorus dopant atoms

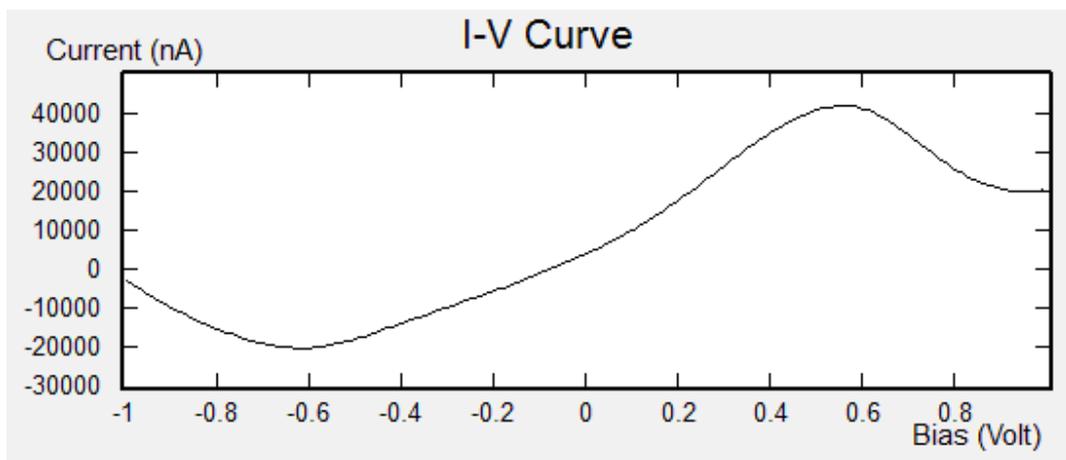

(h) Silicon nanowire with eight boron and eight phosphorus dopant atoms

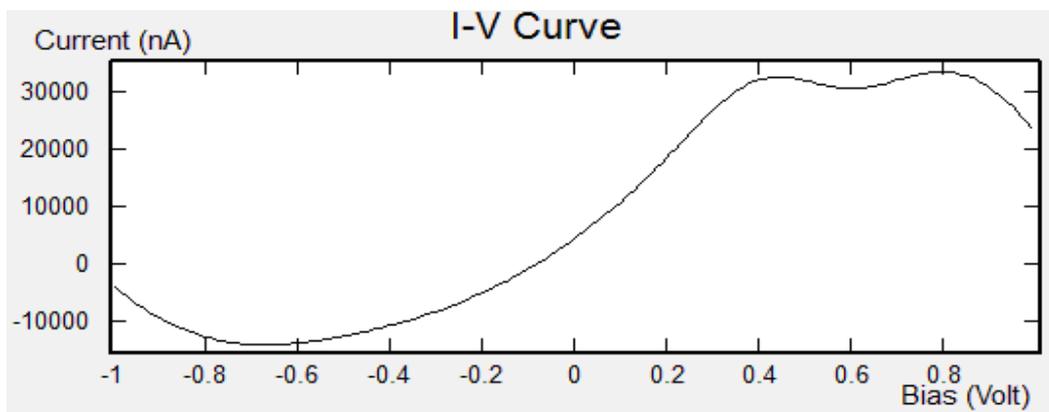

(i) Silicon nanowire with nine boron and nine phosphorus dopant atoms

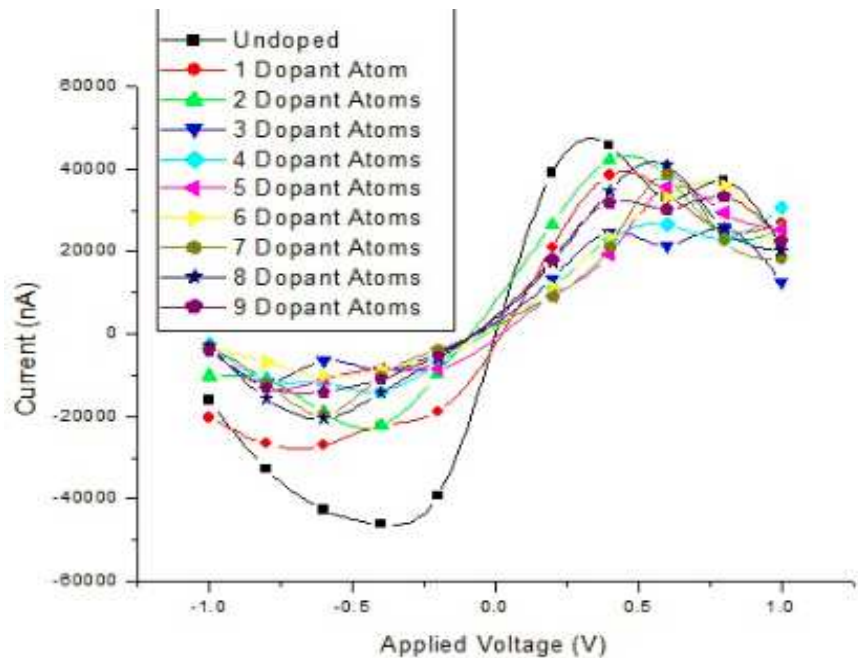

(j) Consolidated Plot

**Fig. 3.2** Electrical Characteristics of Doped Silicon Nanowire

Looking at the consolidated plot above, it is evident that if the device is simulated for a bias voltage greater than 1 V, the breakdown regions can be seen. The conductance of the nanowire modeled as a Zener diode has been computed and plotted against the applied voltage to obtain the plot as seen in Figure 3.3. Though not very high, but a finite conductance can be seen at zero bias condition as the nanowire is very small in terms of diameter and there is no insulating layer in between the P- and N-regions to obstruct the zero bias conduction of carriers. There are many curves that cross each other as the applied voltage exceeds 1.5 V.

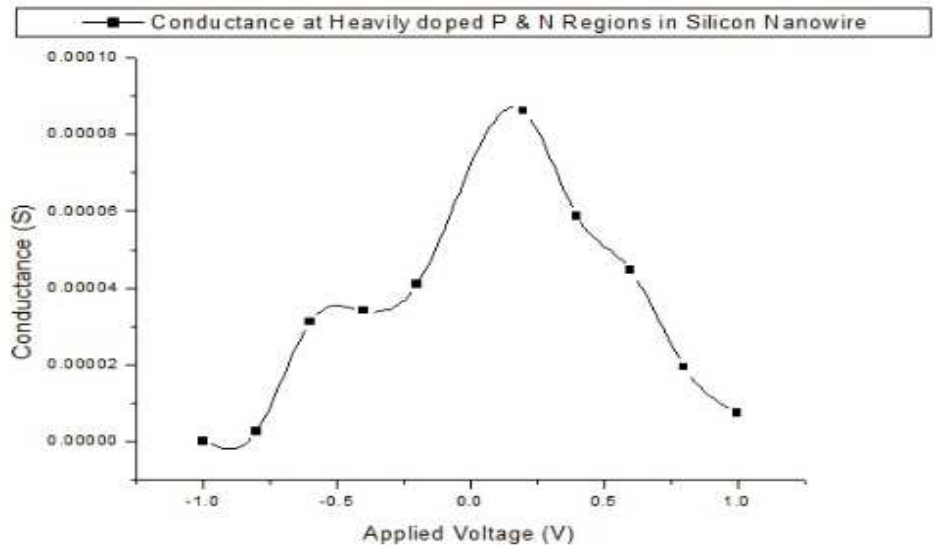

**Fig. 3.3** Variation of conductance with applied voltage

### 3.1.3 Uniform Doping of Silicon Nanowire to obtain PIN Diode

The silicon nanowire has been doped in such a way that a PIN diode is modeled as part of the nanowire. The left and right ends of the nanowire have been doped with boron and phosphorus atoms respectively in a way that there exists a dopant atom (boron or phosphorus) after every two silicon atoms while the central region is left intrinsic/undoped. The doping is done gradually, starting from incorporating one dopant atom of either type at each side of the nanowire, simulating the nanowire, followed by progressively increasing the dopant concentration upto six dopant atoms, atom by atom, on either side of the nanowire with subsequent simulations thereby keeping the central region intrinsic. Once the doping is done to realize a PIN diode, the built structure is biased by sweeping the right end of the structure from -1 V to 1 V while grounding the other end of the nanowire. DFT

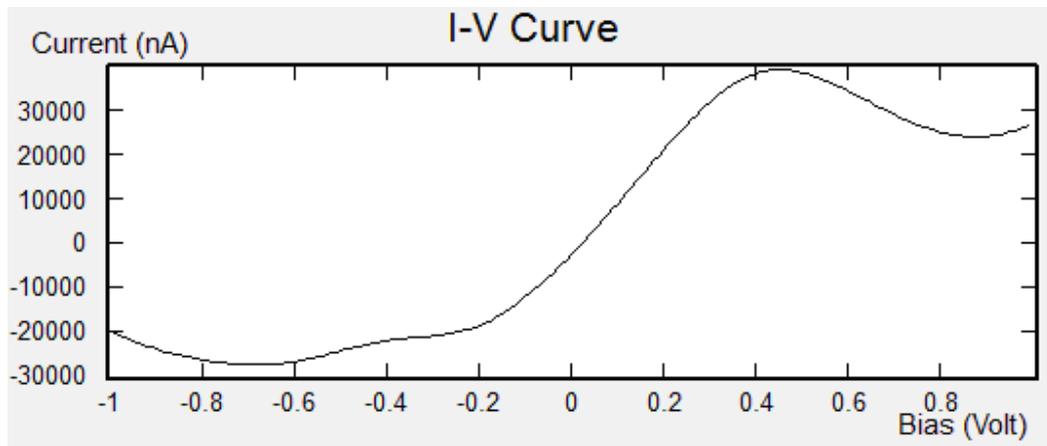

(a) Silicon nanowire with one boron and one phosphorus dopant atom

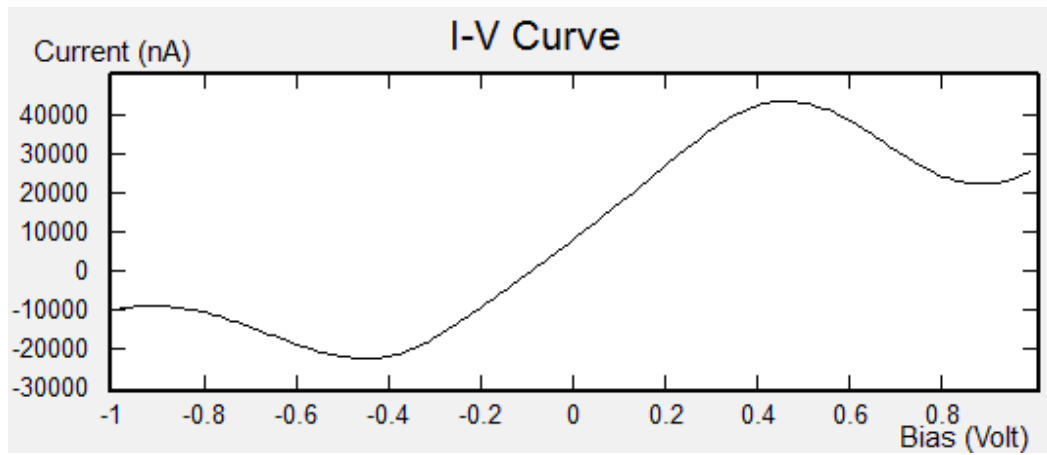

(b) Silicon nanowire with two boron and two phosphorus dopant atoms

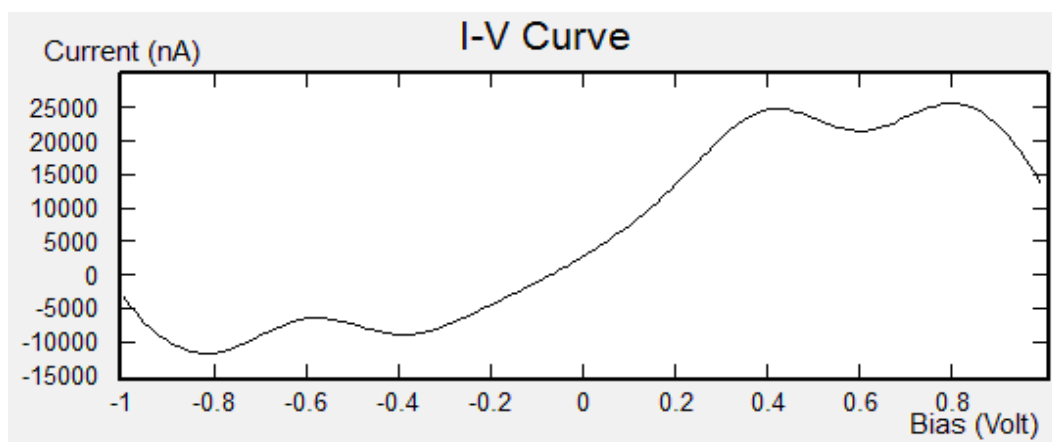

(c) Silicon nanowire with three boron and three phosphorus dopant atoms

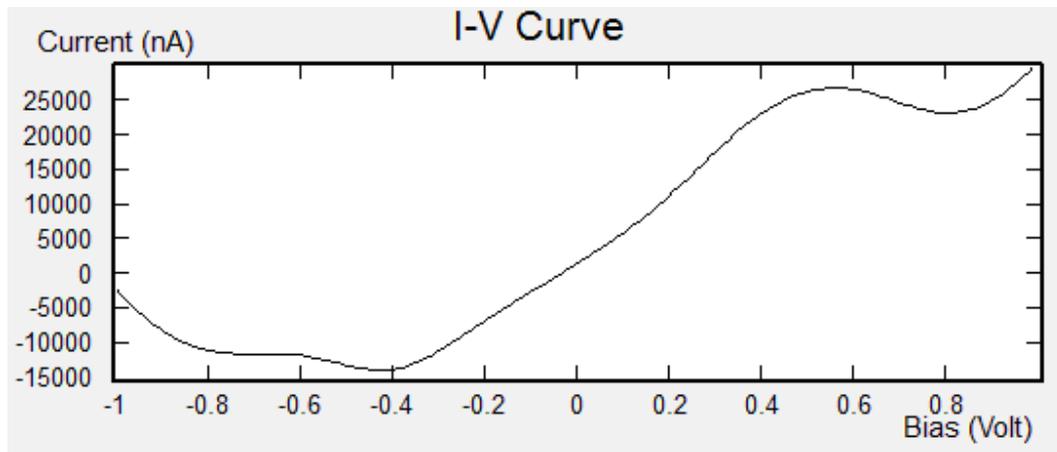

(d) Silicon nanowire with four boron and four phosphorus dopant atoms

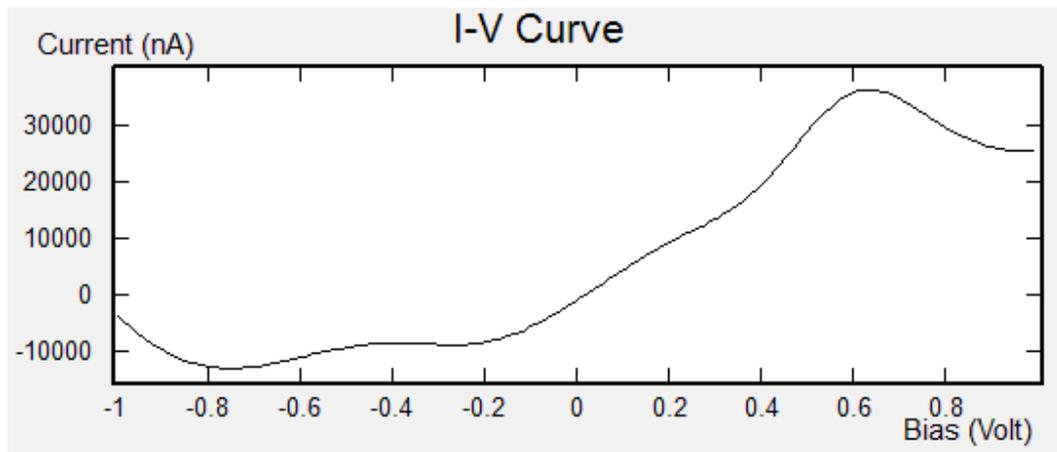

(e) Silicon nanowire with five boron and five phosphorus dopant atoms

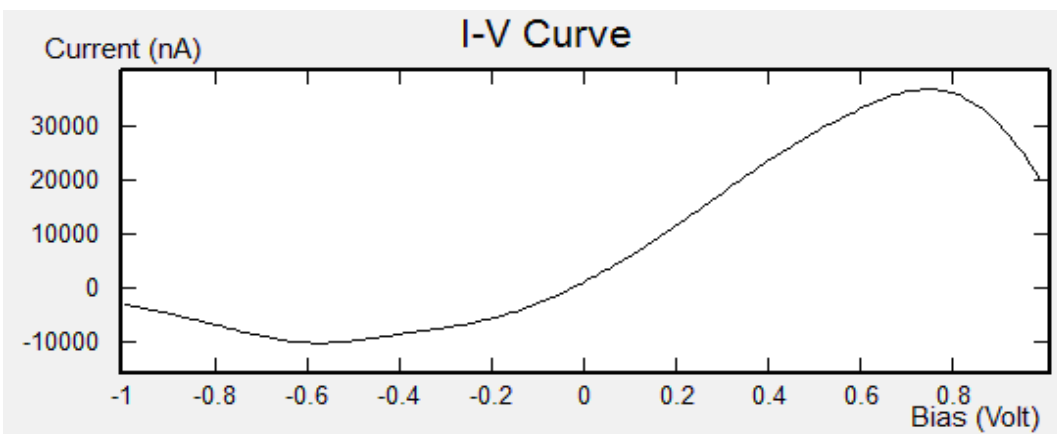

(f) Silicon nanowire with six boron and six phosphorus dopant atoms

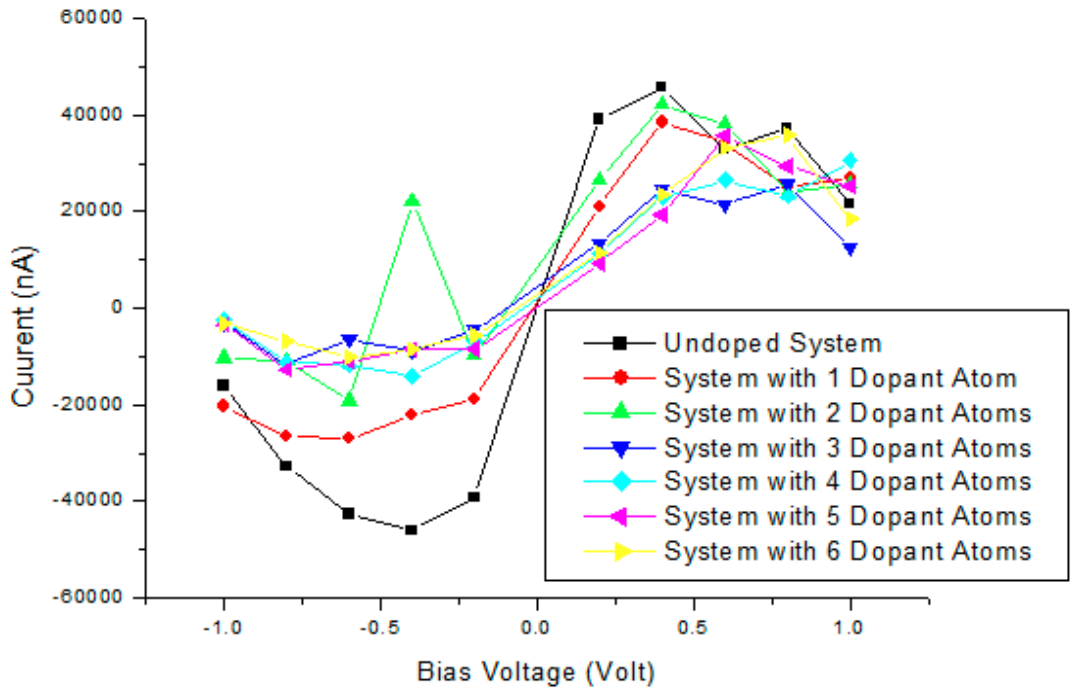

(g) Consolidated Plot

**Fig. 3.4** Electrical Characteristics of Silicon Nanowire Doped in PIN Configuration

simulations based on NEGF formalism yield I-V curves for each of the six cases as depicted in Fig. 3.4. The Fig. 3.4 (g) depicts the consolidated plot.

## 3.1.4 Uniform Doping of Silicon Nanowire to obtain PIP Diode

In this section, the silicon nanowire has been doped in such a way that a PIP diode is modeled as part of the nanowire. The left and right ends of the nanowire have been doped with boron atoms alone in a way that there exists a dopant atom (boron) after every two silicon atoms while the central region is left intrinsic/undoped. The doping is done gradually, starting from incorporating one dopant atom of boron on each side of the nanowire, simulating the nanowire, followed by progressively increasing the dopant concentration upto six dopant atoms, atom by atom, on either side of the nanowire with subsequent simulations thereby keeping the central region intrinsic. Once the doping is done to realize a

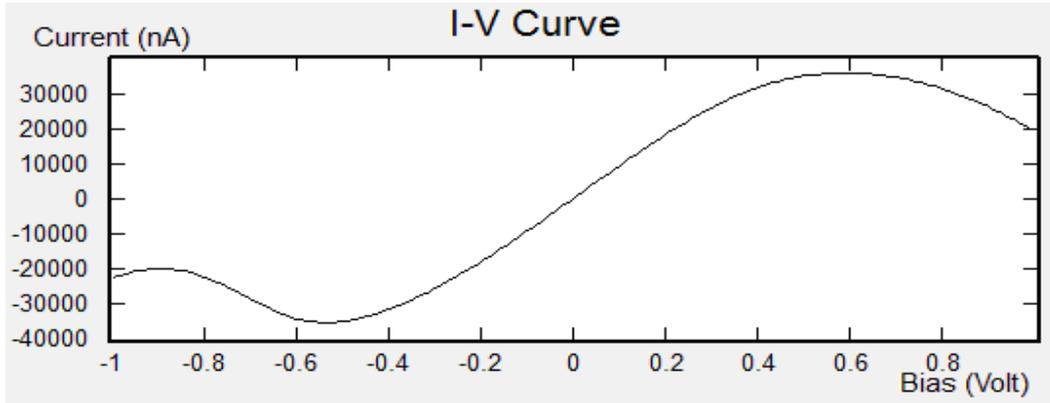

(a) Silicon nanowire with one boron dopant atom on either side of the intrinsic region

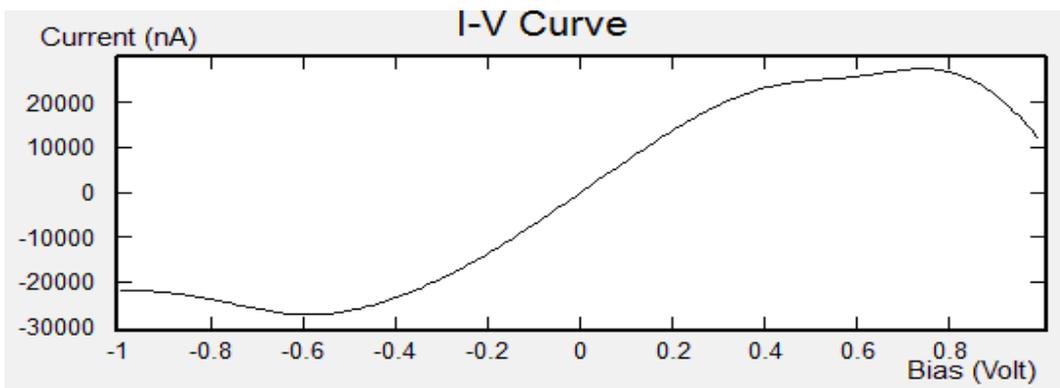

(b) Silicon nanowire with two boron dopant atoms on either side of the intrinsic region

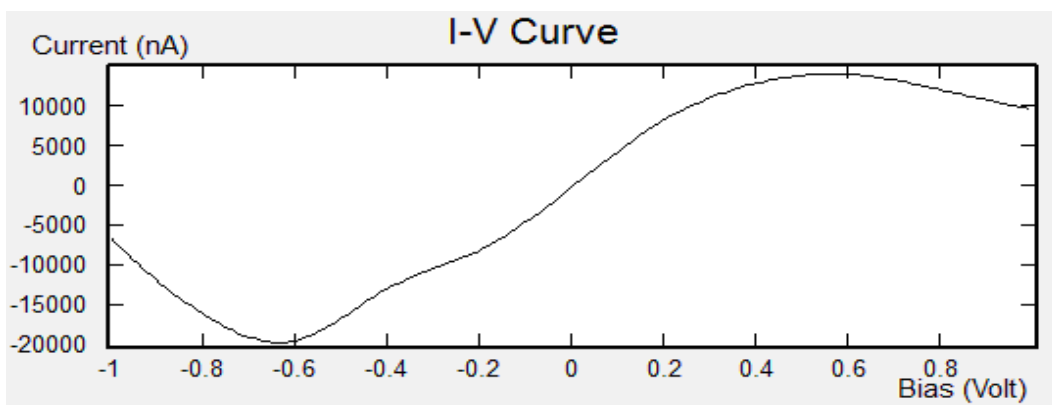

(c)  Silicon nanowire with three boron dopant atoms on either side of the intrinsic region

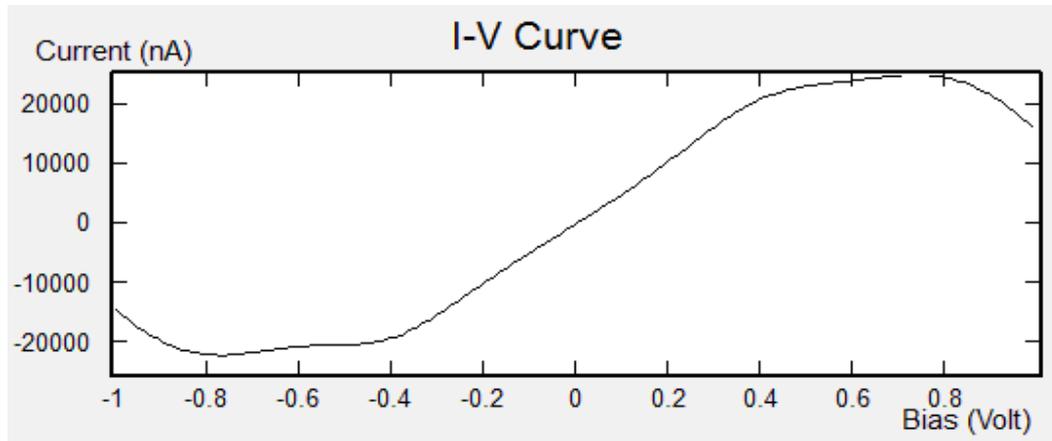

(d) Silicon nanowire with four boron dopant atoms on either side of the intrinsic region

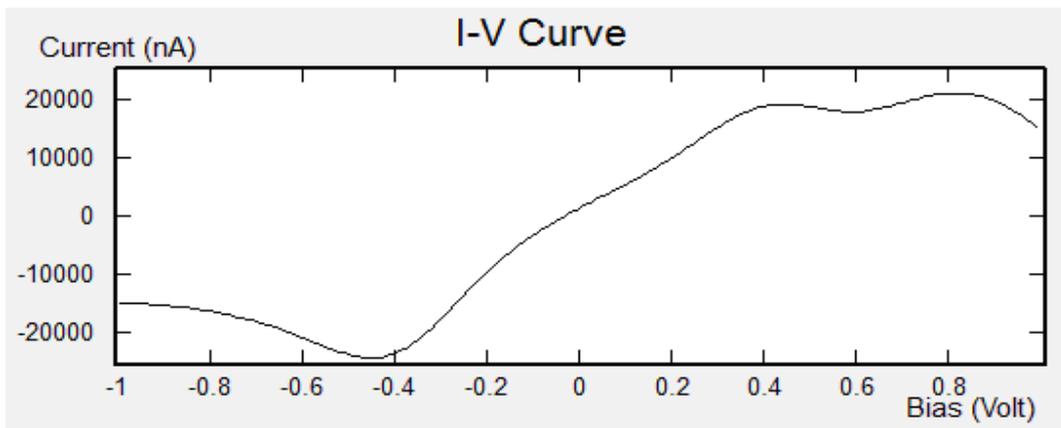

(e) Silicon nanowire with five boron dopant atoms on either side of the intrinsic region

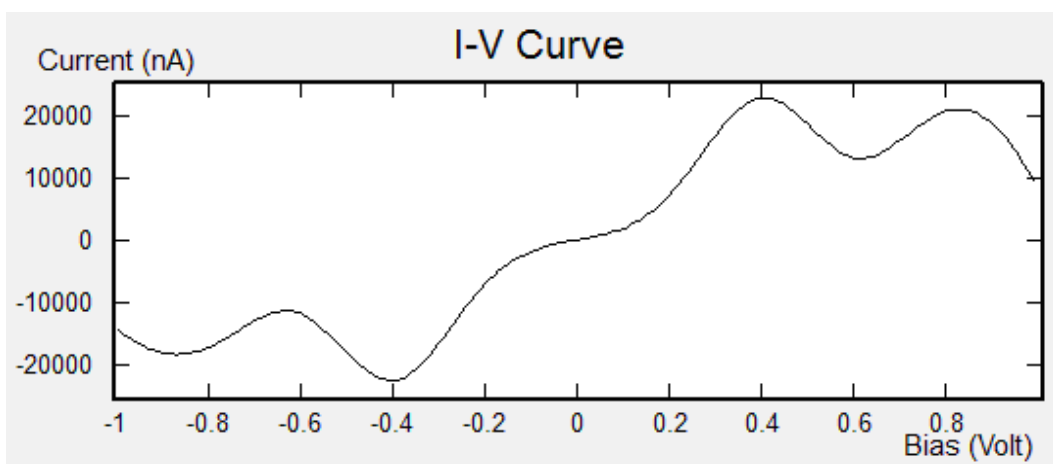

(f) Silicon nanowire with six boron dopant atoms on either side of the intrinsic region

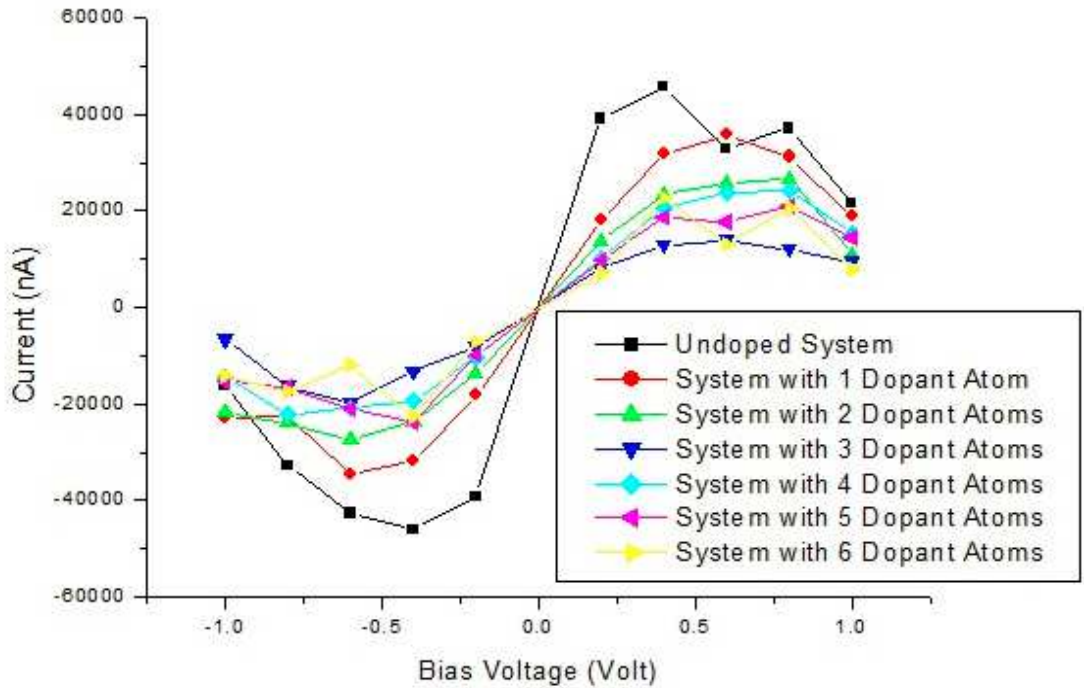

(g) Consolidated Plot

**Fig. 3.5** Electrical Characteristics of Silicon Nanowire Doped in PIP Configuration

PIP diode,the built structure is biased by sweeping the right end of the structure from -1 V to 1 V while grounding the other end of the nanowire. DFT simulations based on NEGF formalism yield I-V curves for each of the six cases as depicted in Fig. 3.5 along with all the I-V curves merged into a single plot to obtain the consolidated plot. As seen in each of the plots, negative resistance is evident, though the extent varies based on the dopant concentration. Therefore, this device can be used for microwave applications.

## 3.1.5 Uniform Doping of Silicon Nanowire to obtain NIN Diode

As done in the previous section, the nanowire is doped with the same type of dopant, the dopant being phosphorus in this case to N-dope both sides of the nanowire while keeping the central region undoped. Here, the silicon nanowire has been doped in such a way that a NIN diode is modeled as part of the nanowire. The left and right ends of the nanowire have been doped with phosphorus atoms alone in a way that there exists a dopant atom (phosphorus) after every two silicon atoms while the central region is left intrinsic/undoped. The doping is done gradually, starting from incorporating one dopant atom of phosphorus on each side of the nanowire, simulating the nanowire, followed by progressively increasing the dopant concentration upto six dopant atoms, atom by atom, on either side of the nanowire with subsequent simulations thereby keeping the central region intrinsic. Once the doping is done to realize a NIN diode, the built structure is biased by sweeping the right end of the structure from -1 V to 1 V while grounding the other end of the nanowire. DFT simulations based on NEGF formalism yield I-V curves for each of the six cases as depicted in Fig. 3.6. As seen in each of the plots, negative resistance is evident in each case, though the extent varies based on the dopant concentration. Therefore, this device can be used for microwave applications.

The I-V curves obtained by progressively adding one dopant each on either side of the nanowire for each of the diode configurations (PIN, PIP and NIN) might look almost the same including the three consolidated plots. But there is in fact marked variation in the values of currents in the individual plots obtained by progressive addition of dopants for each of the three diode configurations which might not be very evident looking at the individual plots as well as in the consolidated plots.

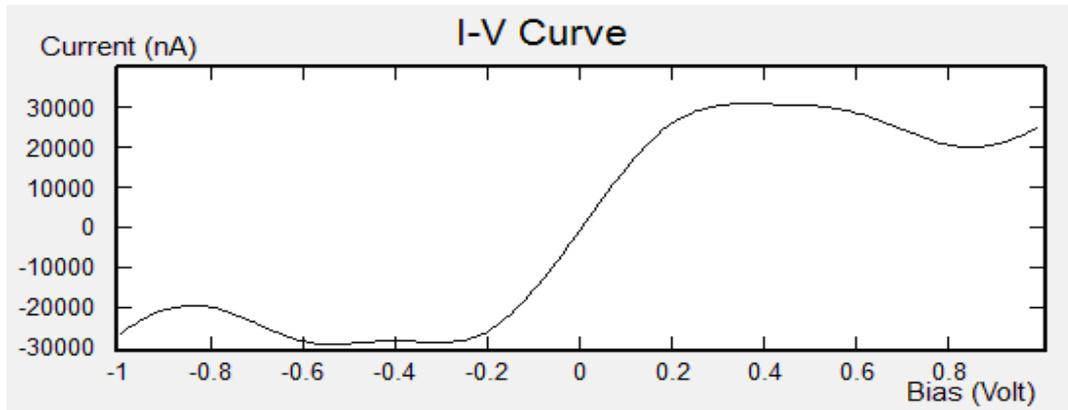

(a) Silicon nanowire with one phosphorus dopant atom on either side of the intrinsic region

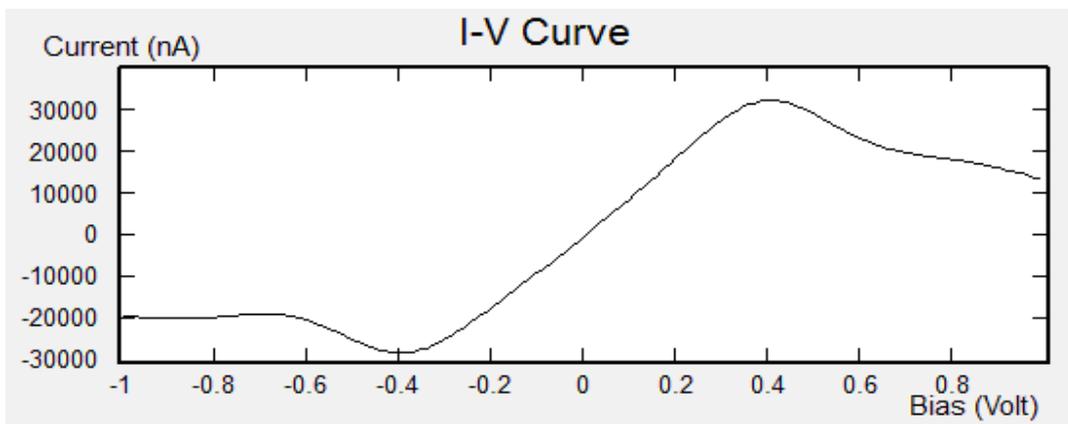

(b) Silicon nanowire with two phosphorus dopant atoms on either side of the intrinsic region

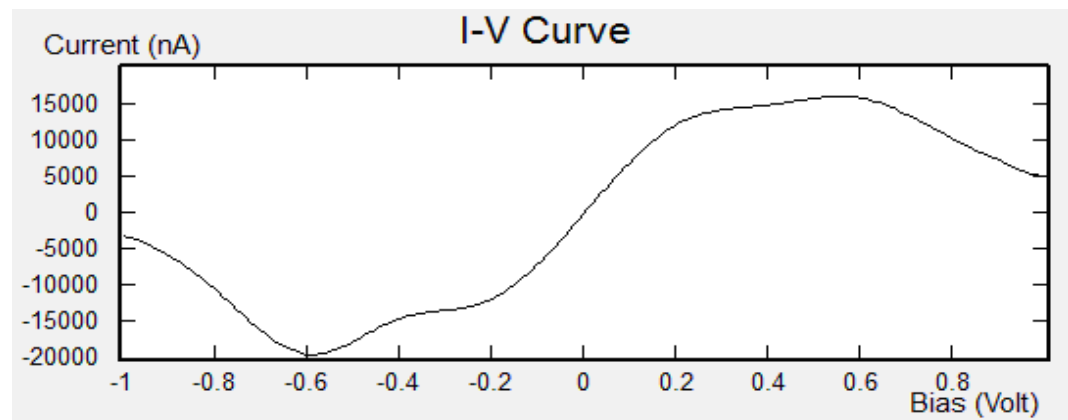

(c) Silicon nanowire with three phosphorus dopant atoms on either side of the intrinsic region

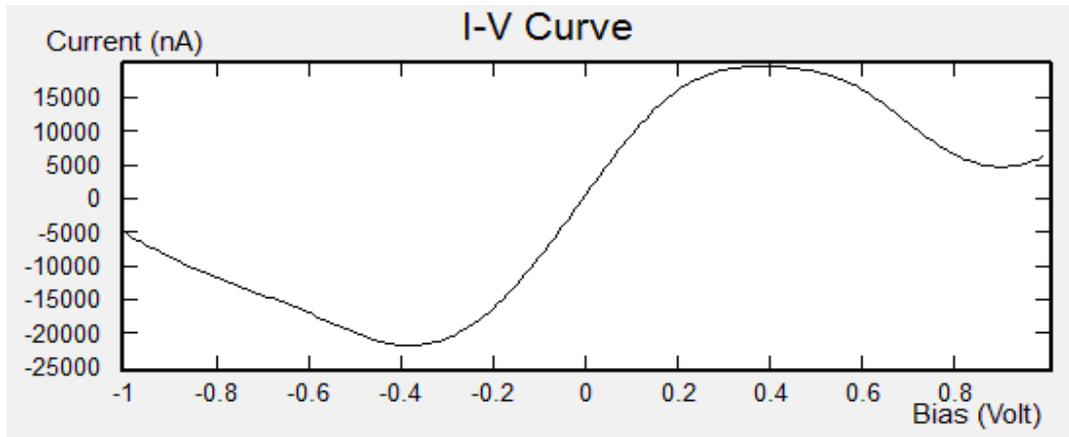

(d) Silicon nanowire with four phosphorus dopant atoms on either side of the intrinsic region

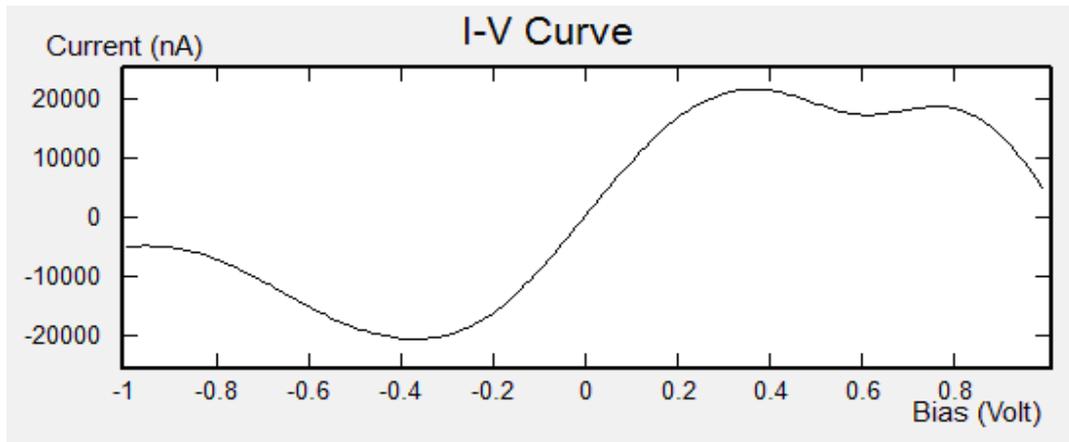

(e) Silicon nanowire with five phosphorus dopant atoms on either side of the intrinsic region

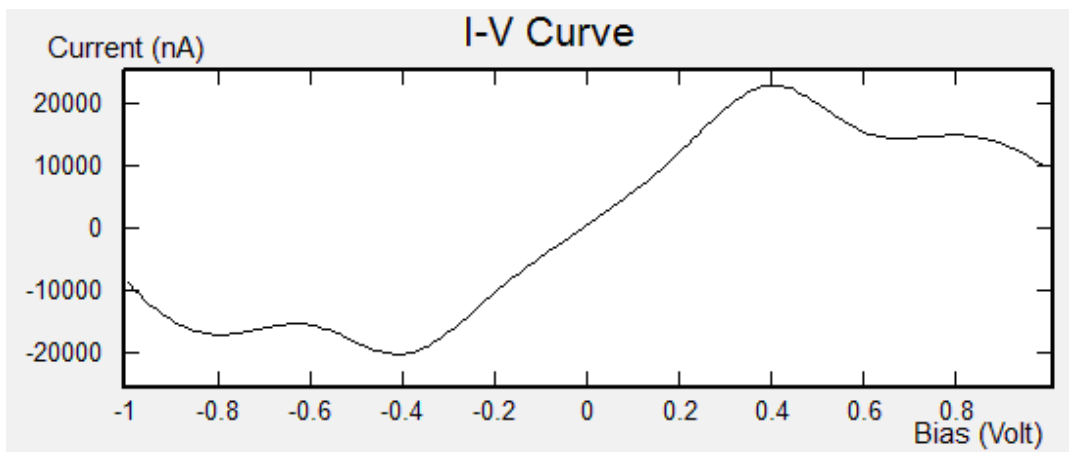

(f) Silicon nanowire with six phosphorus dopant atoms on either side of the intrinsic region

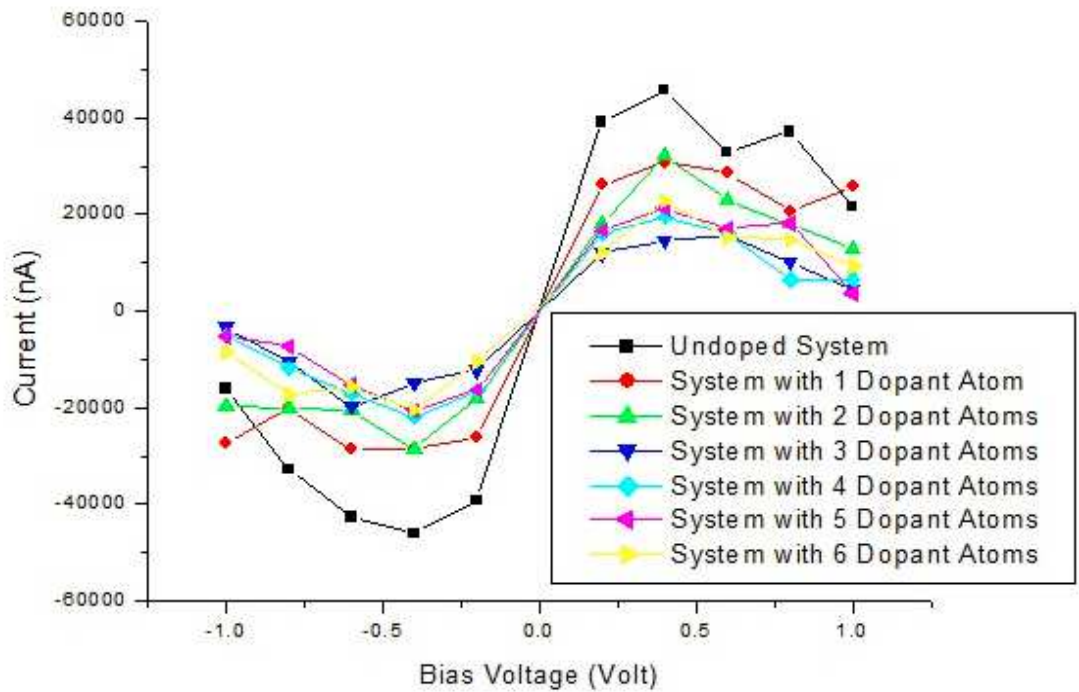

(g) Consolidated Plot

**Fig. 3.6** Electrical Characteristics of Silicon Nanowire Doped in NIN Configuration

## 3.2 Germanium Based Systems

In this section, all the simulations presented are based on germanium nanowire of the same dimension as that of silicon nanowire used in the pervious sections. Firstly, a germanium nanowire of an appropriate geometry has been built using Virtual NanoLab tool present as part of QuantumWise ATK. The total number of atoms comprising the germanium nanowire is again 56. The lattice constant of germanium is taken as 5.65.

### 3.2.1 Simulation of Germanium Nanowire

The germanium nanowire built in the previous section is simulated by applying a voltage sweeping from -1 V to 1 V at the right end of the nanowire while grounding the left end of the silicon nanowire to obtain the I-V curve, as shown in the figure below.

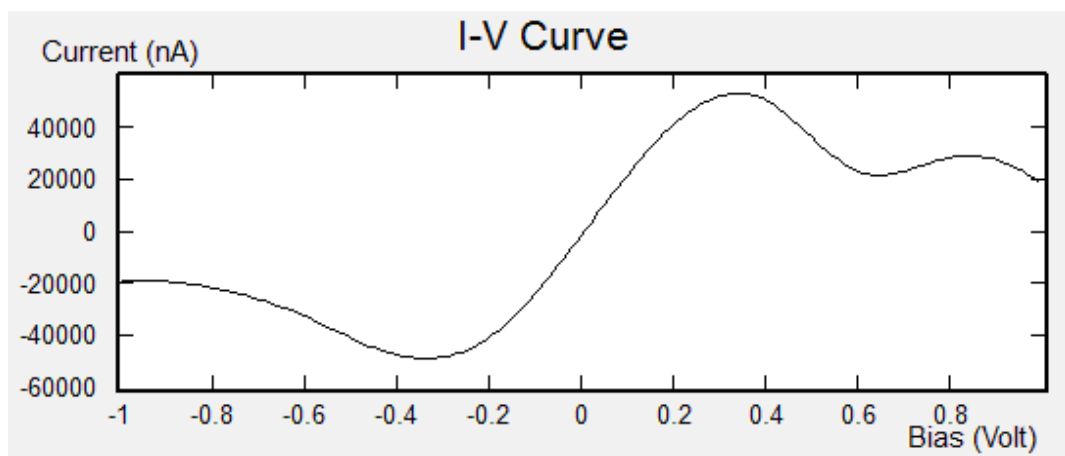

**Fig. 3.7** Electrical characteristics of Germanium Nanowire

### 3.2.2 Uniform Doping of Germanium Nanowire to obtain PN Diode

As done in case of silicon nanowire, the intrinsic germanium nanowire is doped, atom by atom, to investigate the effect of gradual doping on the electrical characteristics taking the characteristics of Fig. 3.7 as reference. To ensure regularity in dopant distribution, the nanowire is doped in a way that after every two germanium atoms, there is a dopant atom sitting on the lattice. For N-doping, phosphorus is used as dopant while for P-doping, boron is used as the dopant, though other N and P-type dopants can also be used.

To start with doped nanowire simulation, one atom of boron and one atom of phosphorus have been incorporated in the nanowire. Once doped, the nanowire is simulated to obtain the electrical characteristics. This is followed by the incorporation of two, three, four, five, six, seven, eight and nine dopant atoms, each of boron and phosphorus to obtain the electrical characteristics. The curves obtained as shown together in Fig. 3.8. Thus, a P-N diode configuration has been obtained where both the P and N regions are heavily doped. This depicts the configuration of a Zener diode for which the electrical characteristics are shown in Fig. 3.8 (i) with the consolidated plot in Fig.3.8 (j). As shown in the figures below, there is a variation in the electrical characteristics with the incorporation of every additional dopant atom of boron and phosphorus. Negative resistance regions are evident in each of the nine plots. But the current levels vary with the addition of dopant. The consolidated plot obtained by simulating the silicon and germanium nanowire based Zener diodes can be compared to observe the variation in the features of both the two plots. Breakdown regions can be seen of the Germanium nanowire is simulated for bias voltages greater than 1 V. As compared to the consolidated plot of Silicon nanowire based Zener diode, the crossing between the different curves is more prominent and more in number in case of Germanium nanowire based Zener diode.

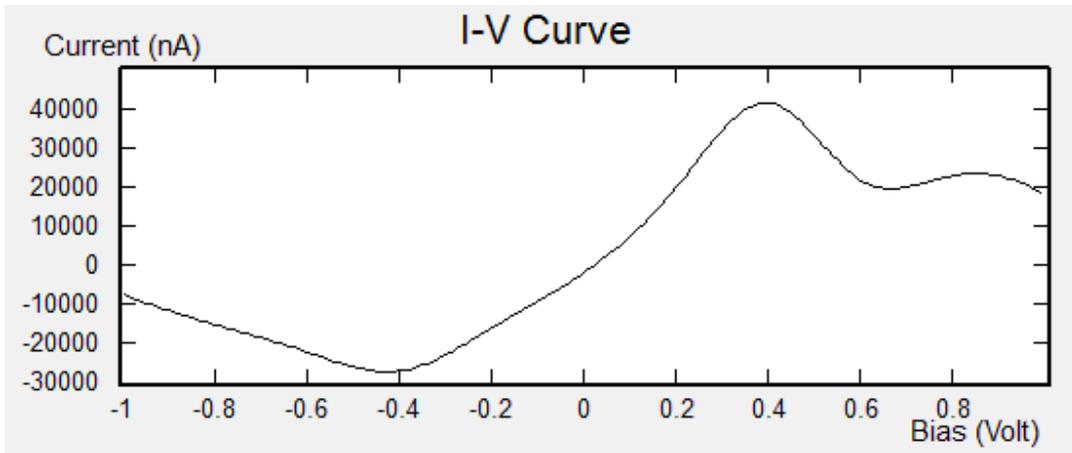

(a) Germanium nanowire with one boron and one phosphorus dopant atom

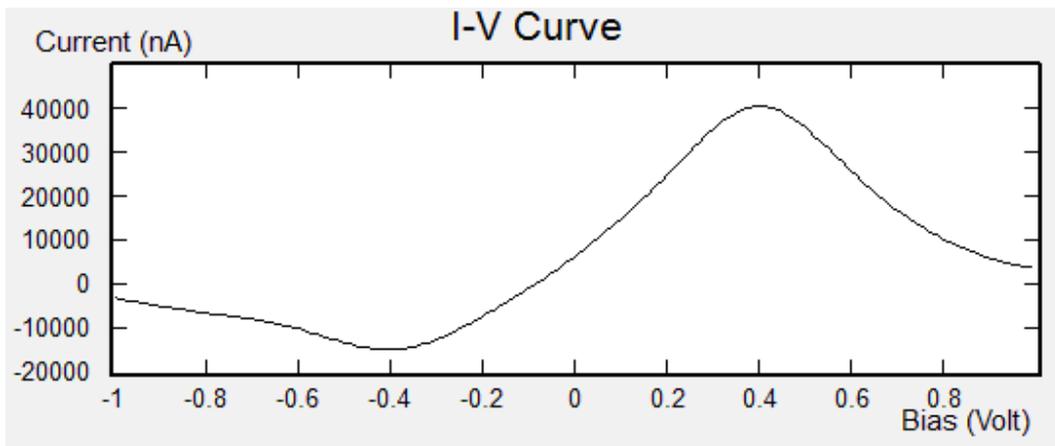

(b) Germanium nanowire with two boron and two phosphorus dopant atoms

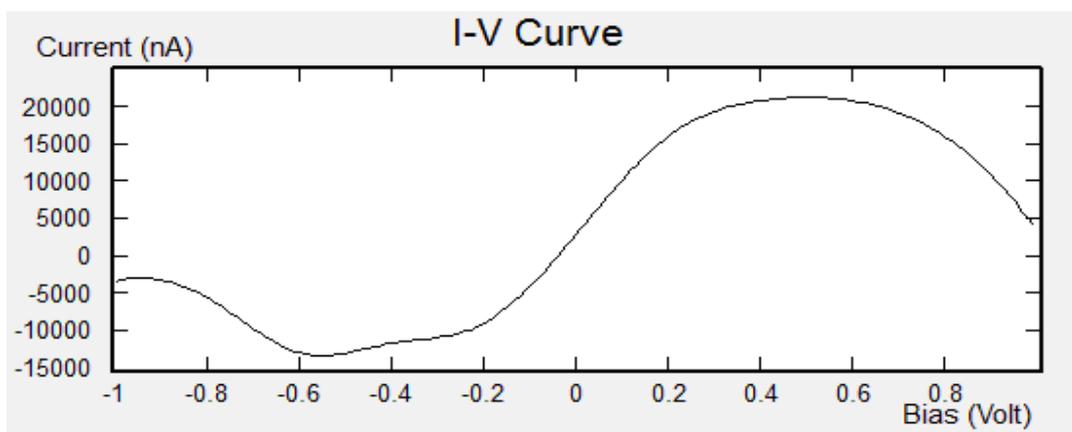

(c) Germanium nanowire with three boron and three phosphorus dopant atoms

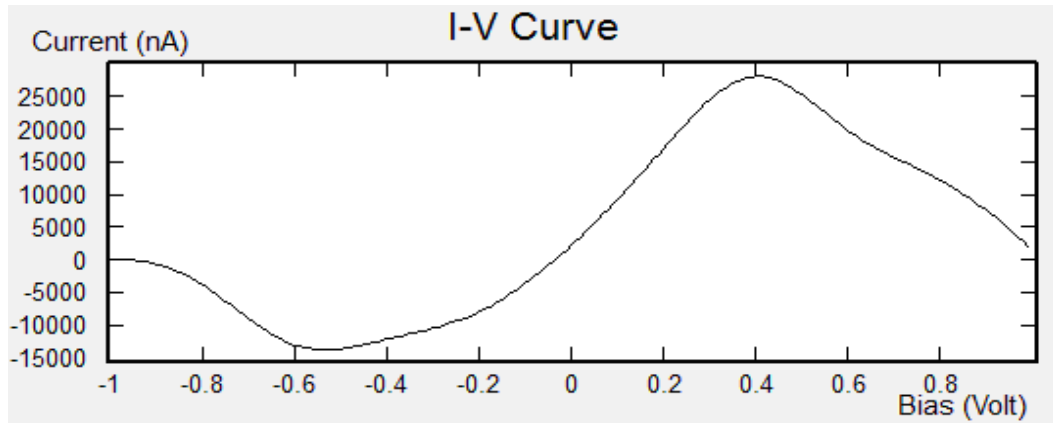

(d) Germanium nanowire with four boron and four phosphorus dopant atoms

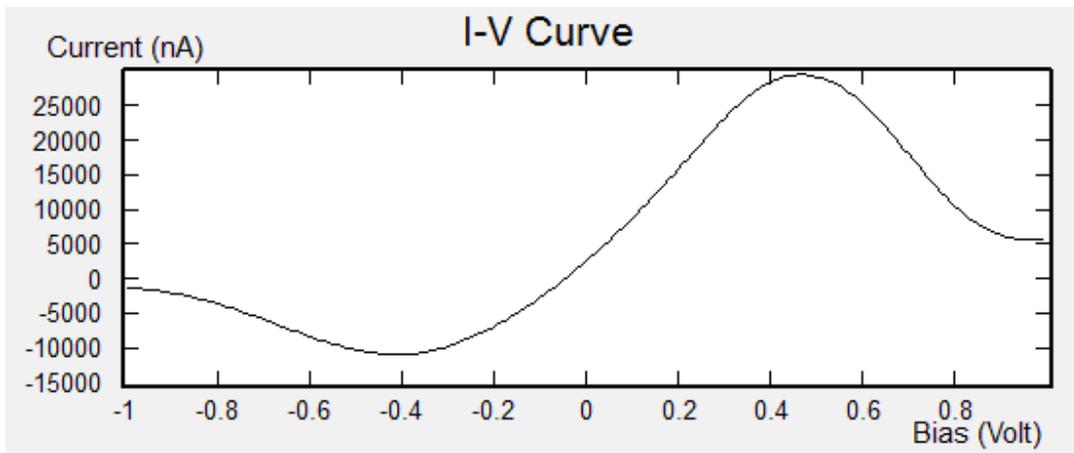

(e) Germanium nanowire with five boron and five phosphorus dopant atoms

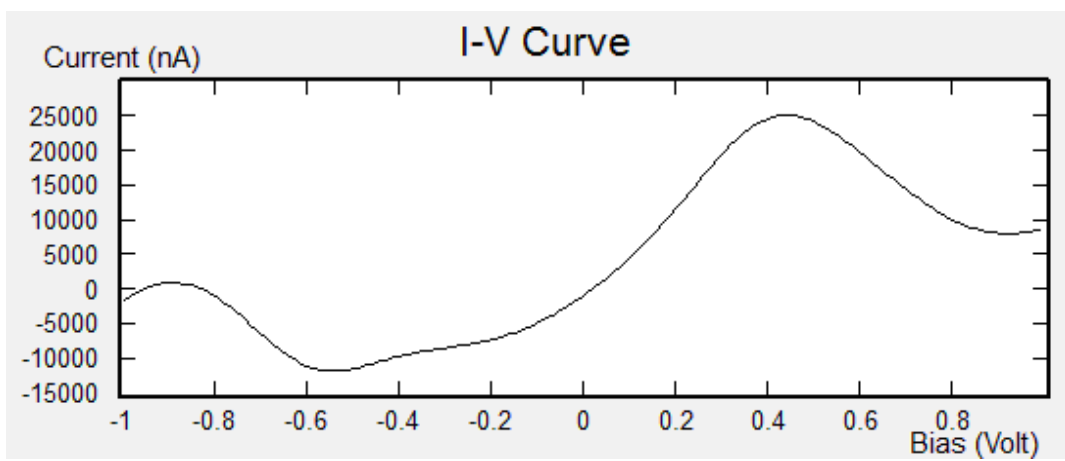

(f) Germanium nanowire with six boron and six phosphorus dopant atoms

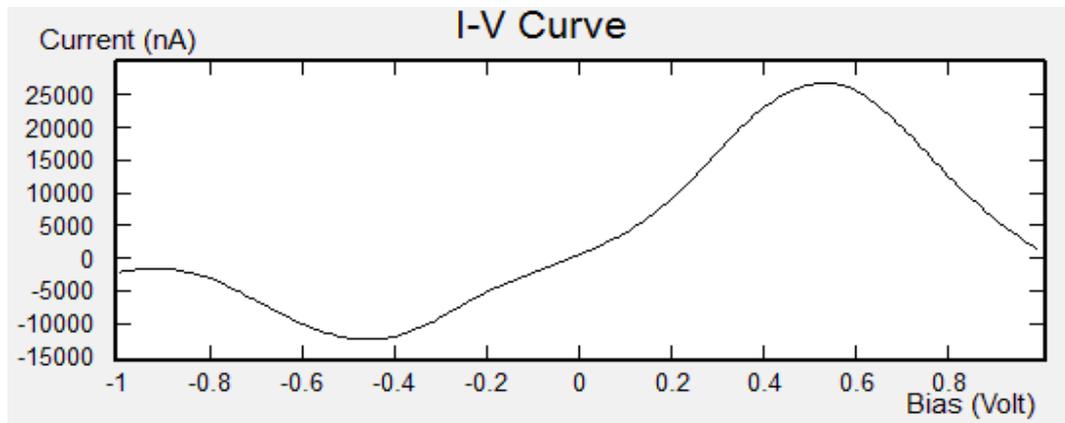

(g) Germanium nanowire with seven boron and seven phosphorus dopant atoms

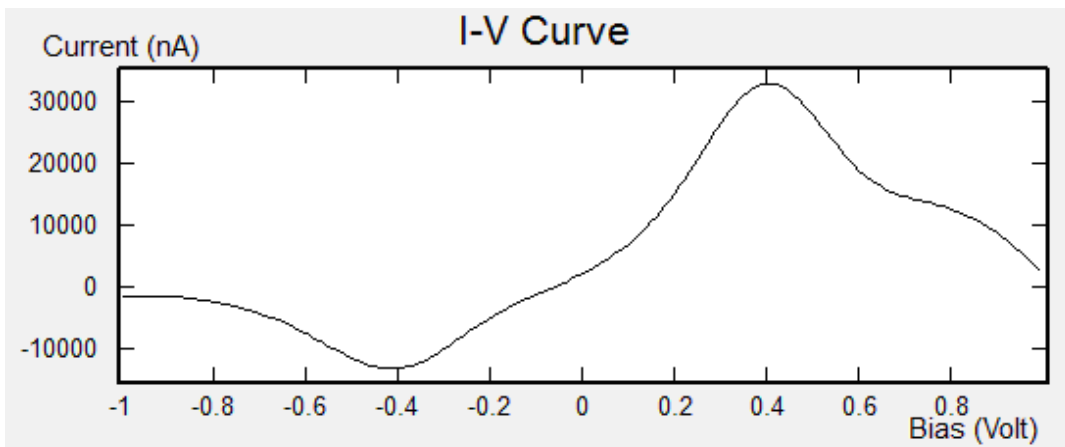

(h) Germanium nanowire with eight boron and eight phosphorus dopant atoms

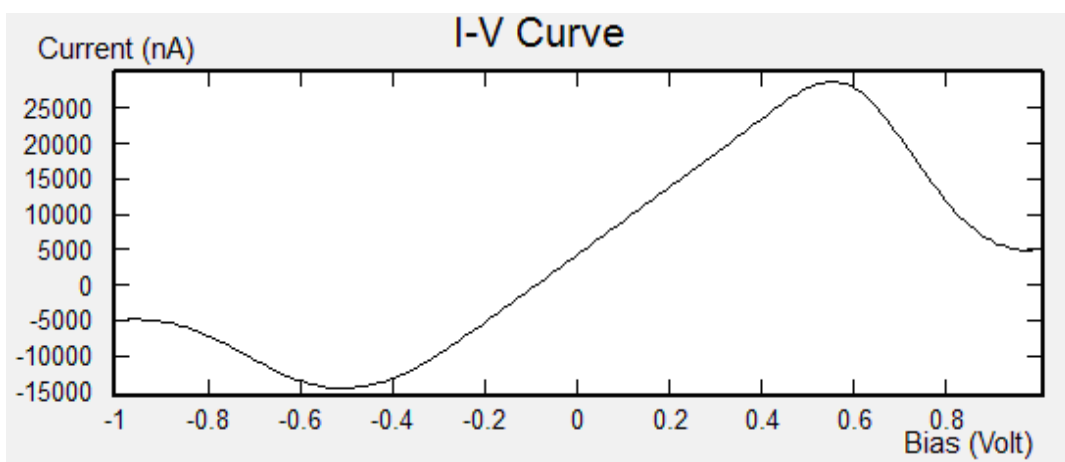

(i)    Germanium nanowire with nine boron and nine phosphorus dopant atoms

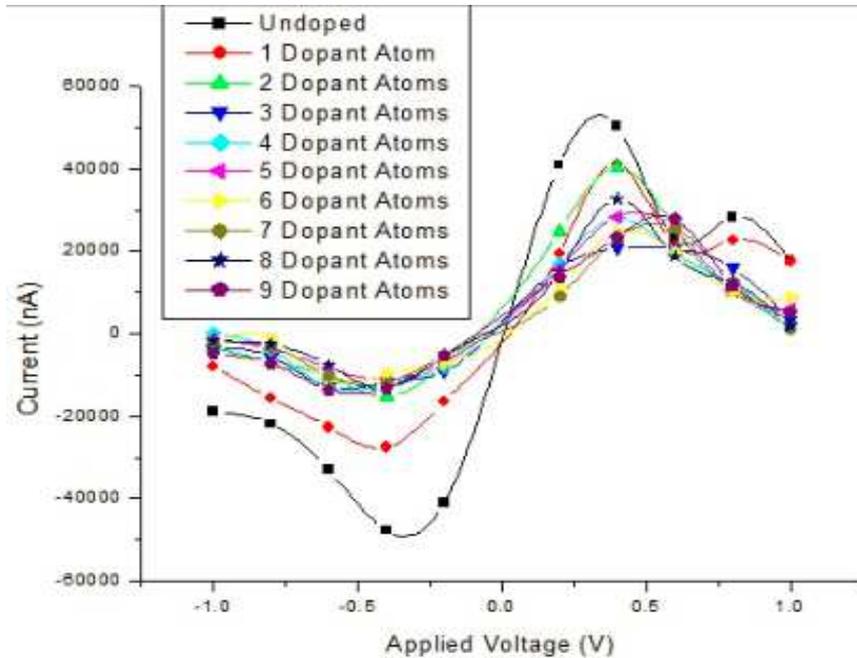

(j) Consolidated Plot

**Fig. 3.8** Electrical Characteristics of Doped Germanium Nanowire

### 3.2.3 Uniform Doping of Germanium Nanowire to obtain PIN Diode

The germanium nanowire has been doped in such a way that a PIN diode is modeled as part of the nanowire. The left and right ends of the nanowire have been doped with boron and phosphorus atoms respectively in a way that there exists a dopant atom (boron or phosphorus) after every two germanium atoms while the central region is left intrinsic/undoped. The doping is done gradually, starting from incorporating one dopant atom of either type at each side of the nanowire, simulating the nanowire, followed by progressively increasing the dopant concentration up to six dopant atoms, atom by atom, on either side of the nanowire with subsequent simulations thereby keeping the central region intrinsic. Once the doping is done to realize a PIN diode, the built structure is biased by sweeping the right end of the structure from -1 V to 1 V while grounding the other end of the nanowire. DFT simulations based on NEGF formalism yield I-V curves for each of the six cases as depicted in Fig. 3.9.

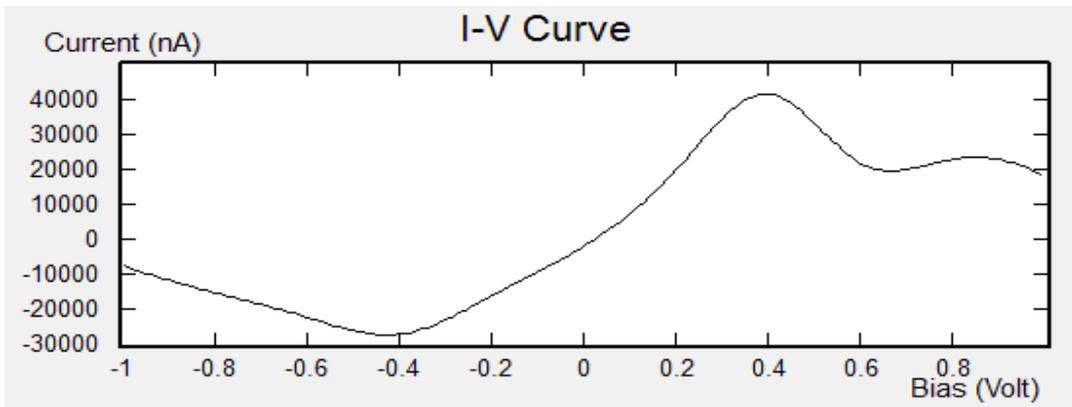

(a) Germanium nanowire with one boron and one phosphorus dopant atom

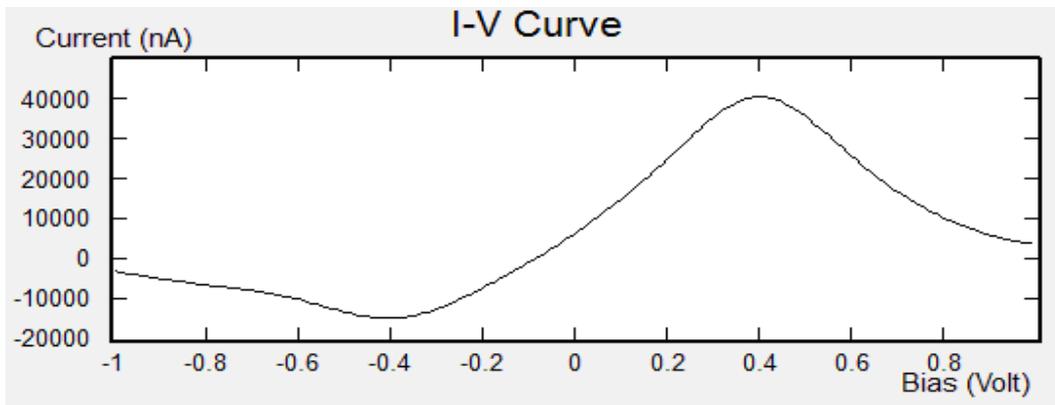

(b) Germanium nanowire with two boron and two phosphorus dopant atoms

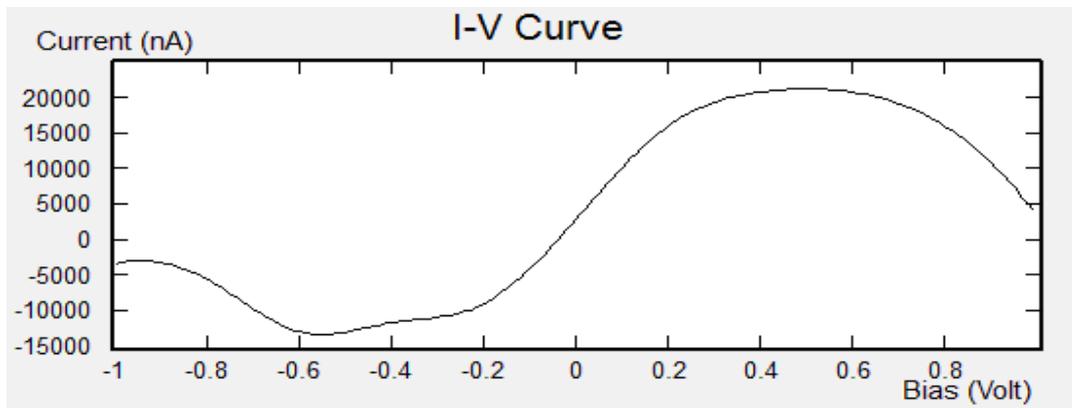

(c) Germanium nanowire with three boron and three phosphorus dopant atoms

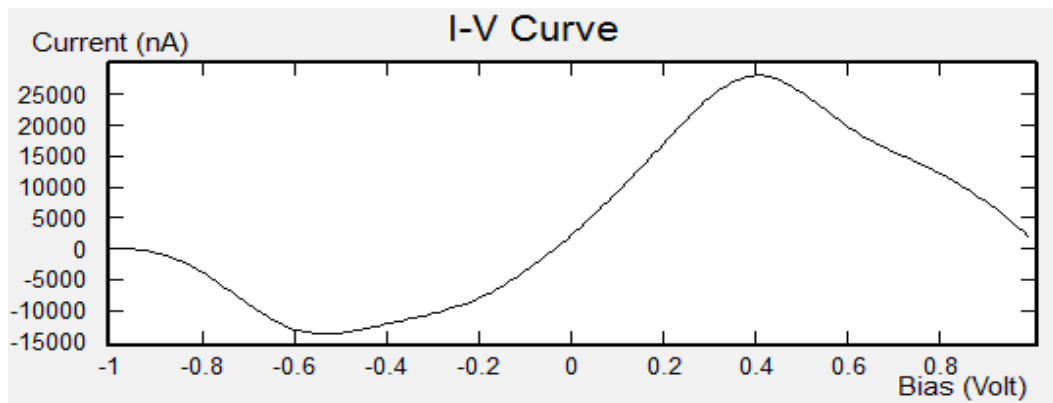

(d) Germanium nanowire with four boron and four phosphorus dopant atoms

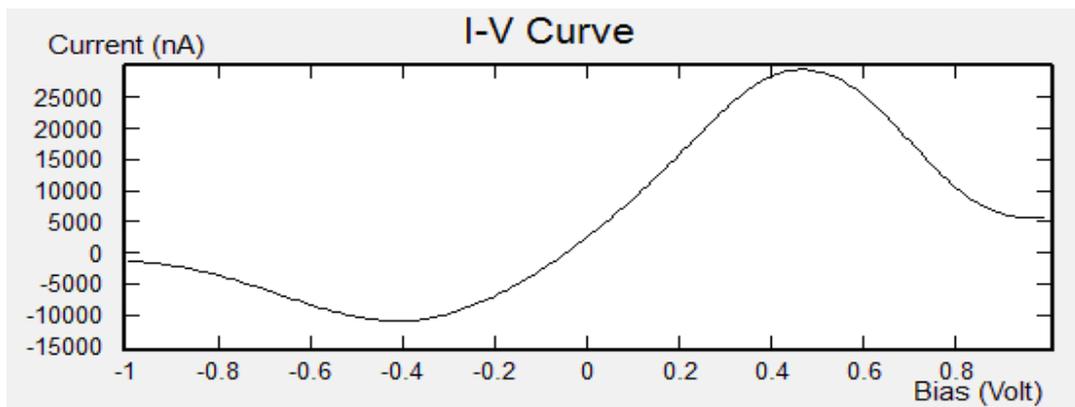

(e) Germanium nanowire with five boron and five phosphorus dopant atoms

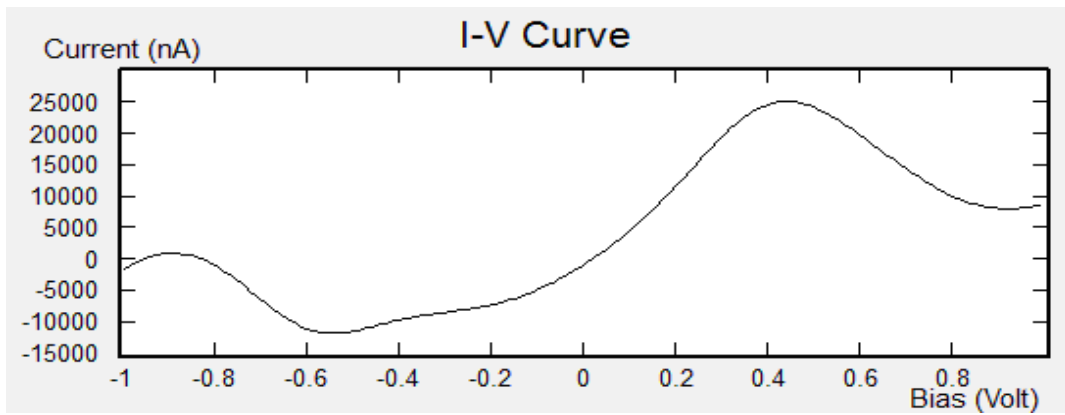

(f) Germanium nanowire with six boron and six phosphorus dopant atoms

**Fig. 3.9** Electrical Characteristics of Germanium Nanowire Doped in PIN Configuration

## 3.2.4 Uniform Doping of Germanium Nanowire to obtain PIP Diode

As done with silicon nanowire, the germanium nanowire has been doped in such a way that a PIP diode is modeled as part of the nanowire. The left and right ends of the nanowire have been doped with boron atoms alone in a way that there exists a dopant atom (boron) after every two silicon atoms while the central region is left intrinsic/undoped. The doping is done gradually, starting from incorporating one dopant atom of boron on each side of the nanowire, simulating the nanowire, followed by progressively increasing the dopant concentration upto six dopant atoms, atom by atom, on either side of the nanowire with subsequent simulations thereby keeping the central region intrinsic. Once the doping is done to realize a PIP diode, the built structure is biased by sweeping the right end of the structure from -1 V to 1 V while grounding the other end of the nanowire. DFT simulations based on NEGF formalism yield I-V curves for each of the six cases as depicted in Fig. 3.10. As seen in each of the plots, negative resistance is evident in each case, though the extent varies based on the dopant concentration. Therefore, this device can be used for microwave applications. There is a variation in the values of currents reported with germanium nanowire as compared to silicon nanowire for each of the input voltages applied.

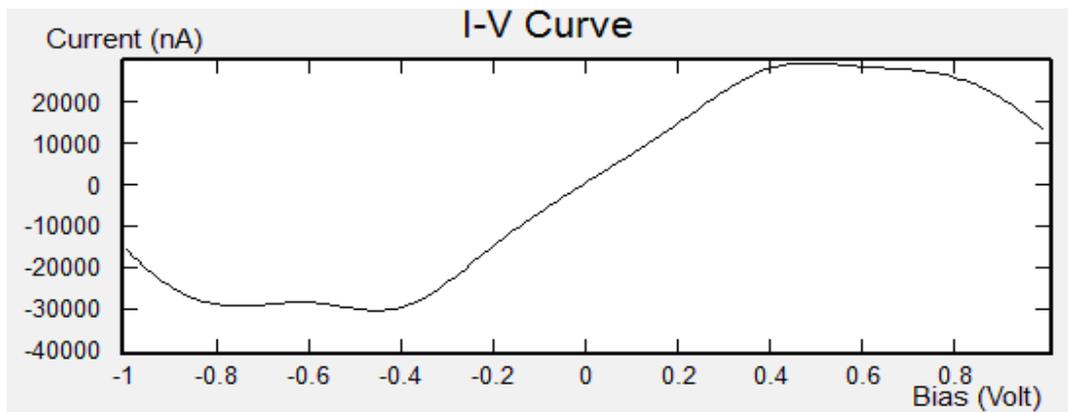

(a) Germanium nanowire with one boron dopant atom on either side of the intrinsic region

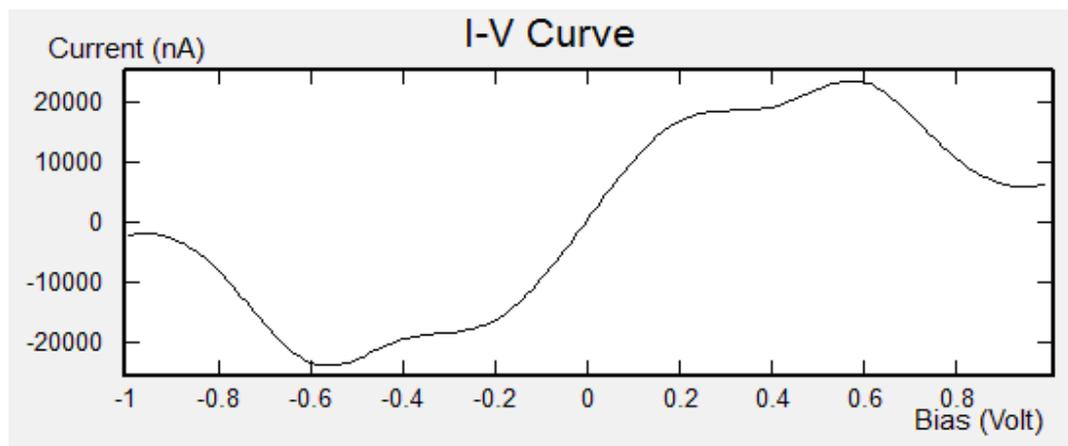

(b) Germanium nanowire with two boron dopant atoms on either side of the intrinsic region

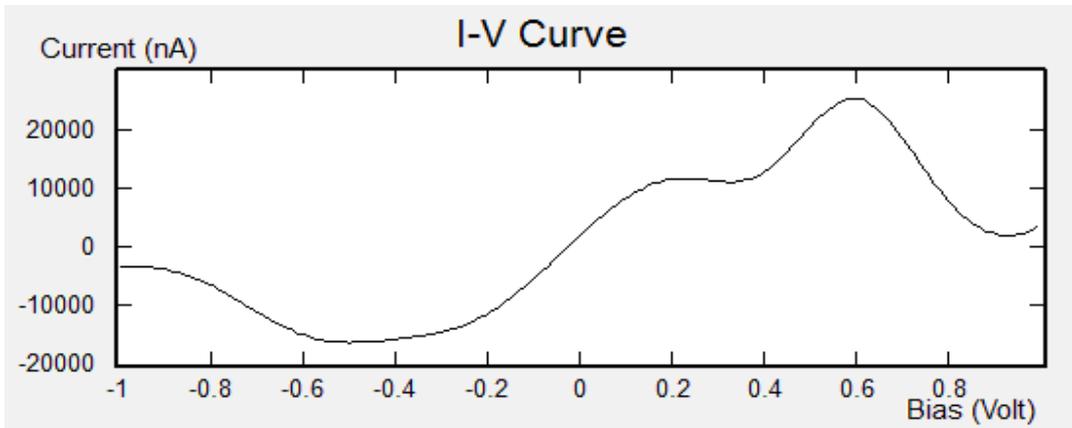

(c) Germanium nanowire with three boron dopant atoms on either side of the intrinsic region

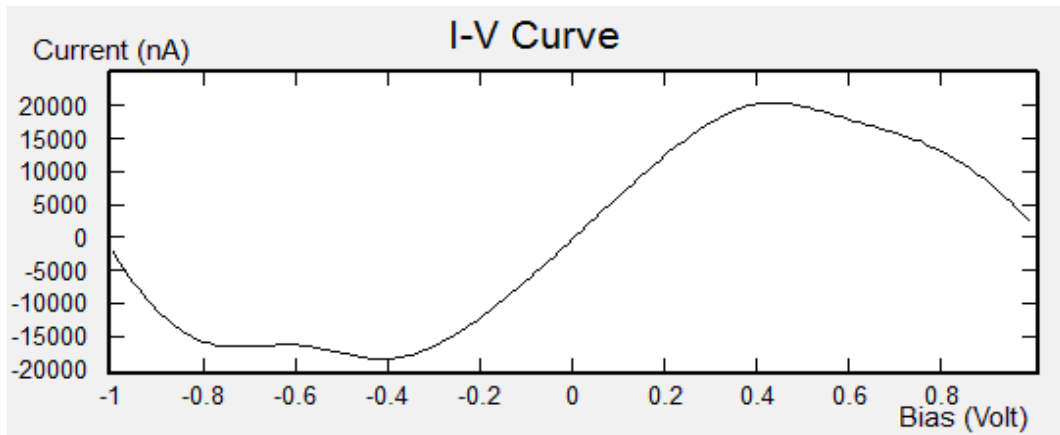

(d) Germanium nanowire with four boron dopant atoms on either side of the intrinsic region

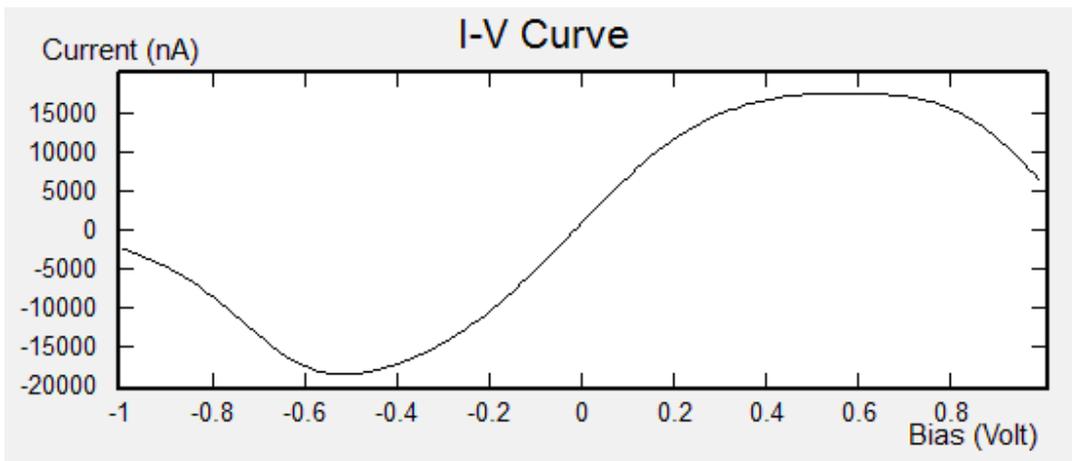

(e) Germanium nanowire with five boron dopant atoms on either side of the intrinsic region

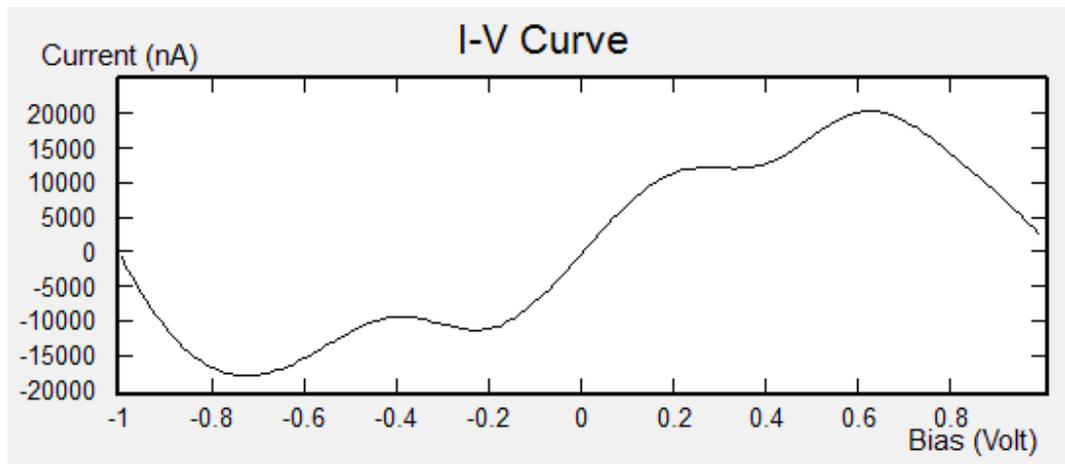

(f) Germanium nanowire with six boron dopant atoms on either side of the intrinsic region

**Fig. 3.10** Electrical Characteristics of Germanium Nanowire Doped in PIP Configuration

## 3.2.5 Uniform Doping of Germanium Nanowire to obtain NIN Diode

As done in the previous section, the nanowire is doped with the same type of dopant, the dopant being phosphorus in this case to N-dope both sides of the nanowire while keeping the central region undoped. Here, the silicon nanowire has been doped in such a way that a NIN diode is modeled as part of the nanowire. The left and right ends of the nanowire have been doped with phosphorus atoms alone in a way that there exists a dopant atom (phosphorus) after every two silicon atoms while the central region is left intrinsic/undoped. The doping is done gradually, starting from incorporating one dopant atom of phosphorus on each side of the nanowire, simulating the nanowire, followed by progressively increasing the dopant concentration upto six dopant atoms, atom by atom, on either side of the nanowire with subsequent simulations thereby keeping the central region intrinsic. Once the doping is done to realize a NIN diode, the built structure is biased by sweeping the right end of the structure from -1 V to 1 V while grounding the other end of the nanowire. DFT simulations based on NEGF formalism yield I-V curves for each of the six cases as depicted in Fig. 3.11. As seen in each of the plots, negative resistance is evident in each case, though the extent varies based on the dopant concentration. Therefore, this device can be used for microwave applications. However, the range of voltages or the exact voltage at which negative resistance regions can be seen differ for each addition of dopant and hence the negative resistance regions are not seen for the same voltage in every plot, there is a shift in the negative resistance region upon addition of an additional dopant atom.

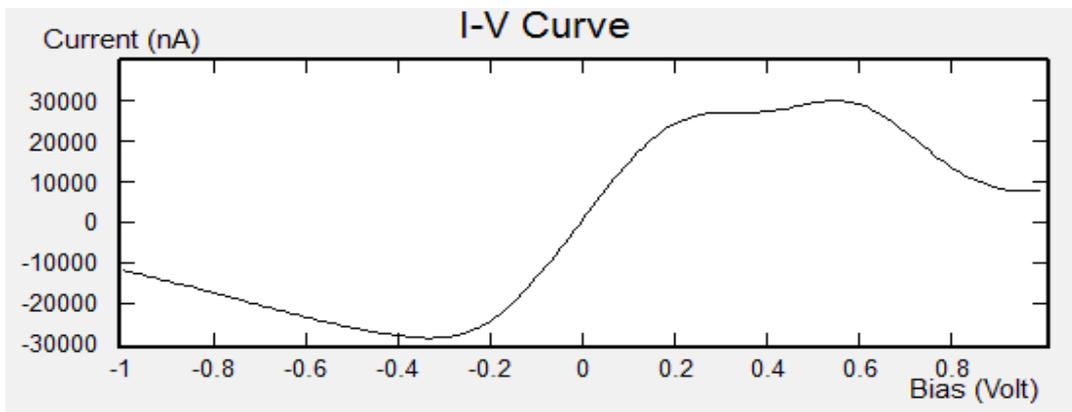

(a) Germanium nanowire with one phosphorus dopant atom on either side of the intrinsic region

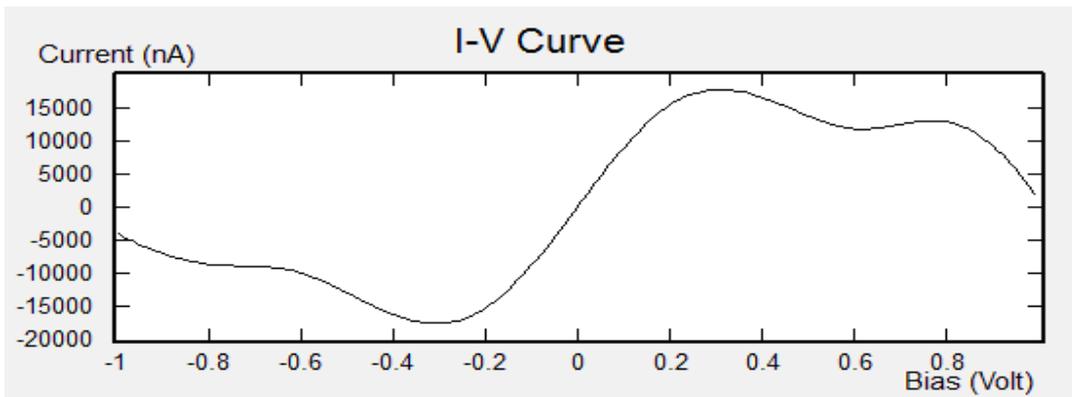

(b) Germanium nanowire with two phosphorus dopant atoms on either side of the intrinsic region

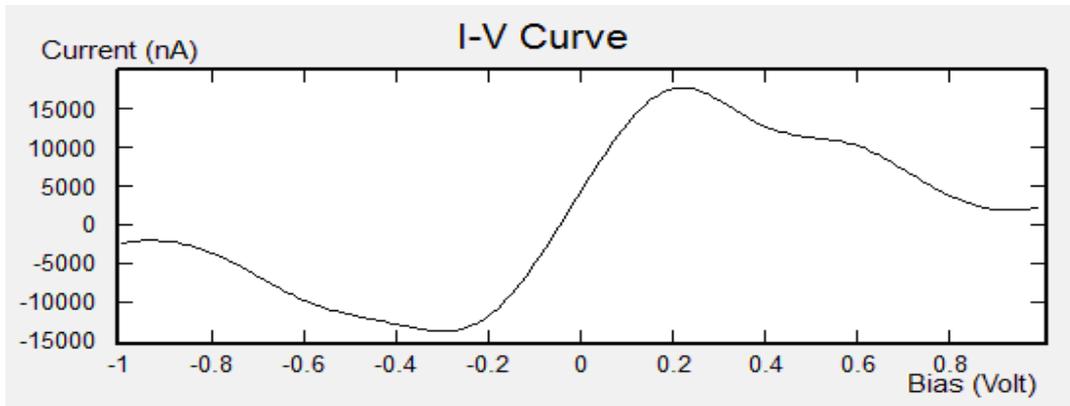

(c) Germanium nanowire with three phosphorus dopant atoms on either side of the intrinsic region

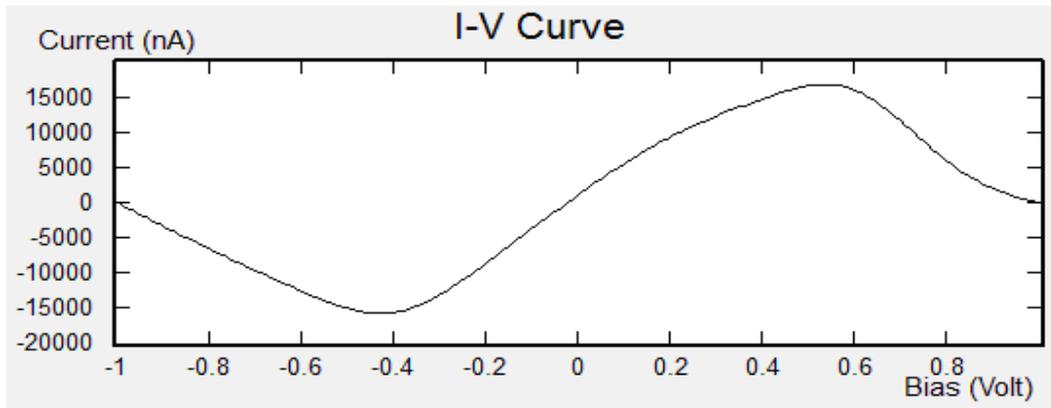

(d) Germanium nanowire with four phosphorus dopant atoms on either side of the intrinsic region

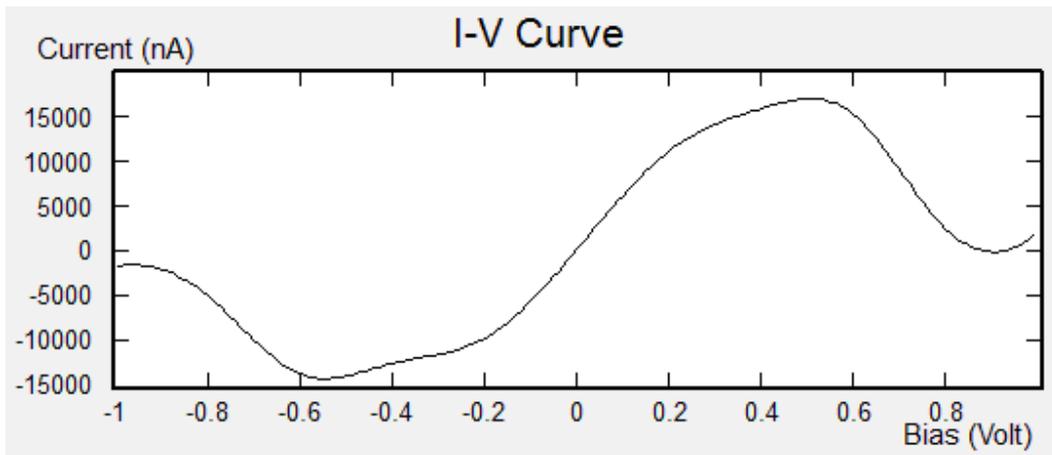

(e) Germanium nanowire with five phosphorus dopant atoms on either side of the intrinsic region

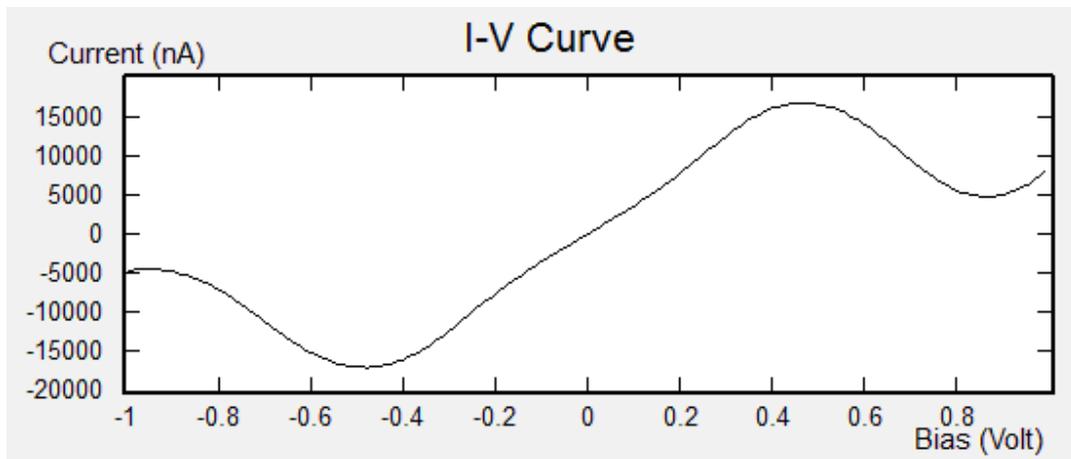

(f) Germanium nanowire with five phosphorus dopant atoms on either side of the intrinsic region

**Fig. 3.11** Electrical Characteristics of Germanium Nanowire Doped in NIN Configuration

## 3.3 Simulation of End Doped Silicon Nanowire

In this section, simulation results have been presented for a silicon nanowire of similar dimensions as discussed in section 3.2 but doped differently as compared to the previous section. Here, the nanowire has been doped in such a way that the dopants have been incorporated towards the two extreme ends such that the intrinsic region at the center of the nanowire widens to a greater extent than the manner the intrinsic region widens as discussed in the previous sections. As done earlier, the silicon nanowire has been dopant towards the two extreme ends in such a way that PIN, PIP and NIN diodes can be realized. Instead of presenting each I-V curve resulting from the addition of each dopant atom on the two sides of the nanowire, this section presents the three consolidated plots pertaining to PIN, PIP and NIN diode realizations by doping the nanowire towards the two extreme ends. Fig. 3.12 depicts the consolidated I-V curves obtained by simulating a silicon nanowire by progressively doping the two ends of the nanowire with dopants of opposite polarity to realize a PIN diode configuration. It can be seen from the plot that for the case with 6 dopant atoms, the current is very low (close to zero) which is not observed in the simulated results of uniformly doped silicon nanowire to realize PIN diode configuration. Also, for the case with 2 dopant atoms, the magnitude of current is less in case of end doped nanowire than in uniformly doped nanowire.

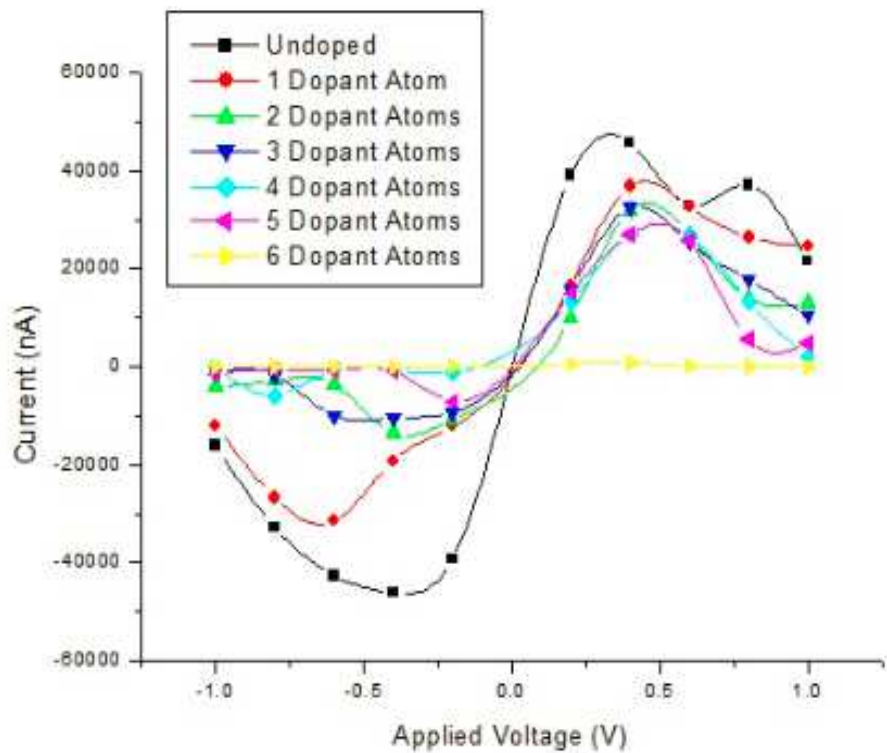

**Fig. 3.12** Consolidated I-V curves of end doped silicon nanowire to realize PIN configuration

The Fig. 3.13 and Fig. 3.14 reports the consolidated I-V curves obtained by simulating the silicon nanowire by doping both the extreme ends with boron and phosphorus respectively, thereby realizing a PIP and an NIN diode respectively.

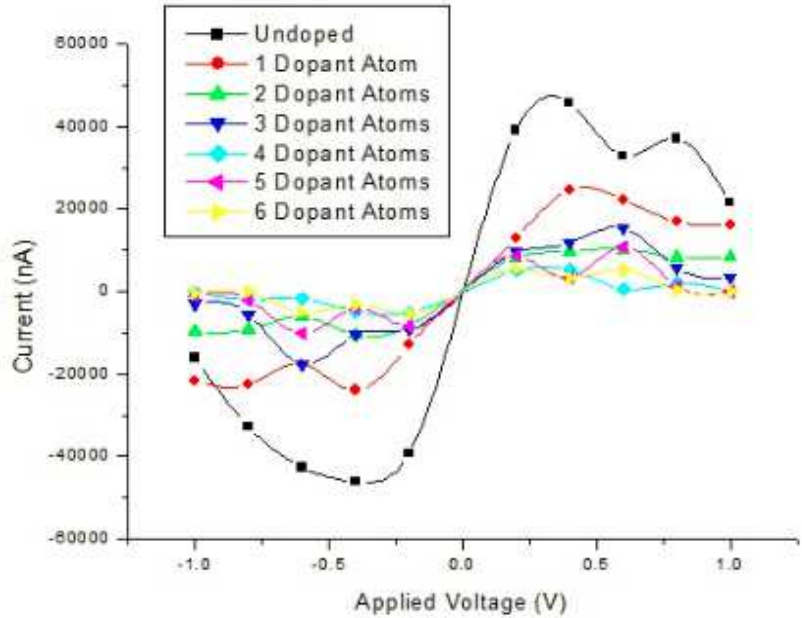

**Fig. 3.13** Consolidated I-V curves of end doped silicon nanowire to realize PIP configuration

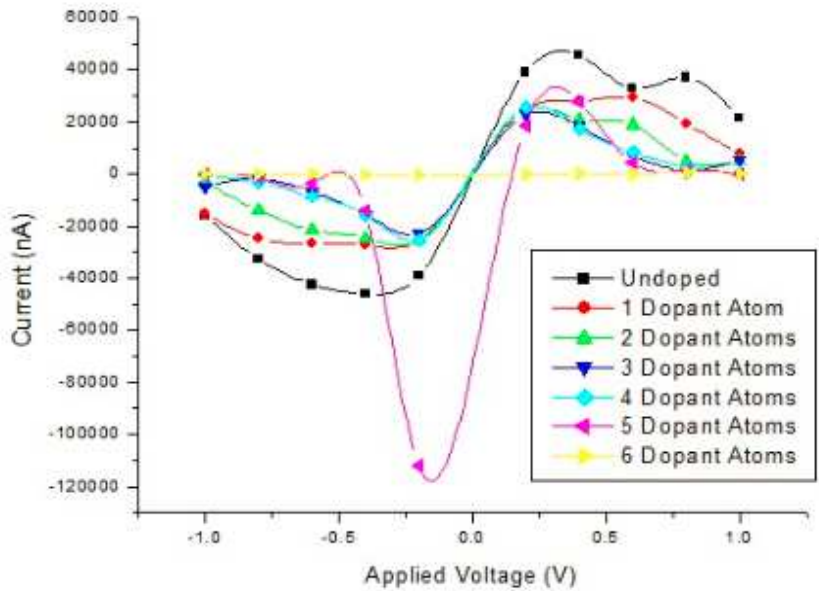

**Fig. 3.14** Consolidated I-V curves of end doped silicon nanowire to realize NIN configuration

When the consolidated I-V curves of the uniformly doped nanowire realizing a PIP diode is compared with the plot of end doped nanowire realizing the same diode, it is observed that the end doped nanowire simulation results in a much flatter characteristics than the ones obtained by simulating the uniformly doped nanowire.

In case of Fig. 3.14, a sudden drop in current is observed for an applied voltage of -0.15 V. The reason for this drop in current has not been reported earlier and needs to be further investigated. As seen in the case of end doped nanowire realizing a PIP diode, near zero current has also been reported when the NIN device has been simulated with six dopants of phosphorus on either side of the intrinsic region, though widely separated from each other.

# 4. Zero Bias Simulation of Silicon Nanowire Transistor with No Gate Dielectric

This section of the thesis presents the modeling and simulation of a silicon nanowire transistor with no separate oxide layer defined for gate dielectric. Through there is no transistor design that would practically be realization without a gate dielectric layer, the purpose of the work presented in this chapter is to study the effect of gate bias on the channel conductance of the FET with no gate dielectric. Here, we define the structure of a H-passivated silicon nanowire along the (100) direction, and set up a field-effect transistor (FET) structure with a cylindrical wraparound gate. We will primarily use the graphical user interface Virtual NanoLab (VNL) for setting up and analyzing the results. The underlying calculation engines for this kind of simulation are **ATK-DFT** and **ATK-SE**. The details pertaining to the Semi Empirical Extended Huckel calculator used and the physics behind the Semi Empirical Extended Huckel Model based calculator used for simulating the device have been presented at the end of this chapter. The screen shots shown below gives a step by step flow to model the device.

## 4.1 Setting up the Si (100) Nanowire Geometry

Start VNL and create a new project and give it a name then click Open. Next launch the Builder via the icon  on the toolbar.

In the builder, click Add → From Database... Type "silicon fcc" in the search field to locate the diamond phase of silicon. Click the icon 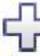 in the lower right-hand corner of the Database window to add the structure to the Stash in the Builder.

Next unfold the Builders panel bar in the right-hand column of the Builder and open the "Surface (Cleave)..." tool.

In the surface cleave tool,

- Keep the default (100) cleave direction, and press "Next >".

- Keep the default surface lattice, and press "Next >".

- Keep the default supercell, this will ensure that the wire direction is perpendicular to the surface, and press "Next >".

Press the "Finish" button to add the cleaved structure to the Stash.

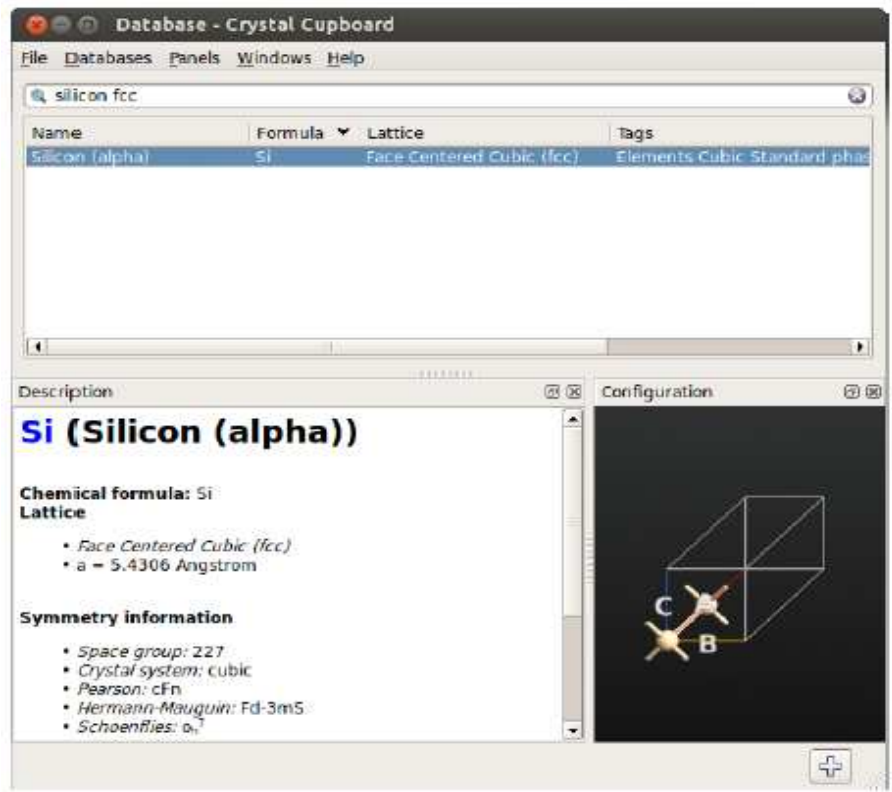

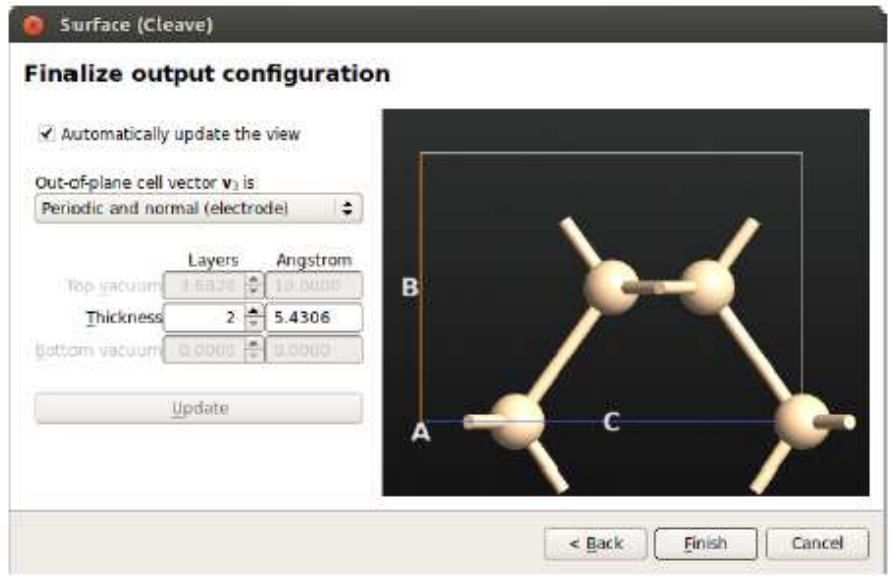

Next open <u>Bulk Tools</u> → <u>Repeat</u> and enter A=2, B=2, C=1, and press **Apply**.

Press Ctrl+R to reset the view in the Builder.

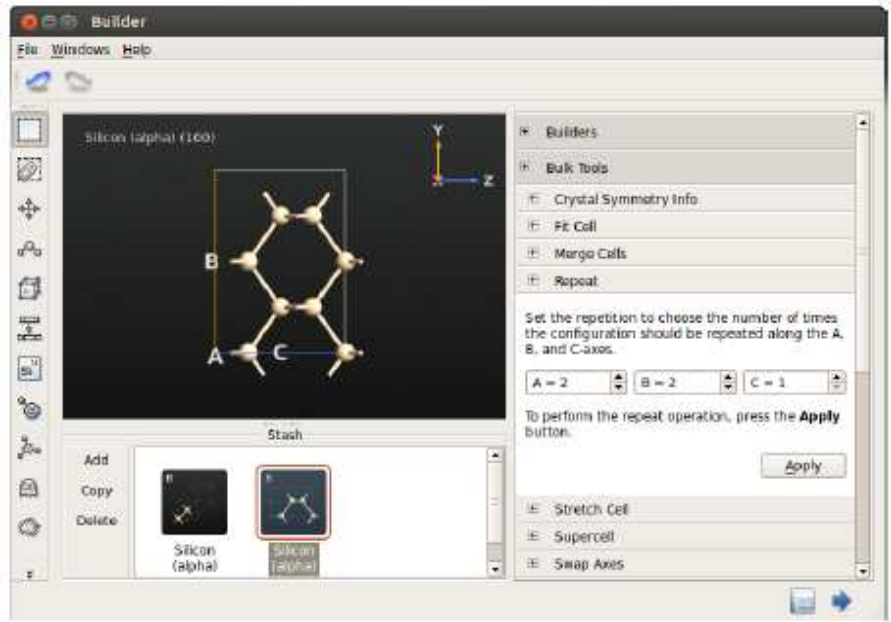

To finalize the setup perform the following steps:

- Open Bulk Tools → Lattice Parameters and set the length of the A and B vectors to 20 Å.

- Open Coordinate Tools → Center and center the structure in all directions.

- Click the H-passivator  in the left-hand tool bar to passivate the structure.

Your structure should now resemble the structure below.

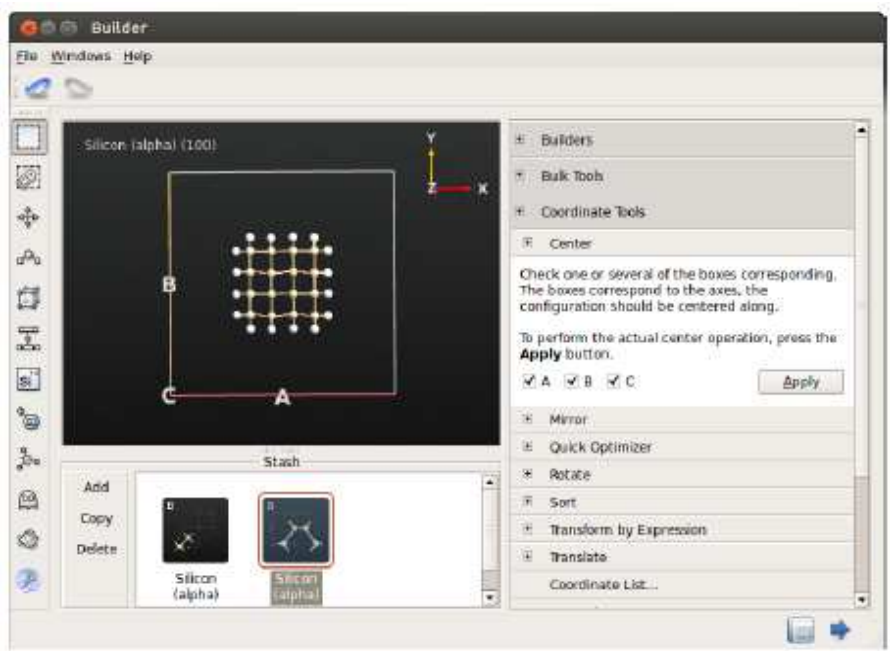

Next send the structure to the **Script Generator**, by using the "Send To" icon  in the lower right-hand corner of the Builder window.

## 4.2 Setting up the FET Device Geometry

To set up the nanowire device geometry you will need the relaxed nanowire geometry. To this end, select the file `si_100_nanowire.nc` in the VNL main window. In the Result Browser you will see that the file contains four bulk configurations. The first one (gID000) is the unrelaxed structure, while the remaining ones correspond to the relaxed structure, calculated with different methods (DFT-GGA, DFT-MGGA and Extended Hückel).

Drag the configuration with Id gID001 onto the Builder icon 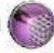 on the VNL toolbar. This will open the Builder with the relaxed nanowire as the active stash item.

Some of the H atoms are located outside of the unit cell. This is not a problem for a purely periodic calculation, but it needs to be fixed for the device, else they can lead to the generation of a wrong electrodes.

To do this, open `Bulk Tools → Wrap`, and press **Apply**.

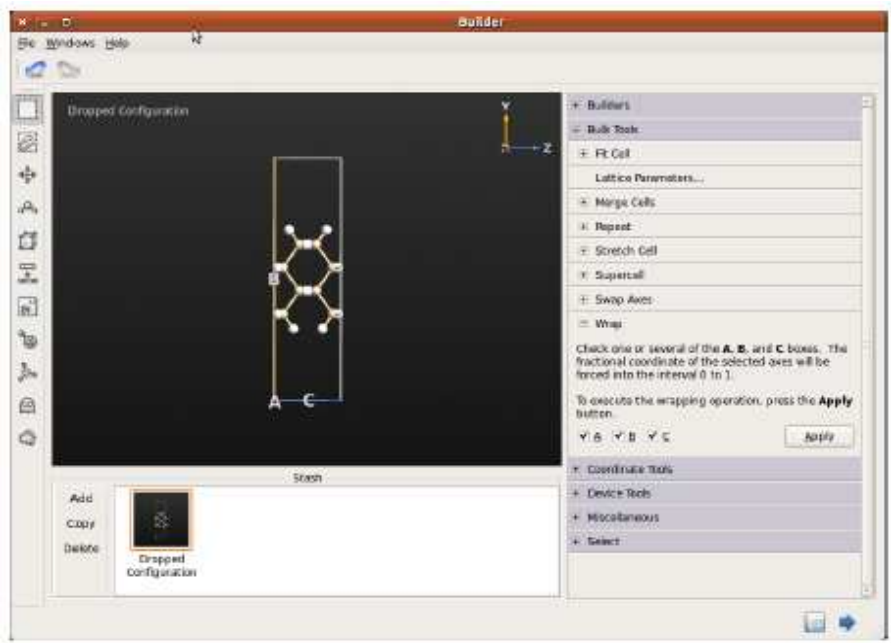

Next, open the `Bulk Tools → Repeat` Panel, set A=1, B=1, C=12, and press **Apply**.

## 4.2.1 Defining the Gate

The next step is to setup a wrap-around metallic gate.

Open <u>Miscellaneous</u> → <u>Spatial Regions...</u>.

1. Right-click in the white area of the widget to insert a new metallic region.

2. Set the value to 0 Volt.

3. Under Geometry, choose **Tube**, to create a cylindrical region.

4. Define the geometry of the tube as shown in the picture below. The cylinder will extend to the edge of the simulation cell, and will cover most of the central part of the nanowire.

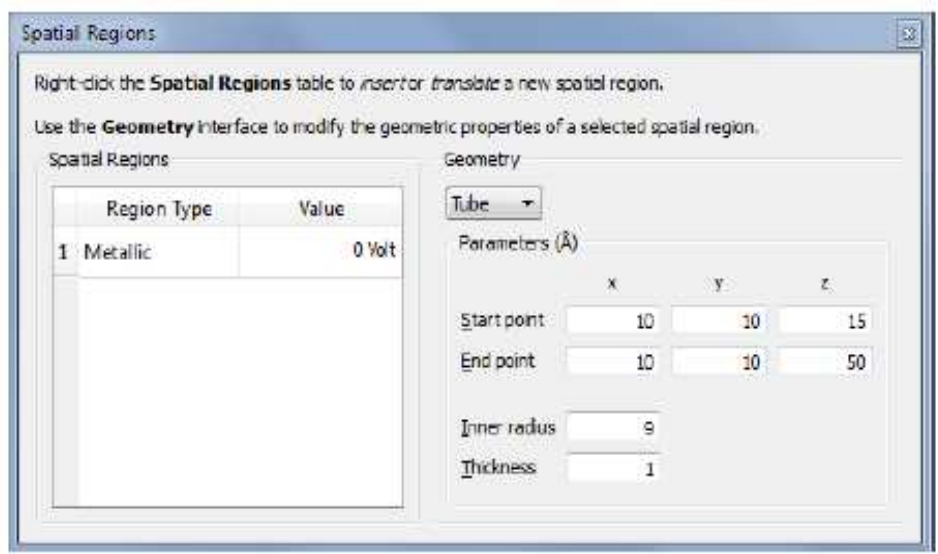

Finally, transform the structure into a device configuration by opening <u>Device Tools</u> → <u>Device From Bulk...</u> and accept the default suggestion for the electrode lengths.

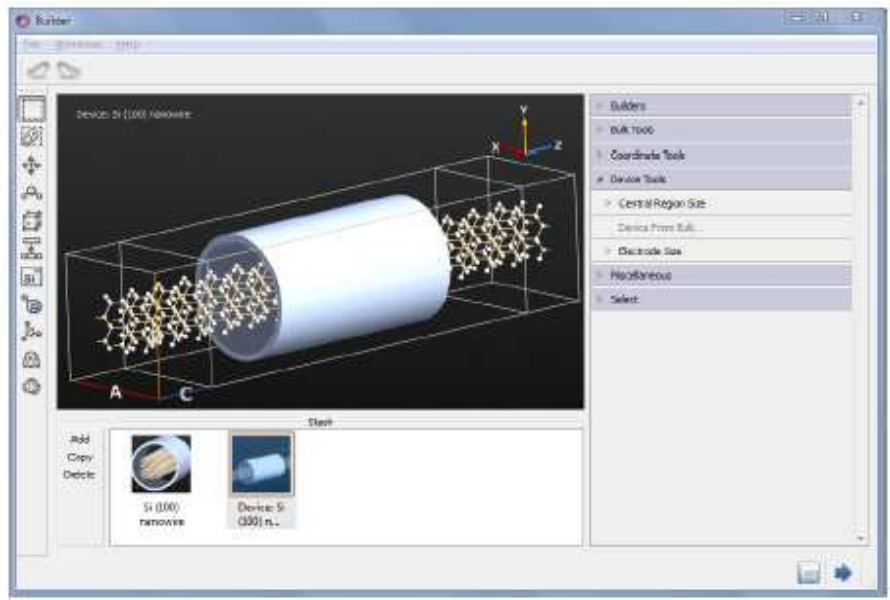

Next send the structure to the **Script Generator**.

## 4.2.2 Zero Gate Voltage Calculation

In the following you will setup a calculation of the transmission spectrum of a p-doped nanowire at zero gate potential using the Extended Hückel model.

For this purpose:

• Add a New Calculator.

• Add Analysis/TransmissionSpectrum.

• Add Analysis/ElectronDifferenceDensity.

- Add Analysis/ElectrostaticDifferencePotential

- Change the default output file to `si_100_nanowire_fet_pin.nc`

Open the **New Calculator** block, Select the "ATK-SE: Extended Hückel (Device)" calculator and make the following settings (similar to before, for the bulk calculation):

- Uncheck "No SCF iteration" to make the calculation selfconsistent.

- Increase the density mesh cut-off to 20 Hartree.

- Go to the **Hückel basis set** tab, and select the Cerda.Hydrogen [C2H4] and Cerda.Silicon [GW diamond] basis sets.

  Set the vacuum level of H to -10 eV.

- Under **Poisson solver**, set **Neumann** boundary conditions in the A and B directions; it would be incorrect to use periodic boundary conditions when there are gates present in the system.

Then open the **Transmission spectrum** block and set the energy range to [-4,4] with 301 points.

## 4.2.3 Doping the wire

To make the calculation more interesting, you will finally introduce doping in the wire. Instead of adding explicit dopants, which would give a very high doping concentration, you will set up the system as a **p-i-n junction**, by adding a certain amount of charge to the electrodes. This will lead to a shift in the Fermi levels of the two electrodes, relative to each other, and in this way a potential drop will be built into the system.

Send the script to the Editor using "Send To"  .

In the script locate the section defining the electrode calculators, and set charge=0.01 and -0.01 for the electrodes, respectively, as illustrated below.

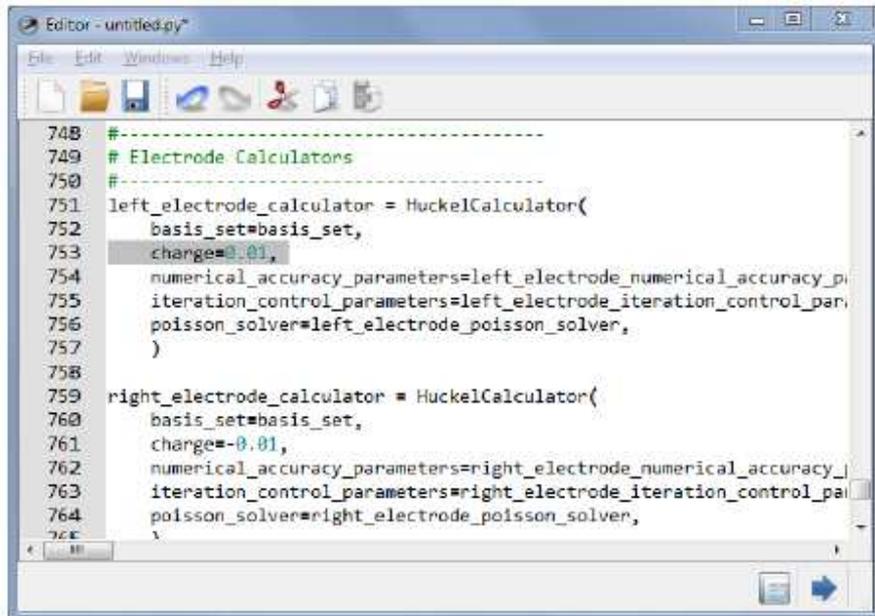

**Note**

The electrode is about 1 nm long and has a cross section of (0.5 nm)x(0.5 nm), so the added charge corresponds to an effective doping level of around $4 \cdot 10^{19}$ cm$^{-3}$.

Save the script, and then send it to the Job Manager and execute the calculation. The calculation will take some time to run, probably between one and two hours on most machines. If you have possibility to run the script in parallel on 4-5 MPI nodes, the time can be reduced to under one hour, however.

## 4.2.4 Analyzing the results

Once the calculation is completed, select the file `si_100_nanowire_fet_pin.nc` in the VNL file browser and plot the transmission spectrum.

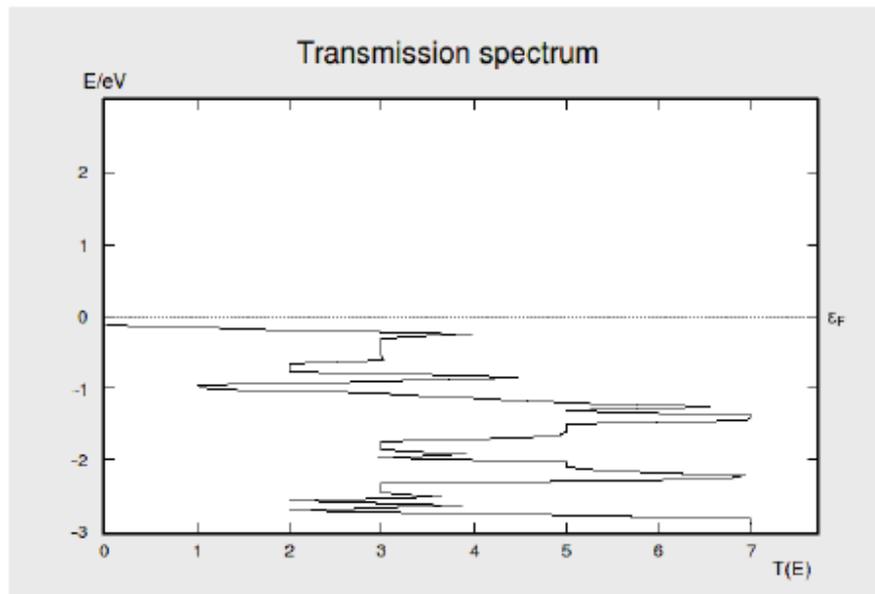

**Fig. 4.1** Transmission spectrum with 0 Volt gate potential. Note how the valence band is located just below the Fermi level due to the p-doping. The conduction band is located above 3 eV and therefore not visible in the plot.

## 4.2.5 Performing a Gate Scan

We will now perform a calculation of the transmission spectrum of the Nanowire FET for different values of the gate bias. From the transmission spectrum we can obtain the conductance as function of gate bias.

The calculation is most conveniently done using scripting. Save the script below in the same directory as the result file `si_100_nanowire_fet_pin.nc`

```python
# Read in the old configuration
device_configuration = nlread("nanodevice_huckel.nc",DeviceConfiguration)[0]
calculator = device_configuration.calculator()
metallic_regions = device_configuration.metallicRegions()

# Define gate voltages
gate_voltage_list=[1.0, 2.0, 3.0, 4.0, 5.0]*Volt
# Define output file name
filename= "nanodevice_huckel.nc"

# Perform loop over gate voltages
for gate_voltage in gate_voltage_list:
    # Set the gate voltages to the new values
    new_regions = [m(value = gate_voltage) for m in metallic_regions]
    device_configuration.setMetallicRegions(new_regions)

    # Make a copy of the calculator and attach it to the configuration
    # Restart from the previous scf state
    device_configuration.setCalculator(calculator(),
        initial_state=device_configuration)
    device_configuration.update()
    nlsave(filename, device_configuration)

    # Calculate analysis objects
    electron_density = ElectronDifferenceDensity(device_configuration)
    nlsave(filename, electron_density)

    electrostatic_potential = ElectrostaticDifferencePotential(device_configuration)
    nlsave(filename, electrostatic_potential)

    transmission_spectrum = TransmissionSpectrum(
        configuration=device_configuration,
        energies=numpy.linspace(-3,3,101)*eV,
        self_energy_calculator=DirectSelfEnergy(),
        )
    nlsave(filename, transmission_spectrum)
    nlprint(transmission_spectrum)
```

Next execute the script by dropping it on the jobmanager or starting it from a command line. The script will take 5 times longer than the previous calculation to finish.

## 4.2.6 Analyzing the Gate Scan

```python
# Read the data
transmission_spectrum_list = nlread("nanodevice_huckel.nc", TransmissionSpectrum)
configuration_list = nlread("nanodevice_huckel.nc", DeviceConfiguration)

conductance = numpy.zeros(len(configuration_list))
gate_bias = numpy.zeros(len(configuration_list))
for i, configuration in enumerate(configuration_list):
    transmission_spectrum = transmission_spectrum_list[i]
    energies = transmission_spectrum.energies().inUnitsOf(eV)
    spectrum = transmission_spectrum.evaluate()
    gate_bias[i] = configuration.metallicRegions()[0].value().inUnitsOf(Volt)
    conductance[i] = transmission_spectrum.conductance()

# Sort the data according to the gate bias
index_list = numpy.argsort(gate_bias)

# Plot the spectra
import pylab
pylab.figure()
ax = pylab.subplot(111)
ax.semilogy(gate_bias[index_list], conductance[index_list])
ax.set_ylabel("Conductance", size=16)
ax.set_xlabel("Gate Bias (Volt)", size=16)
ax.set_ybound(lower=1e-25)

for g,c in zip(gate_bias[index_list], conductance[index_list]):
    print g,c
pylab.show()
```

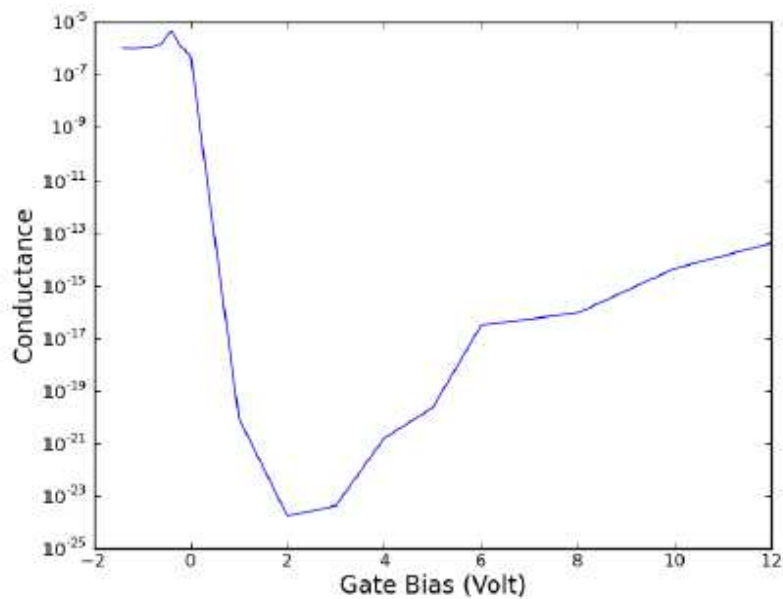

**Fig.4.2** Conductance as a function of the gate bias

```
# Read the data
transmission_spectrum_list = nlread("nanodevice_huckel.nc", TransmissionSpectrum)
configuration_list = nlread("nanodevice_huckel.nc", DeviceConfiguration)

# Plot the spectra
import pylab
pylab.figure()
ax = pylab.subplot(111)
for configuration, transmission_spectrum in zip(configuration_list, transmission_spectrum_list):
    energies = transmission_spectrum.energies().inUnitsOf(eV)
    spectrum = transmission_spectrum.evaluate()
    gate_bias = configuration.metallicRegions()[0].value()
    ax.semilogy( energies,  spectrum, label=str(gate_bias))

ax.set_ylabel("Transmission coefficient", size=16)
ax.set_xlabel("Energy (eV)", size=16)
ax.set_ybound(lower=1e-15)
pylab.show()
```

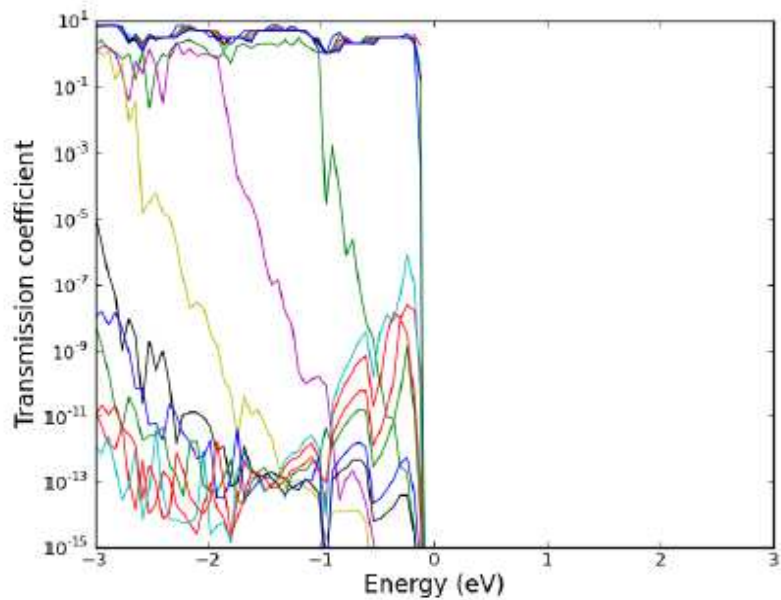

**Fig. 4.3** The transmission spectrum for different gate biases

## 4.3 ATK-SE Package

The ATK-SE package uses the Semi Empirical Extended Huckel Model to calculate the transmission characteristics of the nanowire. The extended Hückel method is a semi empirical quantum chemistry method, developed by Roald Hoffmann since 1963. It is based on the Hückel method but, while the original Hückel method only considers pi orbitals, the extended method also includes the sigma orbitals.

The extended Hückel method can be used for determining the molecular orbitals, but it is not very successful in determining the structural geometry of an organic molecule. It can however determine the relative energy of different geometrical configurations. It involves calculations of the electronic interactions in a rather simple way for which the electron-electron repulsions are not explicitly included and the total energy is just a sum of terms for each electron in the molecule. The off-diagonal Hamiltonian matrix elements are given by an approximation due to Wolfsberg and Helmholz that relates them to the diagonal elements and the overlap matrix element.

$$H_{ij} = KS_{ij} \, (H_{ii} + H_{jj})/2$$

$K$ is the Wolfsberg-Helmholtz constant, and is usually given a value of 1.75. In the extended Hückel method, only valence electrons are considered; the core electron energies and functions are supposed to be more or less constant between atoms of the same type. The method uses a series of parametrized energies calculated from atomic ionization potentials or theoretical methods to fill the diagonal of the Fock matrix. After filling the non-diagonal elements and diagonalizing the resulting Fock matrix, the energies (eigenvalues) and wavefunctions (eigenvectors) of the valence orbitals are found.

It is common in many theoretical studies to use the extended Hückel molecular orbitals as a preliminary step to determining the molecular orbitals by a more sophisticated method such as the CNDO/2 method and ab initio quantum chemistry methods. Since the EHT basis set is fixed, the monoparticle calculated wavefunctions must be projected to the basis set where the accurate calculation is to be done. One usually does this by adjusting the orbitals in the new basis to the old ones by least squares method. As only valence electron wavefunctions are found by this method, one must fill the core electron functions by orthonormalizing the rest of the basis set with the calculated orbitals and then selecting the ones with less energy. This leads to the determination of more accurate structures and electronic properties, or in the case of ab initio methods, to somewhat faster convergence.

The method was first used by Roald Hoffmann who developed, with Robert Burns Woodward, rules for elucidating reaction mechanisms (the Woodward–Hoffmann rules). He used pictures of the molecular orbitals from extended Hückel theory to work out the orbital interactions in these cycloaddition reactions.

A closely similar method was used earlier by Hoffmann and William Lipscomb for studies of boron hydrides. The off-diagonal Hamiltonian matrix elements were given as proportional to the overlap integral.

$$H_{ij} = K \, S_{ij}.$$

This simplification of the Wolfsberg and Helmholz approximation is reasonable for boron hydrides as the diagonal elements are reasonably similar due to the small difference inelectronegativity between boron and hydrogen.

The method works poorly for molecules that contain atoms of very different electronegativity. To overcome this weakness, several groups have suggested iterative schemes that depend on the atomic charge. One such method, that is still widely used in inorganic and organometallic chemistry is the Fenske-Hall method.

## 4.3.1 Introduction to ATK-SE Package

ATK-SemiEmpirical (ATK-SE) can model the electronic properties of molecules, crystals and devices using both self-consistent and non-self-consistent tight-binding models. In this chapter the implemented tight-binding models based on the Slater-Koster model and the extended Hückel model are presented.

In ATK-SE, the non-self-consistent part of the tight-binding Hamiltonian is parametrized using a two-center approximation, i.e. the matrix elements only depend on the distance between two atoms and is independent of the position of the other atoms. In the extended Hückel model, the matrix elements are described in terms of overlaps between Slater orbitals on each site. In this way, the matrix elements can be defined by very few parameters. In the Slater-Koster model, the distance-dependence of the matrix elements is given as a numerical function; this gives higher flexibility, but also makes the fitting procedure more difficult.

The self-consistent part of the calculation is identical for both SE models. The density matrix is calculated from the Hamiltonian using non-equilibrium Green's functions for device systems, while for molecules and crystals it is calculated by diagonalization. The density matrix defines the real-space electron density, and consequently the Hartree potential can be obtained by solving the Poisson equation. The following describes the details of the mathematical formalism behind the implementation.

## 4.3.2. Non-self-consistent Hamiltonian

The Hamiltonian is expanded in a basis of local atomic orbitals (an LCAO expansion)

$$\phi_{nlm}(\mathbf{r}) = R_{nl}(r)Y_{lm}(\hat{r}),$$

The Hamiltonian is expanded in a basis of local atomic orbitals (an LCAO expansion)

$$\phi_{nlm}(\mathbf{r}) = R_{nl}(r)Y_{lm}(\hat{r}),$$

where $Y_{lm}$ is a spherical harmonic and $R_{nl}$ is a radial function. Typically, the atomic orbitals used in the LCAO expansion has a close resemblance to the atomic eigen functions.

### 4.3.2.1 Onsite terms

With this form of the basis set, the onsite elements are given by

$$S_{ij}^{\text{onsite}} = \delta_{ij},$$
$$H_{ij}^{\text{onsite}} = E_i \delta_{ij},$$

where $E_i$ is an adjustable parameter, which often is close to the atomic eigen energy.

### 4.3.2.2 Offsite terms in the extended Hückel model

The central object in the extended Hückel model is the overlap matrix,

$$S_{ij} = \int_V \phi_i(\mathbf{r} - \mathbf{R}_i)\phi_j(\mathbf{r} - \mathbf{R}_j)\,\mathrm{d}\mathbf{r}.$$

To calculate this integral the form of the basis functions must be specified. In the extended Hückel model the basis functions are parametrized by Slater orbitals

$$R_{nl}(r) = \frac{r^{n-1}}{\sqrt{(2n)!}} \left[ C_1 (2\eta_1)^{n+\frac{1}{2}} e^{-\eta_1 r} + C_2 (2\eta_2)^{n+\frac{1}{2}} e^{-\eta_2 r} \right].$$

The LCAO basis is described by the adjustable parameters $\eta_1, \eta_2, C_1,$ and $C_2$. These parameters must be defined for each angular shell of valence orbitals, for each element.

The overlap matrix defines the Hamiltonian

$$H_{ij} = \frac{1}{4}(\beta_i + \beta_j)(E_i + E_j)S_{ij},$$

where $E_i$ is the onsite orbital energy and $\beta_i$ is a Hückel fitting parameter (often chosen to be 1.75).

## 4.3.2.3 Weighting schemes

There are two variants of the weighting schemes of the orbital energies of the offsite Hamiltonian. The scheme used above

$$\frac{1}{2}\beta(E_i + E_j)S_{ij},$$

where $\beta = \frac{1}{2}(\beta_i + \beta_j)$ is due to Wolfsberg, while Hoffmann uses

$$\frac{1}{2}(\beta + \alpha^2 + (1 - \beta)\alpha^4)(E_i + E_j),$$

where $\alpha = (E_i - E_j)/(E_i + E_j)$.

Both variants are available in ATK-SE through the parameters Wolfsberg Weighting and Hoffmann Weighting which can be given to the **HuckelCalculator** and the **DeviceHuckelCalculator** classes.

## 4.3.2.4 Offsite Hamiltonian in the Slater-Koster model

The overlap matrix is given by pairwise integrals between the different basis functions. These integrals can be pre-calculated for all relevant distances and different orbital combinations, and stored in so-called Slater-Koster tables. The Slater-Koster table stores the distance-dependent parameters $s(d, Z_1, Z_2, l_1, l_2, m)$, where $d$ is the distance, $Z_1, Z_2$ the element types, $l_1, l_2$ the angular momentum of the two orbitals, and the index $m \leq \min(l_1, l_2)$.

From the Slater-Koster tables, the overlap matrix elements are given by

$$S_{ij} = \sum_{m \leq \min(l_i, l_j)} \alpha_{l_i, m_i, l_j, m_j, m}(\hat{R}_{ij}) s(d_{ij}, Z_i, Z_j, l_i, l_j, m),$$

where $\alpha$ are the Slater-Koster expansion coefficients.

In the Slater-Koster model it is assumed that also the Hamiltonian has a pairwise form, and a Slater-Koster table is generated for the Hamiltonian matrix elements. This table may be generated by calculating Hamiltonian matrix elements for a set of dimer distances or by simply fitting matrix elements to the band structure for different lattice constants.

In ATK-SE, the Slater-Koster table is constructed either by providing the path to a directory containing compatible Slater-Koster files (see **DFTBDirectory** and **HotbitDirectory**), or directly using the **SlaterKosterTable** class.

Note that the extended Hückel model is a Slater-Koster model too, with a special fitting procedure for the Hamiltonian matrix elements.

## 4.3.3 Self-consistent Hamiltonian

In the self-consistent semi-empirical models in ATK, the electron density is computed using the tight-binding model as described above. Th density gives rise to a Hartree potential $V_H$. The calculation of the Hartree potential is described in detail in the section called "The Hartree Potential"

The Hartree potential is included through an additional term in the Hamiltonian

$$H_{ij}^{SCF} = \frac{1}{2}(V_H(\mathbf{R}_i) + V_H(\mathbf{R}_j))S_{ij}.$$

### 4.3.3.1 Electron density

The electron density is given by the occupied eigen functions

$$n(\mathbf{r}) = \sum_\alpha |\psi_\alpha(\mathbf{r})|^2 f\left(\frac{\varepsilon_\alpha - \varepsilon_F}{kT}\right),$$

where $f(x) = 1/(1 + e^x)$ is the Fermi function, $\varepsilon_F$ the Fermi energy, $T$ the electron temperature, and $\varepsilon_\alpha$ the energy of eigen state $\psi_\alpha$.

Next write the eigen states in the Slater orbital basis as

$$\psi_\alpha = \sum_i c_{\alpha i} \phi_i,$$

and see that the total number of electrons, $N = \int_V n(\mathbf{r})\,\mathrm{d}\mathbf{r}$ is given by

$$N = \sum_{ij} D_{ij} S_{ij},$$

where $D_{ij} = \sum_\alpha c_{\alpha i}^* c_{\alpha j} f\left(\frac{\varepsilon_\alpha - \varepsilon_F}{kT}\right)$ is the density matrix.

### 4.3.3.2 An approximate atom-based electron density

In practice, a simple approximation is used for the electron density. To this end, introduce the Mulliken population

$$m_\mu = \sum_{i \in \mu} \sum_j D_{ij} S_{ij},$$

of atom number $\mu$, and write the total number of electrons as a sum of atomic contributions, $N = \sum_\mu m_\mu$. The radial dependence of each atomic-like density is represented by a Gaussian function, and the total induced charge in the system is approximated by

$$\delta n(\mathbf{r}) = \sum_\mu \delta m_\mu \sqrt{\frac{\alpha_\mu}{\pi}} e^{-\alpha_\mu |\mathbf{r} - \mathbf{R}_\mu|^2},$$

where $\delta m_\mu = m_\mu - Z_\mu$ is the total charge of atom $\mu$, i.e. the sum of the valence electron charge $m_\mu$ and the ionic charge $-Z_\mu$.

To see the significance of the width $\alpha_\mu$ of the Gaussian orbital, calculate the electrostatic potential from a single Gaussian density at position $\mathbf{R}_\mu$

$$V_{H\mu}(\mathbf{r}) = (m_\mu - Z_\mu) \frac{\mathrm{Erf}(\sqrt{\alpha_\mu}|\mathbf{r} - \mathbf{R}_\mu|)}{|\mathbf{r} - \mathbf{R}_\mu|}.$$

The onsite value of the Hartree potential is $V_{H\mu}(\mathbf{R}_\mu) = (m_\mu - Z_\mu)U_\mu$, where $U_\mu = 2\sqrt{\frac{\alpha_\mu}{\pi}}$ is the onsite Hartree shift. In ATK-SE, it is the value of $U_\mu$ which is specified, and this value is used to determine the width $\alpha_\mu$ of the Gaussian using the above relation.

### 4.3.3.3 Onsite Hartree shift parameters

The shell-dependent onsite Hartree shift ($U_l$) can be obtained from an atomic calculation.

$U_l$ is related to the linear shift of the eigen energy $\varepsilon_l$, of shell $l$, as function of the shell occupation $q_l$:

$$U_l = \frac{d\varepsilon_l}{dq_l}.$$

Thus, $U_l$ can be obtained by performing atomic calculations with different values of $q_l$.

ATK provides a database for $U_l$ calculated using the DFT GGA.PBE functional. Access to the data is through the function **ATK_U**.

Due to backwards compatibility, the HoffmanHuckelParameters and MullerHuckelParameters do not use the **ATK_U** database.

### 4.3.4 Spin Polarization

The following spin dependent term is added to the Hamiltonian

$$H_{ij}^\sigma = \pm \frac{1}{2} S_{ij} \left( dE_{l_i} + dE_{l_j} \right),$$

where the sign in the equation depends on the spin.

The spin splitting $dE_l$ of shell $l$ is calculated from the spin-dependent Mulliken populations $\mu_l$ of each shell at the local site as

$$dE_l = \sum_{l' \in \mu_l} W_{ll'} \left( m_{l'\uparrow} - m_{l'\downarrow} \right).$$

### 4.3.4.1 Onsite spin-split parameters

The shell-dependent spin splitting strength $W_{ll'}$ can be obtained from a spin-polarized atomic calculation,

$$W_{ll'} = \frac{1}{2} \left( \frac{d\varepsilon_{l\uparrow}}{dm_{l'\uparrow}} - \frac{d\varepsilon_{l\uparrow}}{dm_{l'\downarrow}} \right).$$

Since $W_{ll'}$ enters symmetrically in the Hamiltonian, it is convenient to symmetrize it

$$\bar{W}_{ll'} = \frac{1}{2}(W_{ll'} + W_{l'l}).$$

ATK provides a database for $\bar{W}_{ll'}$.

## 4.3.5 Tight-binding total energy

The calculation of the total energy follows [24] and [6]. The total energy has five terms:

$$E = E_{H^0} + E_{\delta H} + E_{ext} + E_{spin} + E_{pp}.$$

The terms in the equation are

- $E_{H^0}$ is the one-electron energy of the non-self-consistent Hamiltonian, given by

$$E_{H^0} = \sum_{ij} D_{ij} H_{ij}^0$$

- $E_{\delta H}$ is the electrostatic difference energy,

$$E_{\delta H} = \frac{1}{2} \int V_0^H(\mathbf{r}) \delta n(\mathbf{r}) d\mathbf{r}$$

- $E_{ext}$ is the electrostatic interaction between the electrons and an external field.

$$E_{ext} = \int V^{ext}(\mathbf{r}) \delta n(\mathbf{r}) d\mathbf{r}$$

- $E_{spin}$ is the spin polarization energy

$$E_{spin} = -\frac{1}{2} \sum_{\mu} \sum_{l \in \mu} \sum_{l' \in \mu} W(Z_\mu, l, l') m_l m_{l'}$$

- $E_{pp}$ is the repulsive energy from a pair-potential between each atom pair, $V^{pp}(Z_\mu, Z_{\mu'}, R_{\mu,\mu'})$.

$$E_{\mathrm{pp}} = \sum_{\mu < \mu'} V^{\mathrm{pp}}(Z_{\mu}, Z_{\mu'}, R_{\mu,\mu'})$$

It is optional to add this term to the tight-binding model, it does not affect the electronic structure. The tight-binding model will, however, not give sensible geometries without a repulsive pair-potential.

### 4.3.6 Parameters for the Slater-Koster method

The Slater-Koster parameters can be provided either through the **SlaterKosterTable** class or through various 3. party formats. The supported 3. party formats are the slater-koster files from the DFTB consortium and the slater-koster files from the Hotbit consortium.

To use parameters in the DFTB or Hotbit format, put the parameter files in a single directory, and setup the basis set and pair potentials using the functions **DFTBDirectory** or**HotbitDirectory**.

### 4.3.7 Shipped DFTB and Hotbit Parameters

The current version of ATK-SE is shipped with a number of DFTB style parameters from the CP2K consortium and the hotbit consortium. It is most easy to setup these basis sets using the Virtual NanoLab (VNL)..

### 4.3.8 Shipped Slater-Koster Table parameters

A number of orthogonal tight-binding parameters are provided. The parameters are from Vogl et. al and Jancu et. al. It is most easy to setup these basis sets using the Virtual NanoLab (VNL).

## 4.3.9 Parameters for the extended Huckel method

The parameters $\eta_1, \eta_2, C_1, C_2,$ and $E$ must be defined for each valence orbital, while $\beta$ and $U$ only depend on the element type. Different parameter sets are provided with ATK-SE, but it is also possible to provide user-defined parameters in the input file using the **HuckelBasisParameters** class.

The tables below provide a mapping between the symbols in the equations and the corresponding keywords.

**Table 4.1**: HuckelBasisParameters

| Symbol | **HuckelBasisParameters** |
|--------|--------------------------|
| $E_i$ | ionization_potential |
| $\beta$ | wolfsberg_helmholtz_constant |
| $U$ | onsite_hartree_shift |
| $W$ | onsite_spin_split |
| $E^{\text{VAC}}$ | vacuum_level |

**Table 4.2**: SlaterOrbital parameters

| Symbol | **SlaterOrbital parameters** |
|--------|------------------------------|
| $n$ | principal_quantum_number |
| $l$ | angular_momentum |
| $\eta$ | slater_coefficients |
| $C$ | weights |

The current version of ATK comes with built-in Hoffmann and Müller parameter sets, which are appropriate for organic molecules. For crystalline structures, both metals and organic materials like graphene, parameters from J. Cerda are provided.

To combine parameters from different sources, it is important to make sure they use the same energy zero level, in order to obtain correct charge transfers. This can be obtained by ensuring that the crystals have the correct work function and molecules the correct ionisation energies. For this purpose, an additional parameter $E^{\mathrm{VAC}}$ is introduced, which shifts the energy of the vacuum level. I.e. if a calculation with $E^{\mathrm{VAC}} = 0$ eV has a work function of 6.5 eV, then by setting $E^{\mathrm{VAC}} = -1.5$ eV all bands shift rigidly upwards by 1.5 eV and the work function becomes 5.0 eV.

**Note**

The Hückel parameters have been fitted for non-self-consistent calculations. To use the parameters in self-consistent calculations, the self-consistent onsite shifts must be compensated by a reverse shift of the vacuum_levels.

## 4.3.10 The Hartree Potential

The Hartree potential is defined as the electrostatic potential from the electron charge density and must be calculated from the Poisson equation

$$\nabla^2 V^H[n](\mathbf{r}) = -4\pi n(\mathbf{r}),$$

The Poisson equation is a second-order differential equation and a boundary condition is required in order to fix the solution. Molecular systems have the boundary condition that the potential asymptotically goes to zero. In bulk systems, the boundary condition is that the potential is periodic.

**Note**

Periodic boundary conditions only determine the Hartree potential up to an additive constant, reflecting the physics that the bulk electrostatic potential does not have a fixed value relative to the vacuum level. Experimentally this can be measured, through the different work functions of different facets of a crystal.

**Solving the Poisson equation using Fourier transform**

For systems with periodic boundary conditions and no dielectric and metallic regions, the Poisson's equation can be solved using a **<u>FastFourierSolver</u>** . The **<u>FastFourierSolver</u>** is the most efficient solver within the ATK package.

**Solving the Poisson equation with a multi-grid solver**

For general systems, the Poisson equation is solved using an algebraic **<u>MultigridSolver</u>**. The system is enclosed in a bounding box, and the Hartree potential is defined on a regular grid inside the bounding box. Different boundary conditions can be imposed on the solution at the bounding box surface.

**DirichletBoundaryCondition**

The Hartree potential is zero at the boundary.

**NeumannBoundaryCondition**

The negative gradient of the Hartree potential, e.g. the electric field, is zero at the boundary.

**PeriodicBoundaryCondition**

The potential has identical values on opposite faced boundaries.

**MultipoleBoundaryCondition**

The potential at the boundary is determined by calculating the monopole, dipole and quadrupole moments of the charge distribution inside the box, and using these moments to extrapolate the value of the electro-static potential at the boundary of the box

It is possible to include an electro-static interaction with a continuum metallic or dielectric material inside the bounding box. The continuum metals are handled by constraining the Hartree potential within the metallic region to a fixed value. Dielectric materials are handled by introducing a spatially dependent dielectric constant, $\epsilon(\mathbf{r})$, where $\epsilon(\mathbf{r}) = \epsilon_K$ inside the dielectric material with dielectric constant $\epsilon_K$, and $\epsilon(\mathbf{r}) = \epsilon_0$ outside the dielectric materials

It is possible to perform calculations of solvents. In this case, the volume of the configuration is defined by inscribing each atom in a sphere with a size given by the van der Waals radius of the element. Inside the volume of the configuration the dielectric constant is 1; outside the volume of the configuration the dielectric constant is equal to the value of **solvent_dielectric_constant**.

# 5. Simulation of Silicon Nanowire FET with Silicon Dioxide as Gate Dielectric

In this chapter, the modeling and simulation of silicon nanowire field effect transistor with silicon dioxide as gate dielectric has been presented. The tool used is Quantumwise ATK ver.13.1 and the calculator used in ATK-SE: Extended Hückel (Device). The device geometry consisting of the thickness of the metal gate and the gate dielectric layers and diameters of the two cylinders (gate dielectric and metal gate) considering the FET has a gate wrap around structure, left and right electrode lengths vary from the dimensions used to carry out the simulation presented in the previous chapter.

## 5.1 Simulation Settings

The nanowire transistor simulated using silicon dioxide as gate dielectric has a silicon nanowire oriented in (100) direction. The nanowire is essentially a single crystal structure in Face Centred Cubic orientation. The two ends of the nanowire are doped to obtain a doping concentration of $4 \times 10^9$ $cm^{-3}$ in the source and drain regions, as the electrode is about 1 nm long and has a cross section of (0.5 nm)x(0.5 nm). As the nanowire transistor simulated has a Gate All Around structure to obtain better control over the carriers in the channel, the gate dielectric layer and the metal gate are essentially cylindrical in shape to wrap over the silicon nanowire, unlike conventional planar MOSFETs where the gate dielectric and the metal gate are just material stacks sitting over the channel area on the substrate. The thickness of the gate dielectric (silicon dioxide) used in this design is 2 Å while the inner radius is 5 Å. The thickness of the metal gate is 3 Å while the inner radius is 7 Å.

The steps involved in modeling the nanowire FET have been presented below in the form of screen shots from the tool:

Step 1 – This step shows the snapshot of the tool once it is invoked.

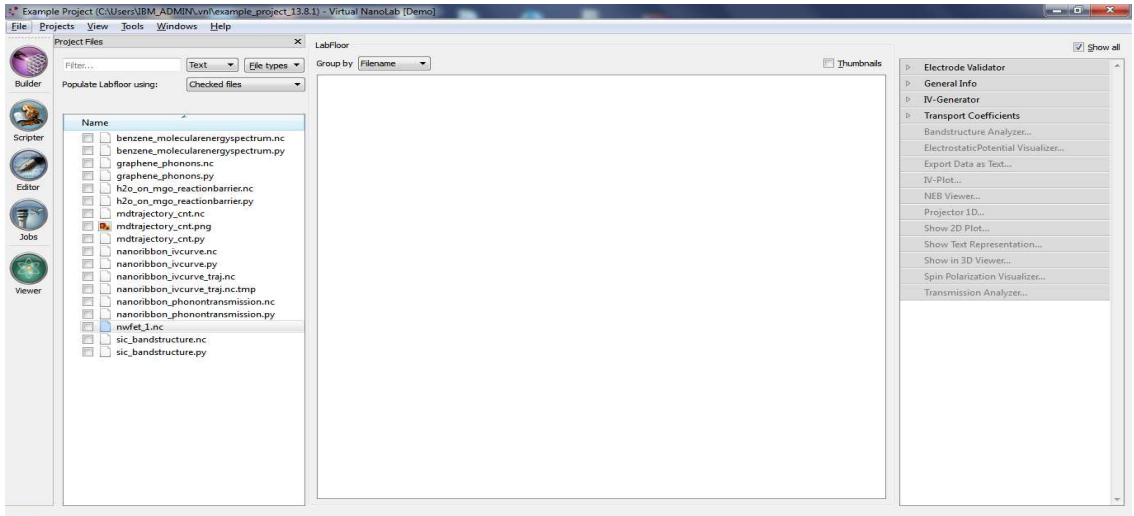

Step 2 – When the Builder tab on the left hand side is clicked, it gives the screen as shown in the snapshot below.

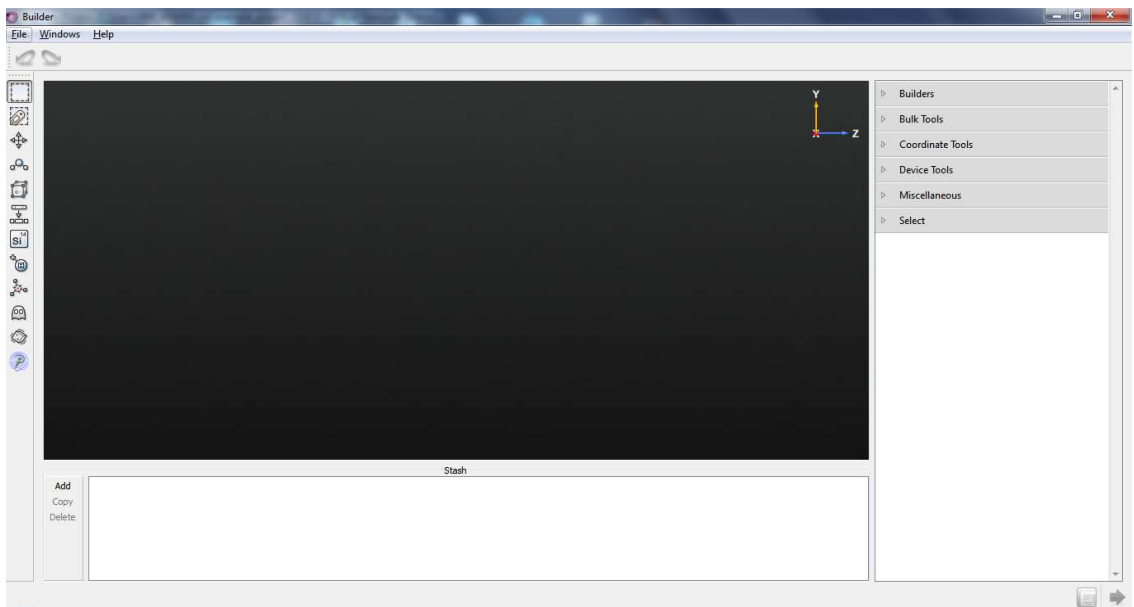

Step 3 – Click on "Add" that can be seen below the black screen to obtain a drop down from which a selection is to be made.

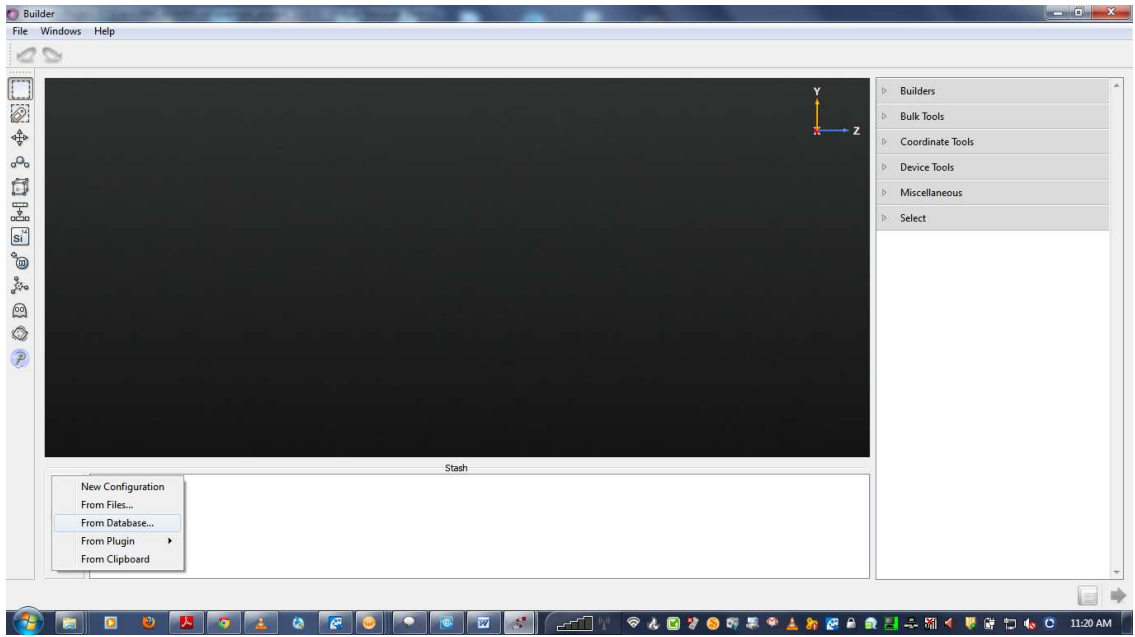

Step 4 – Out of all options, select Database. This operation implies that the designer opts to pull out the silicon atom from the material database of the tool. Once Database is selected, enter "silicon fcc" is the search space at the top of the page. This locates the silicon face centered cubic structured atom which is used to build the nanowire.

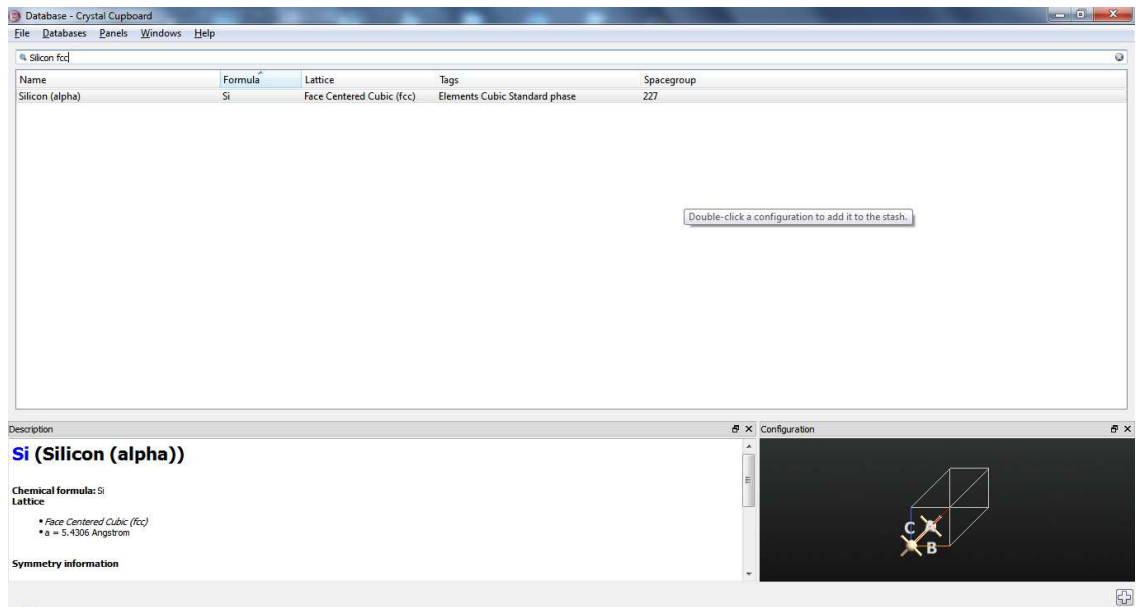

Step 5 – Once the silicon fcc is selected from database, click on Builders in the right hand side to obtain a drop down of many options.

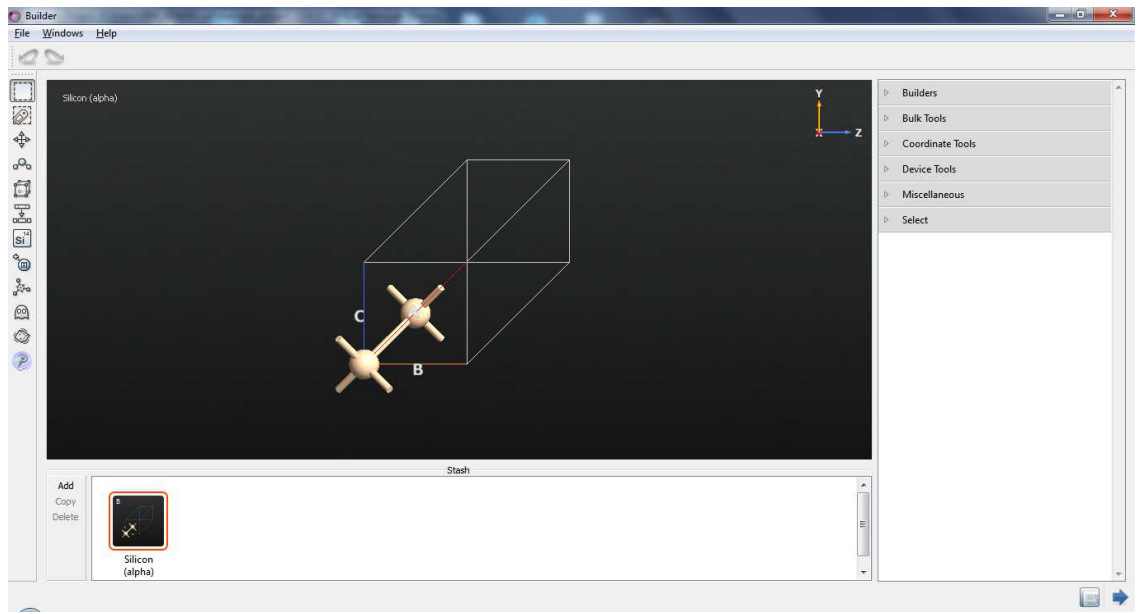

Step 6 – Click on Surface (Cleave)

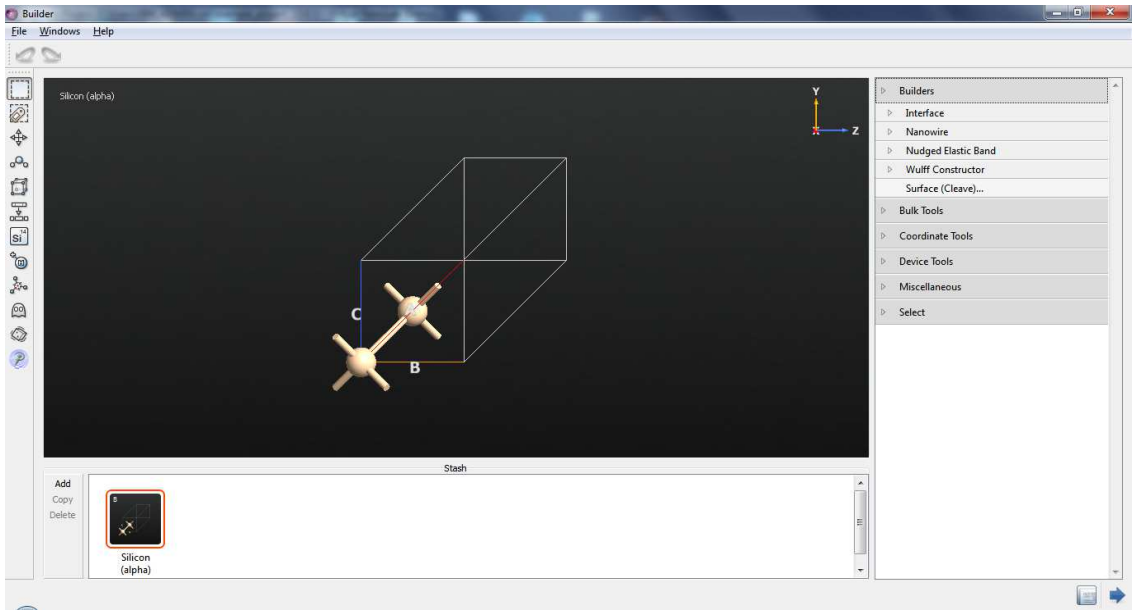

Step 7 – Select the Miller Indices and click on Next.

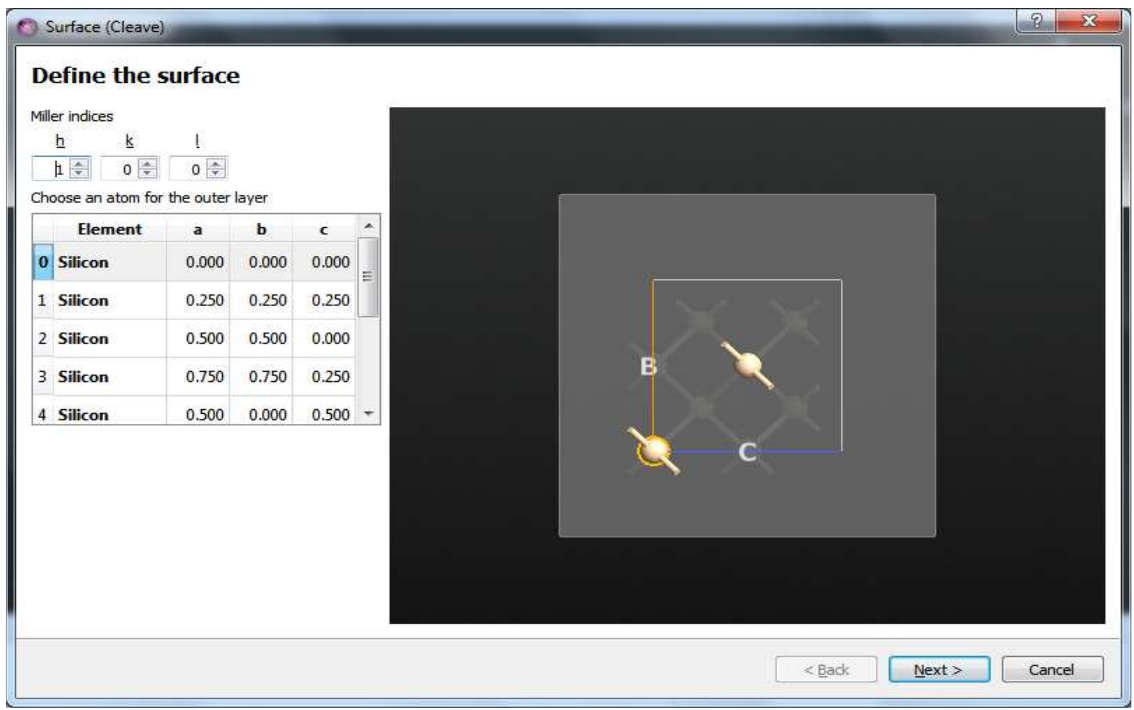

Step 8 - Once the surface lattice is defined, click on Next and then Finish.

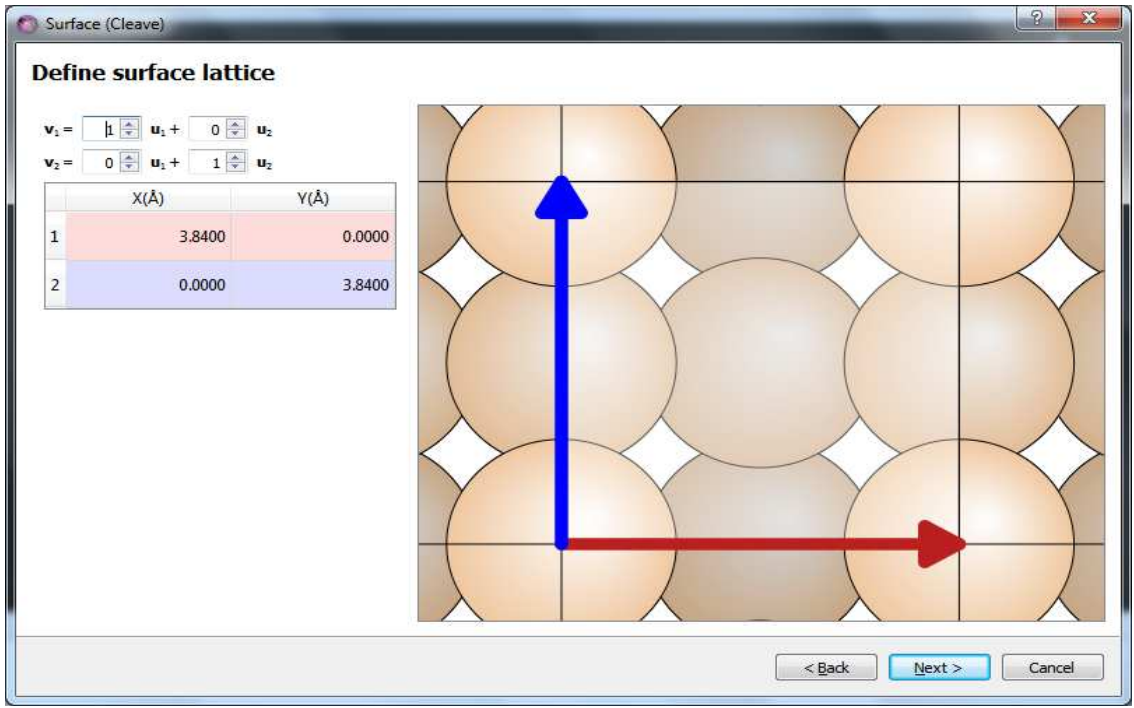

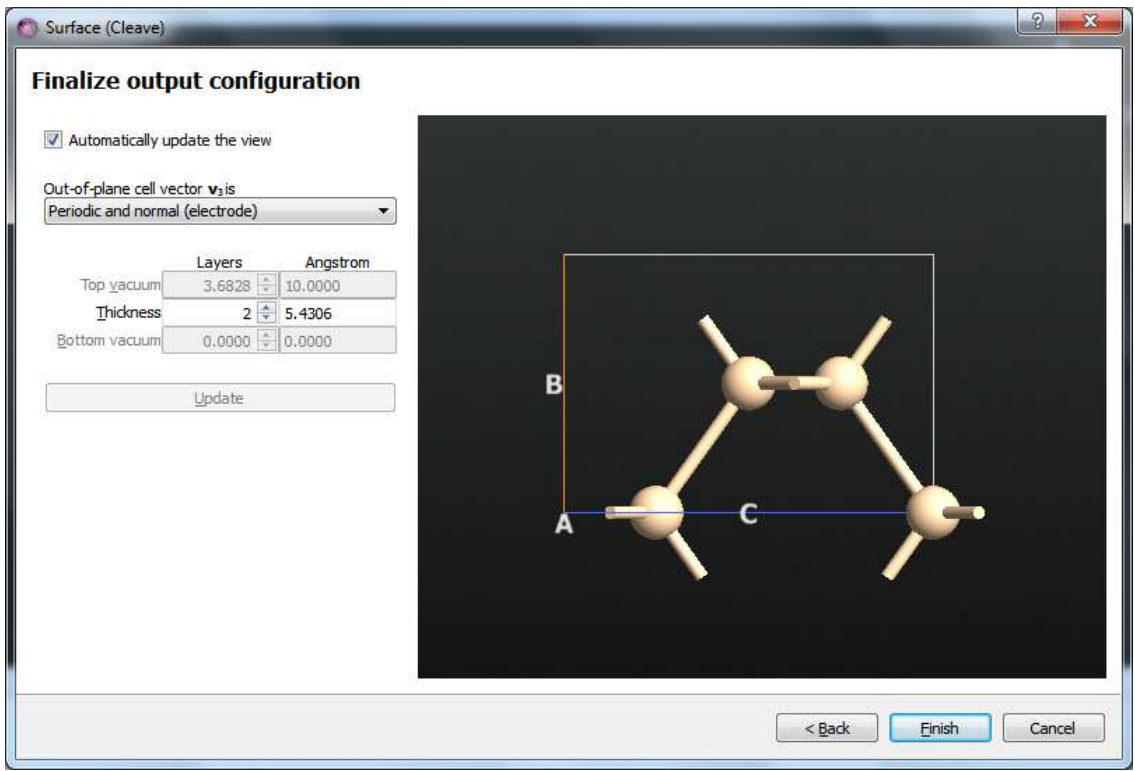

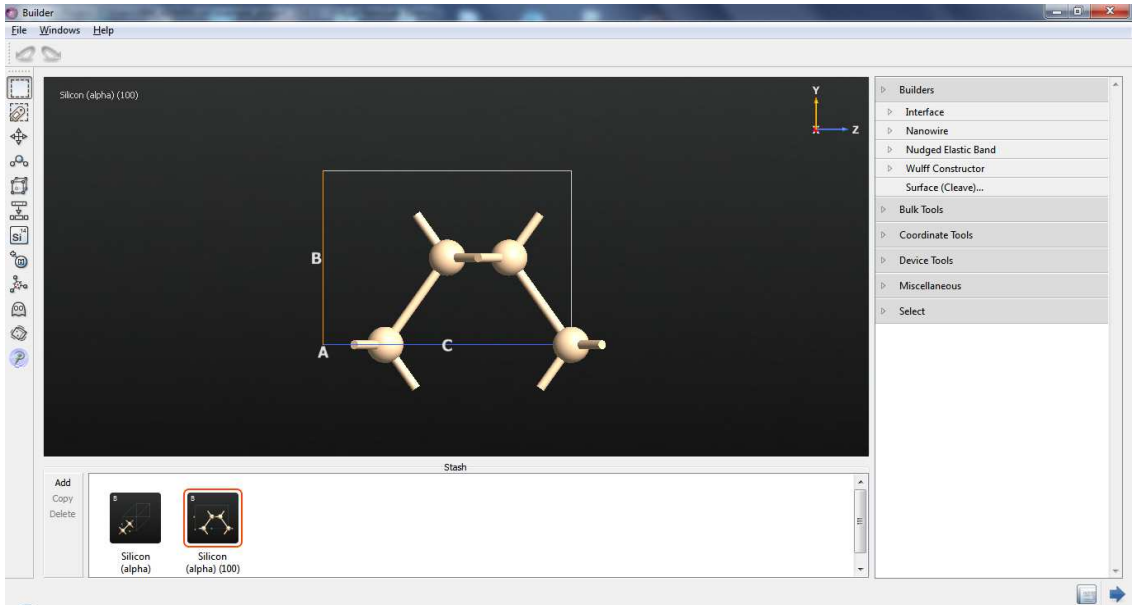

Step 9 – Click on Bulk Tools and select Repeat. Enter (A = 2, B = 2, C = 1) and click on Apply.

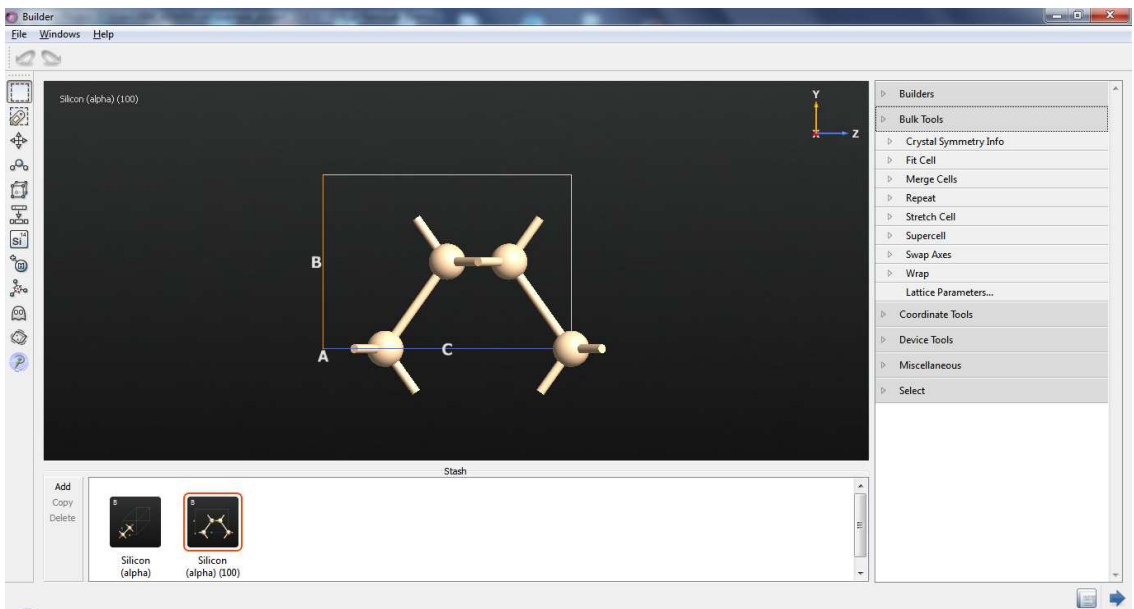

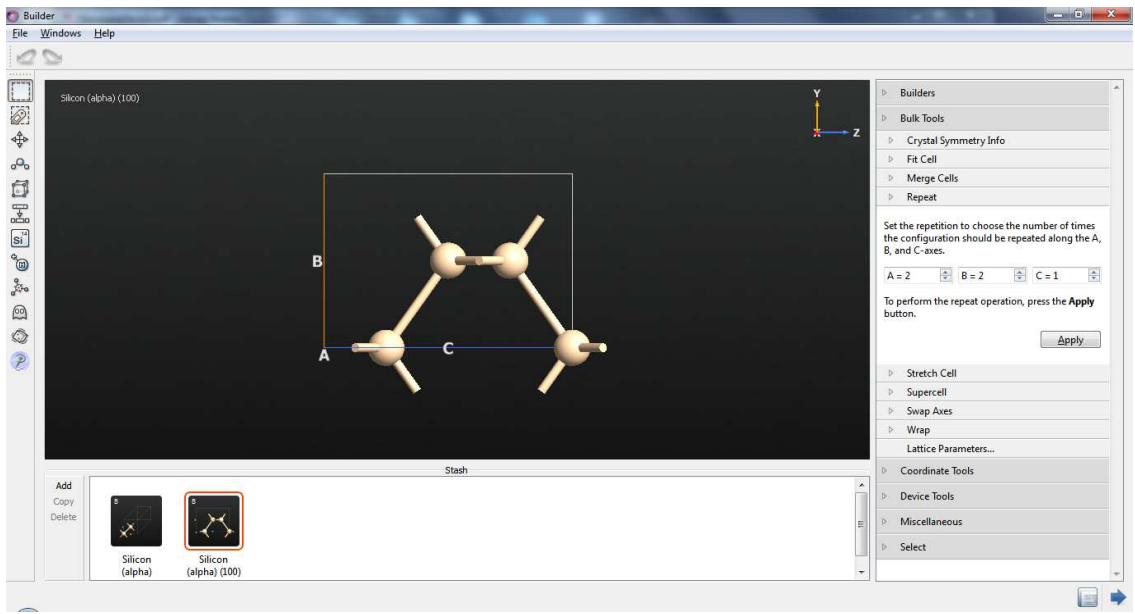

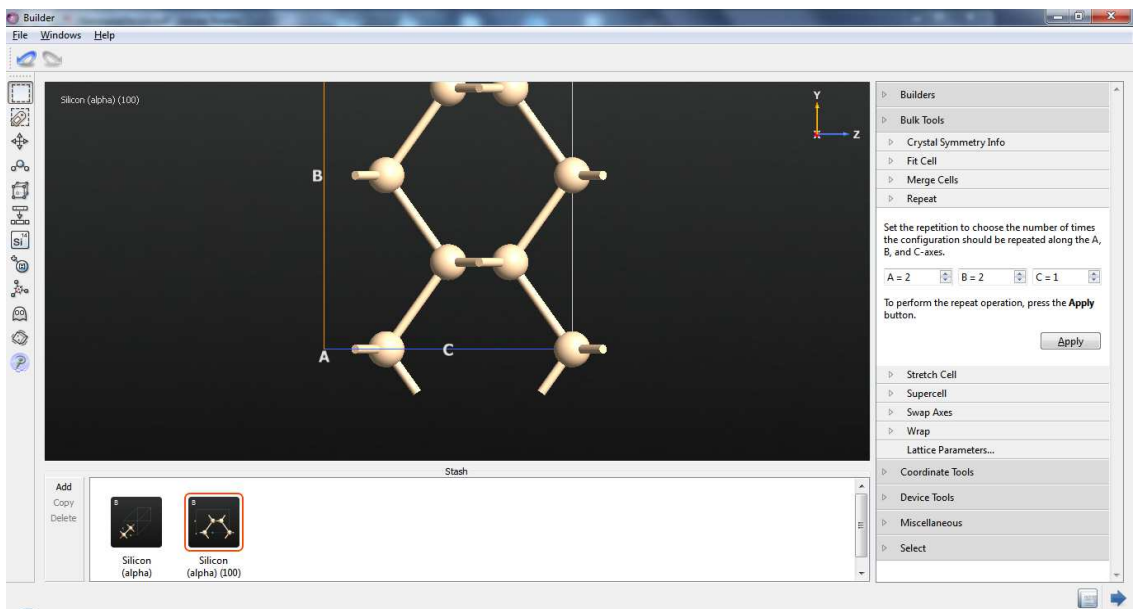

Step 10 -  Press ctrl+R to position the design to the centre of the screen.

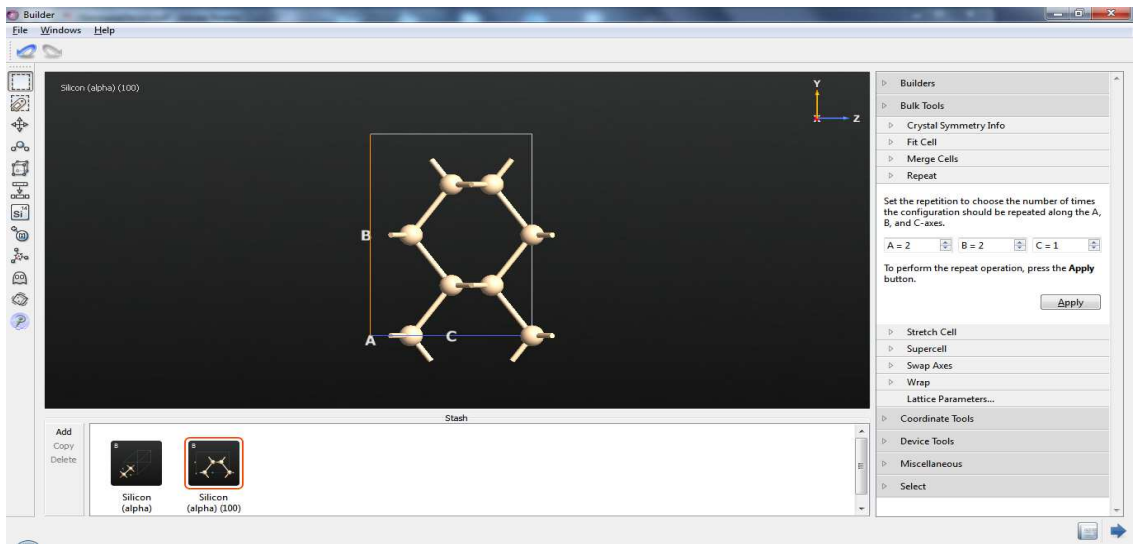

**Step 11 -** Click on Lattice Parameters on the right hand side and enter the lattice parameters as shown in the screen shot below.

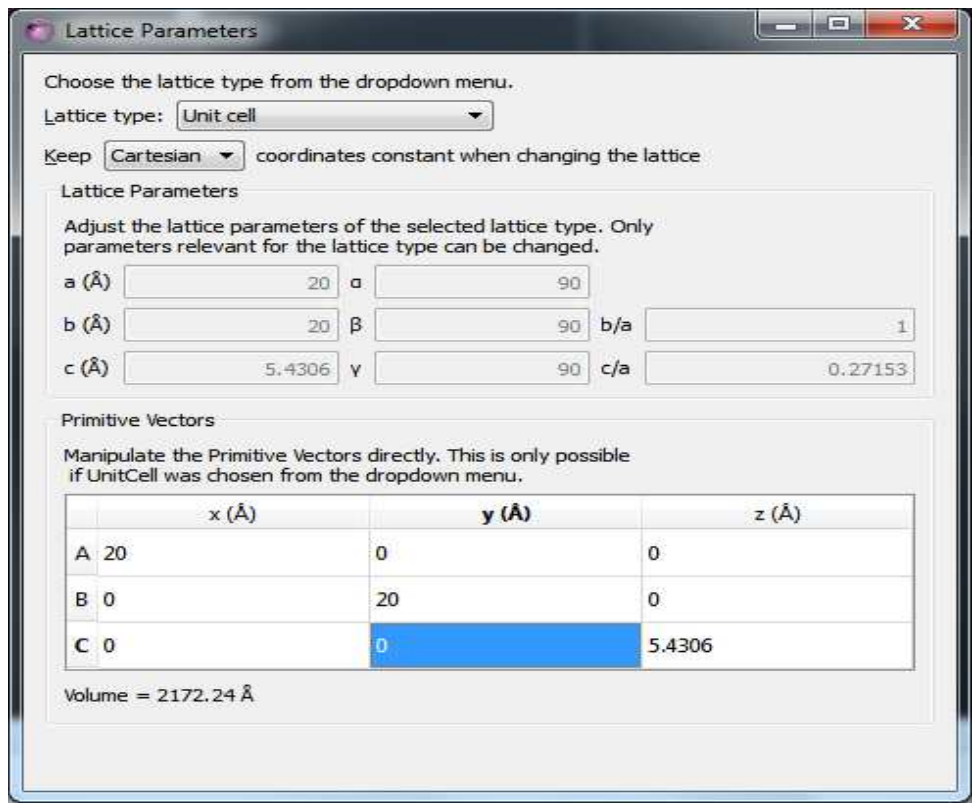

**Step 12** - Under Coordinate tools at the right hand side, click on Center to align the design properly.

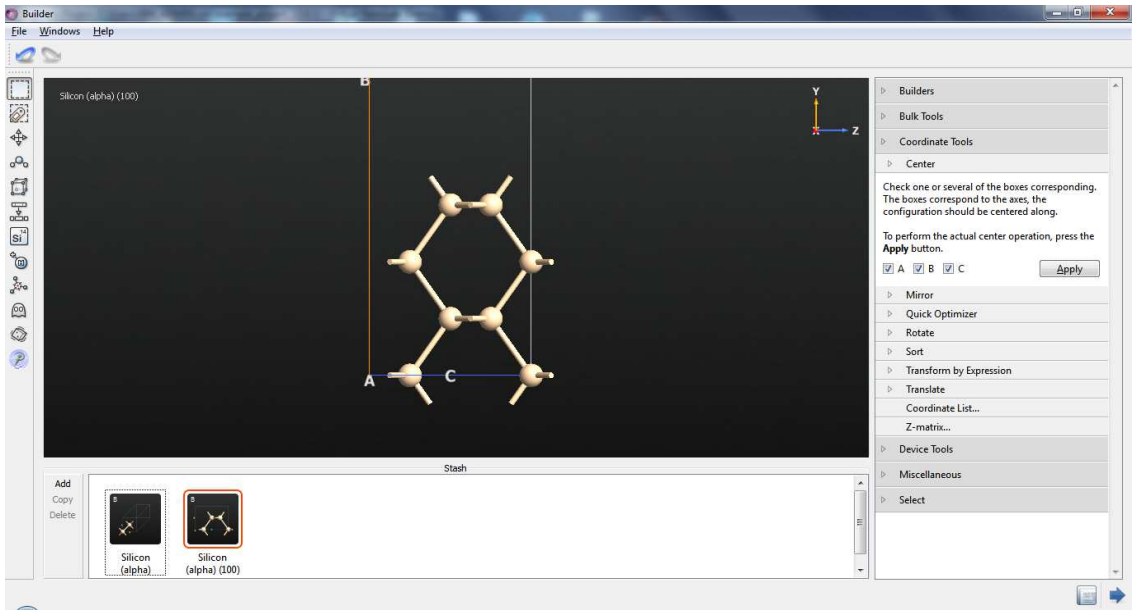

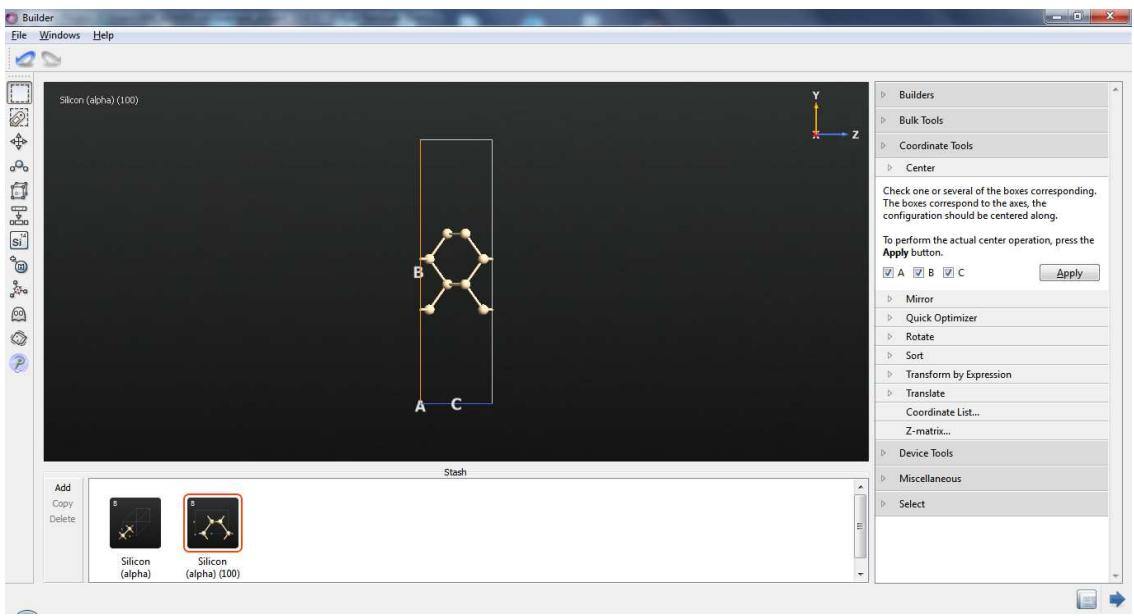

**Step 13** – Click on H-passivation on the left hand side of the screen to attach hydrogen atoms on the silicon atoms.

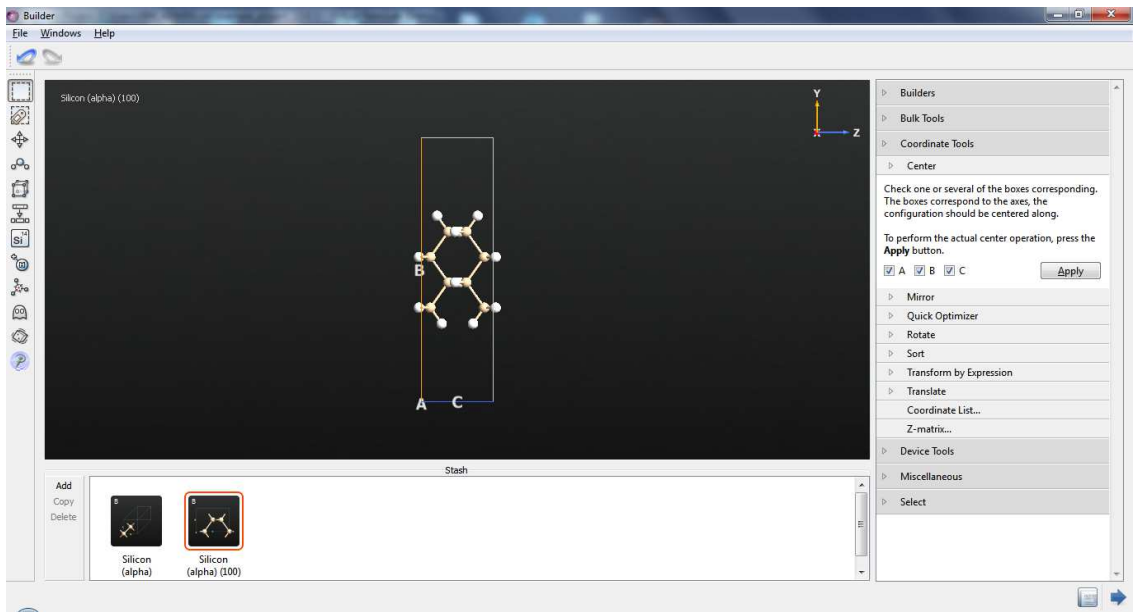

Step 14 -  On the right hand side, under Bulk Tools, click on Wrap. Check A,B and C and click on Apply. This step prepares the structure built so far to form a cylinder once the same structure is repeated.

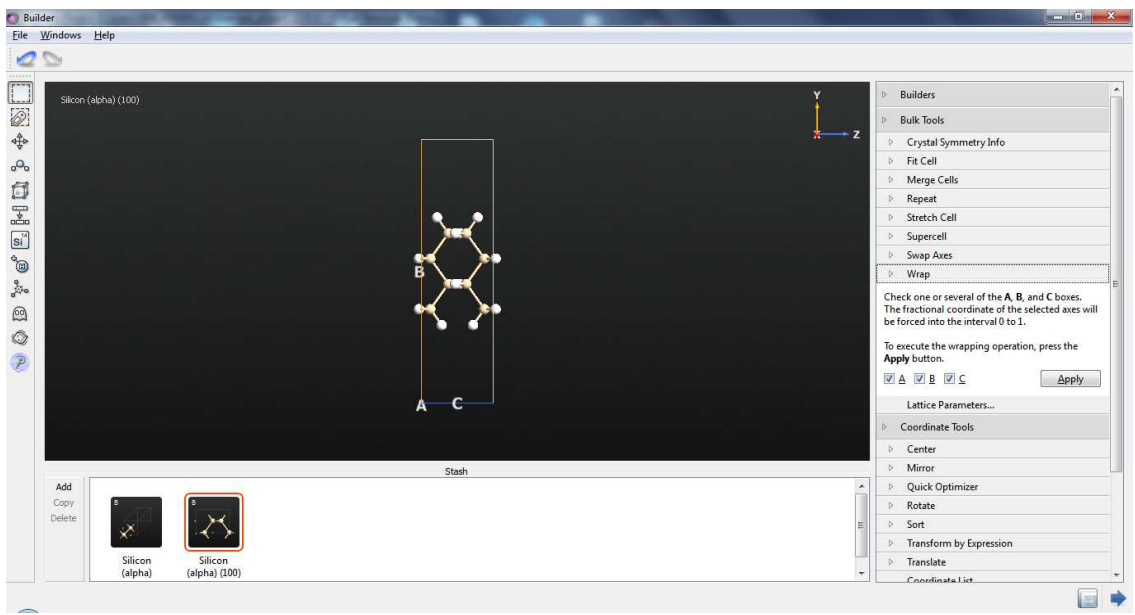

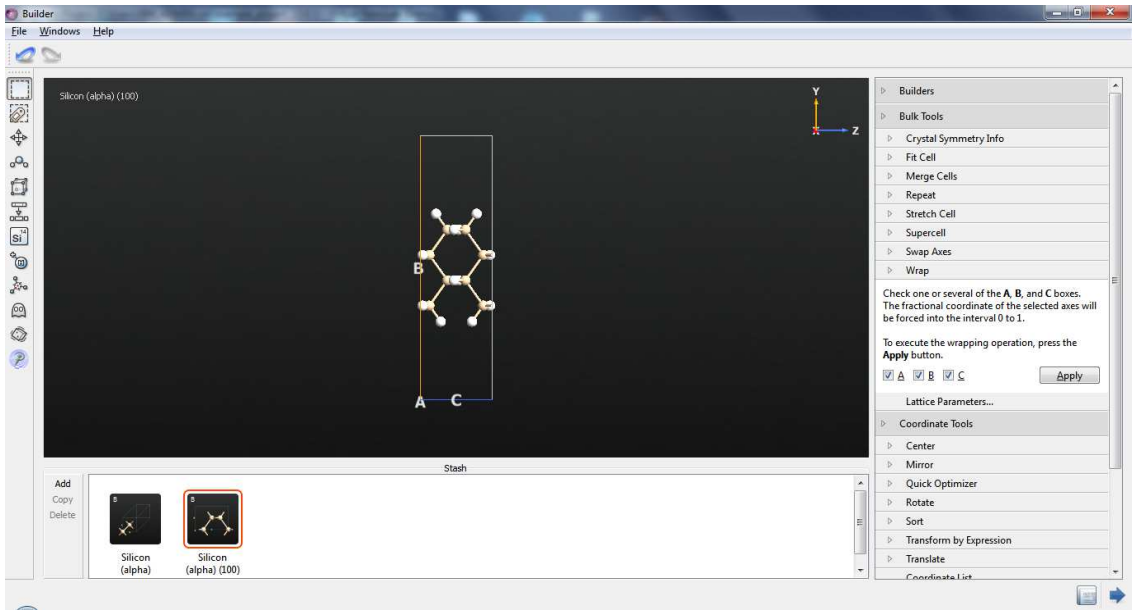

Step 15 – Under Bulk Tools, click on Repeat and enter A=1, =1 and C=12. This gives the silicon nanowire oriented in (100) direction.

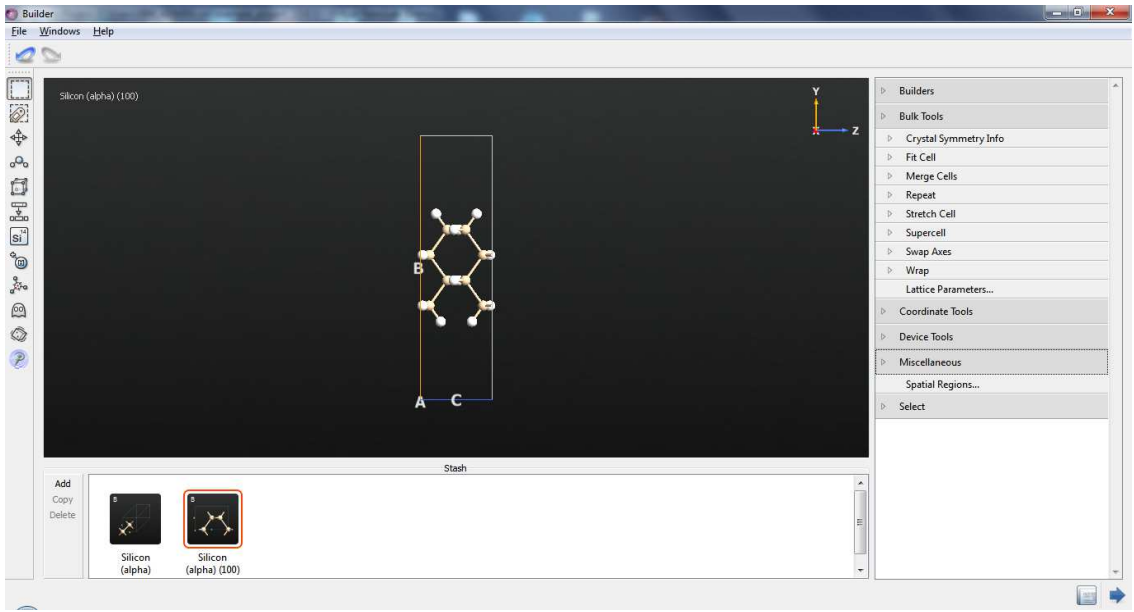

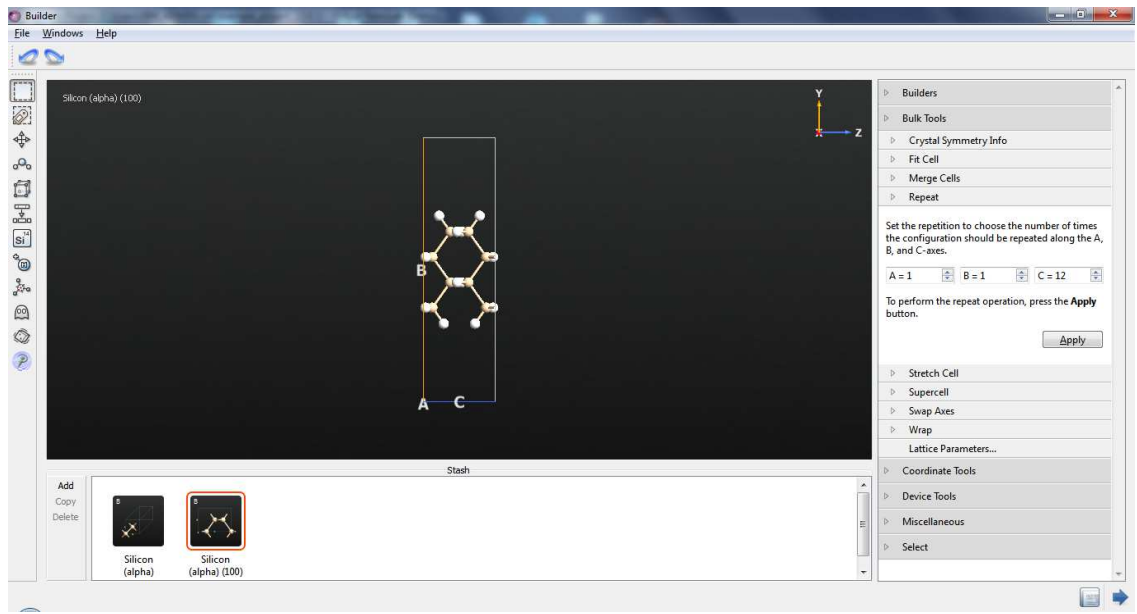

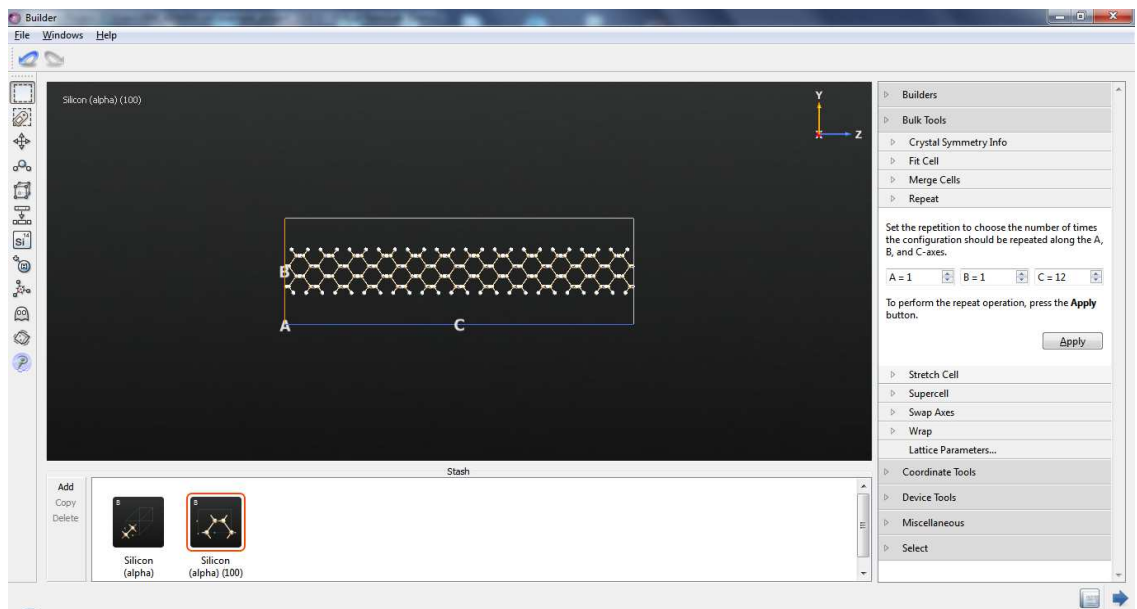

Step 16 – On the right hand side, click on Miscellaneous to obtain the screen shown below. Right click on the left side (under Region Type) and select Dielectric. Enter the dielectric constant value as 3.9. This value defines the gate dielectric to be silicon dioxide. Enter the coordinates as shown in the screen shot, together with the thickness and inner radius.

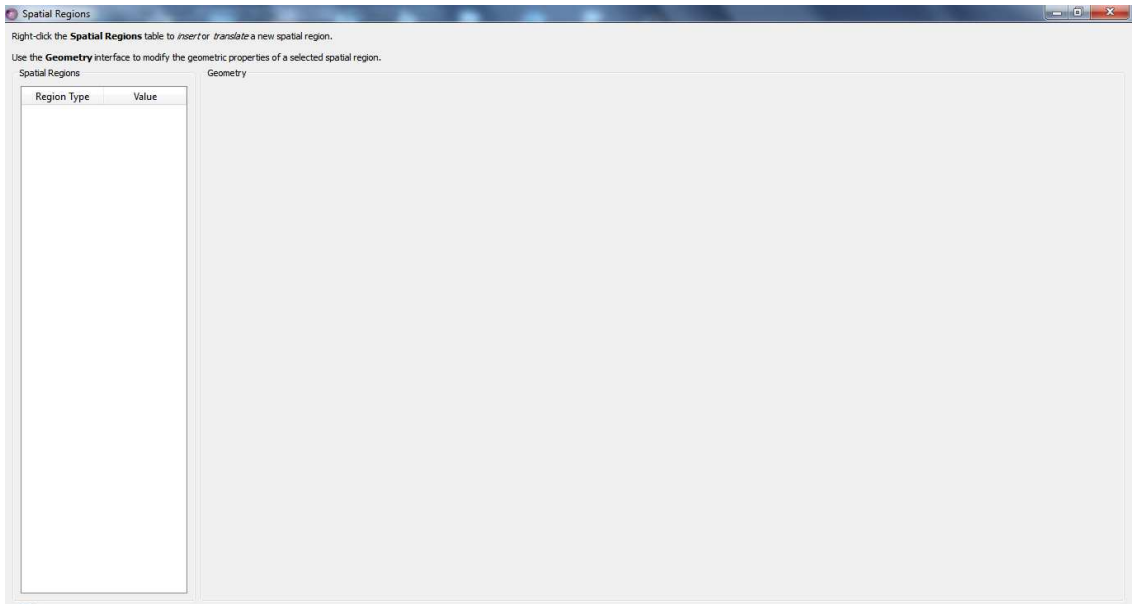

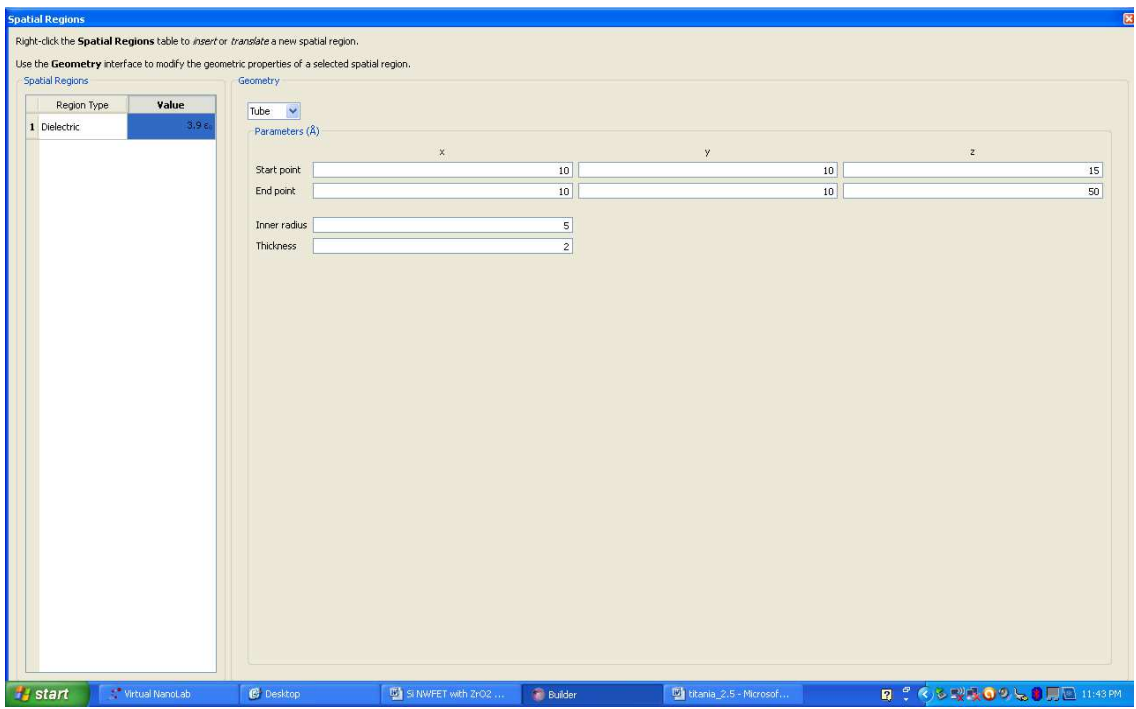

**Step 17 -** Once the dielectric is defined, the layer wraps around the nanowire as shown in the three screen shots below.

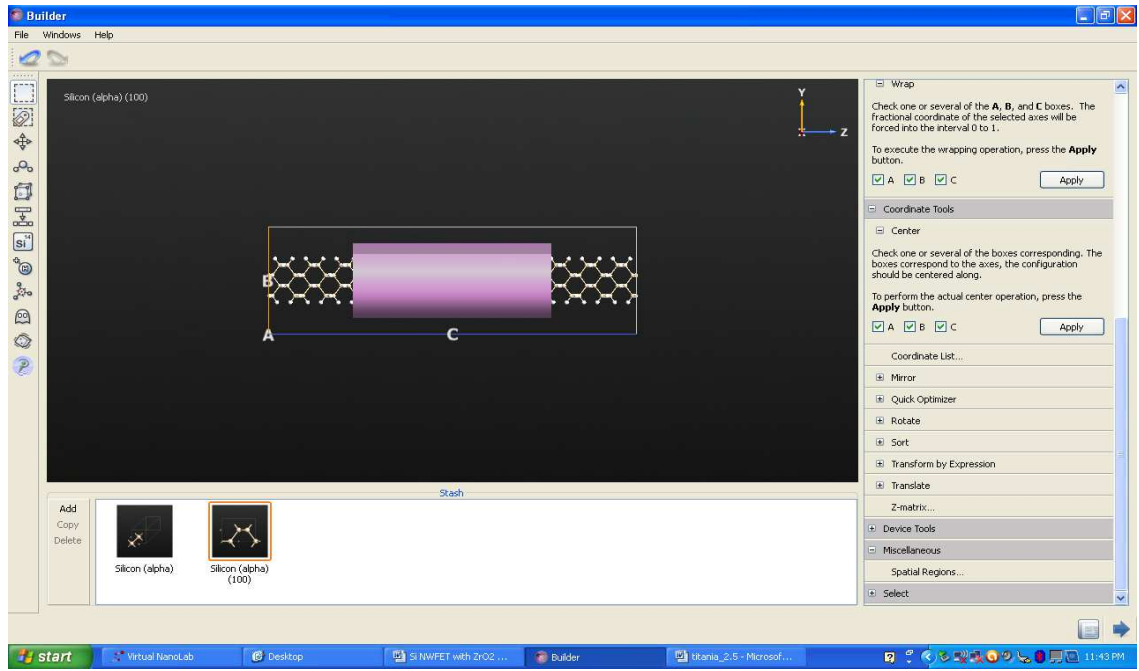

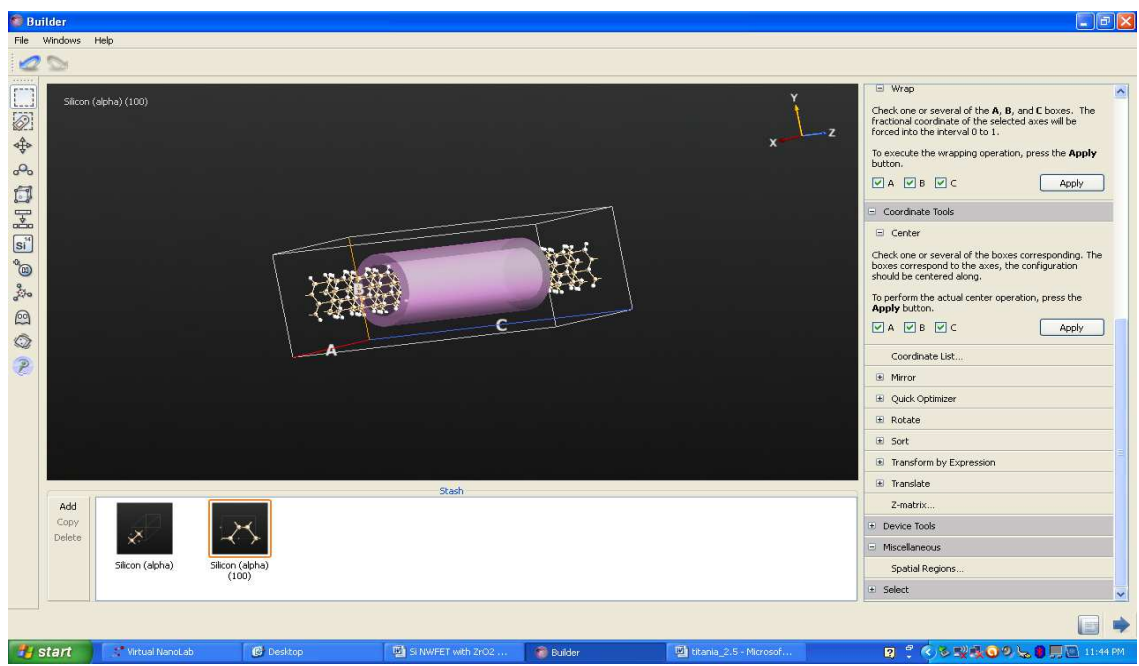

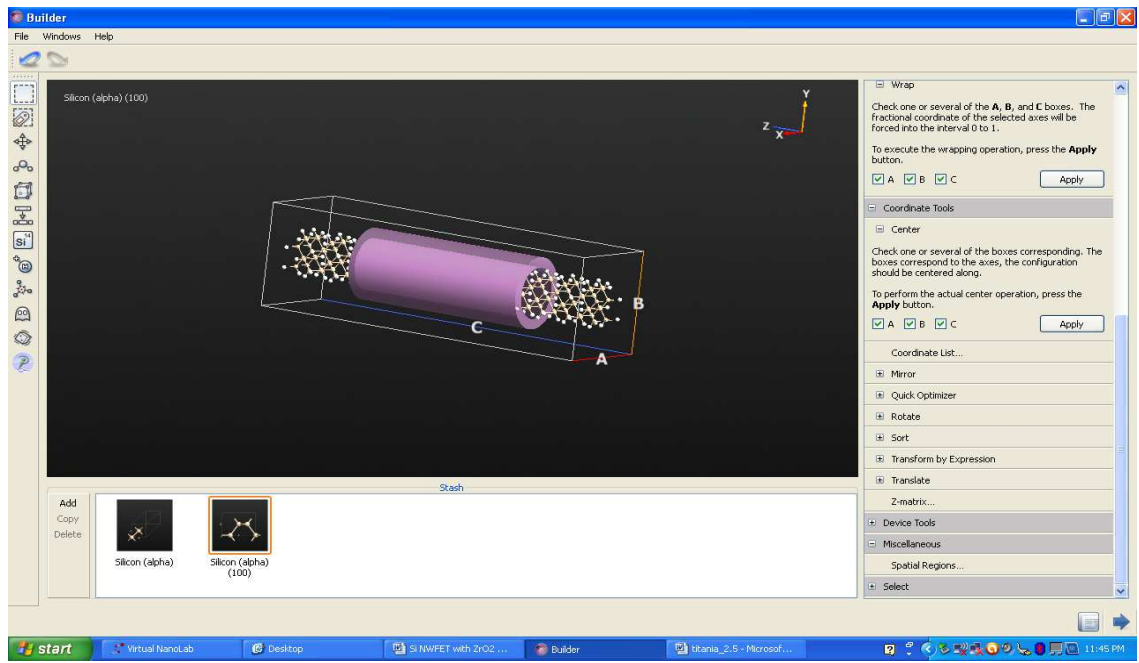

Step 18 - Repeat the same process that is used to define the gate dielectric to define the metal gate. The following screen shot shows the nanowire with the gate dielectric and the metal gate. The gate voltage is varied from 0 V to 2.5 V in steps of 0.5 V

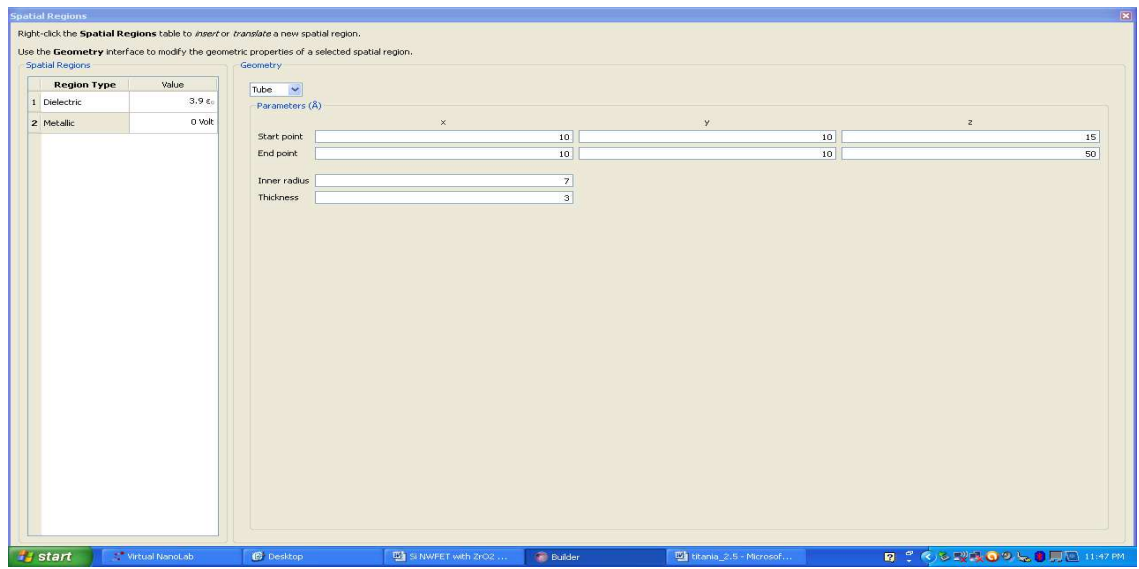

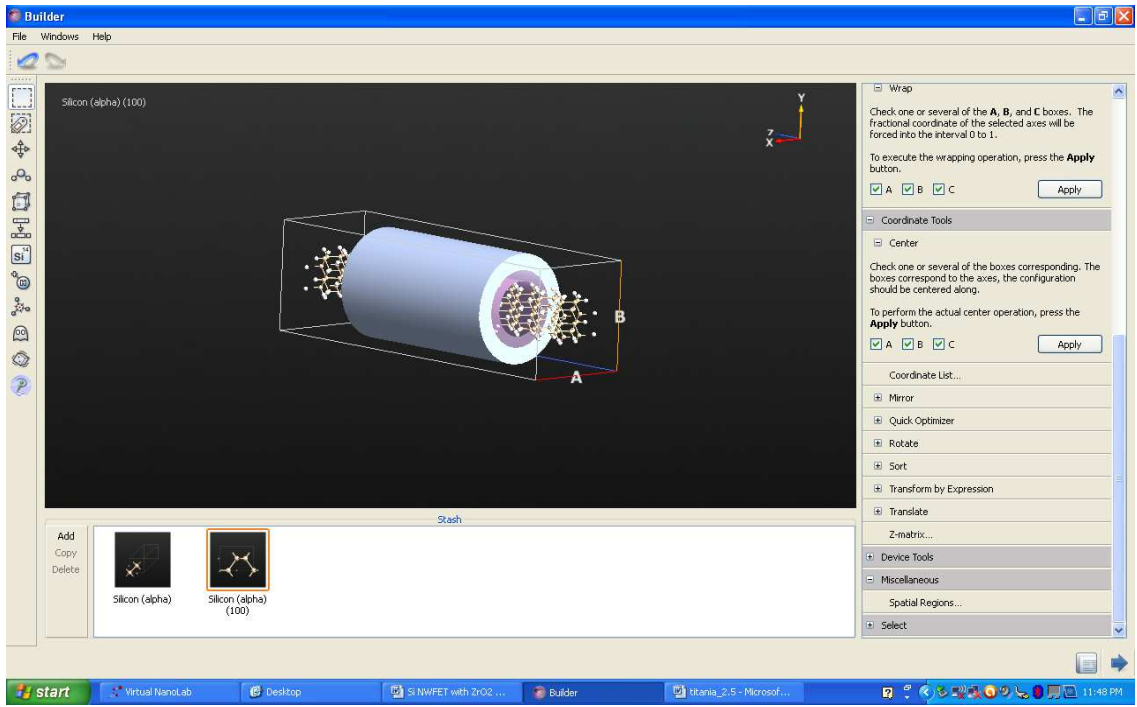

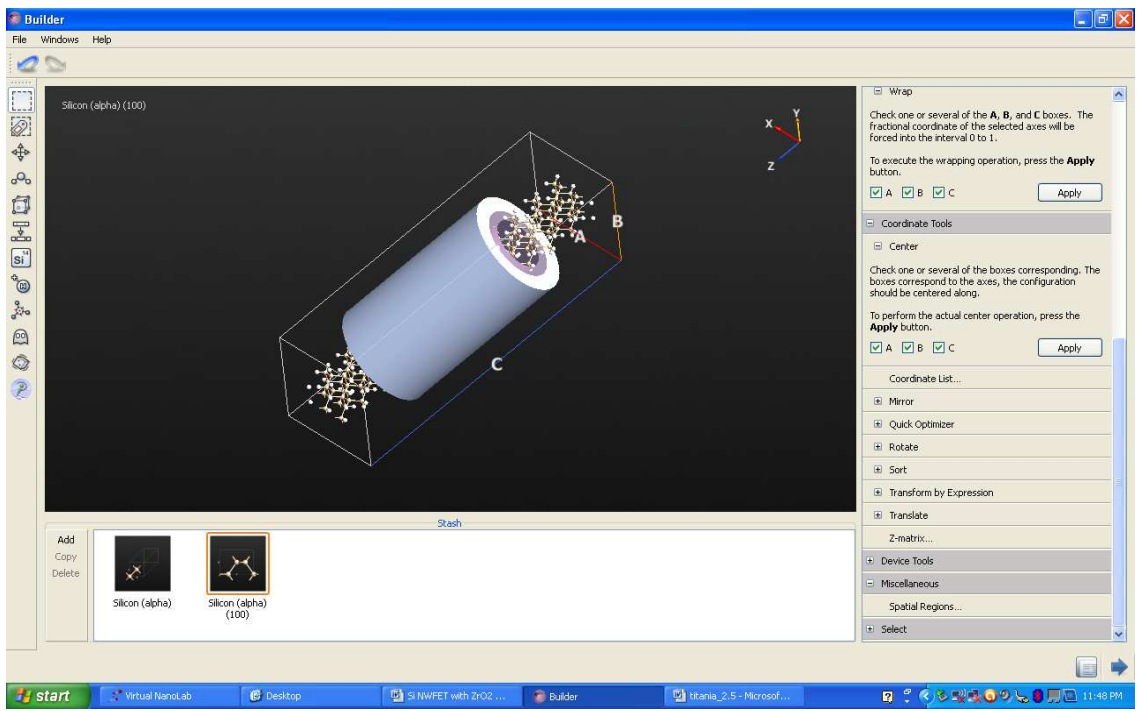

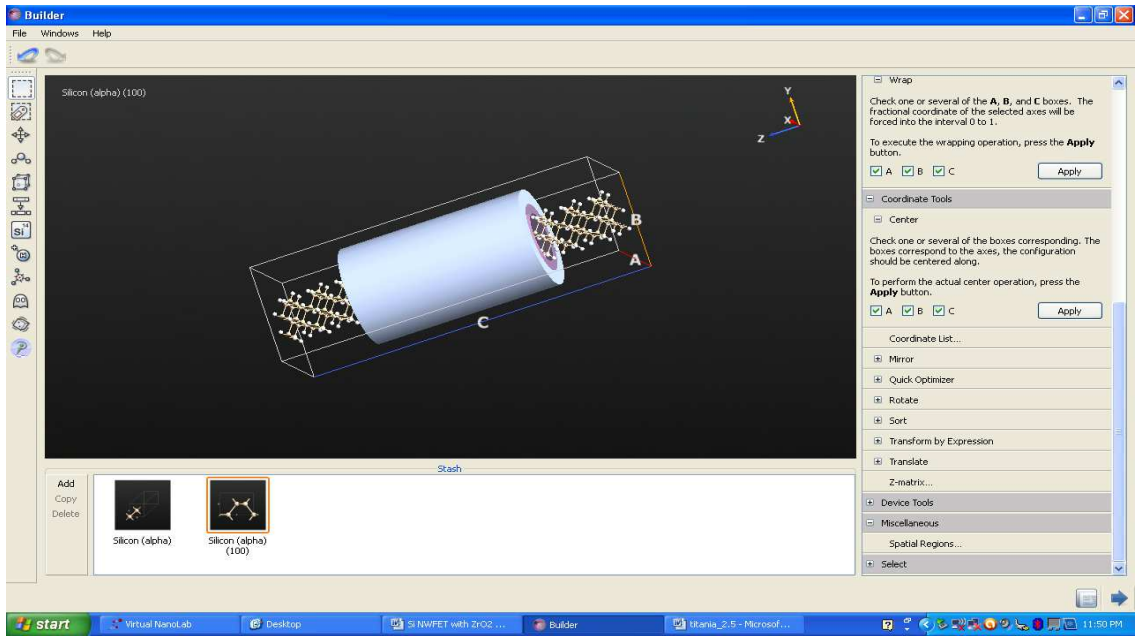

Step 19 – Click on Device Tools on the right hand side to define the electrode lengths as shown in the screen shot below. Click on OK to obtain the final device structure that is to be used for simulation.

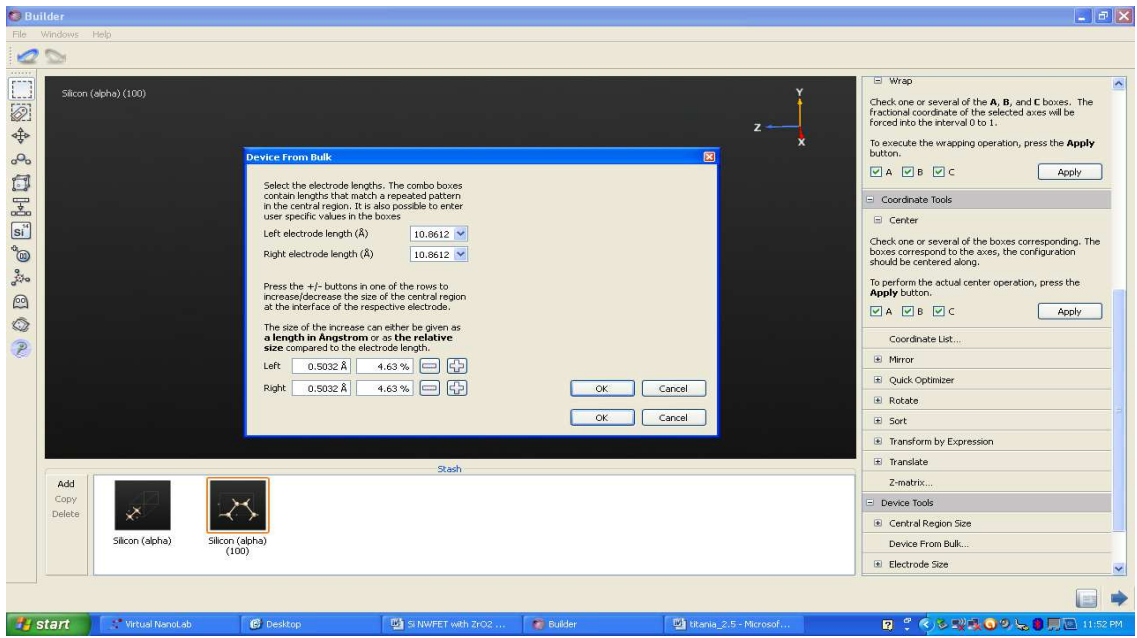

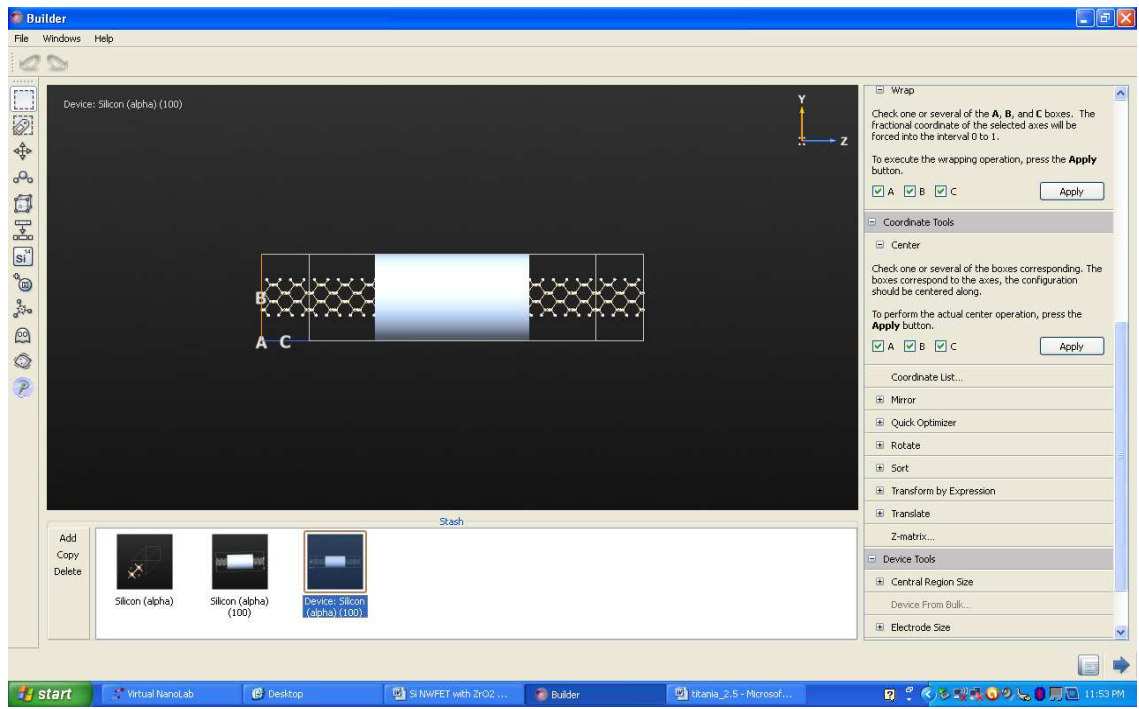

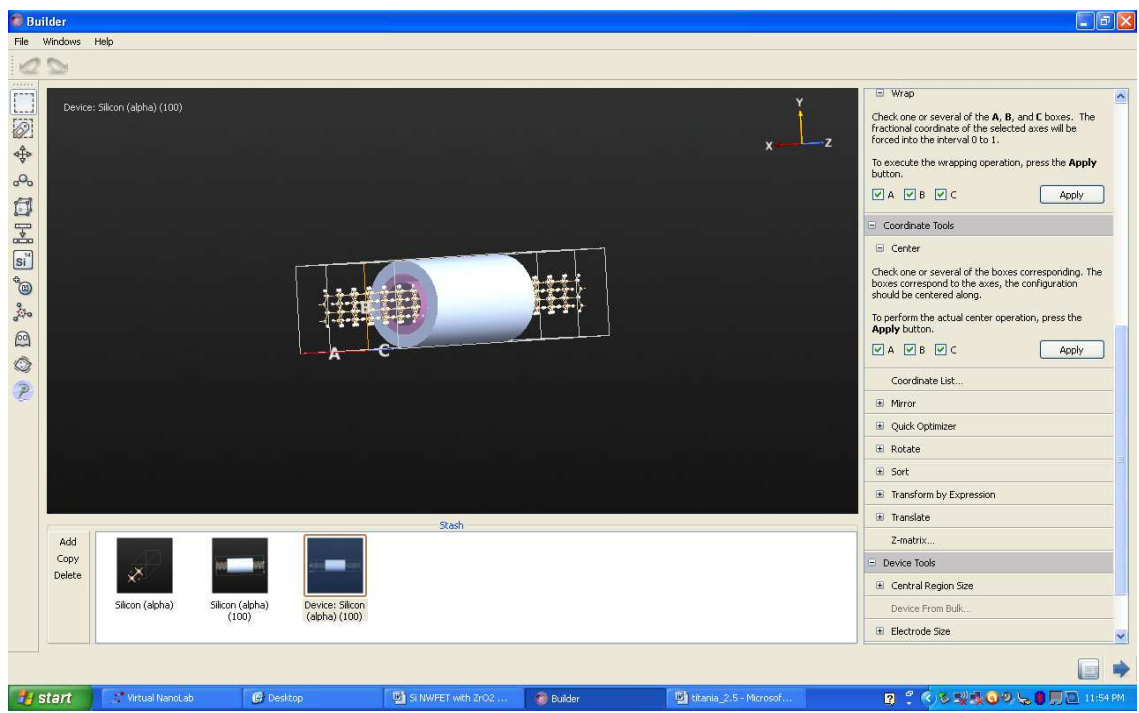

Step 20 – The built structure is then dropped to Script Generator. The options that have to be taken into account for simulating the built device can be seen in the screen shot below, as instantiated from the Blocks section on the left.

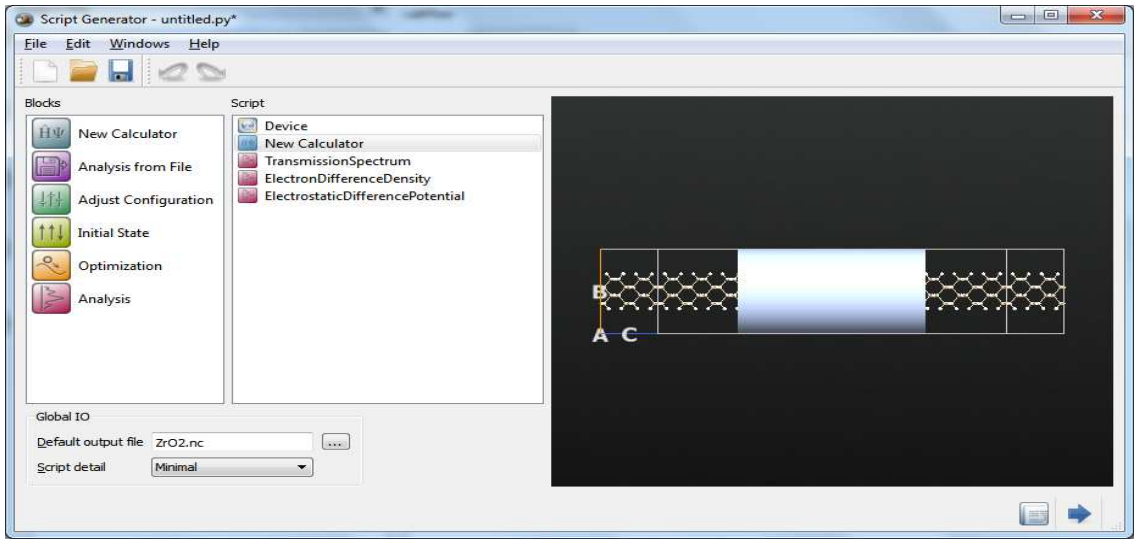

Step 21 – The calculator settings can be seen in the three screen shots below. The right electrode voltage is varied from 0 V to 2 V in steps of 0.25 V. The gate voltage is anyways defined in Step 18.

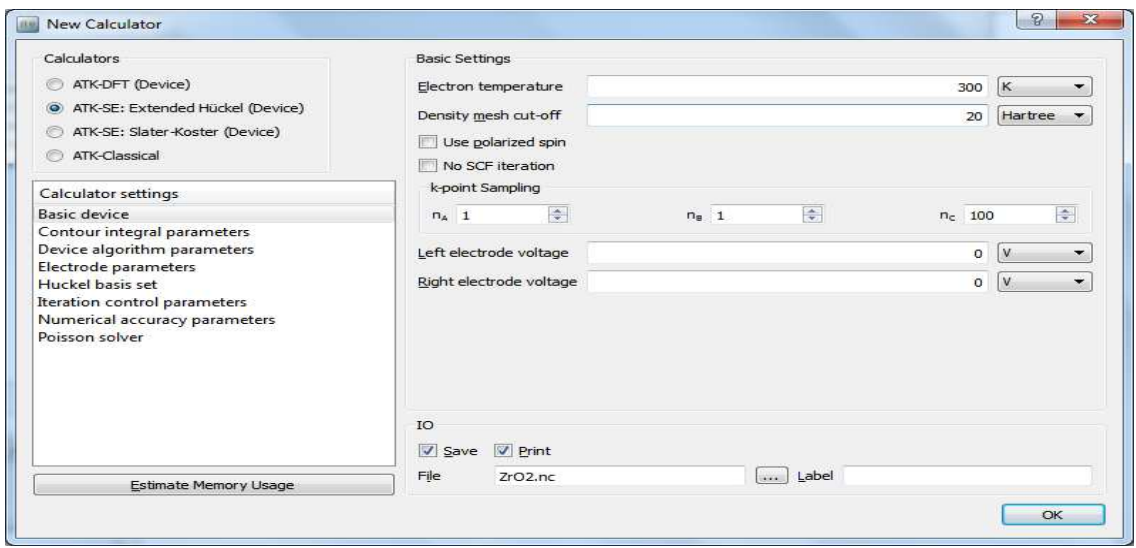

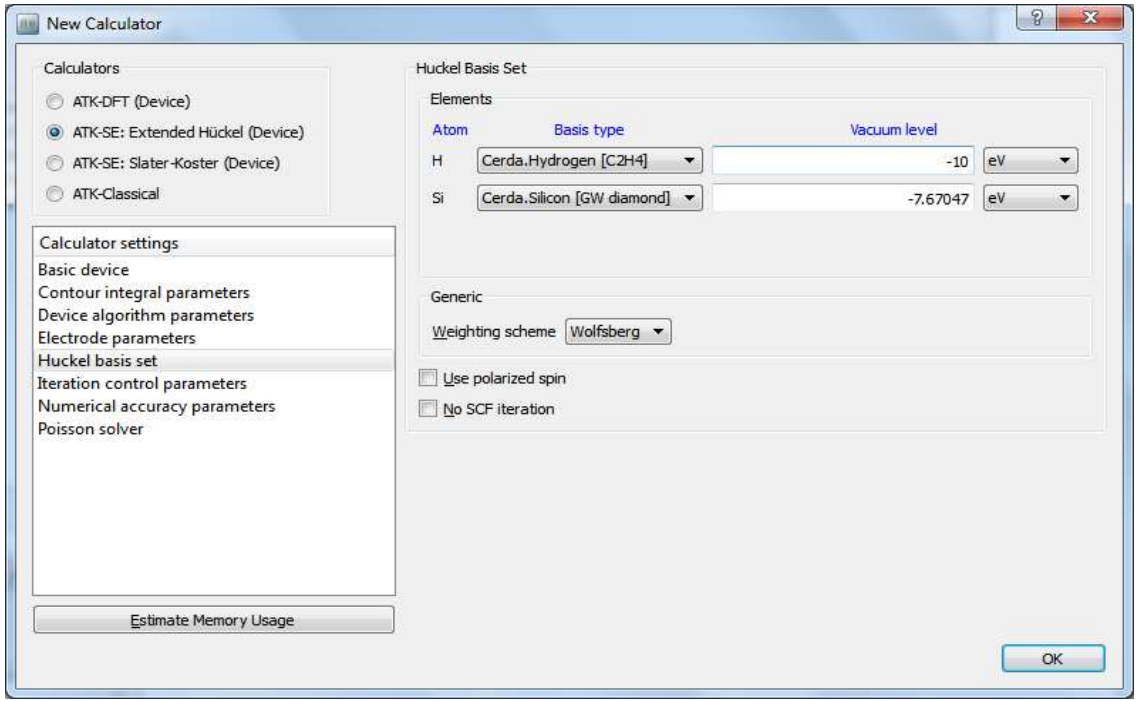

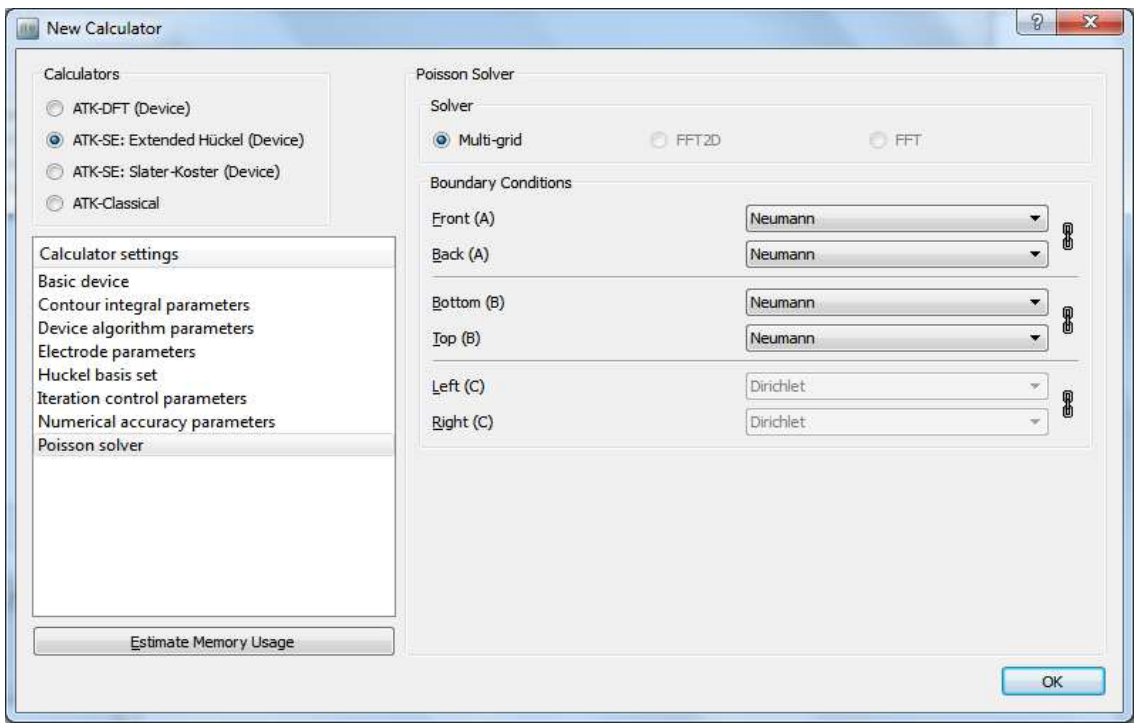

Step 22 - The transmission spectrum settings can be seen below. Enter the settings and click on OK.

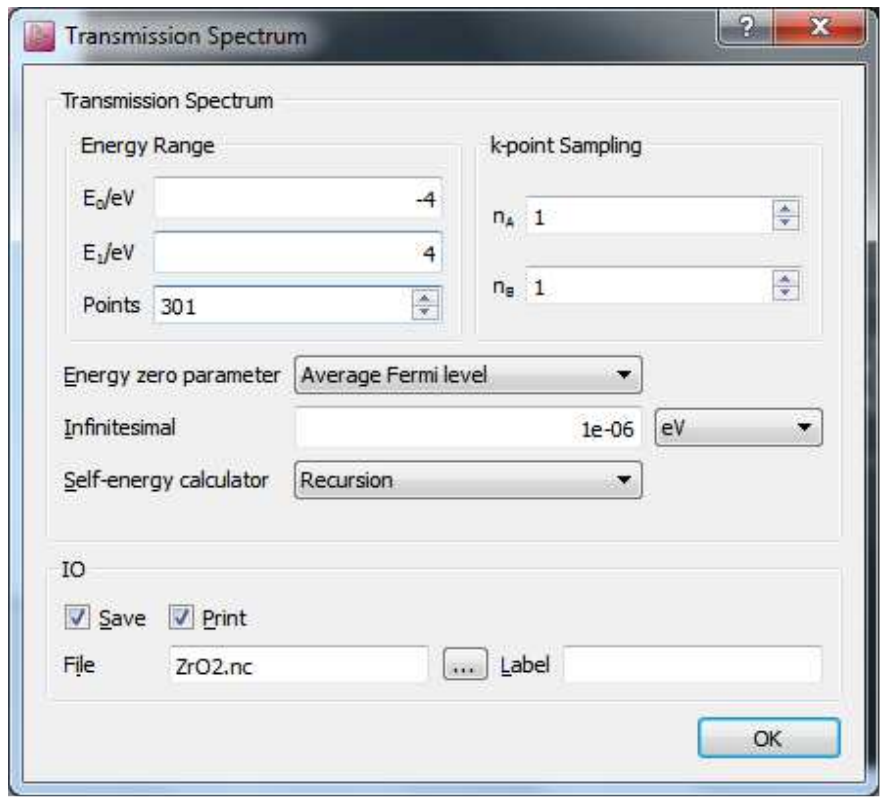

Step 23 – The built device with the calculations defined has to be dropped to the Editor. To specify the doping definitions, the code in the editor needs to be altered in the Electrode Calculator section as shown in the screen shot below. Once done, the script has to be saved and then dropped to Job Manager. Run Queue is clicked to run the jobs.

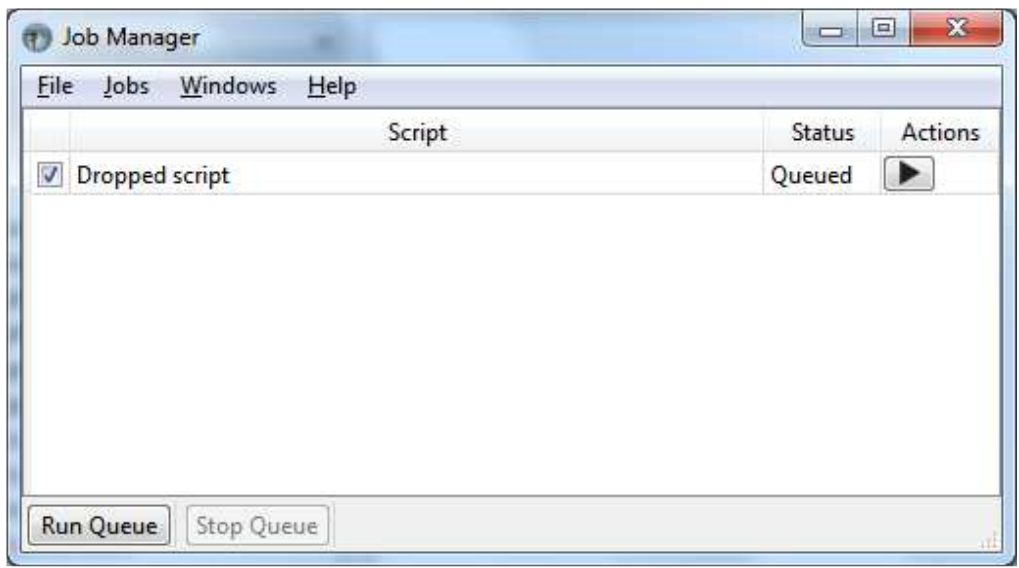

## 5.2 Simulation Results and Analysis

Once the simulation settings are set using the steps discussed in the previous section, iterative simulations have been performed in such a way that for a gate bias of 0 V, drain voltage is swept from 0 V to 2 V in steps of 0.2 V. For each value of drain voltage, a single simulation is run followed by the subsequent simulation with a different drain voltage. Once the drain voltage is swept from 0 V to 2 V for gate bias of 0V, the gate voltage is changed to 0.5 V and simulations have been carried out for different values of drain voltages. In this way, the nanowire field effect transistor has been simulated for gate bias of 0 V to 2.5 V in steps of 0.5 V, drain voltage being swept from 0 V to 2 V for each gate voltage set.

For a particular gate voltage, simulation result gives the I-V curve obtained by sweeping the drain voltage from 0 V to 2 V. The I-V curves obtained for gate voltages of 0 V, 0.5 V, 1 V, 1.5 V, 2 V and 2.5 V have been presented in Fig. 5.1 (a) – (f).

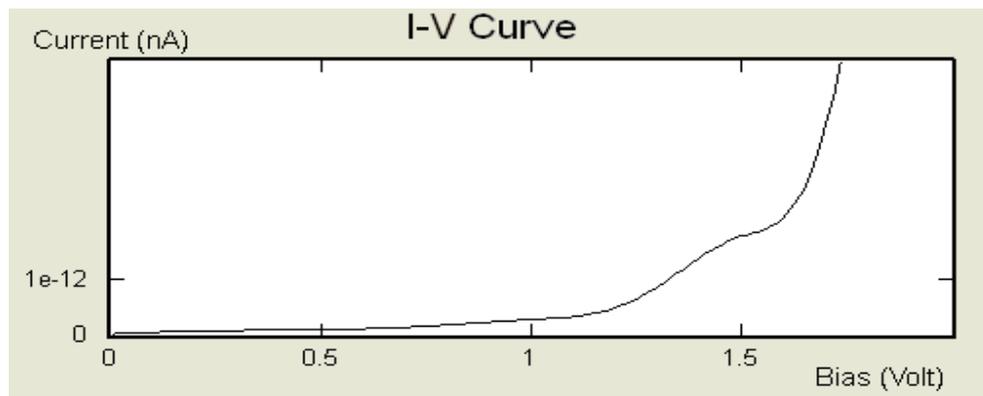

(a) $V_G = 0$ V

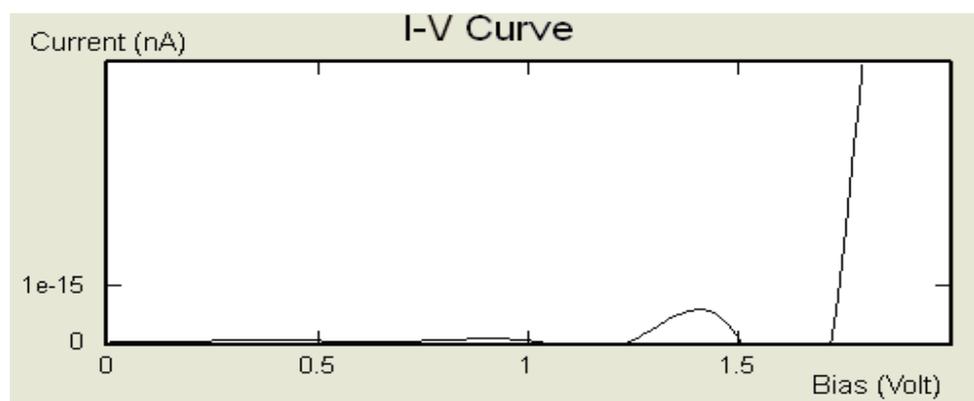

(b) $V_G = 0.5$ V

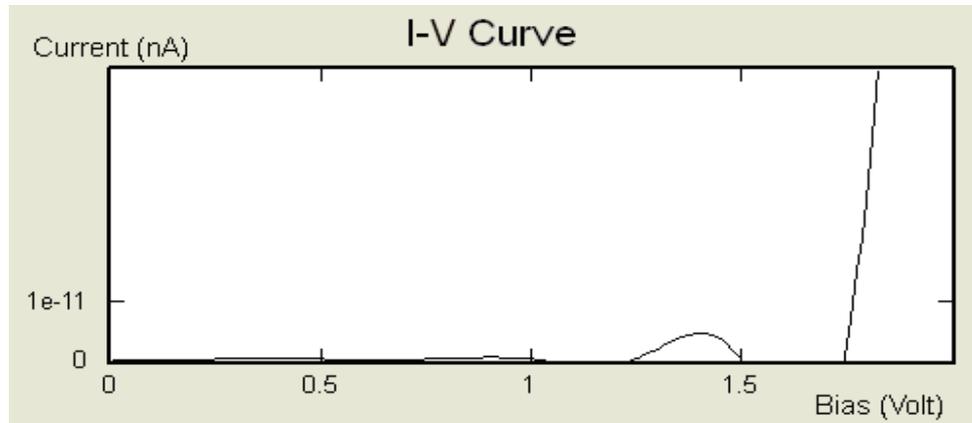

(c) $V_G$ = 1 V

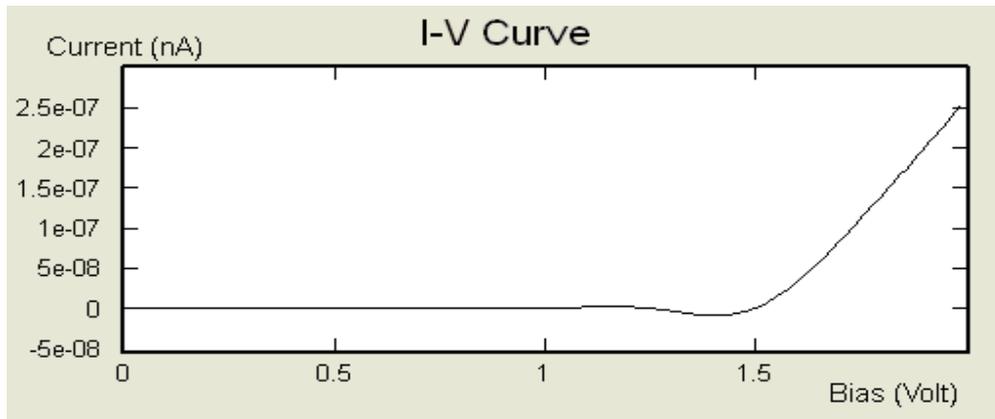

(d) $V_G$ = 1.5 V

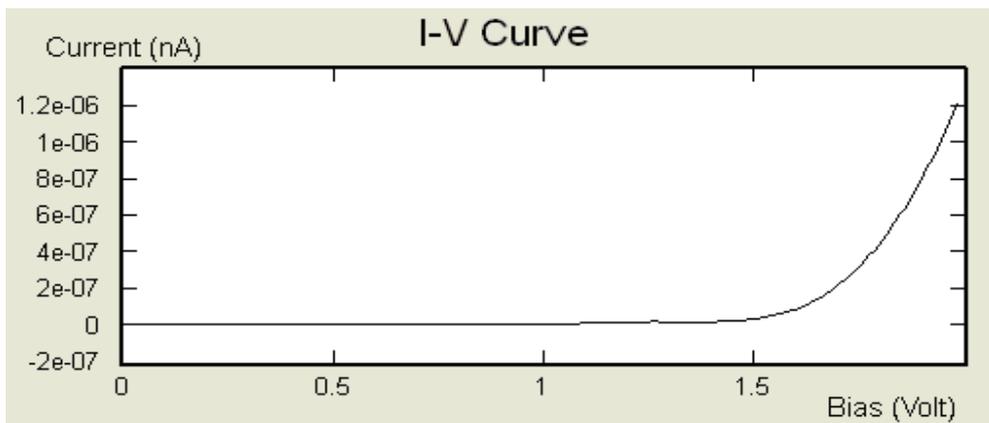

(e) $V_G$ = 2 V

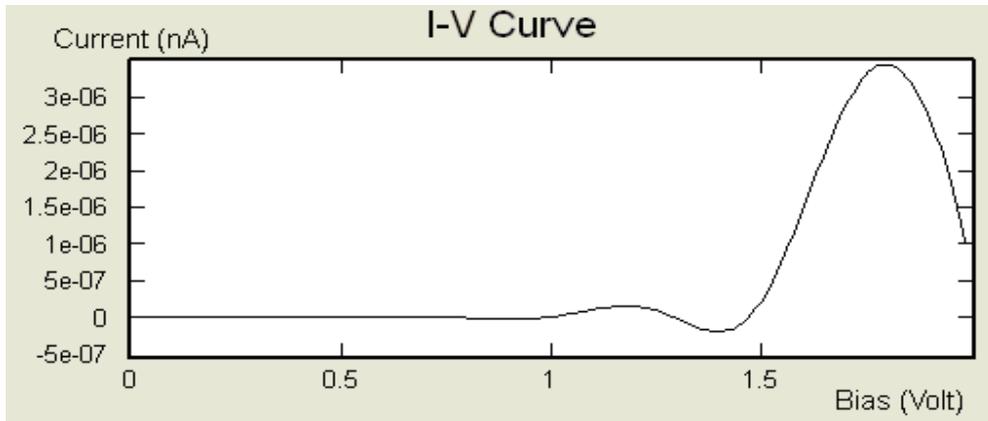

(f) $V_G$ = 2.5 V

**Fig. 5.1** $I_D$-$V_D$ Characteristics of Silicon NWFET with SiO₂ Gate Dielectric

Once the electrical characteristics have been obtained for the NWFET for six different gate voltages, the values of current have been extracted from each I-V curve for each value of drain voltage and gate voltage. Using the extracted data, the transfer characteristics and the drain characteristics have been plotted, as shown in Fig. 5.2 and Fig. 5.3 respectively. In Fig. 5.2, the drain current has been plotted against gate voltage for different values of drain voltage. The threshold voltage, $V_{TH}$ of the NWFET is equal to 0.21 V.

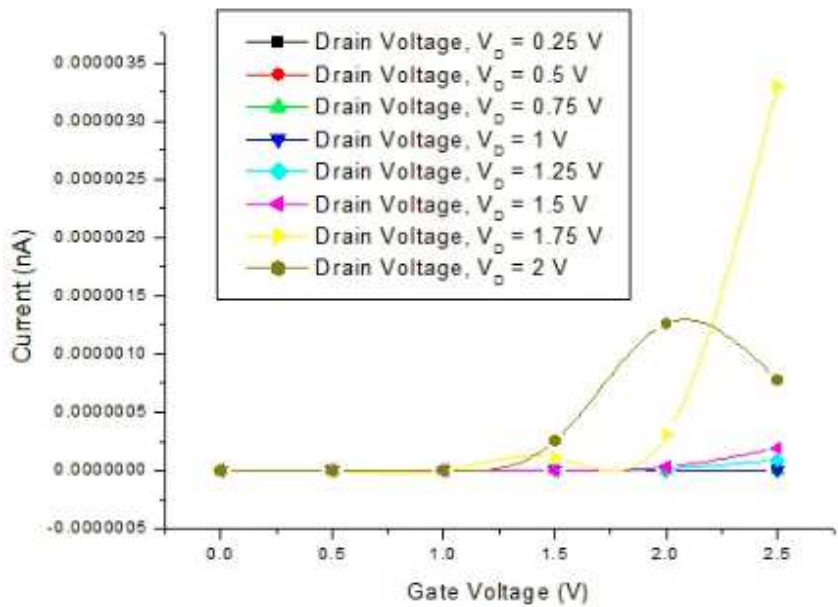

**Fig.5.2** Transfer Characteristics of Silicon NWFET with SiO₂ Gate Dielectric

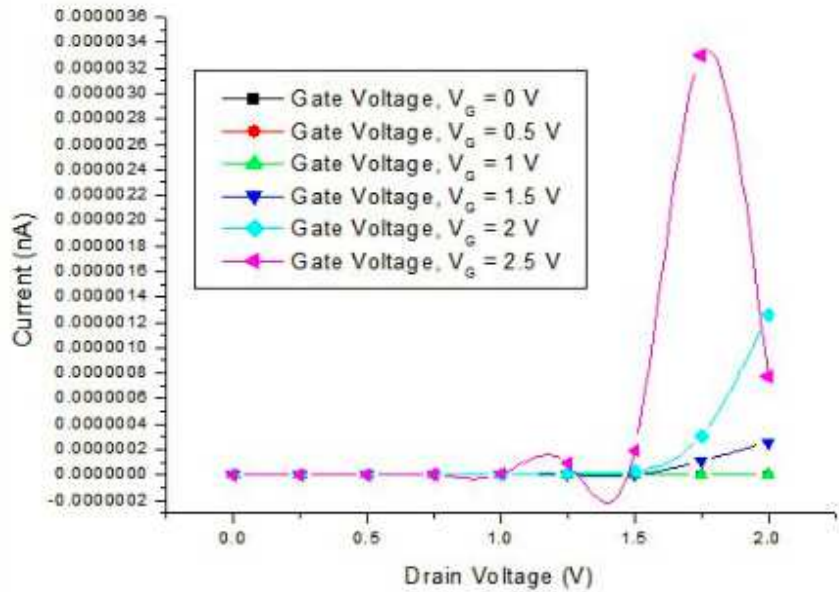

**Fig.5.3** Drain Characteristics of Silicon NWFET with $SiO_2$ Gate Dielectric

The channel conductance has been computed for different gate voltages used for the simulation. Fig. 5.4 depicts the plot of channel conductance plotted against the corresponding gate voltages applied. A similar curve can be obtained by plotting the conductance values against drain voltages applied for different gate voltage values as depicted in Fig. 5.5. As it is evident from the two plots, the channel conductance is low for low values of gate and drain voltages in both the plots. As the voltages are increased, the channel conductance shoots up and is highest for $V_G$ = 2.5 V and $V_D$ = 2 V. Also, from Fig. 5.1, negative resistance regions are evident in a couple of plots, which implies that for the voltage values specified in those plots, the NWFET can be used for microwave applications.

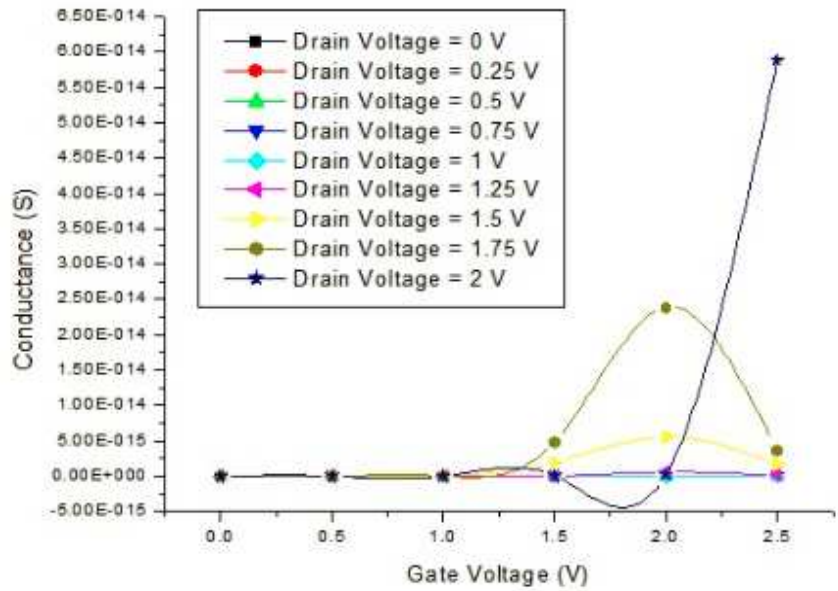

**Fig. 5.4** Conductance plotted against Gate Voltage

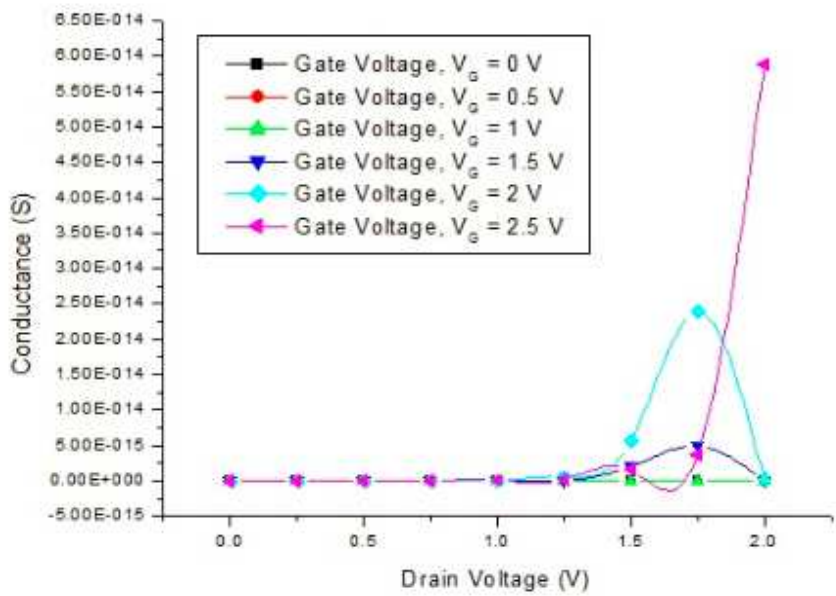

**Fig. 5.5** Conductance plotted against Drain Voltage

## 5.3 Temperature Dependence of Channel Conductance of Si NWFET with SiO$_2$ Gate Dielectric

For operating temperatures of 0 K, 77 K, 300 K and 325 K, the NWFET with SiO$_2$ gate dielectric has been simulated for a gate bias of V$_{TH}$ = 0.21 V. It can be observed from the plot in Fig. 5.6 that the channel conductance shoots up at room temperature, ensuring that the device with SiO$_2$ gate dielectric is well suited to operate at room temperature. Where as, the conductance is low at other operating temperatures.

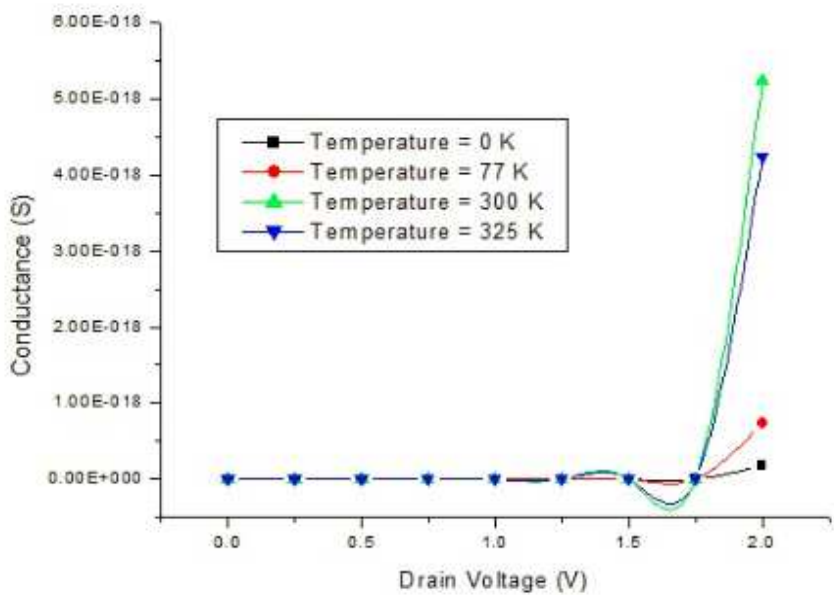

**Fig. 5.6** Temperature Dependence of Channel Conductance with Variation in Drain Voltage

The Fig. 5.7 depicts the temperature dependence of drain current with a variation in drain voltage for a gate voltage, V$_G$ = V$_{TH}$. As evident from the plot above, the drain current shoots up at room temperature, ensuring that the device with SiO$_2$ gate dielectric is well suited to operate at room temperature. Where as, the current is low at other operating temperatures. Therefore, room temperature operation promises higher drive current than other operating temperatures. But when it comes to low drive current leading to lower power dissipation, the silicon NWFET can be operated at lower temperatures.

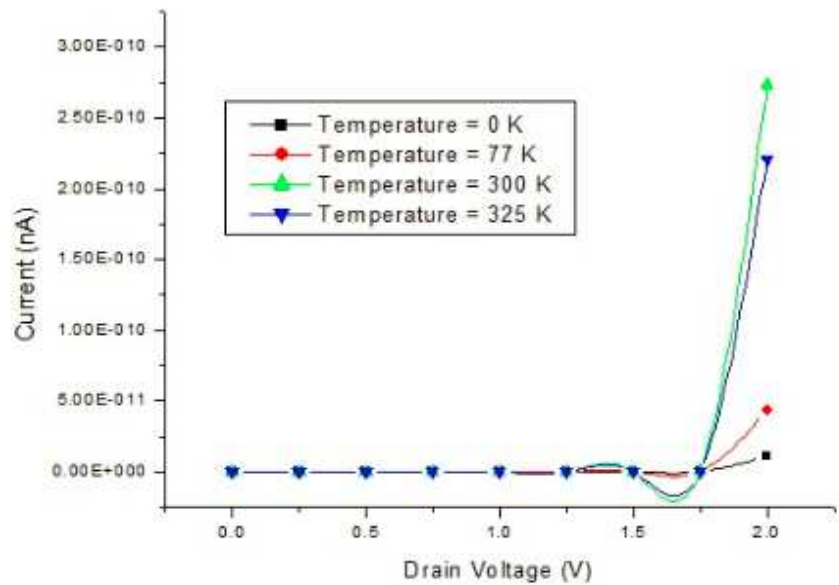

**Fig. 5.7** Temperature Dependence of Drain Current with Variation in Drain Voltage

# 6. Simulation of Silicon Nanowire FET with Zirconium Dioxide as Gate Dielectric

In this chapter, the modeling and simulation of silicon nanowire field effect transistor with zirconium dioxide as gate dielectric has been presented. The tool used is Quantumwise ATK ver.13.1 and the calculator used in ATK-SE: Extended Hückel (Device). The device geometry consisting of the thickness of the metal gate and the gate dielectric layers and radii of the two cylinders (gate dielectric and metal gate) considering the FET has a gate all around structure, left and right electrode lengths vary from the dimensions used to carry out the simulation presented in the previous chapter.

## 6.1 Simulation Settings

The nanowire transistor simulated using zirconium dioxide as gate dielectric has a silicon nanowire oriented in (100) direction, similar to the structure discussed in Chapter 5. The nanowire is essentially a single crystal structure in Face Centred Cubic orientation. The two ends of the nanowire are doped to obtain a doping concentration of 4 X $10^9$ cm$^{-3}$ in the source and drain regions, as the electrode is about 1 nm long and has a cross section of (0.5 nm)x(0.5 nm). As the nanowire transistor simulated has a Gate All Around structure to obtain better control over the carriers in the channel, the gate dielectric layer and the metal gate are essentially cylindrical in shape to wrap over the silicon nanowire, unlike conventional planar MOSFETs where the gate dielectric and the metal gate are just material stacks sitting over the channel area on the substrate. The thickness of the gate dielectric (zirconium dioxide) used in this design is 2 Å while the inner radius is 5 Å. The thickness of the metal gate is 3 Å while the inner radius is 7 Å. The thickness and inner radii of the layers have been intentionally kept the same as used in the structure described in Chapter 5 so that it is convenient to compare the characteristics of both the devices which differ in terms of the gate dielectric material.

The steps involved in modeling the nanowire FET are the same as described in Chapter 5, except that while entering the dielectric constant of the gate dielectric, the value to be entered is 25. Rests of all the steps remain the same.

## 6.2 Simulation Results and Analysis

Once the simulation settings are set as done in Chapter 5 with just a single change with respect to the gate dielectric constant value being equal to 25 instead of 3.9, iterative simulations have been performed in such a way that for a gate bias of 0 V, drain voltage is swept from 0 V to 2 V in steps of 0.2 V. For each value of drain voltage, a single simulation is run followed by the subsequent simulation with a different drain voltage. Once the drain voltage is swept from 0 V to 2 V for gate bias of 0V, the gate voltage is changed to 0.5 V and simulations have been carried out for different values of drain voltages. In this way, the nanowire field effect transistor has been simulated for gate bias of 0 V to 2.5 V in steps of 0.5 V, drain voltage being swept from 0 V to 2 V for each gate voltage set.

For a particular gate voltage, simulation result gives the I-V curve obtained by sweeping the drain voltage from 0 V to 2 V. The I-V curves obtained for gate voltages of 0 V, 0.5 V, 1 V, 1.5 V, 2 V and 2.5 V have been presented in Fig. 6.1 (a) – (f). It can be seen that the plots shown below are very similar in terms of the current values and the features observed. This is because the entire device

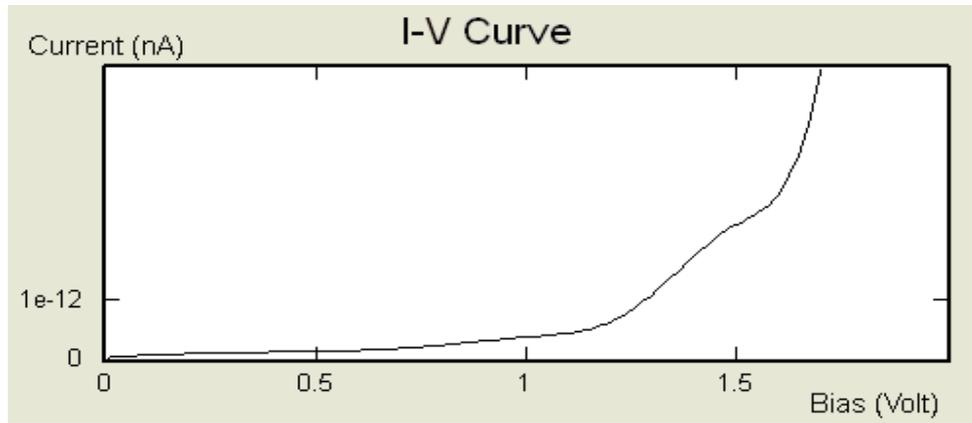

(a) $V_G$ = 0 V

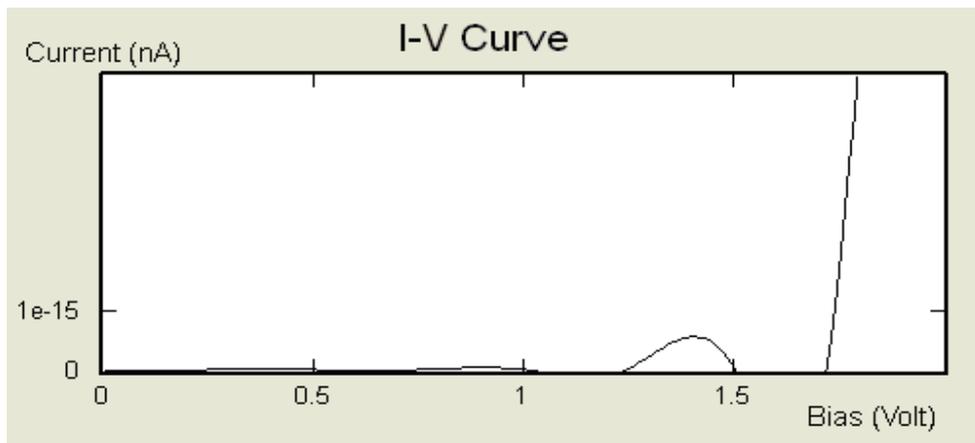

(b) $V_G$ = 0.5 V

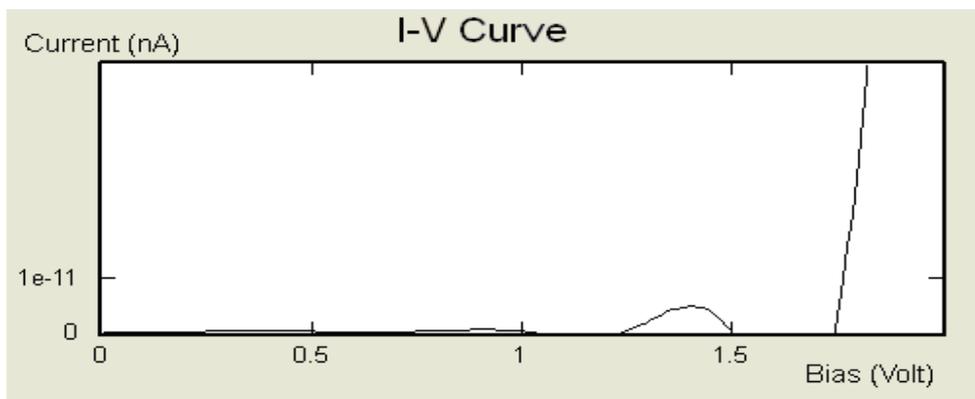

(c) $V_G$ = 1 V

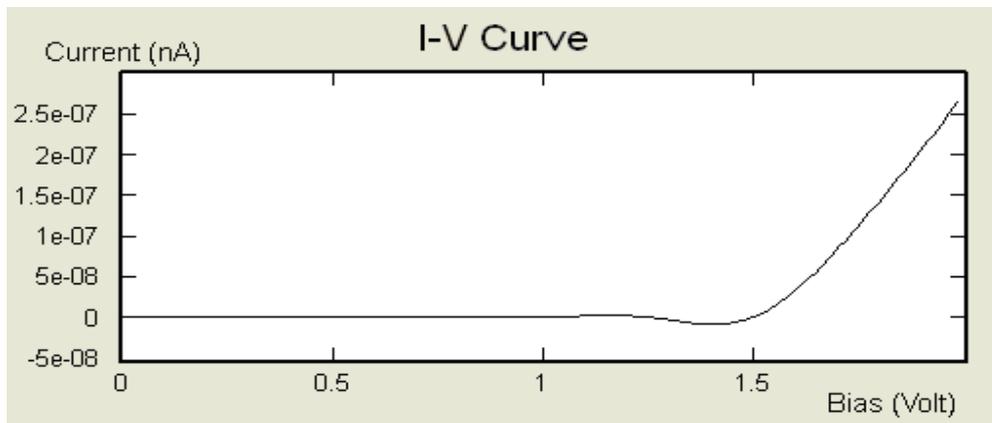

(d) $V_G$ = 1.5 V

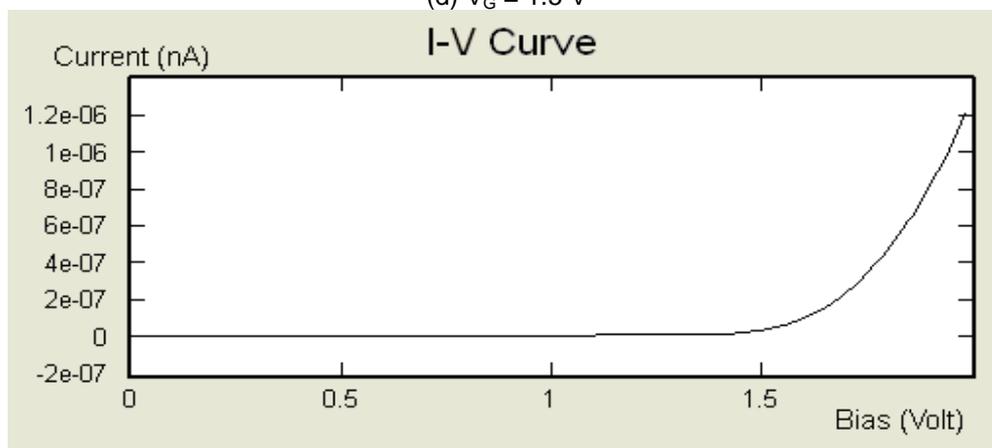

(e) $V_G$ = 2 V

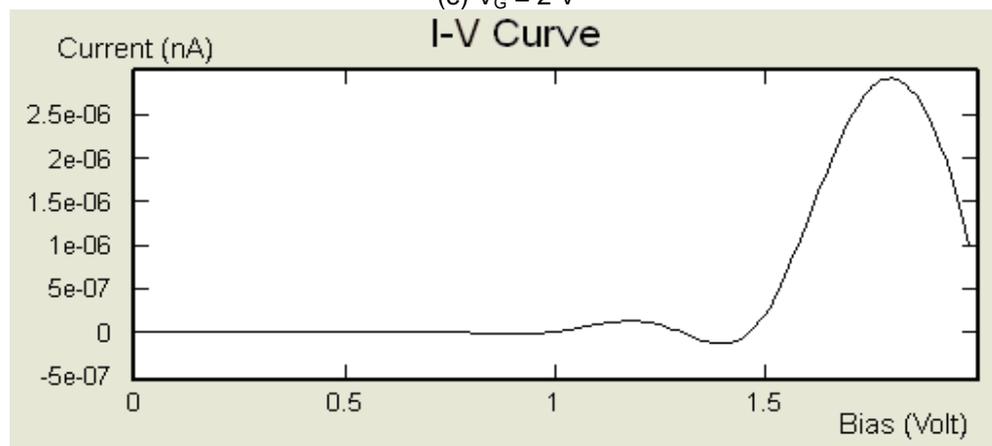

(f) $V_G$ = 2.5 V

**Fig. 6.1** $I_D$-$V_D$ Characteristics of Silicon NWFET with $ZrO_2$ Gate Dielectric

structure remains the same as used in Chapter 5, the only change is the dielectric constant.

Once the electrical characteristics have been obtained for the NWFET for six different gate voltages, the values of current have been extracted from each I-V curve for each value of drain voltage and gate voltage. Using the extracted data, the transfer characteristics and the drain characteristics have been plotted, as shown in Fig. 6.2.

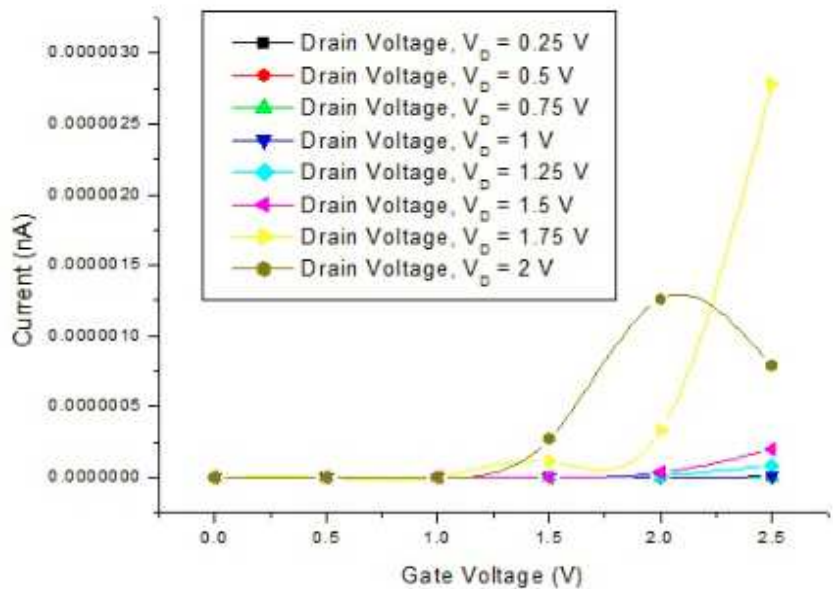

**Fig. 6.2** Transfer Characteristics of Silicon NWFET with ZrO$_2$ Gate Dielectric

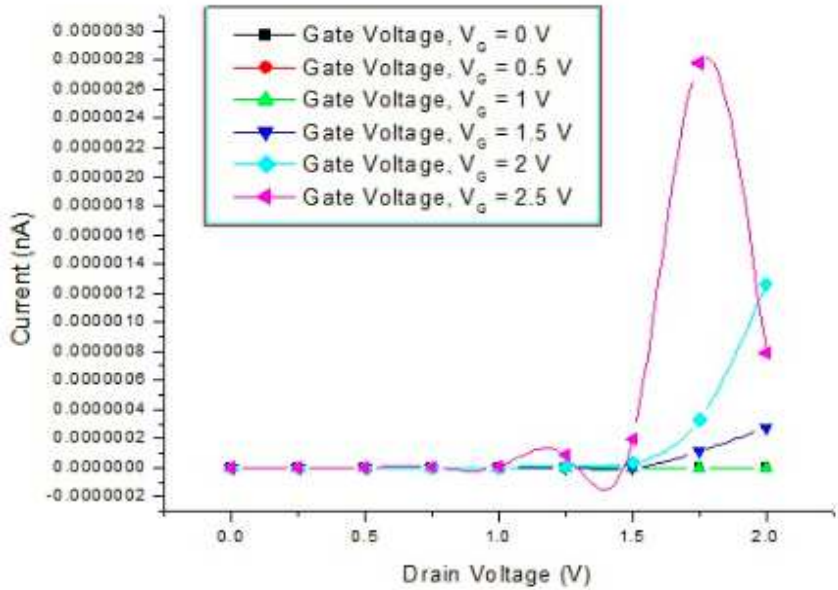

**Fig. 6.3** Drain Characteristics of Silicon NWFET with ZrO$_2$ Gate Dielectric

and Fig. 6.3 respectively. In Fig. 6.2, the drain current has been plotted against gate voltage for different values of drain voltage. The threshold voltage, $V_{TH}$ of the NWFET is equal to 0.17 V.

The channel conductance has been computed for different gate voltages used for the simulation. Figure 6.4 depicts the plot of channel conductance plotted against the corresponding gate voltages applied. A similar curve can be obtained by plotting the conductance values against drain voltages applied for different gate voltage values as depicted in Figure 6.5. As it is evident from the two plots, the channel conductance is low for low values of gate and drain voltages in both the plots. As the voltages are increased, the channel conductance shoots up and is highest for $V_G = 2.5$ V and $V_D = 2$ V. Also, from Figure 6.1, negative resistance regions are evident in a couple of plots, which implies that for the voltage values specified in those plots, the NWFET can be used for microwave applications.

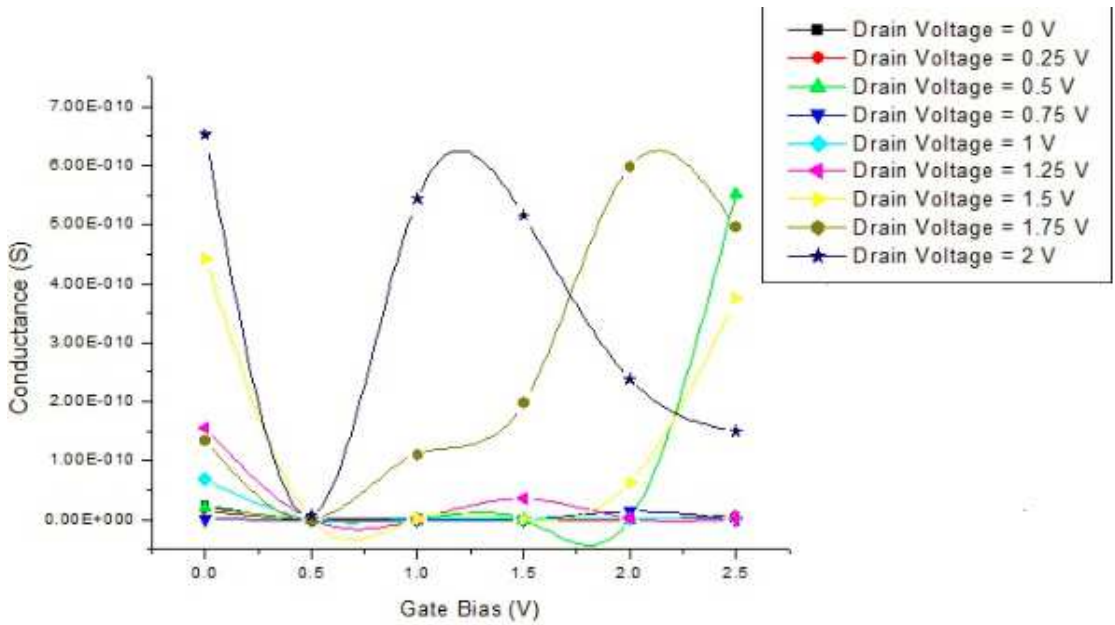

**Fig. 6.4** Conductance plotted against Gate Voltage

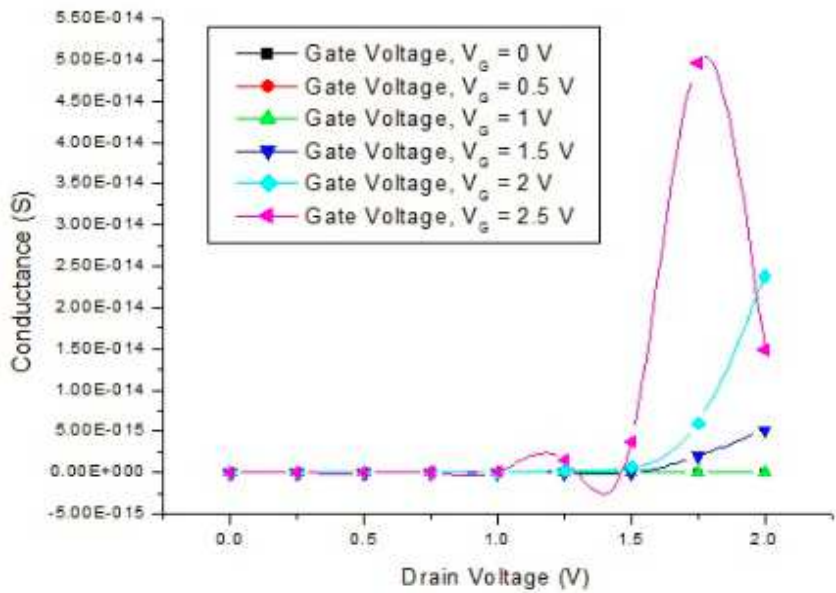

**Fig. 6.5** Conductance plotted against Drain Voltage

It can be seen from the transfer and drain characteristics of the silicon NWFET using $SiO_2$ and $ZrO_2$ as dielectrics respectively, the current values are slightly less in case of NWFET with $ZrO_2$ as gate dielectric. Moreover, due to higher dielectric constant (25>3.9), the gate leakage current is obstructed to a much larger extent in case of $ZrO_2$ than in NWFET with $SiO_2$ gate dielectric. This leads to much lower power consumption in case of $ZrO_2$ gate dielectric based NWFET as compared to $SiO_2$ gate dielectric based NWFET.

## 6.3 Comparison of $SiO_2$ based NWFET with $ZrO_2$ based Silicon NWFET

From the transfer and drain characteristics of the NWFET with $SiO_2$ gate dielectric and $ZrO_2$ gate dielectric respectively, it is evident that the power consumption is more in case of Silicon NWFET with $SiO_2$ gate dielectric owing to gate leakage current, as discussed in the previous section. Also, the comparisons of the off state drain current in the two devices as plotted against the drain voltages with different gate dielectrics have been presented in Fig. 6.6. It is clear from the plot that the off state drain current is slightly higher in case of silicon NWFET with $SiO_2$ gate dielectric as compared to the NWFET with $ZrO_2$ gate dielectric.

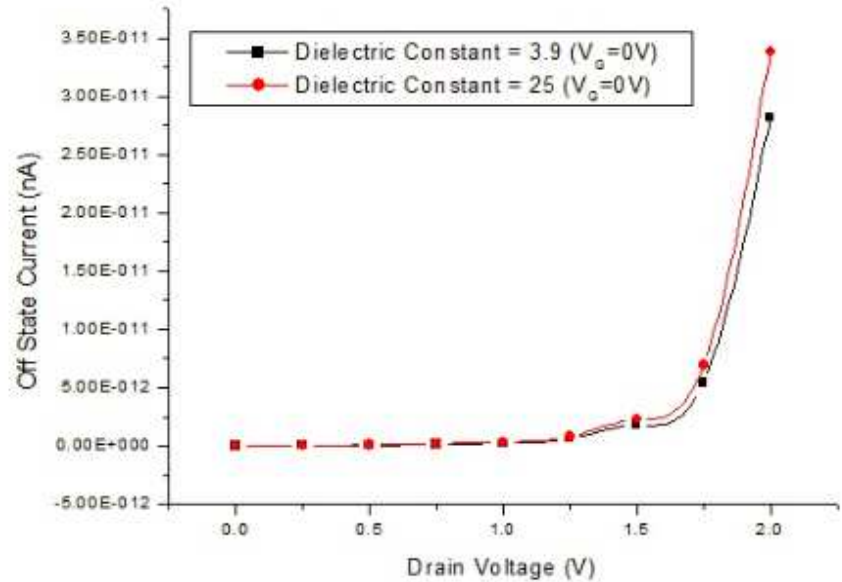

**Fig. 6.6** Comparison of off state drain current of Si NWFET with SiO₂ gate dielectric against ZrO₂ gate dielectric

Similar to the plot above, the channel conductance can be plotted against the drain voltages for the two NWFETs with SiO₂ and ZrO₂ gate dielectrics respectively, as shown in Fig. 6.7. As seen in Fig. 6.6, the off state channel conductance is slightly higher in case of silicon NWFET with SiO₂ gate dielectric as compared to the NWFET with ZrO₂ gate dielectric. This comparison adds to the performance metrics of silicon NWFET with ZrO2 as gate dielectric, owing to its lower power consumption and obstruction to short channel effects owing to higher dielectric constant of ZrO₂.

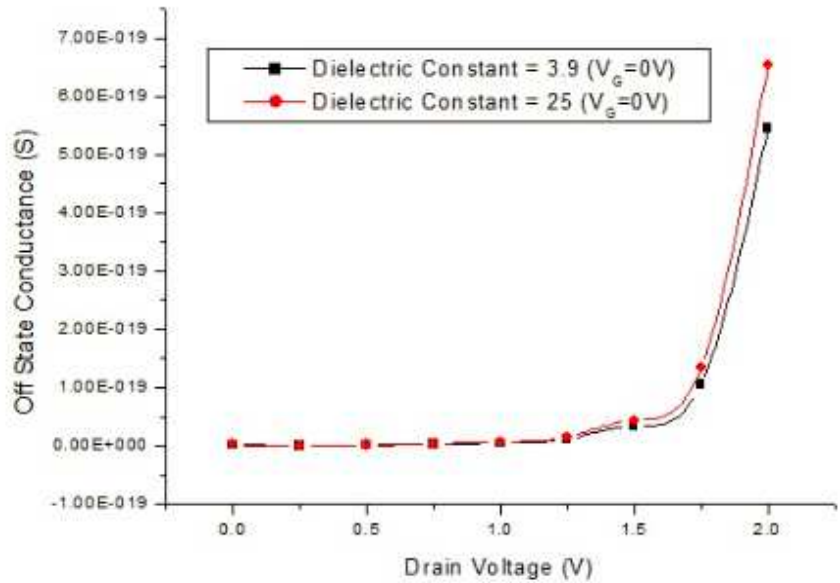

**Fig. 6.7** Comparison of off state channel conductance of Si NWFET with SiO$_2$ gate dielectric against ZrO$_2$ gate dielectric

# 7. Conclusion and Future Work

## 7.1 Conclusions

In this project, we have investigated the electrical characteristics of an n-channel Silicon Nanowire Field Effect Transistor (FET) using the quantum-ballistic transport model. It was found that nanowire FET exhibited superior current characteristics and controlled short channel effects better. In conclusion, the simulation study revealed that the silicon nanowire FET is an attractive candidate for FET device design for future technology nodes. It is expected that a nanowire FET can carry forward MOSFET downscaling to ~ 10 nm gate lengths. However, quantum effects would start playing an important role as device shrinks deep into sub-10nm nanometer regime.

## 7.2 Future Work

The current transport model used for simulations assumes a ballistic transport, i.e. carrier transport without any scattering in the channel. A more accurate transport model could be used bringing into account the surface scattering effects, which start to dominate at very small diameters (<5 nm) [1]. Also the simulation software assumes an ideal wrap around gate for the nanowire FET. Since this ideal structure is not possible in practice, changes can be incorporated into the device structure to more faithfully reproduce the actual device structure. Also, as new materials are being used for fabrication of MOSFETs as discussed in chapter 1, further work could be done by simulating nanowire FET with new materials such as SiGe nanowire channel, compound semiconductor nanowire channel with other high-k dielectrics.

# APPENDIX

## I. Generic Python Code for  Silicon NWFET with $V_D$ = 0 V and $V_G$ = 0 with SiO$_2$ Gate Dielectric:

```python
# ------------------------------------------------------------
# TwoProbe configuration
# ------------------------------------------------------------

# ------------------------------------------------------------
# Left electrode
# ------------------------------------------------------------

# Set up lattice
vector_a = [20.0, 0.0, 0.0]*Angstrom
vector_b = [0.0, 20.0, 0.0]*Angstrom
vector_c = [0.0, 0.0, 10.8612]*Angstrom
left_electrode_lattice = UnitCell(vector_a, vector_b, vector_c)

# Define elements
left_electrode_elements = [Hydrogen, Hydrogen, Silicon, Silicon, Silicon, Silicon,
Hydrogen,
            Hydrogen, Hydrogen, Hydrogen, Silicon, Silicon, Silicon, Silicon,
            Hydrogen, Hydrogen, Hydrogen, Hydrogen, Silicon, Silicon, Silicon, Silicon,
            Silicon, Hydrogen, Hydrogen, Hydrogen, Hydrogen, Silicon, Silicon,
            Silicon, Silicon, Hydrogen, Hydrogen, Hydrogen, Hydrogen, Silicon,
            Silicon, Silicon, Silicon, Hydrogen, Hydrogen, Hydrogen, Hydrogen,
            Silicon, Silicon, Silicon, Hydrogen, Hydrogen, Hydrogen, Hydrogen,
            Hydrogen, Silicon, Silicon, Silicon, Silicon, Hydrogen, Hydrogen,
            Hydrogen, Hydrogen, Silicon, Silicon, Silicon, Silicon, Hydrogen,
            Hydrogen]

# Define coordinates
left_electrode_coordinates = [[ 5.91157506,  7.11998944,  0.175653 ],
            [ 5.91157506, 10.96000352,  0.175653 ],
            [ 9.03999648,  7.11998944,  0.678825 ],
            [12.88001056,  7.11998944,  0.678825 ],
            [ 9.03999648, 10.96000352,  0.678825 ],
            [12.88001056, 10.96000352,  0.678825 ],
            [ 9.03999648, 14.08842494,  1.181997 ],
            [12.88001056, 14.08842494,  1.181997 ],
            [ 9.03999648,  5.91157506,  1.533303 ],
            [12.88001056,  5.91157506,  1.533303 ],
```

[  9.03999648,   9.03999648,   2.036475  ],
[ 12.88001056,   9.03999648,   2.036475  ],
[  9.03999648,  12.88001056,   2.036475  ],
[ 12.88001056,  12.88001056,   2.036475  ],
[  5.91157506,   9.03999648,   2.539647  ],
[  5.91157506,  12.88001056,   2.539647  ],
[ 14.08842494,   9.03999648,   2.890953  ],
[ 14.08842494,  12.88001056,   2.890953  ],
[  7.11998944,   9.03999648,   3.394125  ],
[ 10.96000352,   9.03999648,   3.394125  ],
[  7.11998944,  12.88001056,   3.394125  ],
[ 10.96000352,  12.88001056,   3.394125  ],
[  7.11998944,   5.91157506,   3.897297  ],
[ 10.96000352,   5.91157506,   3.897297  ],
[  7.11998944,  14.08842494,   4.248603  ],
[ 10.96000352,  14.08842494,   4.248603  ],
[  7.11998944,   7.11998944,   4.751775  ],
[ 10.96000352,   7.11998944,   4.751775  ],
[  7.11998944,  10.96000352,   4.751775  ],
[ 10.96000352,  10.96000352,   4.751775  ],
[ 14.08842494,   7.11998944,   5.254947  ],
[ 14.08842494,  10.96000352,   5.254947  ],
[  5.91157506,   7.11998944,   5.606253  ],
[  5.91157506,  10.96000352,   5.606253  ],
[  9.03999648,   7.11998944,   6.109425  ],
[ 12.88001056,   7.11998944,   6.109425  ],
[  9.03999648,  10.96000352,   6.109425  ],
[ 12.88001056,  10.96000352,   6.109425  ],
[  9.03999648,  14.08842494,   6.612597  ],
[ 12.88001056,  14.08842494,   6.612597  ],
[  9.03999648,   5.91157506,   6.963903  ],
[ 12.88001056,   5.91157506,   6.963903  ],
[  9.03999648,   9.03999648,   7.467075  ],
[ 12.88001056,   9.03999648,   7.467075  ],
[  9.03999648,  12.88001056,   7.467075  ],
[ 12.88001056,  12.88001056,   7.467075  ],
[  5.91157506,   9.03999648,   7.970247  ],
[  5.91157506,  12.88001056,   7.970247  ],
[ 14.08842494,   9.03999648,   8.321553  ],
[ 14.08842494,  12.88001056,   8.321553  ],
[  7.11998944,   9.03999648,   8.824725  ],
[ 10.96000352,   9.03999648,   8.824725  ],
[  7.11998944,  12.88001056,   8.824725  ],
[ 10.96000352,  12.88001056,   8.824725  ],
[  7.11998944,   5.91157506,   9.327897  ],

```
                   [ 10.96000352,   5.91157506,   9.327897  ],
                   [  7.11998944,  14.08842494,   9.679203  ],
                   [ 10.96000352,  14.08842494,   9.679203  ],
                   [  7.11998944,   7.11998944,  10.182375  ],
                   [ 10.96000352,   7.11998944,  10.182375  ],
                   [  7.11998944,  10.96000352,  10.182375  ],
                   [ 10.96000352,  10.96000352,  10.182375  ],
                   [ 14.08842494,   7.11998944,  10.685547  ],
                   [ 14.08842494,  10.96000352,  10.685547  ]]*Angstrom

# Set up configuration
left_electrode = BulkConfiguration(
    bravais_lattice=left_electrode_lattice,
    elements=left_electrode_elements,
    cartesian_coordinates=left_electrode_coordinates
    )

# -------------------------------------------------------------
# Right electrode
# -------------------------------------------------------------

# Set up lattice
vector_a = [20.0, 0.0, 0.0]*Angstrom
vector_b = [0.0, 20.0, 0.0]*Angstrom
vector_c = [0.0, 0.0, 10.8612]*Angstrom
right_electrode_lattice = UnitCell(vector_a, vector_b, vector_c)

# Define elements
right_electrode_elements = [Hydrogen, Hydrogen, Silicon, Silicon, Silicon, Silicon,
Hydrogen,
                   Hydrogen, Hydrogen, Hydrogen, Silicon, Silicon, Silicon, Silicon,
                   Hydrogen, Hydrogen, Hydrogen, Hydrogen, Silicon, Silicon, Silicon,
                   Silicon, Hydrogen, Hydrogen, Hydrogen, Hydrogen, Silicon, Silicon,
                   Silicon, Silicon, Hydrogen, Hydrogen, Hydrogen, Hydrogen, Silicon,
                   Silicon, Silicon, Hydrogen, Hydrogen, Hydrogen, Hydrogen, Hydrogen,
                   Silicon, Silicon, Silicon, Silicon, Hydrogen, Hydrogen, Hydrogen,
                   Hydrogen, Silicon, Silicon, Silicon, Silicon, Hydrogen, Hydrogen,
                   Hydrogen, Hydrogen, Silicon, Silicon, Silicon, Silicon, Hydrogen,
                   Hydrogen]

# Define coordinates
right_electrode_coordinates = [[  5.91157506,   7.11998944,   0.175653  ],
                   [  5.91157506,  10.96000352,   0.175653  ],
                   [  9.03999648,   7.11998944,   0.678825  ],
                   [ 12.88001056,   7.11998944,   0.678825  ],
```

```
[  9.03999648, 10.96000352,  0.678825  ],
[ 12.88001056, 10.96000352,  0.678825  ],
[  9.03999648, 14.08842494,  1.181997  ],
[ 12.88001056, 14.08842494,  1.181997  ],
[  9.03999648,  5.91157506,  1.533303  ],
[ 12.88001056,  5.91157506,  1.533303  ],
[  9.03999648,  9.03999648,  2.036475  ],
[ 12.88001056,  9.03999648,  2.036475  ],
[  9.03999648, 12.88001056,  2.036475  ],
[ 12.88001056, 12.88001056,  2.036475  ],
[  5.91157506,  9.03999648,  2.539647  ],
[  5.91157506, 12.88001056,  2.539647  ],
[ 14.08842494,  9.03999648,  2.890953  ],
[ 14.08842494, 12.88001056,  2.890953  ],
[  7.11998944,  9.03999648,  3.394125  ],
[ 10.96000352,  9.03999648,  3.394125  ],
[  7.11998944, 12.88001056,  3.394125  ],
[ 10.96000352, 12.88001056,  3.394125  ],
[  7.11998944,  5.91157506,  3.897297  ],
[ 10.96000352,  5.91157506,  3.897297  ],
[  7.11998944, 14.08842494,  4.248603  ],
[ 10.96000352, 14.08842494,  4.248603  ],
[  7.11998944,  7.11998944,  4.751775  ],
[ 10.96000352,  7.11998944,  4.751775  ],
[  7.11998944, 10.96000352,  4.751775  ],
[ 10.96000352, 10.96000352,  4.751775  ],
[ 14.08842494,  7.11998944,  5.254947  ],
[ 14.08842494, 10.96000352,  5.254947  ],
[  5.91157506,  7.11998944,  5.606253  ],
[  5.91157506, 10.96000352,  5.606253  ],
[  9.03999648,  7.11998944,  6.109425  ],
[ 12.88001056,  7.11998944,  6.109425  ],
[  9.03999648, 10.96000352,  6.109425  ],
[ 12.88001056, 10.96000352,  6.109425  ],
[  9.03999648, 14.08842494,  6.612597  ],
[ 12.88001056, 14.08842494,  6.612597  ],
[  9.03999648,  5.91157506,  6.963903  ],
[ 12.88001056,  5.91157506,  6.963903  ],
[  9.03999648,  9.03999648,  7.467075  ],
[ 12.88001056,  9.03999648,  7.467075  ],
[  9.03999648, 12.88001056,  7.467075  ],
[ 12.88001056, 12.88001056,  7.467075  ],
[  5.91157506,  9.03999648,  7.970247  ],
[  5.91157506, 12.88001056,  7.970247  ],
[ 14.08842494,  9.03999648,  8.321553  ],
```

```
                        [ 14.08842494,  12.88001056,   8.321553  ],
                        [  7.11998944,   9.03999648,   8.824725  ],
                        [ 10.96000352,   9.03999648,   8.824725  ],
                        [  7.11998944,  12.88001056,   8.824725  ],
                        [ 10.96000352,  12.88001056,   8.824725  ],
                        [  7.11998944,   5.91157506,   9.327897  ],
                        [ 10.96000352,   5.91157506,   9.327897  ],
                        [  7.11998944,  14.08842494,   9.679203  ],
                        [ 10.96000352,  14.08842494,   9.679203  ],
                        [  7.11998944,   7.11998944,  10.182375  ],
                        [ 10.96000352,   7.11998944,  10.182375  ],
                        [  7.11998944,  10.96000352,  10.182375  ],
                        [ 10.96000352,  10.96000352,  10.182375  ],
                        [ 14.08842494,   7.11998944,  10.685547  ],
                        [ 14.08842494,  10.96000352,  10.685547  ]]*Angstrom

# Set up configuration
right_electrode = BulkConfiguration(
    bravais_lattice=right_electrode_lattice,
    elements=right_electrode_elements,
    cartesian_coordinates=right_electrode_coordinates
    )

# ------------------------------------------------------------
# Central region
# ------------------------------------------------------------

# Set up lattice
vector_a = [20.0, 0.0, 0.0]*Angstrom
vector_b = [0.0, 20.0, 0.0]*Angstrom
vector_c = [0.0, 0.0, 65.1672]*Angstrom
central_region_lattice = UnitCell(vector_a, vector_b, vector_c)

# Define elements
central_region_elements = [Hydrogen, Hydrogen, Silicon, Silicon, Silicon, Silicon,
Hydrogen,
                        Hydrogen, Hydrogen, Hydrogen, Silicon, Silicon, Silicon, Silicon,
                        Hydrogen, Hydrogen, Hydrogen, Hydrogen, Silicon, Silicon, Silicon,
                        Silicon, Hydrogen, Hydrogen, Hydrogen, Hydrogen, Silicon, Silicon,
                        Silicon, Silicon, Hydrogen, Hydrogen, Hydrogen, Hydrogen, Silicon,
                        Silicon, Silicon, Silicon, Hydrogen, Hydrogen, Hydrogen, Hydrogen,
                        Silicon, Silicon, Silicon, Silicon, Hydrogen, Hydrogen, Hydrogen,
                        Hydrogen, Silicon, Silicon, Silicon, Silicon, Hydrogen, Hydrogen,
                        Hydrogen, Hydrogen, Silicon, Silicon, Silicon, Silicon, Hydrogen,
                        Hydrogen, Hydrogen, Hydrogen, Silicon, Silicon, Silicon, Silicon,
```

Hydrogen, Hydrogen, Hydrogen, Hydrogen, Silicon, Silicon, Silicon, Silicon, Hydrogen, Hydrogen, Hydrogen, Hydrogen, Silicon, Silicon, Silicon, Silicon, Hydrogen, Hydrogen, Hydrogen, Hydrogen, Silicon, Silicon, Silicon, Silicon, Hydrogen, Hydrogen, Hydrogen, Hydrogen, Silicon, Silicon, Silicon, Silicon, Hydrogen, Hydrogen, Hydrogen, Hydrogen, Silicon, Silicon, Silicon, Silicon, Hydrogen, Hydrogen, Hydrogen, Hydrogen, Silicon, Silicon, Silicon, Silicon, Silicon, Hydrogen, Hydrogen, Hydrogen, Hydrogen, Silicon, Silicon, Silicon, Silicon, Hydrogen, Hydrogen, Hydrogen, Silicon, Silicon, Silicon, Silicon, Hydrogen, Hydrogen, Hydrogen, Hydrogen, Silicon, Silicon, Silicon, Silicon, Hydrogen, Hydrogen, Hydrogen, Hydrogen, Silicon, Silicon, Silicon, Silicon, Hydrogen, Hydrogen, Hydrogen, Hydrogen, Silicon, Silicon, Silicon, Silicon, Silicon, Hydrogen, Hydrogen, Hydrogen, Hydrogen, Silicon, Silicon, Silicon, Hydrogen, Hydrogen, Hydrogen, Silicon, Silicon, Silicon, Silicon, Hydrogen, Hydrogen, Hydrogen, Hydrogen, Silicon, Silicon, Silicon, Hydrogen, Hydrogen, Hydrogen, Hydrogen, Silicon, Silicon, Silicon, Silicon, Hydrogen, Hydrogen, Hydrogen, Hydrogen, Silicon, Silicon, Silicon, Silicon, Hydrogen, Hydrogen, Hydrogen, Hydrogen, Silicon, Silicon, Silicon, Silicon, Silicon, Hydrogen, Hydrogen, Hydrogen, Hydrogen, Silicon, Silicon, Silicon, Silicon, Hydrogen, Hydrogen, Hydrogen, Silicon, Silicon, Silicon, Silicon, Hydrogen, Hydrogen, Hydrogen, Hydrogen, Silicon, Silicon, Silicon, Silicon, Hydrogen, Hydrogen, Hydrogen, Hydrogen, Silicon, Silicon, Silicon, Silicon, Hydrogen, Hydrogen, Hydrogen, Hydrogen, Silicon, Silicon, Silicon, Silicon, Silicon, Hydrogen, Hydrogen, Hydrogen, Hydrogen, Silicon, Silicon, Silicon, Hydrogen, Hydrogen, Hydrogen, Silicon, Silicon, Silicon, Silicon, Hydrogen, Hydrogen, Hydrogen, Hydrogen, Silicon, Silicon, Silicon, Hydrogen, Hydrogen, Hydrogen, Hydrogen, Silicon, Silicon, Silicon, Silicon, Hydrogen, Hydrogen, Hydrogen, Hydrogen, Silicon, Silicon, Silicon, Silicon, Hydrogen, Hydrogen, Hydrogen, Hydrogen, Silicon, Silicon, Silicon, Silicon, Silicon, Hydrogen, Hydrogen, Hydrogen, Hydrogen, Silicon, Silicon, Silicon, Silicon, Hydrogen, Hydrogen, Hydrogen, Silicon, Silicon, Silicon, Silicon, Hydrogen, Hydrogen, Hydrogen, Hydrogen, Silicon, Silicon, Silicon, Silicon, Hydrogen, Hydrogen, Hydrogen, Hydrogen, Silicon, Silicon, Silicon, Silicon, Hydrogen, Hydrogen, Hydrogen, Hydrogen, Silicon, Silicon, Silicon, Silicon, Silicon, Hydrogen, Hydrogen, Hydrogen, Hydrogen, Silicon, Silicon, Silicon, Hydrogen, Hydrogen, Hydrogen, Silicon, Silicon, Silicon, Silicon, Hydrogen, Hydrogen, Hydrogen, Hydrogen, Silicon, Silicon, Silicon, Hydrogen, Hydrogen, Hydrogen, Hydrogen, Silicon, Silicon, Silicon, Silicon, Hydrogen, Hydrogen, Hydrogen, Hydrogen, Silicon, Silicon, Silicon, Silicon, Hydrogen, Hydrogen, Hydrogen, Hydrogen, Silicon, Silicon, Silicon, Silicon, Silicon, Hydrogen, Hydrogen, Hydrogen, Hydrogen, Silicon, Silicon, Silicon, Hydrogen, Hydrogen, Hydrogen, Silicon, Silicon, Silicon, Silicon, Hydrogen, Hydrogen, Hydrogen, Hydrogen, Silicon, Silicon, Silicon, Hydrogen, Hydrogen, Hydrogen, Hydrogen, Silicon, Silicon, Silicon, Silicon, Hydrogen, Hydrogen, Hydrogen, Hydrogen, Silicon, Silicon, Silicon, Silicon, Hydrogen, Hydrogen, Hydrogen, Hydrogen, Silicon, Silicon, Silicon, Silicon, Silicon, Hydrogen, Hydrogen]

\# Define coordinates
central_region_coordinates = [[ 5.91157506,  7.11998944,  0.175653 ],
                    [ 5.91157506, 10.96000352,  0.175653 ],
                    [ 9.03999648,  7.11998944,  0.678825 ],
                    [ 12.88001056,  7.11998944,  0.678825 ],
                    [ 9.03999648, 10.96000352,  0.678825 ],
                    [ 12.88001056, 10.96000352,  0.678825 ],
                    [ 9.03999648, 14.08842494,  1.181997 ],
                    [ 12.88001056, 14.08842494,  1.181997 ],
                    [ 9.03999648,  5.91157506,  1.533303 ],
                    [ 12.88001056,  5.91157506,  1.533303 ],
                    [ 9.03999648,  9.03999648,  2.036475 ],
                    [ 12.88001056,  9.03999648,  2.036475 ],
                    [ 9.03999648, 12.88001056,  2.036475 ],
                    [ 12.88001056, 12.88001056,  2.036475 ],
                    [ 5.91157506,  9.03999648,  2.539647 ],
                    [ 5.91157506, 12.88001056,  2.539647 ],
                    [ 14.08842494,  9.03999648,  2.890953 ],
                    [ 14.08842494, 12.88001056,  2.890953 ],
                    [ 7.11998944,  9.03999648,  3.394125 ],
                    [ 10.96000352,  9.03999648,  3.394125 ],
                    [ 7.11998944, 12.88001056,  3.394125 ],
                    [ 10.96000352, 12.88001056,  3.394125 ],
                    [ 7.11998944,  5.91157506,  3.897297 ],
                    [ 10.96000352,  5.91157506,  3.897297 ],
                    [ 7.11998944, 14.08842494,  4.248603 ],
                    [ 10.96000352, 14.08842494,  4.248603 ],
                    [ 7.11998944,  7.11998944,  4.751775 ],
                    [ 10.96000352,  7.11998944,  4.751775 ],
                    [ 7.11998944, 10.96000352,  4.751775 ],
                    [ 10.96000352, 10.96000352,  4.751775 ],
                    [ 14.08842494,  7.11998944,  5.254947 ],
                    [ 14.08842494, 10.96000352,  5.254947 ],
                    [ 5.91157506,  7.11998944,  5.606253 ],
                    [ 5.91157506, 10.96000352,  5.606253 ],
                    [ 9.03999648,  7.11998944,  6.109425 ],
                    [ 12.88001056,  7.11998944,  6.109425 ],
                    [ 9.03999648, 10.96000352,  6.109425 ],
                    [ 12.88001056, 10.96000352,  6.109425 ],
                    [ 9.03999648, 14.08842494,  6.612597 ],
                    [ 12.88001056, 14.08842494,  6.612597 ],
                    [ 9.03999648,  5.91157506,  6.963903 ],
                    [ 12.88001056,  5.91157506,  6.963903 ],
                    [ 9.03999648,  9.03999648,  7.467075 ],

```
[ 12.88001056,  9.03999648,  7.467075 ],
[  9.03999648, 12.88001056,  7.467075 ],
[ 12.88001056, 12.88001056,  7.467075 ],
[  5.91157506,  9.03999648,  7.970247 ],
[  5.91157506, 12.88001056,  7.970247 ],
[ 14.08842494,  9.03999648,  8.321553 ],
[ 14.08842494, 12.88001056,  8.321553 ],
[  7.11998944,  9.03999648,  8.824725 ],
[ 10.96000352,  9.03999648,  8.824725 ],
[  7.11998944, 12.88001056,  8.824725 ],
[ 10.96000352, 12.88001056,  8.824725 ],
[  7.11998944,  5.91157506,  9.327897 ],
[ 10.96000352,  5.91157506,  9.327897 ],
[  7.11998944, 14.08842494,  9.679203 ],
[ 10.96000352, 14.08842494,  9.679203 ],
[  7.11998944,  7.11998944, 10.182375 ],
[ 10.96000352,  7.11998944, 10.182375 ],
[  7.11998944, 10.96000352, 10.182375 ],
[ 10.96000352, 10.96000352, 10.182375 ],
[ 14.08842494,  7.11998944, 10.685547 ],
[ 14.08842494, 10.96000352, 10.685547 ],
[  5.91157506,  7.11998944, 11.036853 ],
[  5.91157506, 10.96000352, 11.036853 ],
[  9.03999648,  7.11998944, 11.540025 ],
[ 12.88001056,  7.11998944, 11.540025 ],
[  9.03999648, 10.96000352, 11.540025 ],
[ 12.88001056, 10.96000352, 11.540025 ],
[  9.03999648, 14.08842494, 12.043197 ],
[ 12.88001056, 14.08842494, 12.043197 ],
[  9.03999648,  5.91157506, 12.394503 ],
[ 12.88001056,  5.91157506, 12.394503 ],
[  9.03999648,  9.03999648, 12.897675 ],
[ 12.88001056,  9.03999648, 12.897675 ],
[  9.03999648, 12.88001056, 12.897675 ],
[ 12.88001056, 12.88001056, 12.897675 ],
[  5.91157506,  9.03999648, 13.400847 ],
[  5.91157506, 12.88001056, 13.400847 ],
[ 14.08842494,  9.03999648, 13.752153 ],
[ 14.08842494, 12.88001056, 13.752153 ],
[  7.11998944,  9.03999648, 14.255325 ],
[ 10.96000352,  9.03999648, 14.255325 ],
[  7.11998944, 12.88001056, 14.255325 ],
[ 10.96000352, 12.88001056, 14.255325 ],
[  7.11998944,  5.91157506, 14.758497 ],
[ 10.96000352,  5.91157506, 14.758497 ],
```

```
[  7.11998944,  14.08842494,  15.109803  ],
[ 10.96000352,  14.08842494,  15.109803  ],
[  7.11998944,   7.11998944,  15.612975  ],
[ 10.96000352,   7.11998944,  15.612975  ],
[  7.11998944,  10.96000352,  15.612975  ],
[ 10.96000352,  10.96000352,  15.612975  ],
[ 14.08842494,   7.11998944,  16.116147  ],
[ 14.08842494,  10.96000352,  16.116147  ],
[  5.91157506,   7.11998944,  16.467453  ],
[  5.91157506,  10.96000352,  16.467453  ],
[  9.03999648,   7.11998944,  16.970625  ],
[ 12.88001056,   7.11998944,  16.970625  ],
[  9.03999648,  10.96000352,  16.970625  ],
[ 12.88001056,  10.96000352,  16.970625  ],
[  9.03999648,  14.08842494,  17.473797  ],
[ 12.88001056,  14.08842494,  17.473797  ],
[  9.03999648,   5.91157506,  17.825103  ],
[ 12.88001056,   5.91157506,  17.825103  ],
[  9.03999648,   9.03999648,  18.328275  ],
[ 12.88001056,   9.03999648,  18.328275  ],
[  9.03999648,  12.88001056,  18.328275  ],
[ 12.88001056,  12.88001056,  18.328275  ],
[  5.91157506,   9.03999648,  18.831447  ],
[  5.91157506,  12.88001056,  18.831447  ],
[ 14.08842494,   9.03999648,  19.182753  ],
[ 14.08842494,  12.88001056,  19.182753  ],
[  7.11998944,   9.03999648,  19.685925  ],
[ 10.96000352,   9.03999648,  19.685925  ],
[  7.11998944,  12.88001056,  19.685925  ],
[ 10.96000352,  12.88001056,  19.685925  ],
[  7.11998944,   5.91157506,  20.189097  ],
[ 10.96000352,   5.91157506,  20.189097  ],
[  7.11998944,  14.08842494,  20.540403  ],
[ 10.96000352,  14.08842494,  20.540403  ],
[  7.11998944,   7.11998944,  21.043575  ],
[ 10.96000352,   7.11998944,  21.043575  ],
[  7.11998944,  10.96000352,  21.043575  ],
[ 10.96000352,  10.96000352,  21.043575  ],
[ 14.08842494,   7.11998944,  21.546747  ],
[ 14.08842494,  10.96000352,  21.546747  ],
[  5.91157506,   7.11998944,  21.898053  ],
[  5.91157506,  10.96000352,  21.898053  ],
[  9.03999648,   7.11998944,  22.401225  ],
[ 12.88001056,   7.11998944,  22.401225  ],
[  9.03999648,  10.96000352,  22.401225  ],
```

```
[ 12.88001056,  10.96000352,  22.401225  ],
[  9.03999648,  14.08842494,  22.904397  ],
[ 12.88001056,  14.08842494,  22.904397  ],
[  9.03999648,   5.91157506,  23.255703  ],
[ 12.88001056,   5.91157506,  23.255703  ],
[  9.03999648,   9.03999648,  23.758875  ],
[ 12.88001056,   9.03999648,  23.758875  ],
[  9.03999648,  12.88001056,  23.758875  ],
[ 12.88001056,  12.88001056,  23.758875  ],
[  5.91157506,   9.03999648,  24.262047  ],
[  5.91157506,  12.88001056,  24.262047  ],
[ 14.08842494,   9.03999648,  24.613353  ],
[ 14.08842494,  12.88001056,  24.613353  ],
[  7.11998944,   9.03999648,  25.116525  ],
[ 10.96000352,   9.03999648,  25.116525  ],
[  7.11998944,  12.88001056,  25.116525  ],
[ 10.96000352,  12.88001056,  25.116525  ],
[  7.11998944,   5.91157506,  25.619697  ],
[ 10.96000352,   5.91157506,  25.619697  ],
[  7.11998944,  14.08842494,  25.971003  ],
[ 10.96000352,  14.08842494,  25.971003  ],
[  7.11998944,   7.11998944,  26.474175  ],
[ 10.96000352,   7.11998944,  26.474175  ],
[  7.11998944,  10.96000352,  26.474175  ],
[ 10.96000352,  10.96000352,  26.474175  ],
[ 14.08842494,   7.11998944,  26.977347  ],
[ 14.08842494,  10.96000352,  26.977347  ],
[  5.91157506,   7.11998944,  27.328653  ],
[  5.91157506,  10.96000352,  27.328653  ],
[  9.03999648,   7.11998944,  27.831825  ],
[ 12.88001056,   7.11998944,  27.831825  ],
[  9.03999648,  10.96000352,  27.831825  ],
[ 12.88001056,  10.96000352,  27.831825  ],
[  9.03999648,  14.08842494,  28.334997  ],
[ 12.88001056,  14.08842494,  28.334997  ],
[  9.03999648,   5.91157506,  28.686303  ],
[ 12.88001056,   5.91157506,  28.686303  ],
[  9.03999648,   9.03999648,  29.189475  ],
[ 12.88001056,   9.03999648,  29.189475  ],
[  9.03999648,  12.88001056,  29.189475  ],
[ 12.88001056,  12.88001056,  29.189475  ],
[  5.91157506,   9.03999648,  29.692647  ],
[  5.91157506,  12.88001056,  29.692647  ],
[ 14.08842494,   9.03999648,  30.043953  ],
[ 14.08842494,  12.88001056,  30.043953  ],
```

```
[  7.11998944,   9.03999648,  30.547125  ],
[ 10.96000352,   9.03999648,  30.547125  ],
[  7.11998944,  12.88001056,  30.547125  ],
[ 10.96000352,  12.88001056,  30.547125  ],
[  7.11998944,   5.91157506,  31.050297  ],
[ 10.96000352,   5.91157506,  31.050297  ],
[  7.11998944,  14.08842494,  31.401603  ],
[ 10.96000352,  14.08842494,  31.401603  ],
[  7.11998944,   7.11998944,  31.904775  ],
[ 10.96000352,   7.11998944,  31.904775  ],
[  7.11998944,  10.96000352,  31.904775  ],
[ 10.96000352,  10.96000352,  31.904775  ],
[ 14.08842494,   7.11998944,  32.407947  ],
[ 14.08842494,  10.96000352,  32.407947  ],
[  5.91157506,   7.11998944,  32.759253  ],
[  5.91157506,  10.96000352,  32.759253  ],
[  9.03999648,   7.11998944,  33.262425  ],
[ 12.88001056,   7.11998944,  33.262425  ],
[  9.03999648,  10.96000352,  33.262425  ],
[ 12.88001056,  10.96000352,  33.262425  ],
[  9.03999648,  14.08842494,  33.765597  ],
[ 12.88001056,  14.08842494,  33.765597  ],
[  9.03999648,   5.91157506,  34.116903  ],
[ 12.88001056,   5.91157506,  34.116903  ],
[  9.03999648,   9.03999648,  34.620075  ],
[ 12.88001056,   9.03999648,  34.620075  ],
[  9.03999648,  12.88001056,  34.620075  ],
[ 12.88001056,  12.88001056,  34.620075  ],
[  5.91157506,   9.03999648,  35.123247  ],
[  5.91157506,  12.88001056,  35.123247  ],
[ 14.08842494,   9.03999648,  35.474553  ],
[ 14.08842494,  12.88001056,  35.474553  ],
[  7.11998944,   9.03999648,  35.977725  ],
[ 10.96000352,   9.03999648,  35.977725  ],
[  7.11998944,  12.88001056,  35.977725  ],
[ 10.96000352,  12.88001056,  35.977725  ],
[  7.11998944,   5.91157506,  36.480897  ],
[ 10.96000352,   5.91157506,  36.480897  ],
[  7.11998944,  14.08842494,  36.832203  ],
[ 10.96000352,  14.08842494,  36.832203  ],
[  7.11998944,   7.11998944,  37.335375  ],
[ 10.96000352,   7.11998944,  37.335375  ],
[  7.11998944,  10.96000352,  37.335375  ],
[ 10.96000352,  10.96000352,  37.335375  ],
[ 14.08842494,   7.11998944,  37.838547  ],
```

[ 14.08842494,  10.96000352,  37.838547  ],
[ 5.91157506,   7.11998944,  38.189853  ],
[ 5.91157506,  10.96000352,  38.189853  ],
[ 9.03999648,   7.11998944,  38.693025  ],
[ 12.88001056,   7.11998944,  38.693025  ],
[ 9.03999648,  10.96000352,  38.693025  ],
[ 12.88001056,  10.96000352,  38.693025  ],
[ 9.03999648,  14.08842494,  39.196197  ],
[ 12.88001056,  14.08842494,  39.196197  ],
[ 9.03999648,   5.91157506,  39.547503  ],
[ 12.88001056,   5.91157506,  39.547503  ],
[ 9.03999648,   9.03999648,  40.050675  ],
[ 12.88001056,   9.03999648,  40.050675  ],
[ 9.03999648,  12.88001056,  40.050675  ],
[ 12.88001056,  12.88001056,  40.050675  ],
[ 5.91157506,   9.03999648,  40.553847  ],
[ 5.91157506,  12.88001056,  40.553847  ],
[ 14.08842494,   9.03999648,  40.905153  ],
[ 14.08842494,  12.88001056,  40.905153  ],
[ 7.11998944,   9.03999648,  41.408325  ],
[ 10.96000352,   9.03999648,  41.408325  ],
[ 7.11998944,  12.88001056,  41.408325  ],
[ 10.96000352,  12.88001056,  41.408325  ],
[ 7.11998944,   5.91157506,  41.911497  ],
[ 10.96000352,   5.91157506,  41.911497  ],
[ 7.11998944,  14.08842494,  42.262803  ],
[ 10.96000352,  14.08842494,  42.262803  ],
[ 7.11998944,   7.11998944,  42.765975  ],
[ 10.96000352,   7.11998944,  42.765975  ],
[ 7.11998944,  10.96000352,  42.765975  ],
[ 10.96000352,  10.96000352,  42.765975  ],
[ 14.08842494,   7.11998944,  43.269147  ],
[ 14.08842494,  10.96000352,  43.269147  ],
[ 5.91157506,   7.11998944,  43.620453  ],
[ 5.91157506,  10.96000352,  43.620453  ],
[ 9.03999648,   7.11998944,  44.123625  ],
[ 12.88001056,   7.11998944,  44.123625  ],
[ 9.03999648,  10.96000352,  44.123625  ],
[ 12.88001056,  10.96000352,  44.123625  ],
[ 9.03999648,  14.08842494,  44.626797  ],
[ 12.88001056,  14.08842494,  44.626797  ],
[ 9.03999648,   5.91157506,  44.978103  ],
[ 12.88001056,   5.91157506,  44.978103  ],
[ 9.03999648,   9.03999648,  45.481275  ],
[ 12.88001056,   9.03999648,  45.481275  ],

[  9.03999648,  12.88001056,  45.481275  ],
[ 12.88001056,  12.88001056,  45.481275  ],
[  5.91157506,   9.03999648,  45.984447  ],
[  5.91157506,  12.88001056,  45.984447  ],
[ 14.08842494,   9.03999648,  46.335753  ],
[ 14.08842494,  12.88001056,  46.335753  ],
[  7.11998944,   9.03999648,  46.838925  ],
[ 10.96000352,   9.03999648,  46.838925  ],
[  7.11998944,  12.88001056,  46.838925  ],
[ 10.96000352,  12.88001056,  46.838925  ],
[  7.11998944,   5.91157506,  47.342097  ],
[ 10.96000352,   5.91157506,  47.342097  ],
[  7.11998944,  14.08842494,  47.693403  ],
[ 10.96000352,  14.08842494,  47.693403  ],
[  7.11998944,   7.11998944,  48.196575  ],
[ 10.96000352,   7.11998944,  48.196575  ],
[  7.11998944,  10.96000352,  48.196575  ],
[ 10.96000352,  10.96000352,  48.196575  ],
[ 14.08842494,   7.11998944,  48.699747  ],
[ 14.08842494,  10.96000352,  48.699747  ],
[  5.91157506,   7.11998944,  49.051053  ],
[  5.91157506,  10.96000352,  49.051053  ],
[  9.03999648,   7.11998944,  49.554225  ],
[ 12.88001056,   7.11998944,  49.554225  ],
[  9.03999648,  10.96000352,  49.554225  ],
[ 12.88001056,  10.96000352,  49.554225  ],
[  9.03999648,  14.08842494,  50.057397  ],
[ 12.88001056,  14.08842494,  50.057397  ],
[  9.03999648,   5.91157506,  50.408703  ],
[ 12.88001056,   5.91157506,  50.408703  ],
[  9.03999648,   9.03999648,  50.911875  ],
[ 12.88001056,   9.03999648,  50.911875  ],
[  9.03999648,  12.88001056,  50.911875  ],
[ 12.88001056,  12.88001056,  50.911875  ],
[  5.91157506,   9.03999648,  51.415047  ],
[  5.91157506,  12.88001056,  51.415047  ],
[ 14.08842494,   9.03999648,  51.766353  ],
[ 14.08842494,  12.88001056,  51.766353  ],
[  7.11998944,   9.03999648,  52.269525  ],
[ 10.96000352,   9.03999648,  52.269525  ],
[  7.11998944,  12.88001056,  52.269525  ],
[ 10.96000352,  12.88001056,  52.269525  ],
[  7.11998944,   5.91157506,  52.772697  ],
[ 10.96000352,   5.91157506,  52.772697  ],
[  7.11998944,  14.08842494,  53.124003  ],

[ 10.96000352,  14.08842494,  53.124003  ],
[ 7.11998944,   7.11998944,  53.627175  ],
[ 10.96000352,   7.11998944,  53.627175  ],
[ 7.11998944,  10.96000352,  53.627175  ],
[ 10.96000352,  10.96000352,  53.627175  ],
[ 14.08842494,   7.11998944,  54.130347  ],
[ 14.08842494,  10.96000352,  54.130347  ],
[ 5.91157506,   7.11998944,  54.481653  ],
[ 5.91157506,  10.96000352,  54.481653  ],
[ 9.03999648,   7.11998944,  54.984825  ],
[ 12.88001056,   7.11998944,  54.984825  ],
[ 9.03999648,  10.96000352,  54.984825  ],
[ 12.88001056,  10.96000352,  54.984825  ],
[ 9.03999648,  14.08842494,  55.487997  ],
[ 12.88001056,  14.08842494,  55.487997  ],
[ 9.03999648,   5.91157506,  55.839303  ],
[ 12.88001056,   5.91157506,  55.839303  ],
[ 9.03999648,   9.03999648,  56.342475  ],
[ 12.88001056,   9.03999648,  56.342475  ],
[ 9.03999648,  12.88001056,  56.342475  ],
[ 12.88001056,  12.88001056,  56.342475  ],
[ 5.91157506,   9.03999648,  56.845647  ],
[ 5.91157506,  12.88001056,  56.845647  ],
[ 14.08842494,   9.03999648,  57.196953  ],
[ 14.08842494,  12.88001056,  57.196953  ],
[ 7.11998944,   9.03999648,  57.700125  ],
[ 10.96000352,   9.03999648,  57.700125  ],
[ 7.11998944,  12.88001056,  57.700125  ],
[ 10.96000352,  12.88001056,  57.700125  ],
[ 7.11998944,   5.91157506,  58.203297  ],
[ 10.96000352,   5.91157506,  58.203297  ],
[ 7.11998944,  14.08842494,  58.554603  ],
[ 10.96000352,  14.08842494,  58.554603  ],
[ 7.11998944,   7.11998944,  59.057775  ],
[ 10.96000352,   7.11998944,  59.057775  ],
[ 7.11998944,  10.96000352,  59.057775  ],
[ 10.96000352,  10.96000352,  59.057775  ],
[ 14.08842494,   7.11998944,  59.560947  ],
[ 14.08842494,  10.96000352,  59.560947  ],
[ 5.91157506,   7.11998944,  59.912253  ],
[ 5.91157506,  10.96000352,  59.912253  ],
[ 9.03999648,   7.11998944,  60.415425  ],
[ 12.88001056,   7.11998944,  60.415425  ],
[ 9.03999648,  10.96000352,  60.415425  ],
[ 12.88001056,  10.96000352,  60.415425  ],

```
                [ 9.03999648,  14.08842494,  60.918597  ],
                [ 12.88001056,  14.08842494,  60.918597  ],
                [ 9.03999648,  5.91157506,  61.269903  ],
                [ 12.88001056,  5.91157506,  61.269903  ],
                [ 9.03999648,  9.03999648,  61.773075  ],
                [ 12.88001056,  9.03999648,  61.773075  ],
                [ 9.03999648,  12.88001056,  61.773075  ],
                [ 12.88001056,  12.88001056,  61.773075  ],
                [ 5.91157506,  9.03999648,  62.276247  ],
                [ 5.91157506,  12.88001056,  62.276247  ],
                [ 14.08842494,  9.03999648,  62.627553  ],
                [ 14.08842494,  12.88001056,  62.627553  ],
                [ 7.11998944,  9.03999648,  63.130725  ],
                [ 10.96000352,  9.03999648,  63.130725  ],
                [ 7.11998944,  12.88001056,  63.130725  ],
                [ 10.96000352,  12.88001056,  63.130725  ],
                [ 7.11998944,  5.91157506,  63.633897  ],
                [ 10.96000352,  5.91157506,  63.633897  ],
                [ 7.11998944,  14.08842494,  63.985203  ],
                [ 10.96000352,  14.08842494,  63.985203  ],
                [ 7.11998944,  7.11998944,  64.488375  ],
                [ 10.96000352,  7.11998944,  64.488375  ],
                [ 7.11998944,  10.96000352,  64.488375  ],
                [ 10.96000352,  10.96000352,  64.488375  ],
                [ 14.08842494,  7.11998944,  64.991547  ],
                [ 14.08842494,  10.96000352,  64.991547  ]]*Angstrom

# Set up configuration
central_region = BulkConfiguration(
    bravais_lattice=central_region_lattice,
    elements=central_region_elements,
    cartesian_coordinates=central_region_coordinates
    )

# Add metallic region
metallic_region_0 = TubeRegion(
    0*Volt,
    start_point = [10.0, 10.0, 15.0]*Ang,
    end_point = [10.0, 10.0, 50.0]*Ang,
    inner_radius = 7.0*Ang,
    thickness = 3.0*Ang,
)

metallic_regions = [metallic_region_0]
central_region.setMetallicRegions(metallic_regions)
```

```
# Add dielectric region
dielectric_region_0 = TubeRegion(
    3.9,
    start_point = [10.0, 10.0, 15.0]*Ang,
    end_point = [10.0, 10.0, 50.0]*Ang,
    inner_radius = 5.0*Ang,
    thickness = 2.0*Ang,
)

dielectric_regions = [dielectric_region_0]
central_region.setDielectricRegions(dielectric_regions)

device_configuration = DeviceConfiguration(
    central_region,
    [left_electrode, right_electrode]
    )

# -----------------------------------------------------------
# Calculator
# -----------------------------------------------------------
#----------------------------------------
# Basis Set
#----------------------------------------
hydrogen_1s = SlaterOrbital(
    principal_quantum_number=1,
    angular_momentum=0,
    slater_coefficients=[ 0.87223*1/Bohr ],
    weights=[ 0.50494 ]
    )

HydrogenBasis = HuckelBasisParameters(
    element=PeriodicTable.Hydrogen,
    orbitals=[ hydrogen_1s ],
    occupations=[ 1.1988 ],
    ionization_potential=[ -17.83841*eV ],
    onsite_hartree_shift=[ 12.848*eV ],
    onsite_spin_split=[[-1.887]]*eV,
    wolfsberg_helmholtz_constant=2.3,
    vacuum_level=-10.0*eV,
    )

basis_set = [
    HydrogenBasis,
```

```
        CerdaHuckelParameters.Silicon_GW_diamond_Basis,
        ]

#----------------------------------------
# Numerical Accuracy Settings
#----------------------------------------
left_electrode_numerical_accuracy_parameters = NumericalAccuracyParameters(
        k_point_sampling=(1, 1, 100),
        density_mesh_cutoff=20.0*Hartree,
        )

right_electrode_numerical_accuracy_parameters = NumericalAccuracyParameters(
        k_point_sampling=(1, 1, 100),
        density_mesh_cutoff=20.0*Hartree,
        )

device_numerical_accuracy_parameters = NumericalAccuracyParameters(
        density_mesh_cutoff=20.0*Hartree,
        )

#----------------------------------------
# Iteration Control Settings
#----------------------------------------
left_electrode_iteration_control_parameters = IterationControlParameters()

right_electrode_iteration_control_parameters = IterationControlParameters()

device_iteration_control_parameters = IterationControlParameters()

#----------------------------------------
# Poisson Solver Settings
#----------------------------------------
left_electrode_poisson_solver = MultigridSolver(
        boundary_conditions=[[NeumannBoundaryCondition,NeumannBoundaryCondition],
                        [NeumannBoundaryCondition,NeumannBoundaryCondition],
                        [PeriodicBoundaryCondition,PeriodicBoundaryCondition]]
        )

right_electrode_poisson_solver = MultigridSolver(
        boundary_conditions=[[NeumannBoundaryCondition,NeumannBoundaryCondition],
                        [NeumannBoundaryCondition,NeumannBoundaryCondition],
                        [PeriodicBoundaryCondition,PeriodicBoundaryCondition]]
        )

device_poisson_solver = MultigridSolver(
```

```
    boundary_conditions=[[NeumannBoundaryCondition,NeumannBoundaryCondition],
                         [NeumannBoundaryCondition,NeumannBoundaryCondition],
                         [DirichletBoundaryCondition,DirichletBoundaryCondition]]
)

#----------------------------------------
# Electrode Calculators
#----------------------------------------
left_electrode_calculator = HuckelCalculator(
    basis_set=basis_set,
    charge=0.01,
    numerical_accuracy_parameters=left_electrode_numerical_accuracy_parameters,
    iteration_control_parameters=left_electrode_iteration_control_parameters,
    poisson_solver=left_electrode_poisson_solver,
    )

right_electrode_calculator = HuckelCalculator(
    basis_set=basis_set,
    charge=-0.01,
    numerical_accuracy_parameters=right_electrode_numerical_accuracy_parameters,
    iteration_control_parameters=right_electrode_iteration_control_parameters,
    poisson_solver=right_electrode_poisson_solver,
    )

#----------------------------------------
# Device Calculator
#----------------------------------------
calculator = DeviceHuckelCalculator(
    basis_set=basis_set,
    numerical_accuracy_parameters=device_numerical_accuracy_parameters,
    iteration_control_parameters=device_iteration_control_parameters,
    poisson_solver=device_poisson_solver,
    electrode_calculators=
        [left_electrode_calculator, right_electrode_calculator],
    )

device_configuration.setCalculator(calculator)
nlprint(device_configuration)
device_configuration.update()
nlsave('nwfet_1.nc', device_configuration)

# ------------------------------------------------------------
# Transmission spectrum
# ------------------------------------------------------------
transmission_spectrum = TransmissionSpectrum(
```

```
        configuration=device_configuration,
        energies=numpy.linspace(-4,4,301)*eV,
        kpoints=MonkhorstPackGrid(1,1),
        energy_zero_parameter=AverageFermiLevel,
        infinitesimal=1e-06*eV,
        self_energy_calculator=RecursionSelfEnergy(),
        )
nlsave('nwfet_1.nc', transmission_spectrum)
nlprint(transmission_spectrum)

# ------------------------------------------------------------
# Electron difference density
# ------------------------------------------------------------
electron_difference_density = ElectronDifferenceDensity(device_configuration)
nlsave('nwfet_1.nc', electron_difference_density)

# ------------------------------------------------------------
# Electrostatic difference potential
# ------------------------------------------------------------
electrostatic_difference_potential =
ElectrostaticDifferencePotential(device_configuration)

nlsave('nwfet_1.nc', electrostatic_difference_potential)
```

# II. Generic Python Code for Uniformly Doped Silicon Nanowire (PIN Realization)

```
-------------------------------------------------------------
# TwoProbe configuration
# -----------------------------------------------------------

# -----------------------------------------------------------
# Left electrode
# -----------------------------------------------------------

# Set up lattice
vector_a = [5.43, 0.0, 0.0]*Angstrom
vector_b = [0.0, 10.86, 0.0]*Angstrom
vector_c = [0.0, 0.0, 16.29]*Angstrom
left_electrode_lattice = UnitCell(vector_a, vector_b, vector_c)

# Define elements
left_electrode_elements = [Silicon, Silicon, Silicon, Silicon, Silicon, Silicon, Silicon,
                Silicon, Silicon, Silicon, Silicon, Silicon]

# Define coordinates
left_electrode_coordinates = [[  1.3575,   1.3575,   1.3575],
                     [ 1.3575,   6.7875,   1.3575],
                     [ 4.0725,   4.0725,   4.0725],
                     [ 4.0725,   9.5025,   4.0725],
                     [ 1.3575,   1.3575,   6.7875],
                     [ 1.3575,   6.7875,   6.7875],
                     [ 4.0725,   4.0725,   9.5025],
                     [ 4.0725,   9.5025,   9.5025],
                     [ 1.3575,   1.3575,  12.2175],
                     [ 1.3575,   6.7875,  12.2175],
                     [ 4.0725,   4.0725,  14.9325],
                     [ 4.0725,   9.5025,  14.9325]]*Angstrom

# Set up configuration
left_electrode = BulkConfiguration(
    bravais_lattice=left_electrode_lattice,
    elements=left_electrode_elements,
    cartesian_coordinates=left_electrode_coordinates
    )

# -----------------------------------------------------------
# Right electrode
```

```
# ------------------------------------------------------------

# Set up lattice
vector_a = [5.43, 0.0, 0.0]*Angstrom
vector_b = [0.0, 10.86, 0.0]*Angstrom
vector_c = [0.0, 0.0, 16.29]*Angstrom
right_electrode_lattice = UnitCell(vector_a, vector_b, vector_c)

# Define elements
right_electrode_elements = [Silicon, Silicon, Silicon, Silicon, Silicon, Silicon, Silicon,
                Silicon, Silicon, Silicon, Silicon, Silicon]

# Define coordinates
right_electrode_coordinates = [[  1.3575,   1.3575,   1.3575],
                  [  1.3575,   6.7875,   1.3575],
                  [  4.0725,   4.0725,   4.0725],
                  [  4.0725,   9.5025,   4.0725],
                  [  1.3575,   1.3575,   6.7875],
                  [  1.3575,   6.7875,   6.7875],
                  [  4.0725,   4.0725,   9.5025],
                  [  4.0725,   9.5025,   9.5025],
                  [  1.3575,   1.3575, 12.2175],
                  [  1.3575,   6.7875, 12.2175],
                  [  4.0725,   4.0725, 14.9325],
                  [  4.0725,   9.5025, 14.9325]]*Angstrom

# Set up configuration
right_electrode = BulkConfiguration(
    bravais_lattice=right_electrode_lattice,
    elements=right_electrode_elements,
    cartesian_coordinates=right_electrode_coordinates
    )

# ------------------------------------------------------------
# Central region
# ------------------------------------------------------------

# Set up lattice
vector_a = [5.43, 0.0, 0.0]*Angstrom
vector_b = [0.0, 10.86, 0.0]*Angstrom
vector_c = [0.0, 0.0, 51.295]*Angstrom
central_region_lattice = UnitCell(vector_a, vector_b, vector_c)

# Define elements
central_region_elements = [Silicon, Silicon, Boron, Silicon, Silicon, Boron, Silicon,
```

---

```
                    Silicon, Boron, Silicon, Silicon, Boron, Silicon, Silicon,
                    Boron, Silicon, Silicon, Boron, Silicon, Silicon, Boron,
                    Silicon, Silicon, Boron, Silicon, Silicon, Boron, Silicon,
                    Silicon, Phosphorus, Silicon, Silicon, Phosphorus, Silicon, Silicon,
                    Phosphorus, Silicon, Silicon, Phosphorus, Silicon, Silicon, Phosphorus,
                    Silicon, Silicon, Phosphorus, Silicon, Silicon, Phosphorus, Silicon,
                    Silicon, Phosphorus, Silicon, Silicon, Phosphorus, Silicon, Silicon]

# Define coordinates
central_region_coordinates = [[ 1.3575,   1.3575,   1.3575],
                               [ 1.3575,   6.7875,   1.3575],
                               [ 4.0725,   4.0725,   4.0725],
                               [ 4.0725,   9.5025,   4.0725],
                               [ 1.3575,   1.3575,   6.7875],
                               [ 1.3575,   6.7875,   6.7875],
                               [ 4.0725,   4.0725,   9.5025],
                               [ 4.0725,   9.5025,   9.5025],
                               [ 1.3575,   1.3575,  12.2175],
                               [ 1.3575,   6.7875,  12.2175],
                               [ 4.0725,   4.0725,  14.9325],
                               [ 4.0725,   9.5025,  14.9325],
                               [ 1.3575,   1.3575,  17.6475],
                               [ 1.3575,   6.7875,  17.6475],
                               [ 4.0725,   4.0725,  18.8475],
                               [ 4.0725,   9.5025,  18.8475],
                               [ 1.3575,   1.3575,  20.6475],
                               [ 4.0725,   4.0725,  20.6475],
                               [ 1.3575,   6.7875,  20.6475],
                               [ 4.0725,   9.5025,  20.6475],
                               [ 1.3575,   1.3575,  22.6475],
                               [ 4.0725,   4.0725,  22.6475],
                               [ 1.3575,   6.7875,  22.6475],
                               [ 4.0725,   9.5025,  22.6475],
                               [ 1.3575,   1.3575,  24.6475],
                               [ 4.0725,   4.0725,  24.6475],
                               [ 1.3575,   6.7875,  24.6475],
                               [ 4.0725,   9.5025,  24.6475],
                               [ 1.3575,   1.3575,  26.6475],
                               [ 4.0725,   4.0725,  26.6475],
                               [ 1.3575,   6.7875,  26.6475],
                               [ 4.0725,   9.5025,  26.6475],
                               [ 1.3575,   1.3575,  28.6475],
                               [ 4.0725,   4.0725,  28.6475],
                               [ 1.3575,   6.7875,  28.6475],
                               [ 4.0725,   9.5025,  28.6475],
```

```
          [ 1.3575,   1.3575,  30.6475],
          [ 4.0725,   4.0725,  30.6475],
          [ 1.3575,   6.7875,  30.6475],
          [ 4.0725,   9.5025,  30.6475],
          [ 1.3575,   1.3575,  32.4475],
          [ 1.3575,   6.7875,  32.4475],
          [ 4.0725,   4.0725,  33.6475],
          [ 4.0725,   9.5025,  33.6475],
          [ 1.3575,   1.3575,  36.3625],
          [ 1.3575,   6.7875,  36.3625],
          [ 4.0725,   4.0725,  39.0775],
          [ 4.0725,   9.5025,  39.0775],
          [ 1.3575,   1.3575,  41.7925],
          [ 1.3575,   6.7875,  41.7925],
          [ 4.0725,   4.0725,  44.5075],
          [ 4.0725,   9.5025,  44.5075],
          [ 1.3575,   1.3575,  47.2225],
          [ 1.3575,   6.7875,  47.2225],
          [ 4.0725,   4.0725,  49.9375],
          [ 4.0725,   9.5025,  49.9375]]*Angstrom

# Set up configuration
central_region = BulkConfiguration(
    bravais_lattice=central_region_lattice,
    elements=central_region_elements,
    cartesian_coordinates=central_region_coordinates
    )

device_configuration = DeviceConfiguration(
    central_region,
    [left_electrode, right_electrode]
    )
```